\documentclass[longauth]{aa}

\usepackage{graphicx,txfonts}
\usepackage[flushleft]{threeparttable}

\newcommand{\almaproj}[1]{\mbox{ADS/JAO.ALMA\#{#1}}}

\usepackage[colorlinks=true,linkcolor=blue,citecolor=blue,urlcolor=blue,
  pdftitle={VAPOLA survey of AGNs and Sgr A* at mm wavelengths with ALMA. II. Spectropolarimetric properties and their evolution from 2017 to 2023},
  pdfauthor={Douglas F. Carlos, Ciriaco Goddi, Alejandro Mus, Nicola Marchili, Kazi L. J. Rygl, Ezequiel Albentosa-Ruiz, Ivan Marti-Vidal, et al. (Event Horizon Telescope Collaboration)}
]{hyperref}

\begin{document}

\title{VAPOLA -- A multiyear, multiband polarization survey of AGNs and Sgr~A* at millimeter wavelengths with ALMA}
\subtitle{II. Spectropolarimetric properties and their evolution from 2017 to 2023}

\author{
        Douglas Carlos\inst{1,2}\fnmsep\thanks{Corresponding author: \texttt{medoug7@gmail.com}\\ $\dagger$~Deceased}
        \and Ciriaco Goddi\inst{2,3,4,5}
        \and Alejandro Mus\inst{3,4,6,7,8}
        \and Nicola Marchili\inst{1,9}
        \and Kazi L. J. Rygl\inst{1}
        \and Ezequiel Albentosa-Ru\'iz\inst{10}
        \and Iv\'an Mart\'{\i}-Vidal\inst{10,11}
        \and Kazunori Akiyama\inst{12,13,14,15}
        \and Antxon Alberdi\inst{16}
        \and Walter Alef\inst{17}
% ...
        }

\institute{
        INAF -- Istituto di Radioastronomia, Via P. Gobetti 101, 40129 Bologna, Italy
        \and Instituto de Astronomia, Geof\'{\i}sica e Ci\^encias Atmosf\'ericas, Universidade de S\~ao Paulo, S\~ao Paulo, SP 05508-090, Brazil
        \and Dipartimento di Fisica, Universit\`a degli Studi di Cagliari, SP Monserrato-Sestu km 0.7, 09042 Monserrato, Italy
        \and INAF -- Osservatorio Astronomico di Cagliari, via della Scienza 5, 09047 Selargius (CA), Italy
        \and INFN, Sezione di Cagliari, Cittadella Univ., 09042 Monserrato (CA), Italy
        \and Soft Computing, Image Processing and Aggregation Research Group (SCOPIA) \& Modelling and Imaging Radio Astronomical Data (MIRADA), Math and Comp. Sci. Department, University of Balearic Islands, Spain
        \and Artificial Intelligence Research, Institute of the Balearic Islands (IAIB), Spain
        \and Health Research Institute of the Balearic Islands (IdISBa), Spain
        \and Max-Planck-Institut f\"ur Radioastronomie, Auf dem H\"ugel 69, 53121 Bonn, Germany
        \and Departament d'Astronomia i Astrof\'isica, Universitat de Val\`encia, C. Dr. Moliner 50, 46100 Burjassot, Val\`encia, Spain
        \and Observatori Astron\`omic, Universitat de Val\`ncia, C. Catedr\'atico Jos\'e Beltr\'an 2, 46980 Paterna, Val\`encia, Spain
        \and Institute of Sensors, Signals and Systems, Heriot-Watt University, Edinburgh EH14 4AS, UK
        \and Massachusetts Institute of Technology Haystack Observatory, 99 Millstone Road, Westford, MA 01886, USA
        \and Mizusawa VLBI Observatory, National Astronomical Observatory of Japan, 2-12 Hoshigaoka, Mizusawa, Oshu, Iwate 023-0861, Japan
        \and School of Astronomy, Institute for Research in Fundamental Sciences (IPM) P.O. Box 19395-5746, Tehran, Iran
        \and Instituto de Astrofísica de Andalucía-CSIC, Glorieta de la Astronomía s/n, 18008 Granada, Spain
        \and Max-Planck-Institut für Radioastronomie, Auf dem Hügel 69, 53121 Bonn, Germany
% ...
}

\date{Received 14 August 2026 / Accepted 18 October 2026}

\titlerunning{VAPOLA II: Spectropolarimetric properties and their evolution}
\authorrunning{Carlos et al.}

\abstract 
   {} {We aim to present a systematic analysis of the spectropolarimetric properties of a sample of 39 active galactic nuclei (AGNs) and Sagittarius~A* observed with the Atacama Large Millimeter/submillimeter Array during five VLBI campaigns between 2017 and 2023.}
   {We characterized the compact cores in total intensity and polarization, focusing on the behavior of the linear-polarization (LP) fraction, electric vector position angle (EVPA), and Faraday rotation measure (RM) over time and spectral domains. We investigated both individual objects -- such as M87, Sgr~A*, 3C273, and 3C279 -- and ensemble properties of different source classes, including flat-spectrum radio quasars, BLLac objects, and other active galaxies.}
   {While total intensity and spectral index are generally stable on weekly timescales, polarization properties often exhibit strong variability, with significant day-to-day changes in LP, EVPA, and RM. Several sources display large EVPA rotations accompanied by variations in LP and RM, in some cases coinciding with flaring activity.}
   {We observe that the magnitude of RM increases with observing frequency for all sources for which we have reliable multiband measurements, consistent with Faraday rotation arising in a magnetized sheath surrounding the relativistic jet, although an origin in the accretion flow -- particularly in the case of Sgr~A* -- cannot be ruled out.}


\maketitle

\nolinenumbers

\section{Introduction}\label{sec:intro}

Temporal and spectral analyses of active galactic nucleus (AGN) polarization offer a multidimensional view of the physical processes taking place in the environments around supermassive black holes (SMBHs).
Short-term variability is often associated with localized changes in the emitting region and can be used to place constraints on its characteristic size \citep{Jorstad2005, MartiVidal2021}, while long-term variations, on timescales of months to years, provide insight into the behavior of more persistent and large-scale structures \citep{Agudo2018b}.
Additionally, the spectral evolution of the polarization resulting from the Faraday effect can help us constrain the plasma properties, and systematic changes over extended periods may reflect the evolution of the magnetic-field topology, shock propagation along the jet, jet precession, or variations in the accretion flow \citep{Pacholczyk1970,  1989Hughes, 1995Pohl, Abraham_1999, Jorstad_2001}.
\\

The Atacama Large Millimeter/submillimeter Array (ALMA) provides a uniquely powerful instrument for investigating the polarization properties of AGNs, owing to its exceptional sensitivity, broad frequency coverage, and high spatial resolution at millimeter and submmillimeter wavelengths.
Moreover, the inclusion of ALMA as a phased-array element \citep{APPPaper,QA2Paper,APP2B7} in global very-long-baseline-interferometry (VLBI) networks -- such as the Global-mm VLBI Array \citep[GMVA;][]{GMVApaper} and the Event Horizon Telescope \citep[EHT;][]{Akiyama_2019_II} -- has dramatically enhanced the sensitivity of long-baseline observations \citep{Goddi2019}.
This capability has enabled landmark results, including the imaging of the shadows of the supermassive black holes M87* \citep{Akiyama_2019_I} and Sgr~A* \citep{EHTC_2022}.

In addition to its role in enhancing the sensitivity of VLBI campaigns, the ALMA observatory provides a rich, full-polarization interferometric dataset with significant standalone scientific value. These data serve multiple purposes: they enable the refinement and validation of VLBI calibration strategies \citep{EHTC_2021a}; provide observational constraints for theoretical models \citep{EHTC_2021b}; and allow detailed studies of millimeter emission, polarization, and Faraday properties of AGNs on arcsecond scales \citep{Goddi_2021, Goddi_2025}.
Motivated by this unique combination of capabilities, we initiated the VLBI AGN POLarization with ALMA (VAPOLA) project, the first online, multi-epoch, multiband repository of high-level data products derived from ALMA observations of AGN and Sgr~A* obtained during global VLBI campaigns. The high-level products are generated using an automated pipeline that processes fully calibrated ALMA data following the procedures established for the analysis of the 2017 observations \citep{Goddi_2021} and subsequently standardized for later epochs in \cite{VAPOLA_I} (VAPOLA I hereafter). For a detailed description of the automated pipeline and the data products publicly available through the repository, we refer the reader to VAPOLA~I. This study represents the first scientific exploitation of the high-level products released through the VAPOLA archive. We analyzed the spectropolarimetric properties of a sample of 39 AGNs and Sgr~A* observed with ALMA during five VLBI campaigns spanning 2017--2023. We performed a comprehensive characterization of the compact cores in both total intensity and polarization. The analysis is carried out in the temporal and spectral domains, allowing us to track variability on weekly and annual timescales and to compare polarization behavior across millimeter and submillimeter wavelengths.
\

This paper is the second in the VAPOLA series and is organized as follows. In Sect. \ref{sec:Obs} we provide an overview of the survey, including the observational epochs, spectral coverage, and the list of observed sources. In Sect. \ref{sec:Data} we summarize the analysis methods, with particular emphasis on the polarization properties of the compact cores and their behavior as a function of frequency and time. The results are presented in Sect. \ref{sec:Results}, where we give a detailed breakdown of the most frequently observed sources, as well as a comparison between the different classes.
In Sect. \ref{sec:Disc} we examine differences in the parameters of different classes, discuss the short- and long-term trends in polarization variability, and analyze the trend of increasing rotation measure (RM) magnitude with frequency. Our conclusions are summarized in Sect. \ref{sec:conc}.
Throughout this work, we adopted a $\Lambda$ cold dark matter cosmology with $H_0 = 73$ km s$^{-1}$ Mpc$^{-1}$, $\Omega_\Lambda = 0.7$, and $\Omega_m = 0.3$.

\section{Survey and sample} \label{sec:Obs}

The dataset presented in this work was acquired with ALMA during the global VLBI campaigns of the GMVA and EHT spanning 2017, 2018, 2021, 2022, and 2023, corresponding to ALMA Cycles 4, 5, 7, 8, and 9, respectively. The campaigns were typically performed in March or April, lasting one to two weeks.
Observations were performed close in time, though not strictly simultaneously, at ALMA Band 3 (B3, $\lambda 3$ mm, or 93.3 GHz) with the GMVA and Bands 6 (B6, $\lambda 1.3$ mm or 221.1 GHz) and 7 (B7, $\lambda 0.87$ mm or 342.6 GHz) with the EHT.
 Each band comprises four spectral windows (SPWs), each with a bandwidth of 1.875 GHz and 240 channels. The central frequencies of the SPWs are listed in Table~\ref{tab:bands}.

Two distinct types of ALMA data were considered: VLBI scans, acquired while ALMA was phased-up and participating in the global campaigns; and non-VLBI scans, taken in standalone mode between VLBI observations for calibration purposes. Calibration and analysis were performed separately for the two modes. Detailed procedures for data calibration are described in \citet{QA2Paper} and VAPOLA I.
\begin{table}[h]
\centering
\caption{Central frequencies of the four SPWs in each ALMA band used by VAPOLA.}\label{tab:bands}
\begin{tabular}{cc}
\hline\hline
Band & SPW central frequencies (GHz) \\
\hline
3 (3 mm) & [86.27, 88.27, 98.33, 100.27]\\
6 (1.3 mm) & [213.1, 215.1, 227.1, 229.1] \\
7 (0.87 mm) & [335.6, 337.54, 347.6, 349.6] \\
\hline\hline
\end{tabular}
\end{table}
\\

The sample includes 39 sources: 29 quasars, nine radio galaxies, and the strong radio source at the Galactic center, \object{Sgr A*}. Source classifications were taken from the NASA Extragalactic Database (NED\footnote{\url{https://ned.ipac.caltech.edu/help/class_list.html}}) based on \cite{Veron-Cetty_2006}.

Among the quasars, 22 are flat-spectrum radio quasars (FSRQ; e.g., \object{3C273}, \object{NRAO530}, \object{and PKS1335-127}) defined by a flat spectral index ($-0.5<\alpha<0.5$) between 1.4 and 5 GHz. Seven others are BLLacs (e.g., \object{3C279} and \object{4C01.28}), characterized by featureless optical spectra lacking strong emission lines. Two sources, \object{3C279} and \object{J0510+1800}, are classified as BLLacs in NED but as FSRQs in the Roma-BZCAT \citep{Massaro_2015}; for consistency, we adopted the BLLac classification. One quasar, \object{J1744-3116}, is of unknown type and lacks a reliable redshift estimate.
The active galaxy class includes well-known sources such as \object{M87}, \object{Centaurus A} (Cen A), \object{M84}, \object{3C84}, and \object{NGC4261}, as well as low-luminosity AGNs such as \object{NGC1052}, \object{NGC4594}, \object{NGC4278}, and \object{NGC5232}.
An extensive list of all sources observed during the 2017--2023 campaigns, including both VLBI and non-VLBI scans, is provided in Table \ref{tab:srcs}.

Detailed observing logs are available through the VAPOLA archive\footnote{\url{https://vapola.ia2.inaf.it/home.html}} under \texttt{Tables/observation\_summary}. These provide, for each dataset, the observing date and duration, central frequency, and wavelength, minimum and maximum baseline lengths, angular resolution, and maximum recoverable scale.

\section{Data analysis} \label{sec:Data}

We first focus on the polarization properties of the compact cores in Sect. \ref{sec:data_polpar}, which represent the main focus of this work. In Sects. \ref{sec:data_spectra} and \ref{sec:data_varpol}, we discuss the analysis in the frequency and time domains for selected sources. Finally, in Sect. \ref{sec:data_imgs} we show some representative images of sources exhibiting extended emission at millimeter and submillimeter wavelengths.

\subsection{Polarization parameters of the cores} \label{sec:data_polpar}

To extract the polarization properties of the compact cores, we modeled each object as a point source at the phase center of the visibilities using the external \texttt{casa} library \texttt{UVMULTIFIT} \citep{UVMULTIFIT}. This is justified because the emission of interest is dominated by the compact core. For each SPW, we fit a delta function at the source center, retrieving the Stokes parameters $I$, $Q$, $U$, and $V$.

From the Stokes $Q$ and $U$ we derived the LP and the EVPA ($\chi$) for each SPW:
\begin{equation}\label{eq:LP}
\textnormal{LP} = \frac{\sqrt{Q^2 + U^2}}{I},
\end{equation}
\begin{equation}\label{eq:evpa}
\chi = \frac{1}{2} \mathrm{atan2}\left(U,\,Q\right).
\end{equation}

To summarize the polarization properties of individual sources, we also computed band-averaged values of Stokes $I$, LP, and EVPA across the four SPWs. Both the per-SPW full-Stokes measurements and the band-averaged polarization parameters are available in tabular form through the VAPOLA archive.
Examples of both tables are presented in Appendix~\ref{ap:poltable}, while the adopted uncertainty budget, including statistical and systematic calibration errors, is detailed in Appendix~\ref{ap:errors}.

\subsection{Spectral features} \label{sec:data_spectra}
The VAPOLA archive includes observations across three ALMA bands, and several sources were observed in multiple bands within the same week. This minimizes potential biases from short-term variability (see Sect. \ref{sec:data_varpol}) when performing multiband analyses. By combining data from the different bands, we can study the polarization behavior of the compact cores over the millimeter to submillimeter wavelength range.

\begin{figure*}
    \centering
    \includegraphics[width=0.95\linewidth]{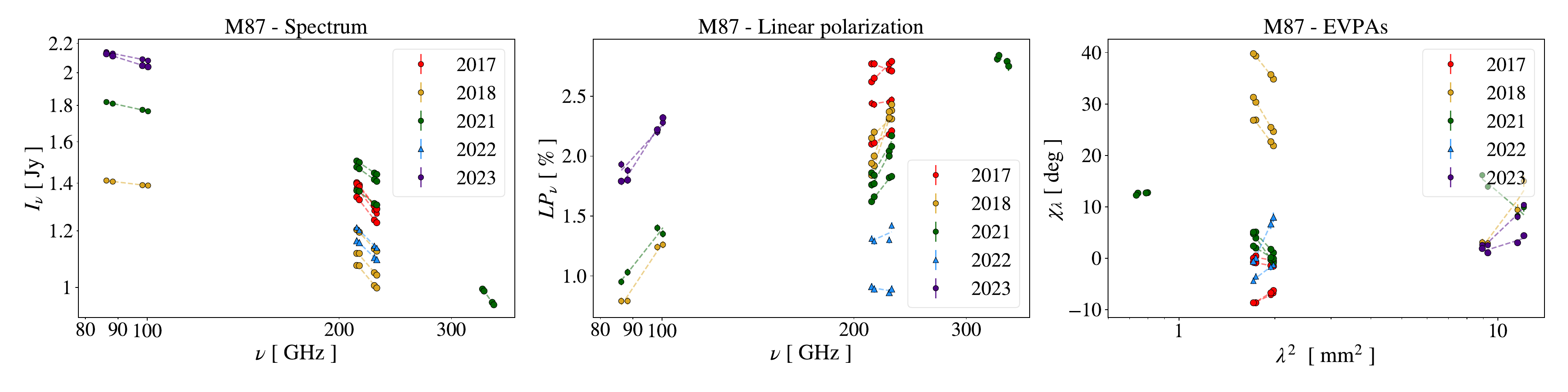}
    
    %\vspace{-6.5mm}
    \includegraphics[width=0.95\linewidth]{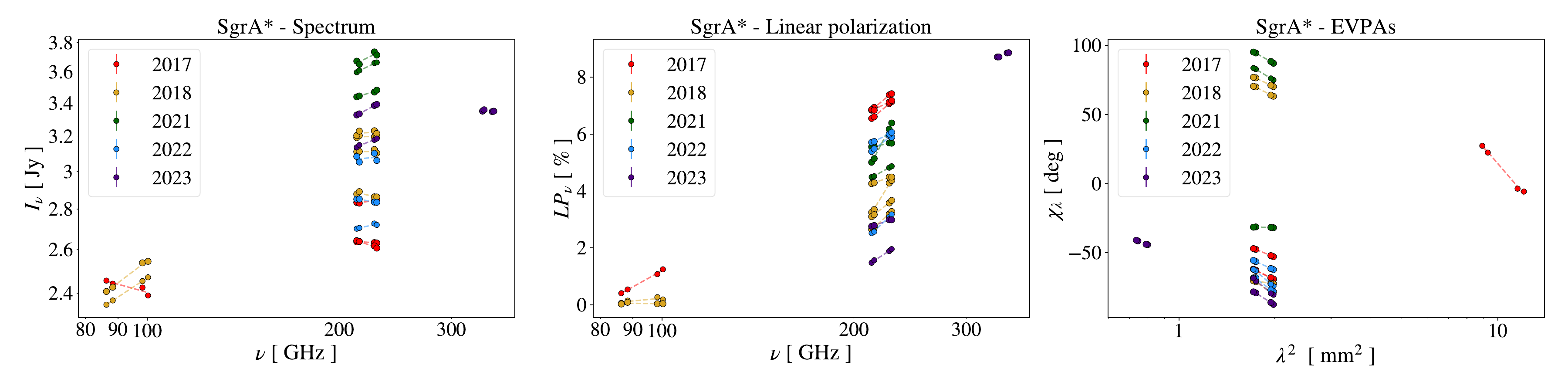}
    %\vspace{-2mm}
    \caption{Spectral distribution of Stokes $I$ (left), LP (center), and EVPA (right) from the compact cores of the most widely observed sources: M87 (top) and Sgr A* (bottom). The remaining sources are shown in Figures \ref{fig:specs2} and \ref{fig:specs3}.}\label{fig:specs}
\end{figure*}

For the best-sampled sources, we summarize these interband spectropolarimetric properties by plotting spectra (Stokes $I$ vs.\ $\nu$), polarization (LP vs.\ $\nu$), and Faraday rotation (EVPA vs.\ $\lambda^2$) in Figs. \ref{fig:specs}, \ref{fig:specs2}, and \ref{fig:specs3}.
To quantify the spectral behavior of LP and EVPA, we computed the depolarization measure ($D$) and the rotation measure (RM); these are simple phenomenological estimates obtained via least-squares linear fits of LP and EVPA across the SPWs of each band. Specifically,
\begin{equation}\label{eq:depol}
D = \frac{d}{d\nu} \text{LP},
\end{equation}
\begin{equation}\label{eq:RM}
\chi = \chi_0 + \text{RM}\ \lambda^2,
\end{equation}
where $\chi_0$ is the intrinsic EVPA extrapolated to zero wavelength, assuming Faraday rotation in an external screen.
For the depolarization measure, $D>0$ corresponds to normal depolarization, in which the degree of linear polarization decreases toward longer wavelengths, whereas $D<0$ indicates inverse depolarization, with polarization increasing toward longer wavelengths.
The latter behavior may arise from complex emitting and Faraday-rotating structures \citep{Sokoloff_1998,Homan_2012,Sullivan2012}.

We discard $D$ and RM measurements from low polarization states (LP $<0.5\%$), as well as measurements for which one or more SPWs deviate by more than $4\sigma$ from the best-fitting linear relation.
The procedure for identifying outliers is described in Appendix \ref{ap:like}, and examples of the table products are shown in Appendix \ref{ap:poltable}.

\subsection{Variability} \label{sec:data_varpol}
The dataset analyzed here spans one to two weeks of observations per year over six years, providing an opportunity to study source variability on both weekly and yearly timescales. As Table \ref{tab:srcs} shows, only a few sources were observed in more than five epochs, and most of them are in   B6. Therefore, we focused the time variability analysis on B6 and for a selected subset of sources of interest: M87, Sgr A*, 3C273, 3C279, QSOB1921-293, NRAO530, 4C01.28, OJ287, and PKS1335-127.

We adopted a two-fold approach, (a) constructing time series of the measured parameters and (b) evaluating their statistical distributions, computing averages, standard deviations, and variability indices. Details of these approaches are given in the following subsections.

\subsubsection{Time series from ALMA measurements}

To visualize the day-to-day and year-to-year evolution of the sources, we plot Stokes $I$, LP, EVPA, and RM measurements over time (Appendix \ref{ap:histplots}). This is shown for M87, Sgr A*, 3C273, and 3C279 in Figs. \ref{fig:hist_M87}, \ref{fig:hist_Sgra}, \ref{fig:hist_3C273}, and \ref{fig:hist_3C279}.

To identify potential outliers and distinguish them from long-term trends, we compared our measurements to the ALMA Grid Survey (GS) light curves spanning 2016--2026. The GS, conducted with the ALMA Compact Array (ACA), regularly monitors bright radio quasars in full-polarization mode alongside Solar System calibrators. Data reduction uses the Analytic Matrix for ALMA POLArimetry (AMAPOLA\footnote{\url{https://www.alma.cl/~skameno/AMAPOLA/}}), providing near-uniform coverage of many targets at millimeter wavelengths since April 2012 \citep{Kameno2023b}.

The GS dataset serves as an essential reference for assessing the calibration quality of our measurements. It overlaps with many VLBI targets, allowing us to place our observations into a broader temporal context. Only M87 and Sgr A* are not covered by the GS.
The evolution of $I$, LP, and EVPA for the most frequently observed sources, binned into seven-day intervals, is shown in the figures of Appendix \ref{ap:amaplots}.
\\

\subsubsection{Variability statistics}
\label{var_stat}
We quantified variability using the following index:

\begin{equation}\label{eq:Valler}
V = \frac{p_\textnormal{max}-p_\textnormal{min}}{p_\textnormal{max}+p_\textnormal{min}},
\end{equation}
where $p_\textnormal{max}$ and $p_\textnormal{min}$ are the maximum and minimum values of the parameter (Stokes $I$ or LP).

Table \ref{tab:var_analysis} reports the median and scatter of Stokes $I$, spectral index $\alpha$, LP, and EVPA, along with variability indices for $I$ and LP, combining VLBI and AMAPOLA measurements when available. Sources observed in fewer than five VLBI epochs but with $\ge$10 AMAPOLA measurements are also included.
Individual cases are detailed in Sect. \ref{sec:Results}, while a comprehensive discussion about the short- and long-term trends is given in Sect. \ref{sec:Disc}

\subsection{Extended source images}\label{sec:data_imgs}
After the full-Stokes imaging of all SPWs, we generated various maps to study the structure of extended sources. In this work, we show maps of Stokes $I$ and of LP fraction with overlaid EVPAs for a selection of sources. The uncertainties quoted for the Stokes $I$ maps include both statistical noise and the systematic uncertainty associated with ALMA's absolute flux calibration, as detailed in Appendix~\ref{ap:errors}.
Additional maps -- including spectral index, polarized intensity, depolarization, and RM -- are available in the VAPOLA archive.

\begin{figure*}
    \centering
    %\vspace{0cm}
    \includegraphics[width=\linewidth]{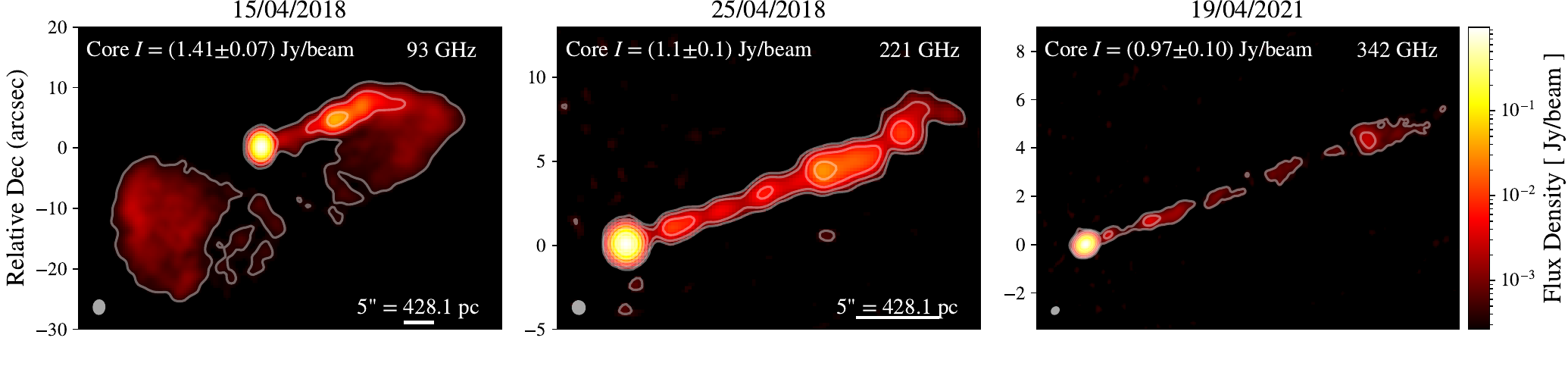}
    
    %\vspace{-0.65cm}
    \includegraphics[width=\linewidth]{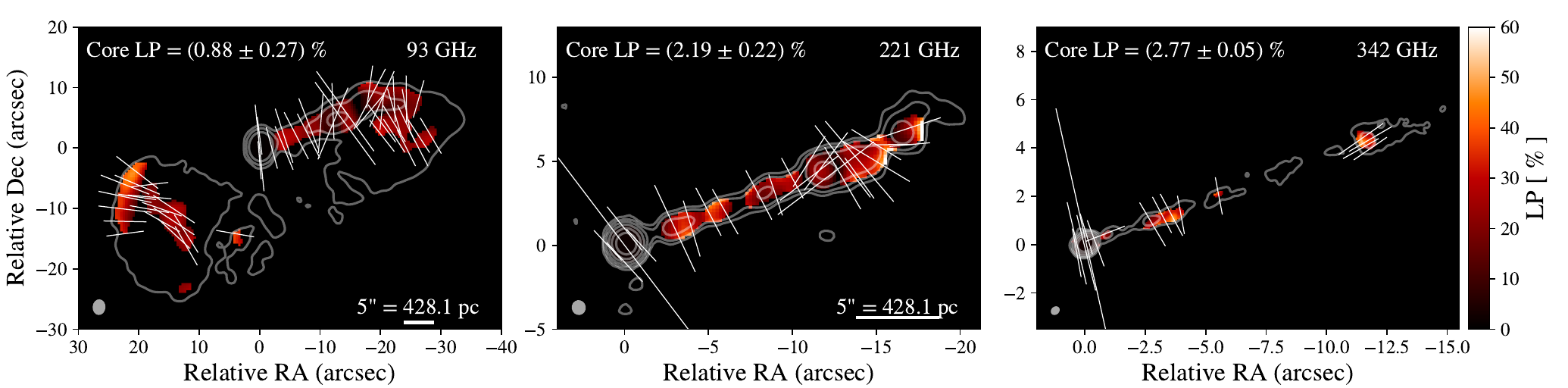}
    \caption{M87 jet images at representative frequencies for each ALMA band. Top: Stokes I. Bottom: LP fraction with overlaid EVPAs (length proportional to polarized flux). Top, middle, and bottom rows correspond to B3, B6, and B7, respectively. Contours start at $4\times$RMS of Stokes $I$, except for B3, which starts at $5\times$RMS and increase by $4\times$RMS at each level; RMS values are 1.12, 0.09, 0.09 mJy for B3, B6, and B7, respectively. Beam dimensions are 2.61" $\times$ 2.13" ($-4.0^\circ$), 0.63" $\times$ 0.61" ($69.8^\circ$), and 0.40" $\times$ 0.33" ($-48.6^\circ$), respectively.}\label{fig:M87_b3b6}
    %\vspace{0.3cm}
    \includegraphics[width=0.95\linewidth]{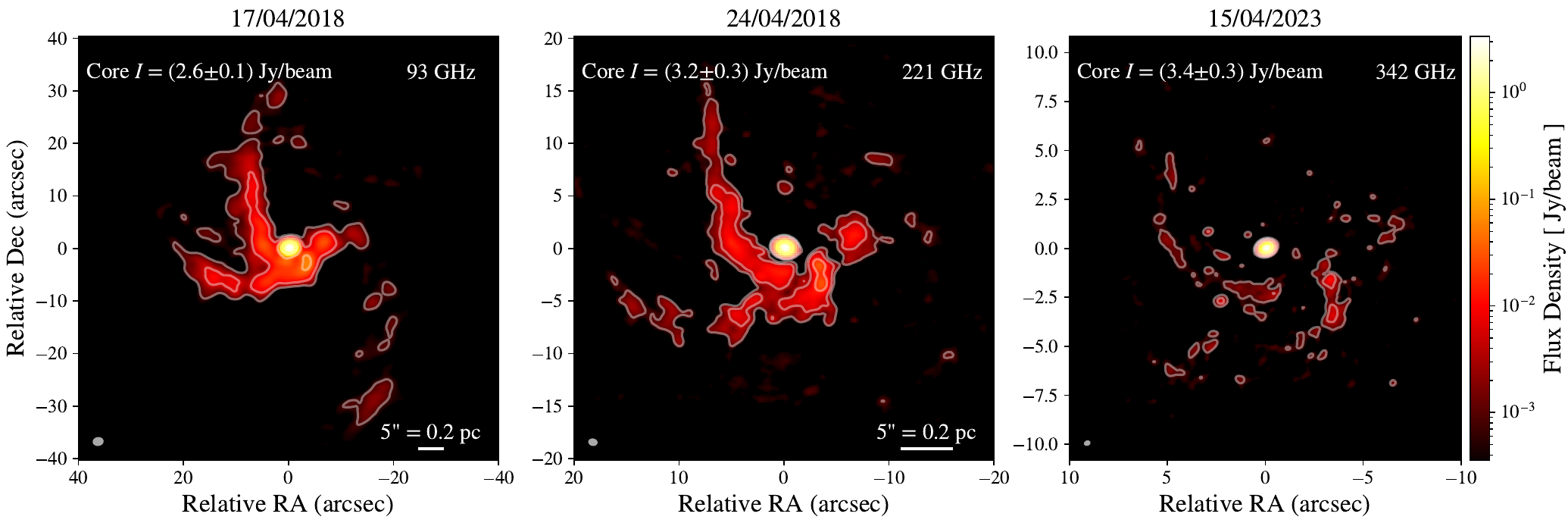}
    \caption{Stokes $I$ images of Sgr A* and the Galactic Center minispiral at representative frequencies for each band. Left: B3, middle: B6, right: B7. Beam dimensions are $2.11''\times1.71''$ ($-81.9^\circ$), $0.90''\times0.72''$ ($79.9^\circ$), and $0.36''\times0.28''$ ($-69.6^\circ$) for B3, B6, B7, respectively. Contours start at $4\times$RMS of Stokes $I$  and increase by $4\times$RMS at each level, with RMS values of 1.6, 0.14, and 0.12 mJy for B3, B6, and B7, respectively. 
    Polarization maps are not shown because the detected polarized emission is confined to the unresolved core of Sgr A*, with no significant polarized emission detected from the extended minispiral structure.}\label{fig:SgrA_b3b6}
\end{figure*}

\begin{figure*}
    \centering
    \includegraphics[width=0.9\linewidth]{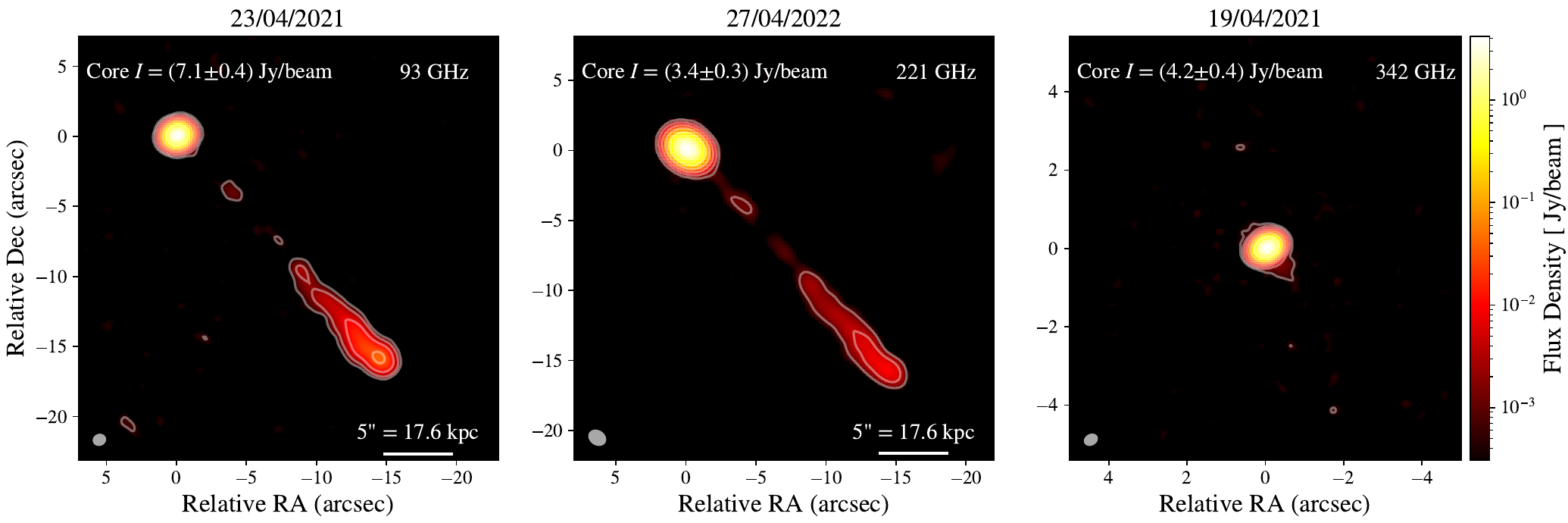}
    
    %\vspace{-0.7cm}
    \includegraphics[width=0.9\linewidth]{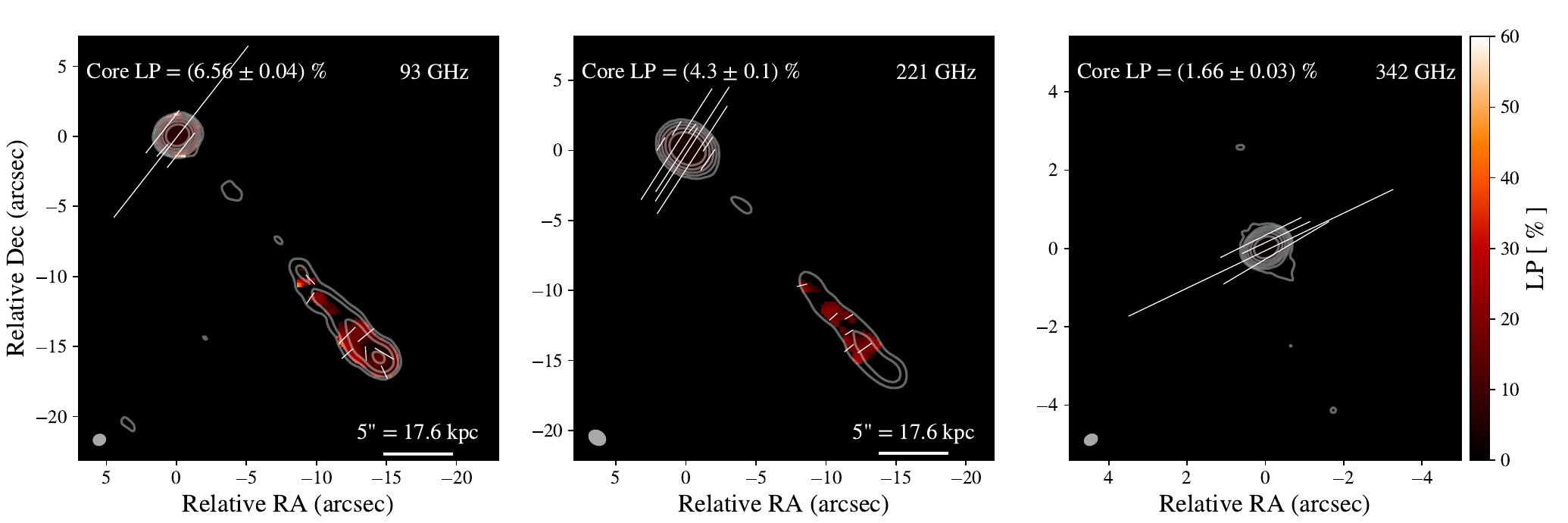}
    \caption{Polarization images of 3C273 at representative frequencies. Top row: Stokes I. Bottom row: LP fraction with overlaid EVPAs. Left, middle, and right: B3, B6, and B7. Beam sizes: B3 0.83" $\times$ 0.69" ($27.3^\circ$), B6 1.31" $\times$ 0.95" ($89.1^\circ$), and B7 0.38" $\times$ 0.30" ($-60.8^\circ$). Contours start at $4\times$RMS of Stokes $I$, except B3 at $5\times$RMS, but all increase by $4\times$RMS at each level; RMS values of 0.19, 0.11, and 0.10 mJy for B3, B6, and B7, respectively.}\label{fig:3c273_b3b6}
    %\vspace{0.5cm}
    \includegraphics[width=0.9\linewidth]{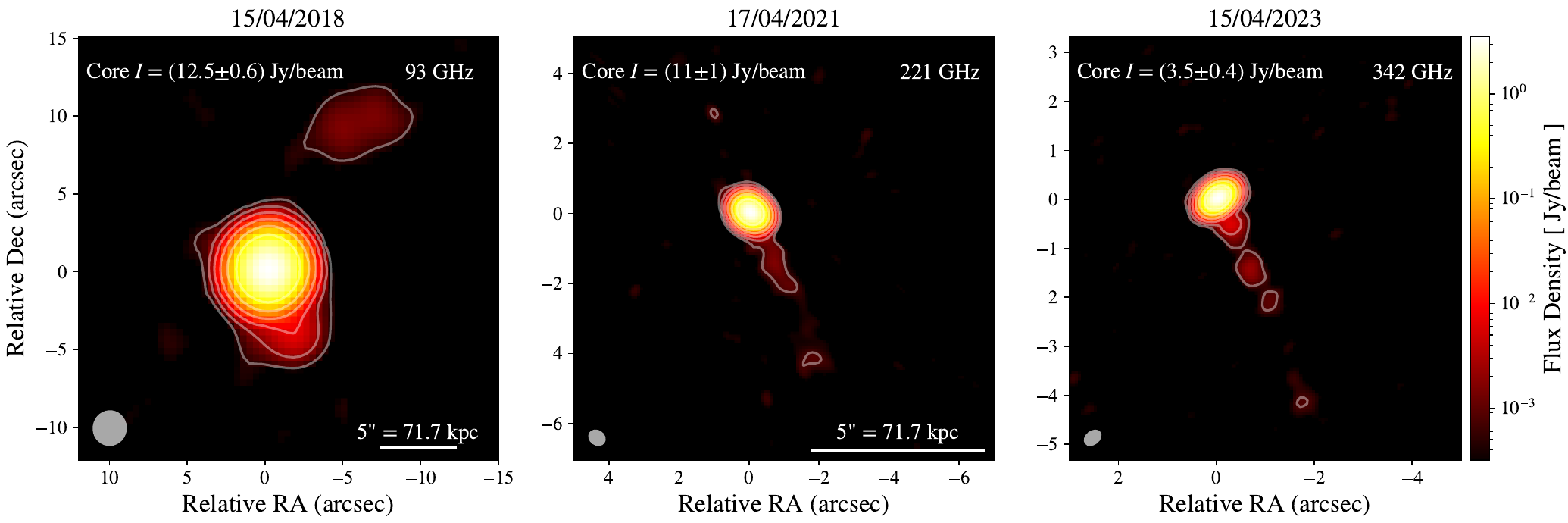}
    
    %\vspace{-0.7cm}
    \includegraphics[width=0.9\linewidth]{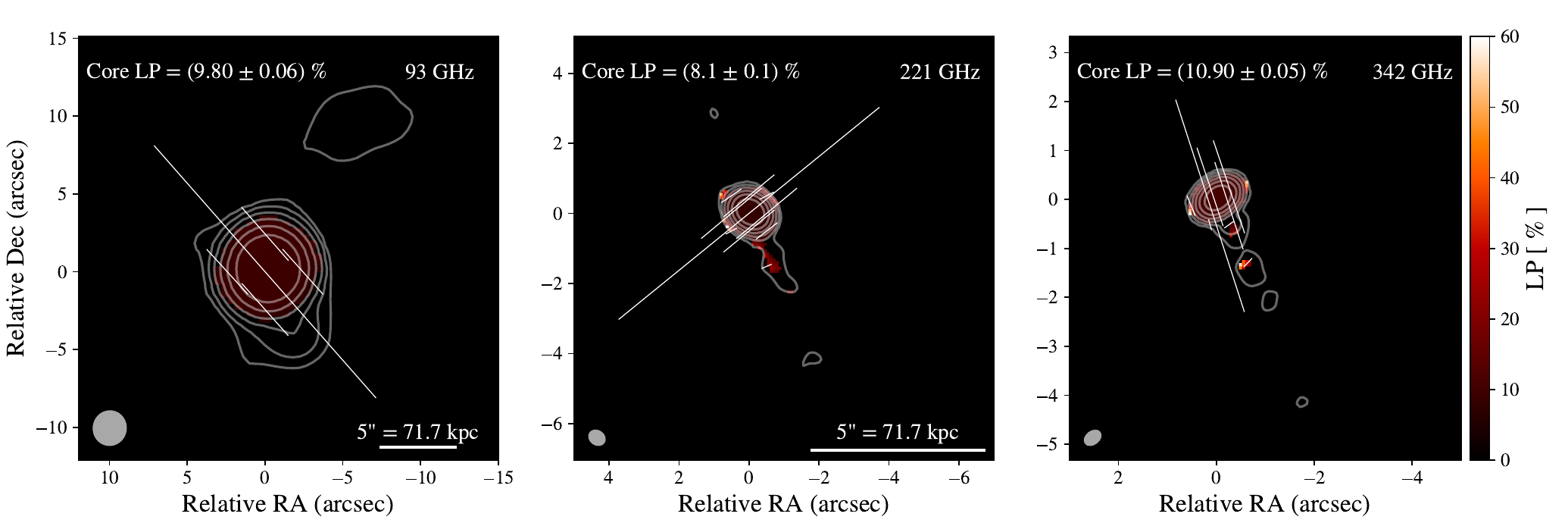}
     \caption{Polarization images of 3C279 at representative frequencies. Top: Stokes $I$. Bottom: LP with overlaid EVPAs. Left, middle, and right: B3, B6, and B7. Beam sizes: B3 2.3" $\times$ 2.2" ($-8.4^\circ$), B6 0.53" $\times$ 0.44" ($54.6^\circ$), B7 0.40" $\times$ 0.29" ($-52.3^\circ$). Contours start at $4\times$RMS and increase by $4\times$RMS at each level; RMS values of 0.23, 0.28, and 0.11 mJy for B3, B6, and B7, respectively.}\label{fig:3c279_b3b6}
\end{figure*}

Polarization maps provide important constraints for probing the magnetic-field configuration in large-scale jets.
We discuss the images of M87 (Fig. \ref{fig:M87_b3b6}), Sgr A* (Fig. \ref{fig:SgrA_b3b6}), 3C273 (Fig. \ref{fig:3c273_b3b6}), and 3C279 (Fig. \ref{fig:3c279_b3b6}) in Sect. \ref{sec:Results}, together with the observed trends in time and frequency of their core properties.

\section{Results}\label{sec:Results}

In this section, we present the main results of the analyses described in Sect. \ref{sec:Data}.
In Sect. \ref{sec:individual}, we summarize the spectropolarimetric behavior of selected well-sampled sources, considering both their temporal and spectral evolution.
In Sect. \ref{sec:class}, we explore systematic differences between object classes across the full sample.

\subsection{Analysis of individual sources}\label{sec:individual}
Here we focus on a few individual sources. They are M87 (Sect. \ref{sec:Res_M87}), Sgr A* (Sect. \ref{sec:Res_SgrA}), 3C273 (Sect. \ref{sec:Res_3C273}), 3C279 (Sect. \ref{sec:Res_3c279}), NRAO530 (Sect. \ref{sec:Res_NRAO530}), QSOB1921-293 (Sect. \ref{sec:Res_QSOB1921-293}), 4C01.28 (Sect. \ref{sec:Res_4C}), OJ287 (Sect. \ref{sec:res_OJ287}), PKS1335-127 (Sect. \ref{sec:res_PKS1335}), and J1744-3116  (Sect. \ref{sec:res_J1744}).

\subsubsection{M87} \label{sec:Res_M87}
In Fig. \ref{fig:M87_b3b6} we show Stokes $I$ maps of the M87 radio galaxy at the three ALMA bands. The maps also include the corresponding LP fraction maps with EVPA orientation overlaid.
The images reveal the kiloparsec-scale jet of M87 extending from the core by $\sim17''$ (1.4 kpc projected) at 343\,GHz to $\sim25''$ (2.1 kpc) at 93\,GHz. The various jet knots are well resolved in the B6 image, while HST-1, located $\sim0.9''$ (74.2 pc) from the core, is seen only in the high-resolution 343\,GHz image \citep{Goddi_2025}.  The inner lobes are prominent at 93\,GHz but are resolved out at higher frequencies. At centimeter and meter wavelengths, observations reveal much more extended, diffuse radio lobes, reaching scales approximately an order of magnitude larger \citep{Owen_2000}.

From the LP maps, it is evident that the core is much less polarized ($\sim \mathrm{few}\ \%$) than the jet ($\sim10$--$60\%$); it is also less polarized than what is seen at horizon scales $\sim10$--$15\%$ \citep{EHTC_2021a, EHTC_2021b}, likely owing to the large amount of structure that is averaged inside the ALMA beam.
The EVPAs are mostly perpendicular to the jet axis, except in knots A and HST-1, where they align with the jet; this is consistent with the presence of a large-scale helical magnetic field in the jet \citep[e.g.,][]{Pasetto_2021_M87}. For a helical field, this can be interpreted as the poloidal component dominating in regions with perpendicular EVPAs, while the toroidal component dominates where the EVPAs are parallel, possibly due to shocks compressing the field \citep{Lyutikov_2005}. Alternatively, EVPA swings along the jet could be due to the kinematics of the different jet components, as relativistic aberration can cause components ejected at different angles to the line of sight to have distinct observed EVPAs.
A more detailed study of the kiloparsec-scale jet will be presented in a follow-up publication.
\\

The evolution of Stokes I, LP, EVPA, and RM is reported in Table \ref{tab:M87resumed_avg} and plotted in Fig. \ref{fig:hist_M87}.
The 1.3 mm flux density is relatively stable across years. With a median of $1.3\pm0.1$ Jy, it has the second lowest variability index for $I$ of $V_I=17.3\%$. The minimum flux was $1.08$ Jy in 2018, and the maximum was $1.47$ Jy in 2021.
M87 displays significant spectral curvature in the wavelength range covered by VAPOLA (see upper panel of Fig. \ref{fig:specs}), going from slightly flat at 93 GHz to increasingly steeper at higher frequencies. The spectral index $\alpha$ also varies over time; it was around $-1$ at 221 GHz in 2017 and 2018, flattening to $-0.65$ in 2021, and steepening again to $-0.9$ in 2022.

M87 exhibits more variability in LP than in $I$, ranging from $1.0\%$ to $2.8\%$ (although it shows the lowest $V_\textnormal{LP}=51\%$). A slight decrease in LP at 221 GHz is observed over the years, from $2.5\%$ in 2017 to $1.1\%$ in 2022. The LP rises with frequency from $\sim1.2\%$ at 93 GHz to $2.8\%$ at 343 GHz in 2021, with a weak depolarization within B7 of $D=(-4.5\pm2.8)\times10^{-5}$ GHz$^{-1}$, suggesting a possible plateau.

The EVPAs at 221 GHz have a median of $-0.4^\circ$ with a scatter of $4.4^\circ$, excluding 2018 when they were $\sim30^\circ$.
The frequency dependence of the EVPAs follows the expected $\lambda^2$ relation within individual bands, but the magnitude of the RM tends to grow with frequency, and sometimes sign flips are observed across bands. For instance, in 2018, the RM at 93 and 221 GHz was $(6.0\pm0.6)\times10^4$~rad~m$^{-2}$ and $(-4.3\pm0.3)\times10^5$~rad~m$^{-2}$, respectively. These variations may reflect different Faraday depths in the source or could be due to intrinsic variability, given the six-day interval between observations in the two bands.

In fact, short-term RM variability on day-to-week timescales, often correlated with LP and EVPA changes, is ubiquitous at 221 GHz. In 2017 \citep{Goddi_2021} the RM in this band flipped sign over four days, while LP rose from $2.1\%$ to $2.7\%$ and the EVPA shifted from $-8^\circ$ to $-0.1^\circ$. In 2018, the EVPA rotated by $12^\circ$ over three days while the RM magnitude decreased by $\sim20\%$ and LP remained stable at $\sim2.2\%$. The opposite trend occurred in 2022; LP decreased from $1.33\%$ to $0.94\%,$ while the EVPA remained stable and the RM magnitude increased over five days.
Notably, EHT horizon-scale imaging shows comparable variability, as the ring-average LP dropped from $\sim15\%$ in 2017 to $\sim5\%$ in 2018 and 2021 and the spiral EVPA pattern changed helicity between epochs \citep{EHTC_2025}.

\subsubsection{Sgr A*} \label{sec:Res_SgrA}
Stokes $I$ images of Sgr A* at ALMA bands B3, B6, and B7, corresponding to April 17, 2018, April 24, 2018, and April 15, 2023, respectively, are shown in Fig. \ref{fig:SgrA_b3b6}. The extended minispiral is clearly visible at 93\,GHz and 221\,GHz but largely resolved out at 343\,GHz, with only its brightest patches above the noise. The diameter of the minispiral decreases from $\sim40''$ (1.6 pc) at 93\,GHz to $\sim10''$ (0.8 pc) at 221\,GHz, while the compact core is always detectable and brightens with frequency.
\\

The averaged properties are reported in Table \ref{tab:SgrAresumed_avg} and plotted in Fig. \ref{fig:hist_Sgra}.
Although Sgr A* is known to vary on timescales of minutes to hours \citep[][]{Baganoff2001, Genzel2003, EHTC_2022,EHT_2022_4, Wielgus_2025}, the day-averaged fluxes are among the most stable in the sample ($V_I=16.9\%$ and a median of $3.1\pm0.3$\,Jy at 221 GHz). Peak and minimum fluxes are 3.7 Jy in 2021 and 2.6 Jy in 2017, respectively. Unlike most other objects in the sample, Sgr A* exhibits a flat or slightly rising millimeter spectral index, with a median and scatter at 221 GHz of $\alpha=0.07\pm0.15$. The millimeter spectrum appears to peak between 200 and 340 GHz, which is consistent with the broad-band ($84$--$720$ GHz) observations by \cite{Liu_2016}. At ALMA scales, contamination from the minispiral complicates the determination of the intrinsic Sgr A* spectrum \citep{Albentosa2025}.

Linear polarization is moderately variable ($V_\textnormal{LP}=62.9\%$, with a median of $4.7\pm1.6\%$ at 221 GHz) and rises steeply with frequency, from $<1\%$ at 93 GHz to $8.8\%$ at 343 GHz in 2023. The highest LP was of $7.1\%$ in 2017, and the lowest was of $1.7\%$ in 2023.
These values are much lower than the $24$--$28\%$ from horizon-scale images \citep{EHT_2024_sgra}.
EVPA measurements show relative stability over the years, with a median and scatter of $-70.7\pm17.3^\circ$, excluding 2018 and 2021.
In 2018, the EVPA shifted from $\sim-73^\circ$ on April 21--22 to $\sim+70^\circ$ on April 24--25. Accounting for the $n\pi$ ambiguity, the change is more likely to have been $\sim40^\circ$, reaching $\sim-110^\circ$ on April 24 \citep{Albentosa2025}. This transition coincided with a rise in LP from 3.8\% to 4.4\%, and a decrease in $|\textnormal{RM}|$ from $(-6.3\pm0.2)\times10^5$ to $(-1.6\pm0.2)\times10^5$~rad~m$^{-2}$ on April 22, before returning to $(-4.6\pm0.2)\times10^5$~rad~m$^{-2}$ on April 24--25.
A similar event occurred in 2021, when the EVPA went from $+79^\circ$ on April 14 to $-89^\circ$ on April 17, passing through an intermediate state of $-32^\circ$ on April 15\footnote{Given the $n\pi$ ambiguity, the EVPA could have been $-101^\circ$ on April 14, fluctuated to $-32^\circ$ on the 15, and gone back to $-89^\circ$ on April 17.}. During this period, Sgr A* reached its highest flux density (3.7 Jy), LP increased from 4.7\% to 5.7\%, and $|\textnormal{RM}|$ dropped temporarily by an order of magnitude from $(-5.7\pm0.1)\times10^5$~rad~m$^{-2}$ on April 14 to $(-2.8\pm1.1)\times10^4$~rad~m$^{-2}$ on April 15, before returning to $(-5.3\pm0.1)\times10^5$~rad~m$^{-2}$ on April 17. The depolarization reached its maximum during these jumps, $(7.24 \pm 0.24)\times10^{-4}$ GHz$^{-1}$ on April 22, 2018, and $(8.62 \pm 0.23)\times10^{-4}$ GHz$^{-1}$ on April 15, 2021, in a trend opposite to that of RM (see Sect. \ref{sec:variab}).
In 2022, the LP rose from 2.8\% on April 20 to 5.7\% after two days, while the EVPA oscillated to $\pm7^\circ$ at around $-66^\circ$, and the $|\textnormal{RM}|$ decreased by half over four days.
In 2023, at 343 GHz, LP was $8.8\%$, considerably higher than the 2.9\% at 221 GHz, while the flux remained similar. The EVPAs differed across bands, but the $|\textnormal{RM}|$ remained high, with the 343 GHz measurement being the largest ever recorded for this source: $(-9.1\pm0.3)\times10^5$~rad~m$^{-2}$.

Despite these variations, the millimeter RM has remained consistently negative ($\sim-5\times10^5$~rad~m$^{-2}$) for decades \citep{Bower_2018}, suggesting a mostly stable magnetic-field configuration in the Faraday screen.
We discuss the apparent tension between this long-term stability and the rapid variability predicted by GRMHD simulations in Sect. \ref{sec:variab}. Additionally, $|\textnormal{RM}|$ also increases with frequency.

\subsubsection{3C273} \label{sec:Res_3C273}
Images of the FSRQ 3C273 at ALMA bands B3, B6, and B7, corresponding to April 23, 2021, April 27, 2022, and April 19, 2021, respectively, are shown in Fig. \ref{fig:3c273_b3b6}. The jet extends up to $\sim23''$ (87.5 kpc) from the core in the southwest direction, with a hot spot with a peak brightness of $\sim 200$ mJy/beam about $1''$ upstream of the end of the jet.
At 343 GHz, only the core and a faint jet of $\sim0.5''$ (1.8 kpc) are visible; the extended emission is largely resolved out.
The morphology is consistent with images at lower frequencies \citep{Perley_2017, Harwood_2022}.
Our maps show that the core is less polarized than the jet, which reaches $\sim30\%$ in B3 and $\sim17\%$ in B6. The EVPA in the core is generally counter-aligned with the jet ($\sim-50^\circ$), while in the jet there are two distinct regions: an upstream section inheriting the core EVPA and a downstream section aligned with the jet. Such a change is consistent with shocks amplifying the toroidal magnetic-field component.
\\

Averaged quantities are reported in Table \ref{tab:3C273resumed} and plotted in Fig. \ref{fig:hist_3C273}.
This is among the most variable sources in terms of flux, with $V_I=62.3\%$ and a median of $4.7\pm 1.7$ Jy.
3C273 showed roughly annual flares from 2017--2022 (Fig. \ref{fig:amaplots}). Since 2022, it has remained in a relatively quiescent state.
The spectral index at 221 GHz steepens over the years, from $\alpha=-0.68$ in 2017 to $-0.85$ in 2021 and $-0.9$ in 2022. At 93 GHz, $\alpha$ was flatter: $-0.34$ in 2017 and $-0.45$ in 2018. Observations across three bands in 2021 show a consistent slope, $\alpha\sim-0.81\pm0.03$, similar to B3 and B7 measurements in 2023 ($\alpha\sim-0.9$).

Linear polarization is variable ($V_\textnormal{LP}=73.4\%$), hovering at $\sim2.3\%$ during the oscillatory phase and rising to $\sim5.1\%$ in quiescence. Narrow-band observations frequently show inverse depolarization, with the LP increasing toward longer wavelengths. For instance, in 2017 the LP decreased from $3.9\%$ at 93 GHz to $2.4\%$ at 221 GHz, with consistent depolarization measures of $(-2.8\pm0.5)\times10^{-4}$ and $(-2.6\pm0.1)\times10^{-4}$ GHz$^{-1}$, respectively.
Observations in 2022 (221~GHz) and 2023 (343~GHz) again show inverse depolarization in the $-[1.6,\,0.4]\times10^{-4}$~GHz$^{-1}$ range.
While inverse depolarization was observed in several sources in this survey, 3C273 exhibits it most frequently.
3C273 is the source that most frequently shows inverse depolarization, which appears to be its natural polarization state during quiescence, as confirmed by AMAPOLA from 2022 onward.

The measurements indicate core EVPA values of $\sim-50^\circ$ in 2017 and $\sim-32^\circ$ in 2018, 2022, and 2023. AMAPOLA data show a mean EVPA of $-37.9^\circ$ with a scatter of $24.8^\circ$. During the rising-flux week of April, 2021, the EVPA appeared to flip from $+76^\circ$ on April 13 and 15 to $-85^\circ$ on April 18, while the 343 GHz measurement on April 19 was $-64^\circ$. Considering the $n\pi$ ambiguity, it is more plausible that the EVPA was $\sim-104^\circ$ on April 13--15 and increased by $19^\circ$ to $-85^\circ$ on April 18, consistently with the AMAPOLA scatter. During this interval, the LP decreased from $2.3\%$ to $1.5\%$, the depolarization dropped by an order of magnitude, and $|$RM$|$ decreased by nearly half on April 15, before returning to $(-2.6\pm0.4)\times10^{5}$ rad/m$^{2}$ on April 18. Possible explanations are discussed in Sect. \ref{sec:variab}.
The RM magnitude increases with frequency, but it can vary with year; it changed sign between the 2017--2018 (positive at low frequencies) and 2021 epochs (negative across all bands).

\subsubsection{3C279} \label{sec:Res_3c279}
In Fig. \ref{fig:3c279_b3b6} we show images of the BLLac 3C279 from April 15, 2018, April 17, 2021, and April 15, 2023, at B3, B6, and B7, respectively. In the 93 GHz image, we see the southwest-pointing jet mostly blended with the core. There is also a less bright feature positioned about $13''$ (190 kpc) northwest of the core; this is a known structure that was already observed at centimeter wavelengths by \cite{1983ApJ...273...64D} and \cite{2002ApJ...581L..15C}. It is visible in all of our 93 GHz maps, but not at higher frequencies, making it likely to be an old inflated radio lobe.
We started to resolve an $\sim5.4''$ (78.5 kpc) long jet at 221 GHz, and it is even better resolved at 343 GHz, where three distinct knots can be seen inside the first $2.2''$ (32.5 kpc) of extension, together with an extra, more distant one located about $4.4''$ (64 kpc) away from the core. The 343 GHz map of the jet is remarkably similar to Figure 2 of \cite{1983ApJ...273...64D} at 5 GHz.
Our maps do not show much of the polarized extended emission due to the low signal of the jet, but they do illustrate how the core's EVPA can change on yearly timescales, as it is oriented along the jet axis in some years and against it in others.
\\

Measurements are reported in Table \ref{tab:3C279resumed} and plotted in Fig. \ref{fig:hist_3C279}.
The 221 GHz flux oscillated around $7.4\pm2.8$ Jy before 2021, peaked at $\sim13$ Jy during 2021--2022, and has since declined to $\sim3$ Jy, making 3C279 the second most variable source in flux ($V_I=61.5\%$).
The spectrum of this BLLac is also nicely characterized by a power law that changes slightly depending on the year.
The spectral index is well described by a power law that steepens slightly with frequency ($\alpha\sim-0.4$ at 93 GHz versus $\sim-0.6$ at 221 GHz), flattens during flaring ($\alpha=-0.5$ in 2021), and steepens in quiescence ($\alpha=-0.9$ in 2023).
3C279 was observed in the three ALMA bands during the flare of 2021 and in 2023.

The LP of 3C279 fluctuated significantly during the 2016--2026 interval, with $V_\textnormal{LP}=87.8\%$. Our measurements show a median and scatter of $(10.1 \pm2.7)\%$ at 221 GHz, but the wider coverage of AMAPOLA shows $(7.3\pm 3.4)\%$.
LP also shows a long-term decline from $\sim15\%$ in 2017 to $\sim6\%$ in 2022, followed by a partial recovery to $\sim10\%$ at the end of the flaring state in 2023; it has remained below $5\%$ since the end of 2024.
There is a clear increase in LP across bands in both 2021 and 2023, with the LP going from $\sim6\%$ at 93 GHz to $\sim10\%$ at 343 GHz, accompanied by a decrease in depolarization toward higher bands.

From 2017 to 2020, the EVPAs of 3C279 remained consistently around $40^\circ$. This changed at some point between 2020 and 2021 (during a gap in monitoring due to the COVID-19 pandemic coinciding with the onset of the flaring state), when it reached around $-60^\circ$ and then drifted to $\sim17^\circ$ in 2023.
Given the $n\pi$ ambiguity, an equivalent scenario involves a $+80^\circ$ rotation to $\sim120^\circ$ in 2021; either way, both scenarios involve changes close to $90^\circ$. We discuss a scenario that could be responsible for this change in Sect. \ref{sec:variab}.
The EVPAs have been unconstrained since late 2024 due to the very low polarization.

The RM of this BLLac is low ($\lesssim$few$\times10^4$ rad m$^{-2}$ at 93 and 221 GHz) and is frequently undetected, but it increases substantially with frequency, reaching $(-1.2\pm0.3)\times10^5$ rad m$^{-2}$ at 343 GHz in 2021. The RM can also change sign depending on the year.

\subsubsection{NRAO530} \label{sec:Res_NRAO530}

The averaged properties of NRAO530 are reported in Table \ref{tab:NRAO530resumed}.
ALMA measurements indicate that the 221~GHz flux density was $1.7\pm0.1$~Jy in 2017--2018, decreasing to $\sim1.4$~Jy in 2021 and further to $0.86$~Jy in 2022.
The AMAPOLA light curve supports this trend and shows more erratic variability between 2016 and 2019, followed by a more quiescent phase from late 2021 onward ($V_I=58.1\%$).
The spectral index at 221~GHz ($\alpha\sim-0.9\pm0.1$) remains fairly stable over the years and is slightly steeper than the values measured at 93~GHz ($\alpha=-0.59$).

According to AMAPOLA, this source is among the most LP-variable in the sample ($V_\textnormal{LP}=96\%$, range 0--7.7\%), and its frequency behavior changed between campaigns: normal depolarization (LP increasing from 93 to 221~GHz) was detected in 2017, giving way to inverse depolarization in 2018.
The LP peaked at $7.7\%$ in 2021, with normal depolarization again detected at 221~GHz.
EVPA and RM are poorly constrained in most epochs due to the generally low LP, with EVPAs oscillating around $(15\pm20)^\circ$ from 2021 onward and negative RMs in the $\sim10^4$--$10^5$~rad~m$^{-2}$ range where detectable.
At sub-parsec scales, EHT imaging resolves polarized emission in multiple components of the inner jet of NRAO530 \citep{Jorstad_2023}; differently oriented components would blend together within the ALMA beam and could help explain why the EVPA and RM are so poorly constrained there.

\subsubsection{QSOB1921--293} \label{sec:Res_QSOB1921-293}
The quasar QSOB1921--293 (J1924--2914) is used as a routine calibrator for Sgr~A*.
ALMA measurements at 221~GHz show that this FSRQ was largely stable ($3.2\pm0.7$~Jy) during 2017--2021 before entering a flaring state ($5.7\pm0.5$~Jy) from late 2021 through late 2023; its flux variability is comparable to that of 3C273 and 3C279, with $V_I=60.7\%$.
The spectral index varies with epoch, becoming flatter ($\alpha\sim-0.6$) during the 2021--2023 flare, and it generally steepens with frequency.

The evolution of averaged quantities is shown in Table \ref{tab:QSOB1921resumed}.
QSOB1921--293 is strongly variable in linear polarization, with $V_\textnormal{LP}=90.5\%$. Both our measurements and AMAPOLA show LP values of $(5\pm2)\%$ during the 2017--2020 period, followed by a sharp decrease to $\sim1\%$ in 2021, just prior to the onset of the flaring state. The LP remained low during the early flare phase before rising to $\sim4.3\%$ by mid 2022.
The LP has weak spectral dependence; depolarization is most clearly detected at 93~GHz and negligible at higher frequencies. There is evidence of inverse depolarization during the 2018 observing week, when $D$ reached $(-1.4\pm0.2)\times10^{-4}$~GHz$^{-1}$ at 221~GHz on April~22.

The EVPAs of QSOB1921--293 are predominantly negative. They ranged between $-50^\circ$ and $-40^\circ$ in 2017--2018, became unconstrained on several days during the low-polarization phase associated with the rising flux in 2021--2022, and settled at around $-30^\circ$ in 2023.
At VLBI (sub-pc) scales, the inner jet is very polarized (up to $25\%$) and has a predominantly toroidal magnetic field \citep{Issaoun_2025}.
Significant RM detections are rare for this source, with the two most robust measurements being of opposite signs: $(7.3\pm1.4)\times10^4$~rad~m$^{-2}$ at 221~GHz in 2018 and $(-9.2\pm1.3)\times10^4$~rad~m$^{-2}$ in 2023. For the non-detections at 221~GHz, the values fluctuate around zero prior to 2021, shift to moderately positive values ($\sim4\times10^4$~rad~m$^{-2}$) in 2021, and become mildly negative in 2023 ($\sim-3\times10^4$~rad~m$^{-2}$).

\subsubsection{4C01.28} \label{sec:Res_4C}
Our measurements of quasar 4C01.28 are reported in Table \ref{tab:4C01.28resumed}. The 1.3\,mm flux density is typically around $2.8\pm0.5$\,Jy. This BLLac object is among the least variable sources in our sample, with a variability index of $V_I=49.9\%$ derived from the AMAPOLA data.
A modest flaring episode occurred between late 2018 and early 2019, after which the source settled into a stable state at $\sim2.5$\,Jy.
The spectral index was seen to steepen with frequency ($\alpha$ from $-0.47$ at 93 GHz to $-0.9$ at 221 GHz) in 2017 and 2018. In contrast, the spectrum was flatter in 2021, with $\alpha\simeq-0.6$ at 93\,GHz and $\alpha\simeq-0.4$ at 343\,GHz.

Linear polarization is strongly variable ($V_\textnormal{LP}=96.7\%$), dropping from $\sim5.5\%$ in 2017 to a minimum of $0.9\%$ in 2018. Consistent with this, the frequency behavior of LP transitioned from normal depolarization in 2017 (LP rising from 93 to 221\,GHz) to inverse depolarization in 2018 during the low-LP state.
The EVPAs at 221\,GHz are also highly variable, showing significant swings and multiple apparent flips over the 2017--2025 period.
It changed from $-23^\circ$ to $-14.4^\circ$ over six days in 2017.
In 2018, a rapid change of nearly $40^\circ$ (from $-28^\circ$ to $+12^\circ$) was observed over four days, while the RM changed sign from positive to negative, coinciding with an increase in LP by a factor of 1.5.
In 2021 EVPAs were consistently $\sim-38^\circ$ across 93 and 343\,GHz, with a small RM of $(-7.9\pm0.7)\times10^3$\,rad\,m$^{-2}$ at 93\,GHz and a nondetection at 343~GHz around $\pm3.5\times10^4$\,rad\,m$^{-2}$.

\subsubsection{OJ287} \label{sec:res_OJ287}
The averaged properties of OJ287 are reported in Table \ref{tab:OJ287resumed}. This quasar is moderately variable in flux at 221 GHz ($V_I=43.7\%$, median $3.2\pm0.8$\,Jy) but shows many small oscillations in the AMAPOLA light curve.
The spectral index is also variable at 221\,GHz($\alpha\sim -[1.1,\, 0.7]$)  and appears to steepen with frequency ($\alpha\sim-0.6$ at 93\,GHz).

The LP of OJ287 is more variable than its total flux, with $V_\textnormal{LP}=74.9\%$ and typical values of $(9.0\pm2.8)\%$. For example, the LP decreased from $\sim9\%$ on April~5, 2017 to $\sim7\%$ on April~10--11. The highest LP at 221\,GHz, $12.8\%$, was measured on April~27, 2018, coinciding with a period of unusually low flux density (1.85\,Jy).
OJ287 generally shows weak but normal depolarization, both within individual bands and across bands. However, during this high-polarization, low-flux state we detected significant inverse depolarization, with $D=(-2.1\pm0.2)\times10^{-4}$\,GHz$^{-1}$ at 221\,GHz. OJ287 is the only BLLac in our ALMA sample that exhibits inverse depolarization.

The EVPA measured at 221\,GHz was approximately $-62^\circ$ in 2017, increased to $-34^\circ$ in 2018, and was $-46^\circ$ in 2022. The broader AMAPOLA dataset yields a median EVPA of $(-36\pm18)^\circ$ over the 2017--2025 period. The frequency dependence of the EVPA generally results in relatively small RMs ($\lesssim10^4$\,rad\,m$^{-2}$ at 93\,GHz), but it changed sign at 221\,GHz between 2017 (positive) and 2022 (negative).
The first EHT imaging of OJ287 identified three components in the inner jet with significant EVPA evolution across five days \citep{Gomez_2026}, which suggests that part of the variability seen by ALMA originates at VLBI scales.

\subsubsection{PKS1335-127} \label{sec:res_PKS1335}
Our measurements are reported in Table \ref{tab:PKS1335-127resumed}, and they show that the 221\,GHz flux density of PKS1335--127 varies substantially from year to year. For example, it decreased from $\sim3.3$\,Jy in 2021 to $\sim1.2$\,Jy in 2022, before increasing again to $\sim2.1$\,Jy in 2023. The AMAPOLA light curve reveals even stronger variability, with a phase of enhanced activity extending from late 2018 to early 2020, and possibly continuing through the gap in coverage until 2021. Overall, PKS1335--127 is among the most variable sources in total flux (with $V_I=64.3\%$ and a median of $2.3\pm0.8$\,Jy).
The spectral index also varies with epoch ($\alpha$ from $-0.4$ to $-0.9$ at 221\,GHz), with no strong frequency dependence in most years.

PKS1335--127 is similarly variable in polarization, with the LP fluctuating around $(5.3\pm3.0)\%$ over the 2016--2026 interval ($V_\textnormal{LP}=84.5\%$). VLBI campaigns have observed this source in two markedly different polarization states. The LP was relatively stable at $\sim4.8\%$ in 2018 and remained near this level during the 2019 flare, but it reached a maximum of $14.2\%$ at 221\,GHz on April~15, 2021 and increased further to $16.7\%$ at 343\,GHz four days later. In the subsequent years the LP dropped sharply, reaching very low values of $\sim0.8\%$ at both 221 and 343\,GHz in 2023.
Overall, the LP is not strongly dependent on frequency. The clearest exception occurred during the high-polarization state of 2021, when the LP increased by $\sim2.5\%$ from 221 to 343 GHz. We detected inverse depolarization on April~22, 2018, with $D=(-1.3\pm0.2)\times10^{-4}$\,GHz$^{-1}$ at 221\,GHz.

The EVPAs have varied with time, transitioning from positive values around $35^\circ$ between 2018 and 2020 to moderately negative values of $\sim-10^\circ$ between 2021 and 2022, and becoming unconstrained afterward due to the low LP.
Occasional RM detections were obtained at 221\,GHz.
In 2022, a $\sim30^\circ$ change in EVPA was observed between March~19 ($-8.5^\circ$) and March~26 ($-39.9^\circ$), accompanied by a sign reversal in RM from $(-4.8\pm1.2)\times10^{4}$ to $(4.3\pm1.3)\times10^{4}$\,rad\,m$^{-2}$ and a modest decrease in LP from $5.3\%$ to $4.8\%$.
Despite the low-polarization state of 2023, we measured a high RM of $(-3.85\pm0.94)\times10^{5}$\,rad\,m$^{-2}$ at 221\,GHz on April~16, which dropped below the noise level ($\sim8\times10^{4}$\,rad\,m$^{-2}$) two days later.

\subsubsection{J1744-3116} \label{sec:res_J1744}
J1744--3116 is a quasar of unknown class located in the direction of the GC and is commonly used as a calibrator for Sgr~A*. It was observed exclusively in non-VLBI mode, primarily during the 2018 campaign at both 93 and 221\,GHz.
Averaged properties are reported in Table \ref{tab:J1744-3116resumed}.
This source is intrinsically weak, with stable 221\,GHz flux densities of $0.36\pm0.02$\,Jy throughout the 2018 observing week. Its spectrum remained stable and is seen to steepen with frequency, going from $-0.45$ at 93 GHz to $-0.9$ at 221 GHz.

The LP was moderate at both bands in 2018, with values in the $\sim5$--$7\%$ range. Notably, J1744--3116 is the only source in our sample that consistently exhibits inverse depolarization, with $D=(-4.4\pm0.3)\times10^{-4}$\,GHz$^{-1}$ at 93\,GHz and a weaker $D\sim(-7.9\pm0.3)\times10^{-5}$\,GHz$^{-1}$ at 221\,GHz. Given the persistent nature of this behavior, in Sect. \ref{sec:class} we argue that J1744--3116 is likely an FSRQ.

The EVPAs show only mild temporal evolution.
At 93\,GHz, the EVPA increased from $31^\circ$ in 2017 to $38^\circ$ in 2018.
The RM is generally small at 93\,GHz ($\sim{-}10^4$\,rad\,m$^{-2}$), but it reached $(1.2\pm0.1)\times10^5$\,rad\,m$^{-2}$ at 221\,GHz on April~21, 2018, before declining over the observing week.

\subsection{Differences between object classes}\label{sec:class}
Using the full set of available ALMA measurements, we can investigate how the millimeter and submillimeter emission properties depend on source class.

\begin{table*}
    \centering
    \caption{Distributions of measurements (medians and standard deviations) for various quantities as a function of source class, combining all available VAPOLA data in the 88--345 GHz range.}\label{tab:classes}
    \begin{tabular}{cccccc}
\hline\hline
Class & $\nu L_\nu$ & $\alpha$ & LP & Dep & $|$RM$|_\textnormal{rest}$ \\
 & (erg/s) & & (\%) & ($10^{-4}$ GHz$^{-1}$) & ($10^5$ rad m$^{-2}$)\\
\hline
FSRQs & $(4.8\pm8.6)\times10^{45}$ & $-0.74\pm0.22$ & $2.73 \pm 2.69$ & $0.79 \pm 2.26$ & $1.34 \pm 2.73$ \\
BL Lacs & $(1.4\pm0.9)\times10^{46}$ & $-0.62\pm0.24$ & $7.99 \pm 3.56$ & $1.50 \pm 1.12$ & $0.69 \pm 2.18$ \\
Galaxies & $(9\pm240)\times10^{40}$ & $-0.62\pm0.51$ & $1.16 \pm 0.93$ & $2.31 \pm 1.06$ & $1.93 \pm 1.93$ \\
SgrA* & $(5.2\pm1.6)\times10^{34}$ & $0.04\pm0.17$ & $4.38 \pm 2.42$ & $3.41 \pm 2.05$ & $5.11 \pm 1.98$ \\
Unknown & $-$ & $-0.88\pm0.24$ & $5.78 \pm 0.63$ & $-4.44 \pm 1.72$ & $0.10 \pm 0.46$\\
\hline\hline
\end{tabular}
\end{table*}

\begin{figure}[h]
    \centering
    \includegraphics[width=0.95\linewidth]{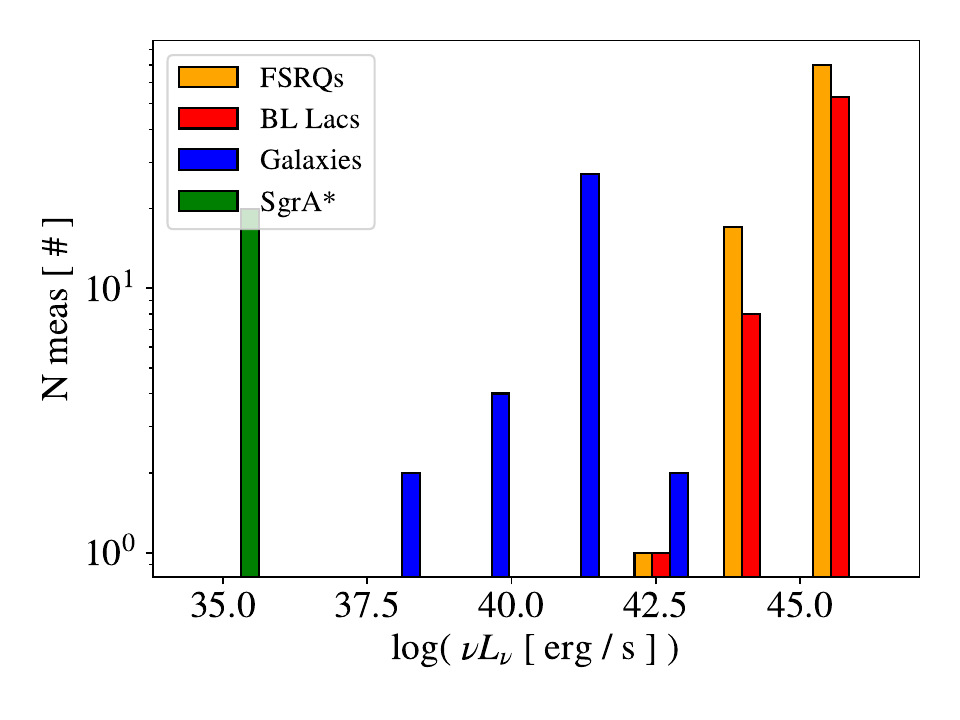}
    %\vspace{-0.5cm}
    \includegraphics[width=0.95\linewidth]{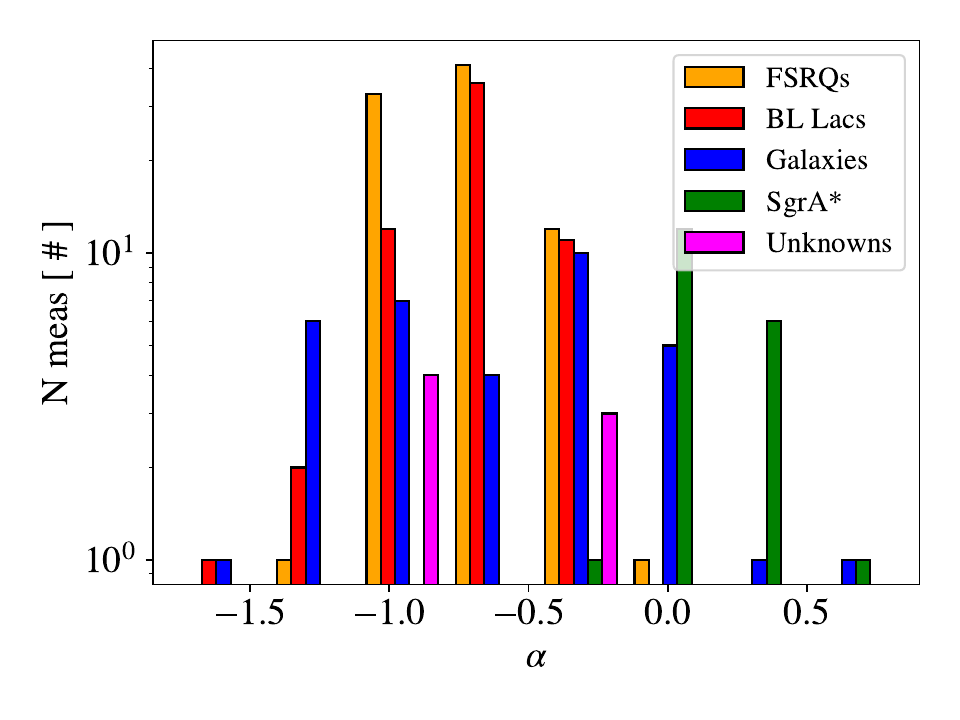}
    %\vspace{-0.5cm}
    \caption{Histogram distribution of spectral properties as a function of source class. Top: millimeter and submillimeter luminosity. Bottom: intraband spectral index. The five main colors (yellow, red, blue, green, and pink) are used to distinguish among the classes (FSRQs, BLLacs, active galaxies, Sgr A*, and unknown, respectively).}
    \label{fig:histograms1}
\end{figure}

\begin{figure}
    \centering
    \includegraphics[width=0.95\linewidth]{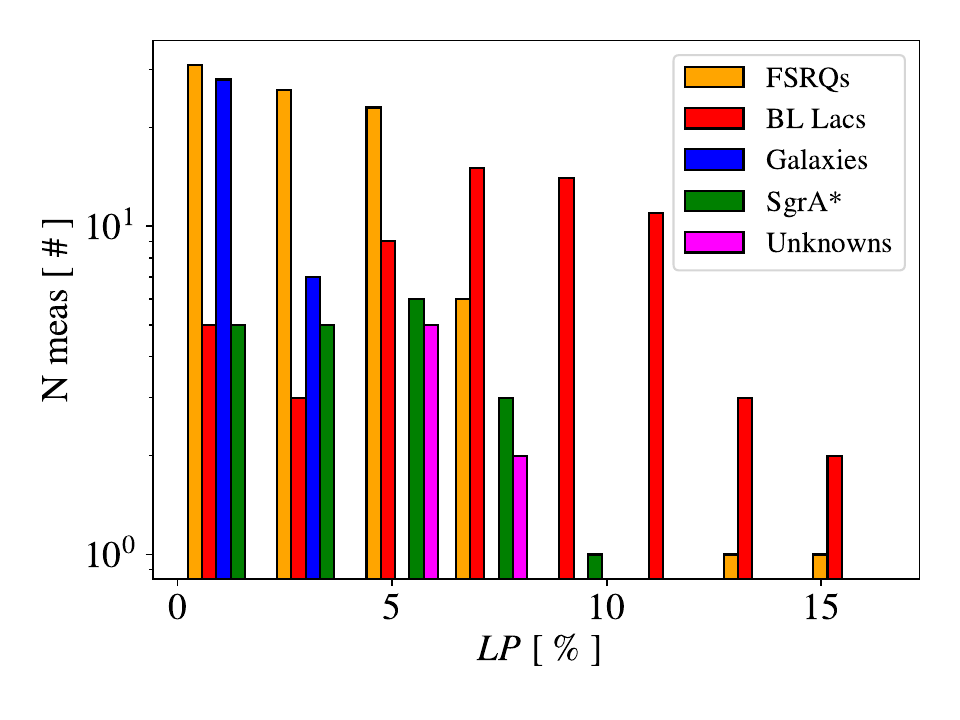}
    %\vspace{-0.5cm}
    \includegraphics[width=0.95\linewidth]{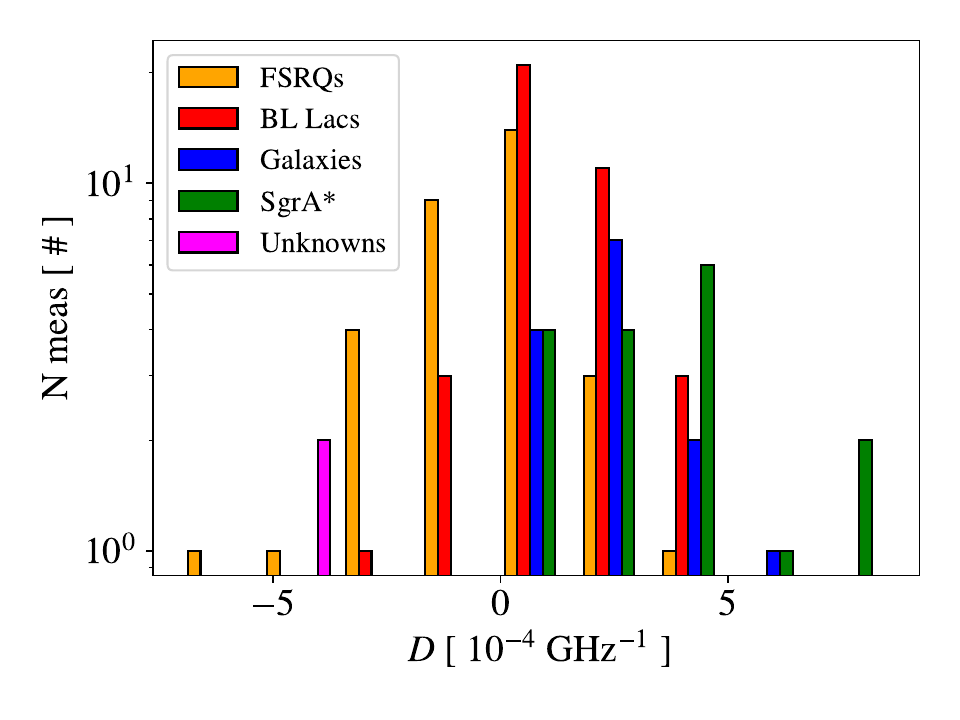}
    
    %\vspace{-0.5cm}
    \includegraphics[width=0.95\linewidth]{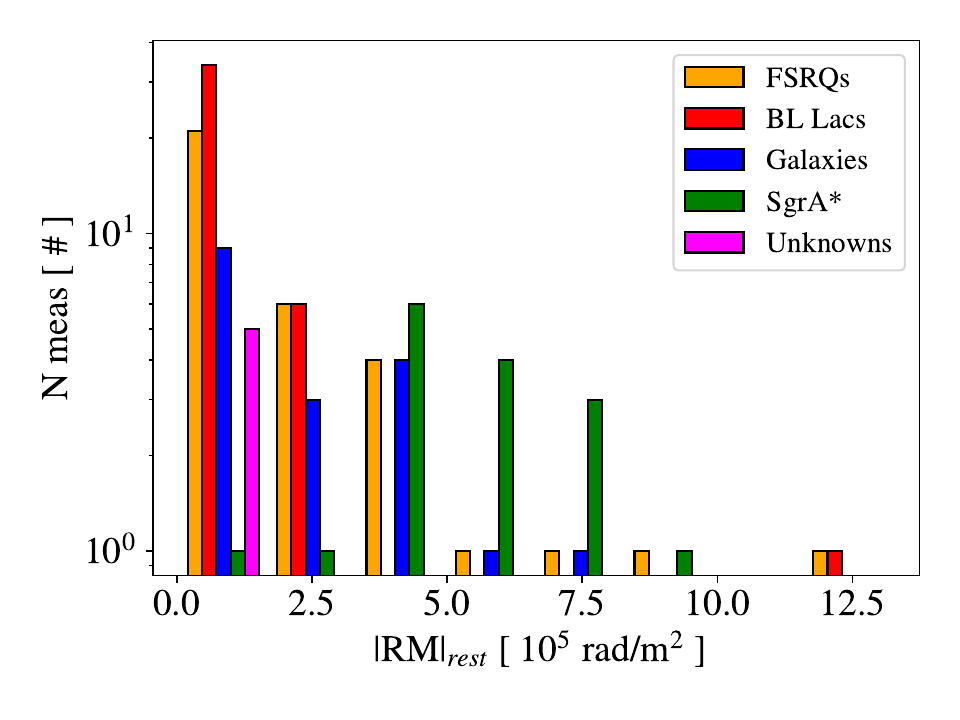}
    %\vspace{-0.25cm}
    \caption{Histogram distribution of polarimetric properties. Top: Linear polarization fraction. Center: Depolarization. Bottom: Magnitude of the rest-frame Faraday rotation measure $|$RM$|_\textnormal{rest}=(1+z)^2|$RM$|$. The color scheme is the same as in Fig. \ref{fig:histograms1}.}
    \label{fig:histograms2}
\end{figure}

We binned the data into eight bins for each quantity. Figure~\ref{fig:histograms1} shows the distributions of luminosities ($\nu L_\nu$) and spectral indices ($\alpha$), while Figure~\ref{fig:histograms2} presents the distributions of LP, $D$, and RM.
To ensure that the histograms of the derived polarization quantities ($D$ and RM) are dominated by reliable measurements, we imposed a signal-to-noise ratio threshold of S/N$>3$ and required LP$>0.5\%$.
When comparing different source classes, we also considered the rest-frame RM, defined as $|$RM$|_\mathrm{rest}=(1+z)^2|$RM$|$. For J1744--3116, whose redshift is unknown, we used the observed RM values.
For sources such as 3C279 and QSOB1921--293, B7 observations probe rest-frame frequencies as high as $\sim500$\,GHz ($\sim0.6$\,mm).

Table~\ref{tab:classes} summarizes the distributions of all spectropolarimetric parameters, listing the median and scatter of each quantity combining measurements from all bands.
As a proxy for the millimeter and submillimeter luminosities, we report the source's luminosity per logarithmic frequency as computed by Equation~(\ref{eq:lum}):
\begin{equation}\label{eq:lum}
    \nu L_\nu = 4\pi \nu I_\nu D_L^2.
\end{equation}

FSRQs and BLLacs span similar $\nu L_\nu$ ranges of $10^{42}$--$10^{46}$\,erg\,s$^{-1}$. The strongest objects in each class are PKS1741--03 among FSRQs, with $(3.7\pm0.4)\times10^{46}$\,erg\,s$^{-1}$, and 4C01.28 among BLLacs, with $(2.8\pm0.2)\times10^{46}$\,erg\,s$^{-1}$, while the weakest are Mrk~501 with $(2.0\pm0.2)\times10^{42}$\,erg\,s$^{-1}$ and AP~Librae with $(2.9\pm0.3)\times10^{43}$\,erg\,s$^{-1}$.
The two classes show modest spectral differences: FSRQs are on average slightly steeper -- with $\alpha=-0.7\pm0.2$ -- than BLLacs, which have $\alpha=-0.6\pm0.2$.

BLLacs are more polarized than FSRQs, with LP fractions of $(8.0\pm3.6)\%$ compared to $(2.7\pm2.7)\%$ for FSRQs, and they span a wider range of values. This is similar to the results from Mojave \citep{Hodge_2018} for longer wavelengths, indicating that this trend extends to the VLBI cores.
The highest LP values in our sample are observed during a brief episode in the FSRQ PKS1335--127, which reached $16\%$ at 343\,GHz in 2021, and in the BLLac 3C279, which reached $14.9\%$ at 221\,GHz in 2017.
The LP of FSRQs also shows less evidence of increasing with frequency compared to other classes.

The median depolarization of FSRQs, $(0.8\pm2.3)\times10^{-4}$\,GHz$^{-1}$, is lower than that of BLLacs, $(1.5\pm1.1)\times10^{-4}$\,GHz$^{-1}$, because FSRQs more frequently exhibit episodes of inverse depolarization; they would occupy similar ranges otherwise. Such episodes are observed in OC--150 at 221\,GHz in 2017, NRAO~530 at 93\,GHz in 2018, QSOB1921--293 and J1744--3116 at both 93 and 221\,GHz in 2018, and 3C273 across multiple epochs (see Sect. \ref{sec:Res_3C273}).
OJ287 is the only BLLac showing inverse depolarization, observed at 221\,GHz on April~27, 2018. However, with only seven objects in this class, the BLLac statistics are dominated by 3C279.
Both the median and the scatter of $|$RM$|_\textnormal{rest}$ are larger for FSRQs $(1.34\pm2.73)\times10^5$~rad~m$^{-2}$ than for BLLacs $(0.69\pm2.18)\times10^5$~rad~m$^{-2}$, and both classes show evidence of $|$RM$|_\textnormal{rest}$ increasing with frequency (see Sect. \ref{sec:RMfreq}).

The active galaxies in our sample are systematically much closer than the quasars, with the furthest being NGC5232 at $z=0.0225$ ($94.2$ Mpc), while the furthest FSRQ and BLLac are, respectively, PKS1243-072 at $z=1.286$ ($8.6$ Gpc) and 4C01.28 at $z=0.894$ ($5.5$ Gpc).
The active-galaxies class also spans the widest range of $\nu L_\nu \sim10^{38}-10^{42}$ erg/s, and their spectra are flatter ($\alpha=-0.2$) at 93 GHz and clearly steepen with frequency; however, the class contains only nine objects and is dominated by the observations of M87.
The strongest and weakest of the active galaxies are, respectively, 3C84 with $(6.6\pm0.8)\times10^{42}$ and NGC4278 with $(7.6\pm0.8)\times10^{38}$ erg/s.
These objects are much less polarized than quasars, showing a median and scatter of only $(1.2\pm0.9)\%$; the highest recorded LP comes from M87, with $2.8\%$ at 343 GHz in 2021.
In terms of depolarization, galaxies have larger $D=(2.3\pm1.1)\times10^{-4}$ GHz$^{-1}$ than both types of quasars.
The $|$RM$|_\textnormal{rest}$ distribution of galaxies has a larger median than that of quasars but a similar spread, with $(2.93\pm1.93)\times10^{5}$~rad~m$^{-2}$.
The observation that active galaxies have a lower LP and higher RM in the millimeter wavelength than quasars was already noted by the works of \citep{Bower_2017, Plambeck_2014, Goddi_2021}.
In fact, the nuclear regions of objects such as Cen A and NGC1052 display evidence of being obscured by a torus of cold gas that could act as a thick Faraday screen, increasing the RM and depolarizing the radiation \citep{Espada_2017, Kameno_2020}.
Obscuration is further supported by recent EHT results for 3C84 from the 2017 campaign \citep{Paraschos_2024}, which reveal a $0.2$ Jy flux density with a very high LP of $17\%$ at VLBI scales; this is in contrast to the $10$ Jy with $1.3\%$ that ALMA measured in the same band.

The quasar of unknown optical class, J1744-3116, is a weak source at 93 GHz ($\sim500$ mJy) located toward the GC; it was observed 7 times by ALMA, exclusively in non-VLBI mode.
The spectrum of this object steepens with frequency. Its polarization is similar to that of BLLacs $(5.8\pm0.6)\%$, shows only inverse depolarization $(-4.44\pm1.72)\times10^{-4}$ GHz$^{-1}$, and has a small (observed) $|$RM$|=(1.0\pm4.6)\times 10^4$~rad~m$^{-2}$. These properties suggest an FSRQ classification, although confirmation requires dedicated optical spectroscopy.

Finally, Sgr~A* is substantially weaker than all other sources in our sample at millimeter and submillimeter wavelengths. With $\nu L_\nu$ of $10^{34}$--$10^{35}$\,erg\,s$^{-1}$, it is roughly three orders of magnitude weaker than the weakest active galaxies and nine orders of magnitude weaker than quasars.
Its spectrum is also much flatter, with $\alpha=0.04\pm0.17$, and can reach values as high as $+0.3$ (Sect. \ref{sec:Res_SgrA}).
The LP of Sgr~A*, $(4.4\pm2.4)\%$, is on average higher than that of active galaxies and FSRQs, and at 343\,GHz it can reach values comparable to BLLacs ($8.8\%$).
It also exhibits stronger depolarization, $(3.41\pm2.05)\times10^{-4}$\,GHz$^{-1}$, particularly at lower frequencies, but it has never shown inverse depolarization.
All available measurements show a consistently negative RM of $(-5.11\pm1.98)\times10^{5}$\,rad\,m$^{-2}$, likely reflecting a stable magnetic-field configuration in the Faraday screen (Sect. \ref{sec:variab}), with a clear increase in magnitude toward higher frequencies (Sect. \ref{sec:RMfreq}).
\\

As a final remark, we stress that the most frequently observed sources may dominate their respective classes, such as 3C279 for BLLacs, QSOB1921-293 for FSRQs, and M87 for active galaxies.
We checked that without them, the results for quasars would not be affected too much; the only major difference is that the LP of BLLacs would be a little lower: $(5.97\pm 3.54)\%$.
The active-galaxies class would be the most affected. Without M87, the median $\alpha$ becomes much flatter, $-0.35\pm0.54$, and the only mildly polarized ($\sim 1\%$) sources left would be NGC4278 and 3C84, bringing the LP of this class down to $(0.25 \pm 0.44)\%$ and leaving high values of RM: $(3.58 \pm 2.76)\times 10^5$~rad~m$^{-2}$.

\section{Discussion}\label{sec:Disc}
In this section, we summarize the variability trends, discuss physical scenarios that could account for the observed changes (Sect. \ref{sec:variab}), examine the RM versus frequency behavior, and investigate whether it can provide insights into jet geometry (Sect. \ref{sec:RMfreq}).

\subsection{Short- and long-term time evolution}\label{sec:variab}
For sources observed multiple times within the same week, we generally do not detect significant changes in total intensity or spectral shape on daily to weekly timescales. Instead, substantial variations in flux density and spectral index predominantly occur over yearly timescales. With the broader temporal coverage provided by AMAPOLA, we interpret the ALMA measurements as snapshots capturing the sources at different stages between low- and high-activity states.

Spectral evolution is most evident in sources that experienced flaring episodes. These events are typically characterized by a spectral flattening during the flare, followed by a steepening in subsequent years. For instance, both 3C279 and QSOB1921--293 exhibited relatively flat spectra during their flares, with $\alpha\sim-0.53$ and $\alpha\sim-0.57$, respectively\footnote{The spectrum of QSOB1921--293 was already flatter in 2021, prior to the flare peak observed in 2022.}. In contrast, 3C273 showed an unusually steep spectrum during its 2021 flare, with a consistent $\alpha=-0.8$ across all three bands. In all cases, a spectral steepening is observed after the flare, which can be attributed to either synchrotron cooling or the fading of the flaring component \citep{Marscher1985, Turler_2000}.

On short (weekly) timescales, polarization properties exhibit substantially stronger variability than total intensity or spectral index. The LP can vary by factors of 0.5 to two from one day to the next, often accompanied by abrupt changes in EVPA and RM.
In several sources, increases in LP appear to correlate with decreases in $|$RM$|$, and vice versa.
This is observed in M87 (2017, 2022) and Sgr~A* (2018, 2021). Conversely, other quasars such as 3C273 during its 2021 flare, 4C01.28 in 2018, and PKS1335--127 in 2022 show either opposite trends or no clear correlation between LP and RM on short timescales.
For sources such as 3C279, QSOB1921--293, and OJ287 polarization variability on weekly timescales is comparatively modest. However, these objects typically exhibit weak RM signals that can drop below detectability from one day to the next, and their RM can change sign across different years. This behavior is also observed in M87 and 3C273 in the ALMA data and was previously reported for 3C273 and 3C279 using VLBA observations by \cite{Zavala_2001}.
\\

On long timescales, the most pronounced polarization changes tend to coincide with flaring activity. In particular, large EVPA rotations are observed in 3C273 during its 2021 flare and in 3C279 between 2020 and its major flare in 2021. One plausible interpretation is that these variations are associated with the emergence or evolution of unresolved jet components, as suggested by \cite{Anderson_2019}.
Such behavior is expected in shock-in-jet models developed to explain polarization variability in VLBI cores at centimeter wavelengths \citep{BK_1979, Lind1985, Marscher1985, Hughes_1989, Qian_2002}. The ejection of new components with polarization properties distinct from those of the core could naturally explain the simultaneous increase in flux, as well as changes in polarization and RM observed in these sources.
Contemporaneous VLBI polarization imaging could be used to test this scenario by constraining the spatial scales at which the variability originates. Arrays such as the GMVA ($\sim60~\mu$as) and the EHT ($\sim20~\mu$as) probe sub-pc scales in 3C273 and 3C279, while the VLBA, EVN, and KVN ($\sim0.7$~mas) are sensitive to structures on parsec scales \citep{Lister_2018, Okino_2022, EHTC_2025}.

In the case of Sgr~A*, the daily polarization variability appears decoupled from the long-term evolution of flux and spectral index, although the limited statistics should be noted.
A long-standing tension exists between the consistently negative RM measured for Sgr~A* \citep{Bower_2018, Wielgus_2024} and horizon-scale GRMHD simulations, which predict highly variable RMs that frequently change sign on hourly timescales \citep{Ricarte_2020, Ressler_2023, EHT_2024_sgra}.
However, the polarization light curves presented in \cite{Albentosa2025} covering the 2018 EHT campaign show that the RM of SgrA* can vary significantly and even take on positive values for short periods, with a characteristic variability timescale of $\sim30$ minutes. If confirmed, this would constitute strong evidence of internal Faraday rotation within the accretion flow itself.
Given these considerations, we expect the intraday RM curve from April 15, 2021 to be marked by strong and/or frequent sign flips on hourly timescales, explaining the very low $|$RM$|$ we measured from the time-averaged data (see Sect. \ref{sec:Res_SgrA}); this will be explored in  a forthcoming publication.
%\vspace{-0.3cm}
\begin{figure}[h]
    \centering
    \includegraphics[width=0.92\linewidth, height=0.65\linewidth]{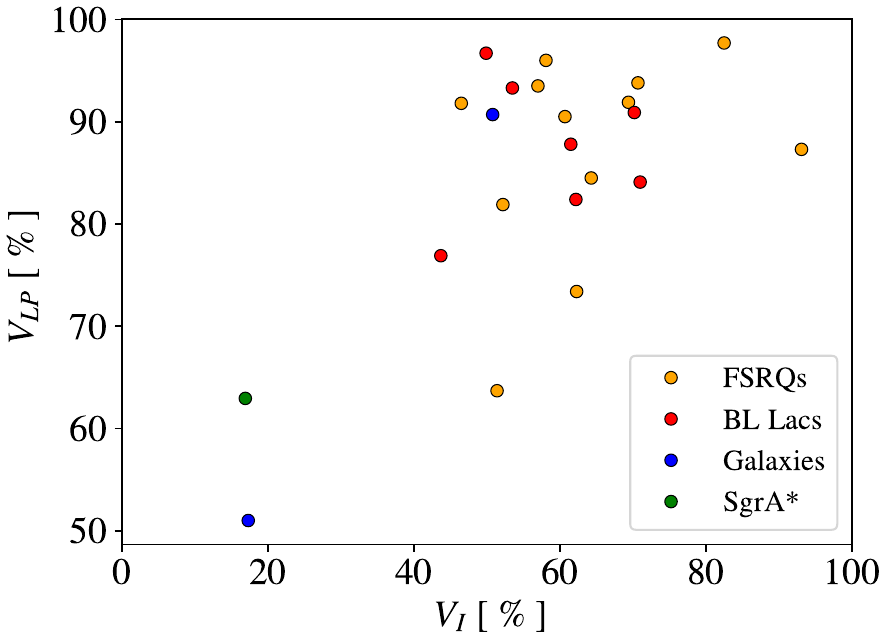}
    
    %\vspace{-0.1cm}
    \includegraphics[width=0.92\linewidth, height=0.65\linewidth]{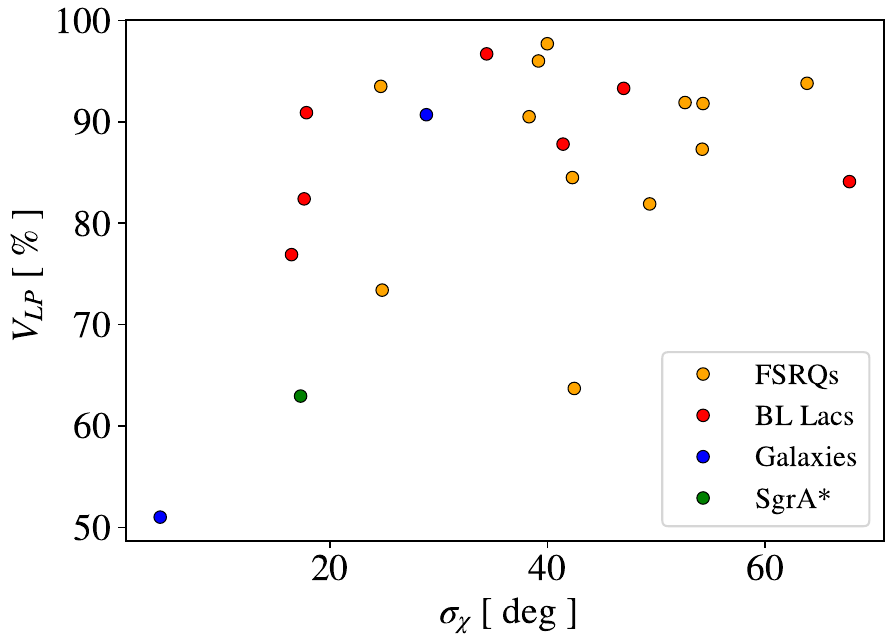}
    %\vspace{-0.2cm}
    \caption{Plots relating different variability indices. Top: LP variability $V_\textnormal{LP}$ versus intensity variability $V_I$. Bottom: LP variability $V_\textnormal{LP}$ versus EVPA scatter, $\sigma_\chi$.}\label{fig:var_class}
\end{figure}

In an attempt to cross-compare the variability features specific to different classes, we can use the variability study conducted in Sect. \ref{var_stat}.
In Fig. \ref{fig:var_class}, we plot the distributions of the different variability indices derived from the available 221 GHz observations (VLBI+AMAPOLA, see Table \ref{tab:var_analysis}).
The upper panel of Fig. \ref{fig:var_class} relates the variability in LP ($V_\textnormal{LP}$) to the variability in Stokes $I$ ($V_I$), while the lower panel relates the variability in LP to the scatter of EVPAs $\sigma_\chi$.
The only active galaxies included in these plots are M87 and 3C84; the latter shows high LP variability in a manner not dissimilar to quasars, while being moderately variable in EVPA.
M87 and Sgr A* are clearly separated from the other objects, given their low variability scores, demonstrating their consistency across different days and years.
FSRQs and BLLacs follow similar distributions of high variability in LP ($V_{LP}>60\%$) and average-to-high variability in both Stokes $I$ ($V_I>40\%$), while spanning a wide range of variability for the EVPA ($\sigma_\chi\sim15^\circ-70^\circ$), with objects such as 3C273 and OJ287 being surprisingly stable (with $\sigma_\chi=24.8^\circ$ and $17.8^\circ$, respectively).

\subsection{Frequency dependence of the RM}\label{sec:RMfreq}
The Faraday RM is sensitive to both the density of the magnetized medium traversed by the radiation and to the magnetic-field component along the line of sight, following this equation:
\begin{equation}\label{eq:RMint}
    \textnormal{RM} = 8.1\times 10^{5}\int n_e B_\parallel \cdot dl \quad\textnormal{rad m}^{-2},
\end{equation}
where $n_e$ is the electron number density (cm$^{-3}$), $B_\parallel$ is the magnetic-field component parallel to the line of sight (G), and $l$ is the path length through the plasma (pc).

Since synchrotron opacity decreases with increasing frequency \citep{BK_1979, Konigl_1981}, observations at different bands are expected to probe physically distinct regions of the source. In particular, higher frequencies should originate closer to the central engine, where both the electron density and magnetic-field strength are presumably higher.
Under a set of simplifying assumptions, \cite{Jorstad_2007} showed that the RM magnitude should scale with observing frequency as a power law: $|\textnormal{RM}| \propto \nu^{a}$. The main assumptions are that (1) the plasma density decreases with distance from the central engine as $n_e \propto r^{-a}$, with $a=2$ for a conical or spherical outflow and $a<2$ for more collimated flows; (2) the toroidal magnetic field dominates the line-of-sight component, $B_\parallel \sim B_\phi \propto r^{-1}$ \citep{Begelman_1984}, as expected if the Faraday screen is even mildly relativistic; (3) the effective integration length is proportional to the distance, $l \propto r$; and (4) the core-shift effect \citep[discovered by][]{Marcaide_1984}, whereby the distance between the VLBI core and the central engine decreases with increasing frequency due to the medium becoming more optically thin, with $r \propto \nu^{-1}$, as expected under equipartition \citep{Lobanov_1998, Sullivan_2009}.
Combining these assumptions leads directly to a power-law dependence of the core RM on frequency: $|\textnormal{RM}| \propto \nu^{a}$. Using VLBI RM measurements at 8, 15, and 43~GHz, \cite{Jorstad_2007} estimated $a$ for several sources including 3C279 ($a=0.92$), 3C345 ($a=1.6$), and 3C454.3 ($a=1.8$). Later, \cite{Hovatta_2019} analyzed 3C273 over a much broader frequency range (4.7--220~GHz) and obtained $a=2.0\pm0.2$. Other studies have reported values spanning $a \in [0.9,\,4]$ \citep{Sullivan2009, Kravchenko2017, Kam_2025}.
\\

Given our multiband ALMA coverage, we investigated the dependence of the rest-frame RM magnitude, $|\textnormal{RM}|_\textnormal{rest}$, on the rest-frame frequency, $\nu_\textnormal{rest}=(1+z)\nu,$ for the best-covered sources in our sample, and we explored the implications for their jet geometry. We included all tentative RM measurements with S/N$>2$ and performed simple power-law fits to derive a global index of $a_\textnormal{all}$ for each source. Since most sources were observed in only one band per year, we combined data from all epochs, but we also performed year-by-year fits when multiband observations were available in order to minimize the effects of intrinsic variability and to assess temporal changes in $a$.
This analysis was carried out for nine well-sampled sources, shown in Figure~\ref{fig:RM_freq}, with results summarized in Table~\ref{tab:jet}. PKS1335--127 was excluded due to the lack of reliable RM measurements.

\begin{table*}
    \centering
    \caption{Power-law fitting results for $|$RM$|$ versus frequency in the millimeter and submillimeter ranges.}\label{tab:jet}
    \begin{tabular}{cccccccc}
    \hline
    \hline
    Source & Class & $a_\textnormal{2017}$ & $a_\textnormal{2018}$ & $a_\textnormal{2021}$ & $a_\textnormal{2022}$ & $a_\textnormal{2023}$ &
    $a_\textnormal{all}$ \\
    \\
    \hline
    M87 & G & $-$ & $2.10 \pm 0.17$ & $2.17\pm0.48$ & $-$ & $-$ & $2.37 \pm 0.34$\\
    SgrA* & - & $1.06$ & $-$ & $-$ & $-$ & $0.66\pm0.53$ & $1.23 \pm 0.22$\\
    3C273 & F & $4$ & $-$ & $2.44\pm0.25$ & $-$ & $2.13\pm0.40$ & $3.04 \pm 0.34$\\
    3C279 & B & $2.56\pm0.98$ & $1.92\pm0.62$ & $1.61\pm0.45$ & $-$ & $1.36\pm0.42$ & $1.83 \pm 0.35$\\
    QSOB1921-293 & F & $-$ & $2.86\pm0.36$ & $-$ & $-$ & $2.06\pm2.32$ & $2.67 \pm 0.28$\\
    NRAO530 & F & $1.28$ & $-$ & $-$ & $-$ & $-$ & $0.78 \pm 0.59$\\
    4C01.28 & B & $-$ & $4.36\pm0.13$ & $-$ & $-$ & $-$ & $2.93 \pm 0.41$\\
    OJ287 & B & $4.41\pm0.33$ & $-$ & $-$ & $-$ & $-$ & $3.10 \pm 0.62$\\
    J1744-3116 & U & $-$ & $3.39\pm0.30$ & $-$ & $-$ & $-$ & $3.11 \pm 0.48$\\
    \hline
    \hline
\end{tabular}
\end{table*}

\begin{figure*}
    \centering
    \includegraphics[width=0.32\linewidth]{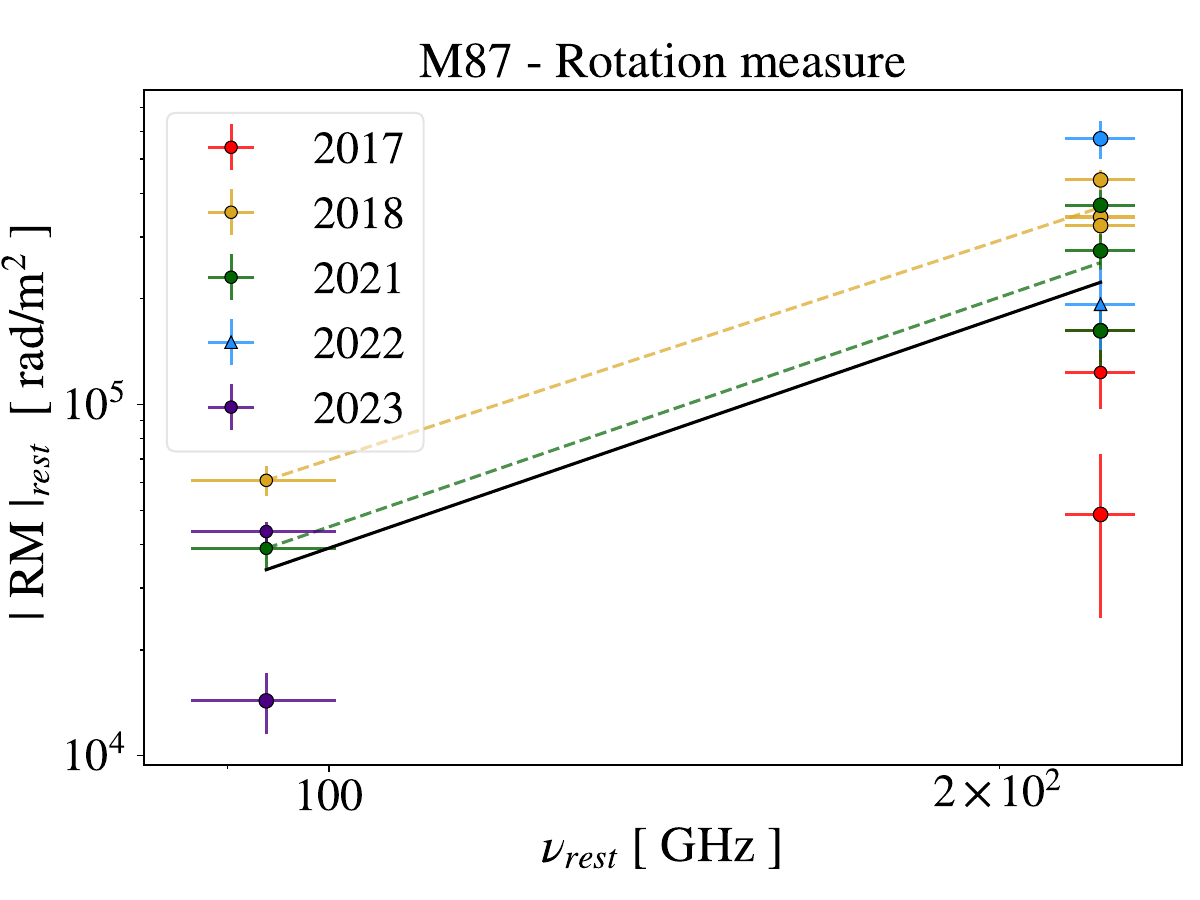}
    \includegraphics[width=0.32\linewidth]{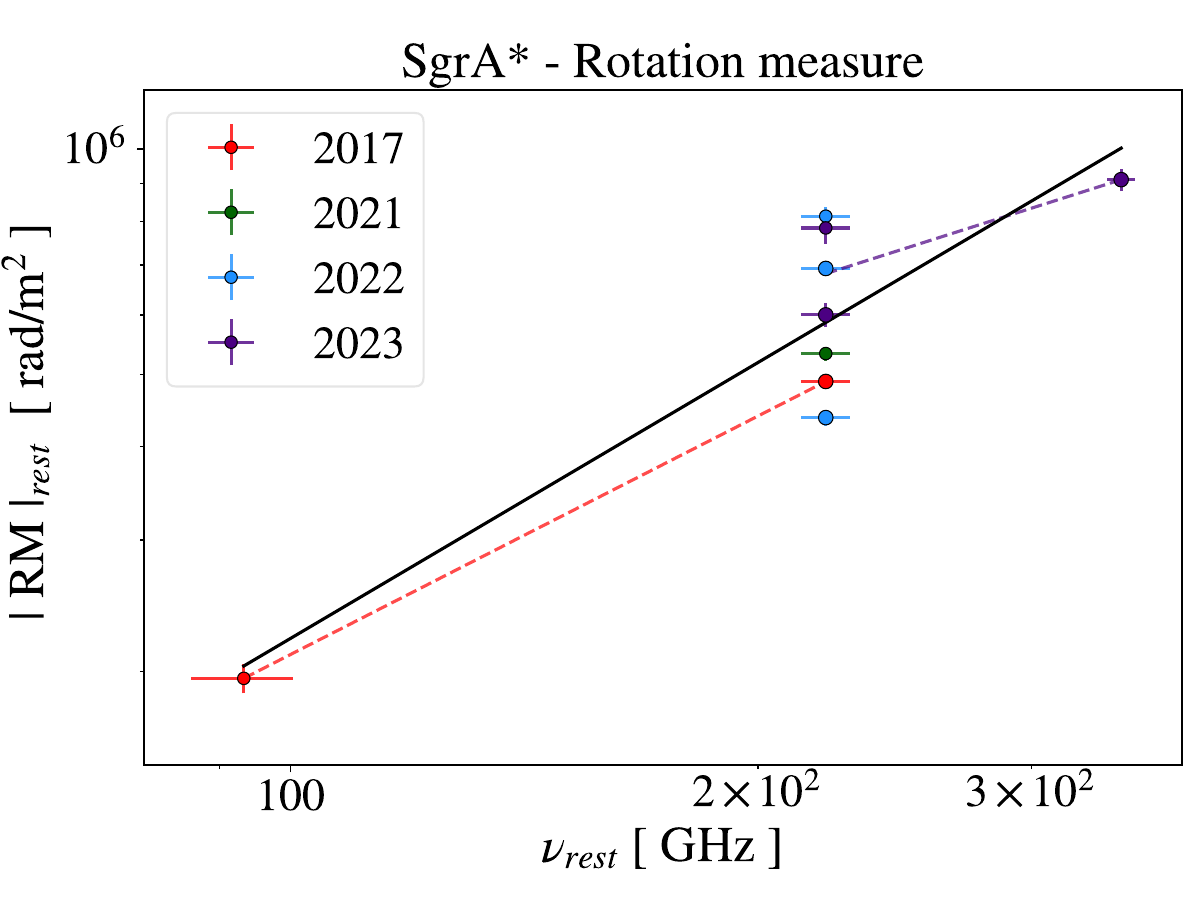}
    \includegraphics[width=0.32\linewidth]{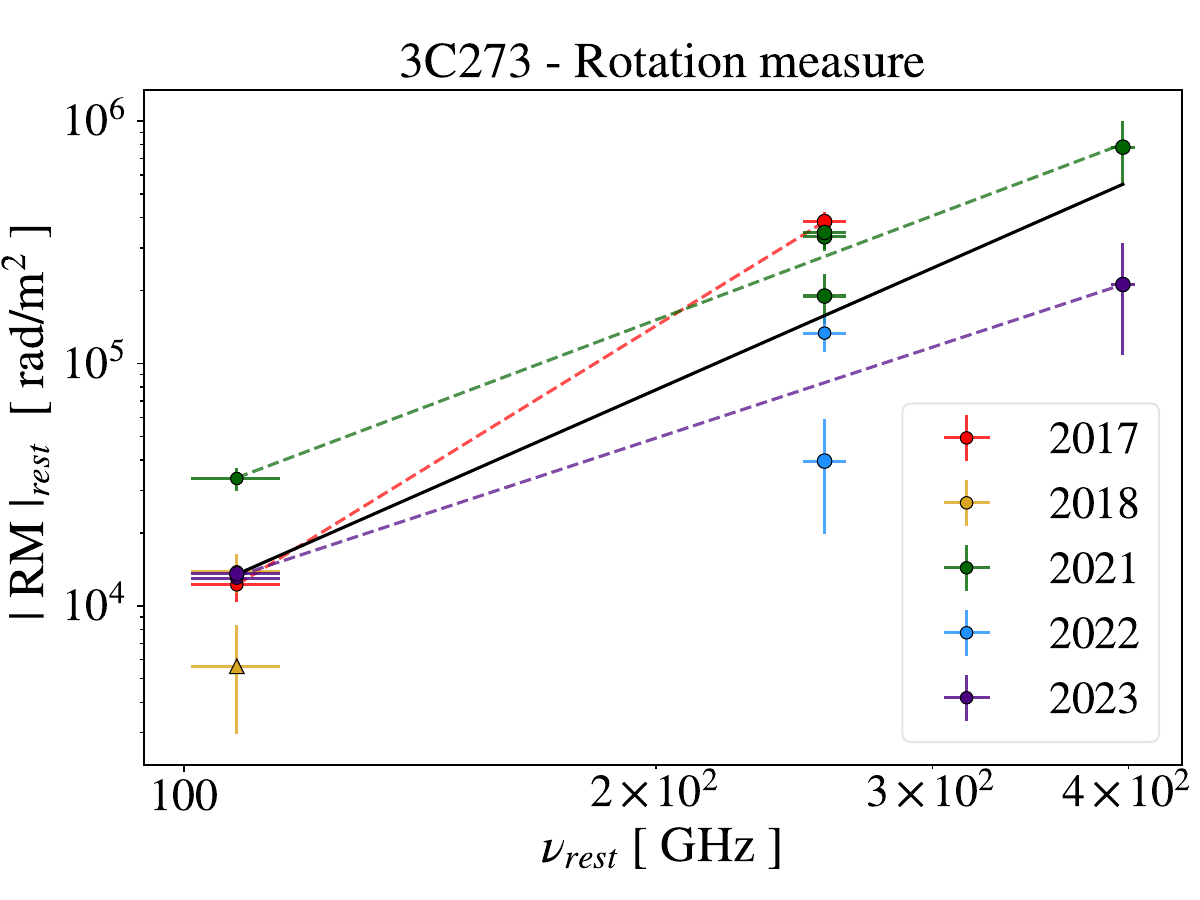}
    
    %\vspace{-5mm}
    \includegraphics[width=0.32\linewidth]{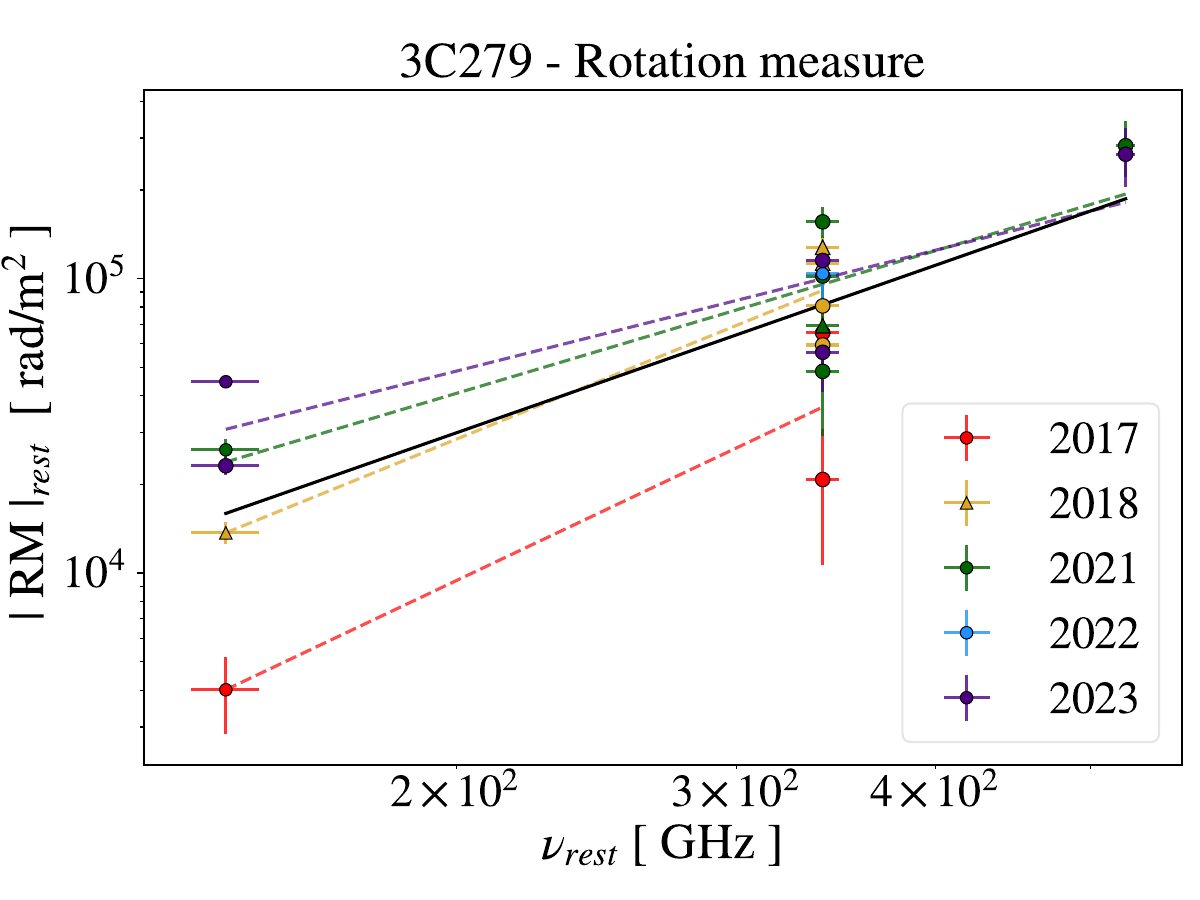}
    \includegraphics[width=0.32\linewidth]{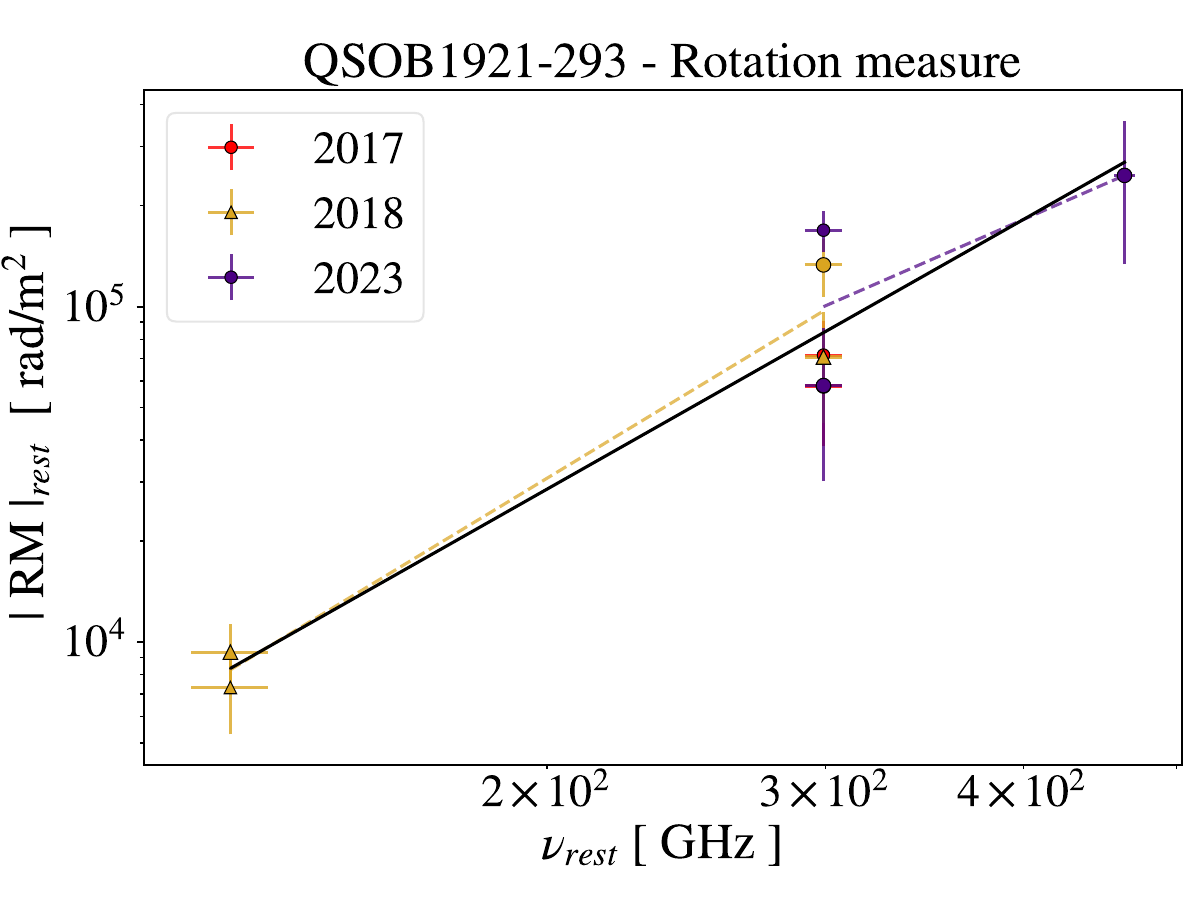}
    \includegraphics[width=0.32\linewidth]{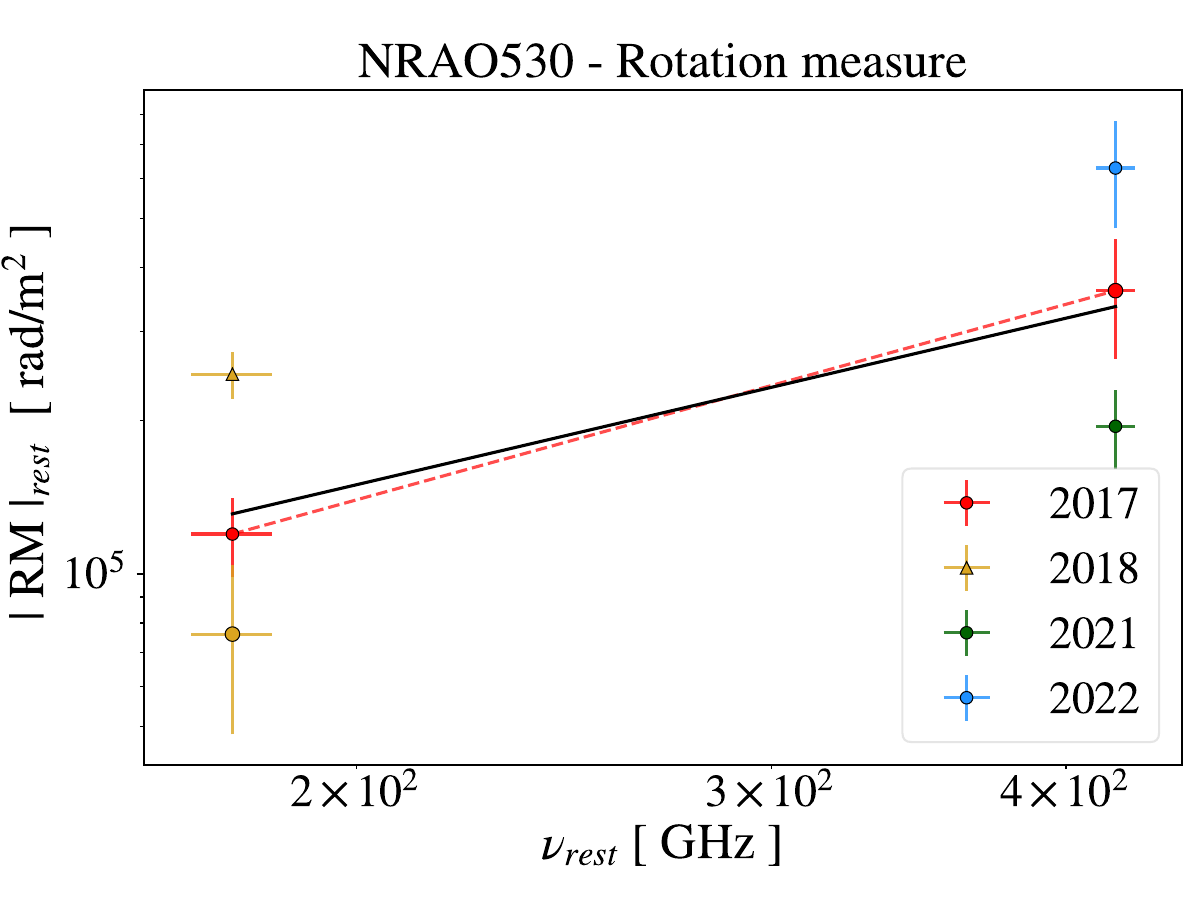}
    
    %\vspace{-5mm}
    \includegraphics[width=0.32\linewidth]{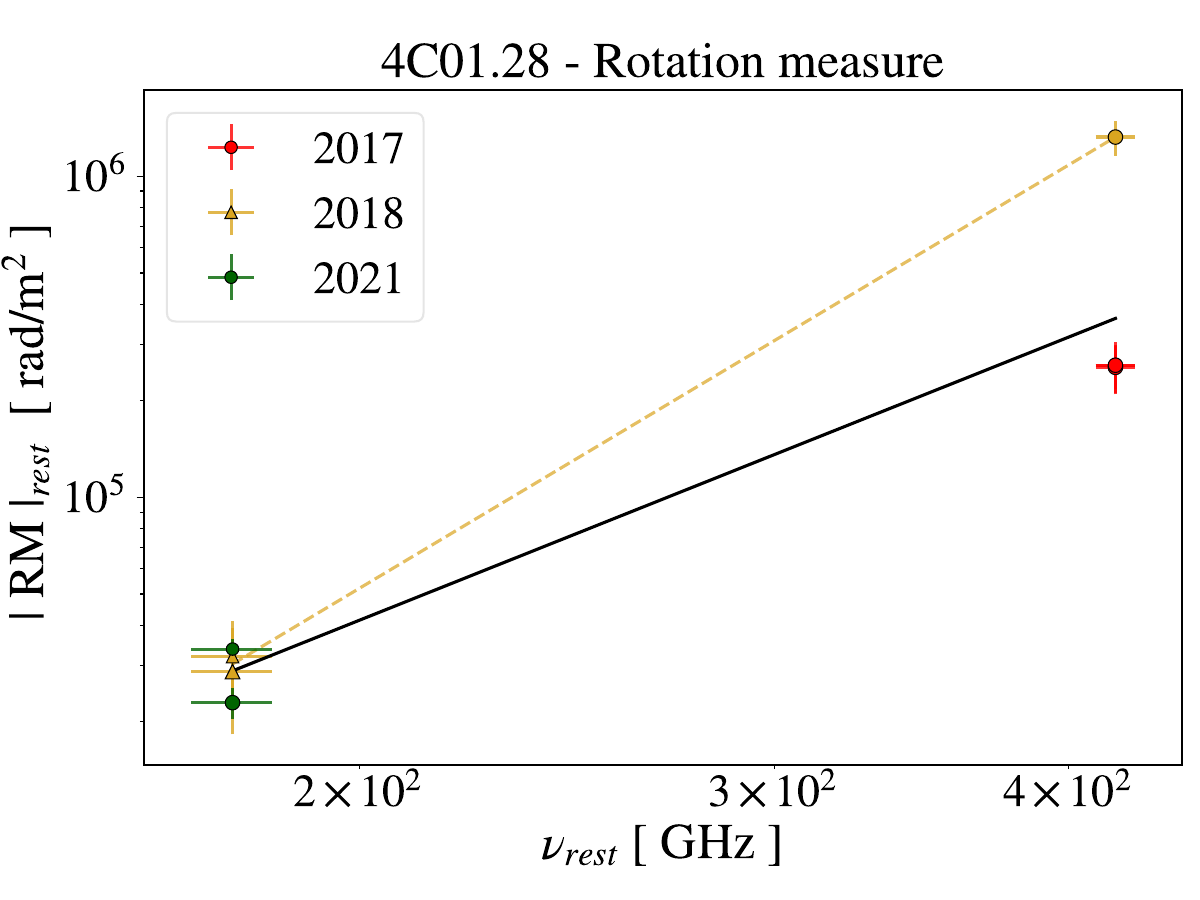}
    \includegraphics[width=0.32\linewidth]{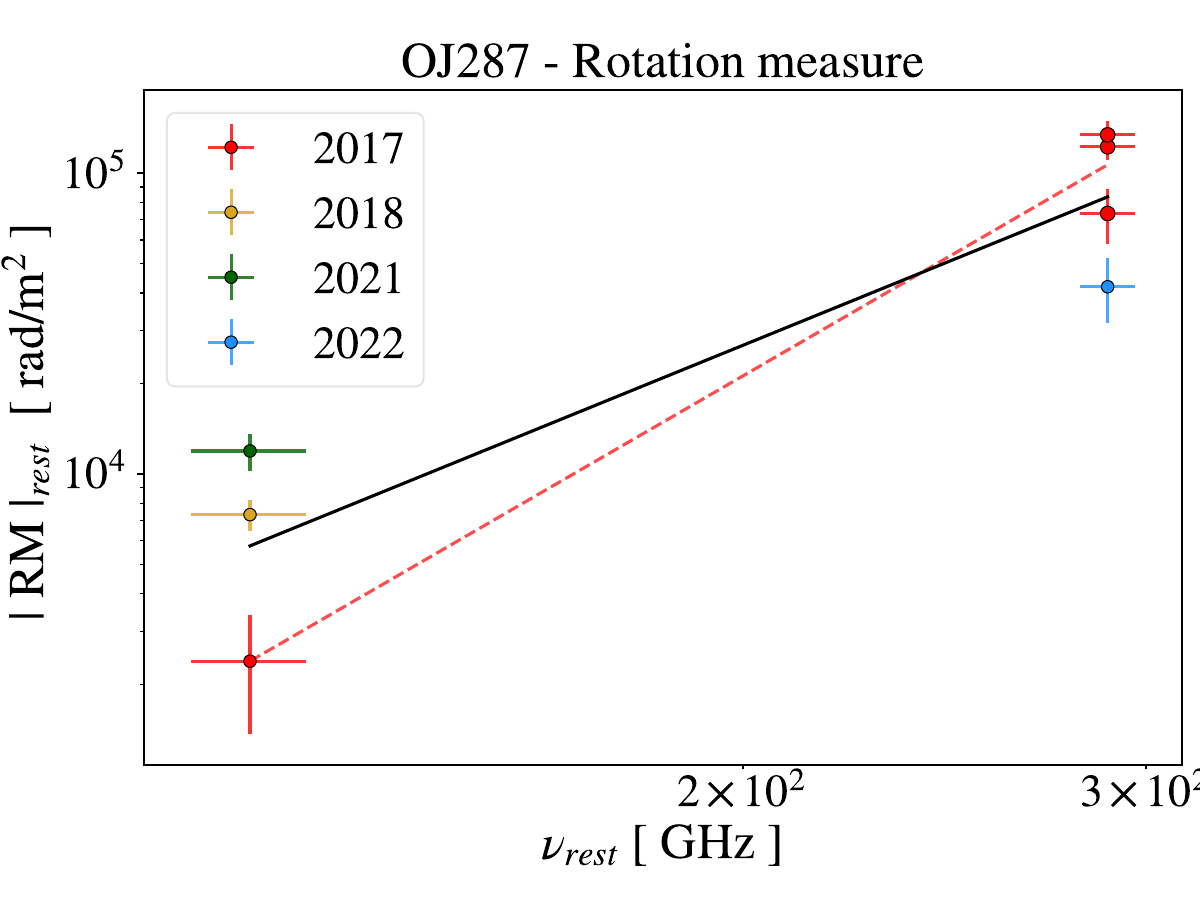}
    \includegraphics[width=0.32\linewidth]{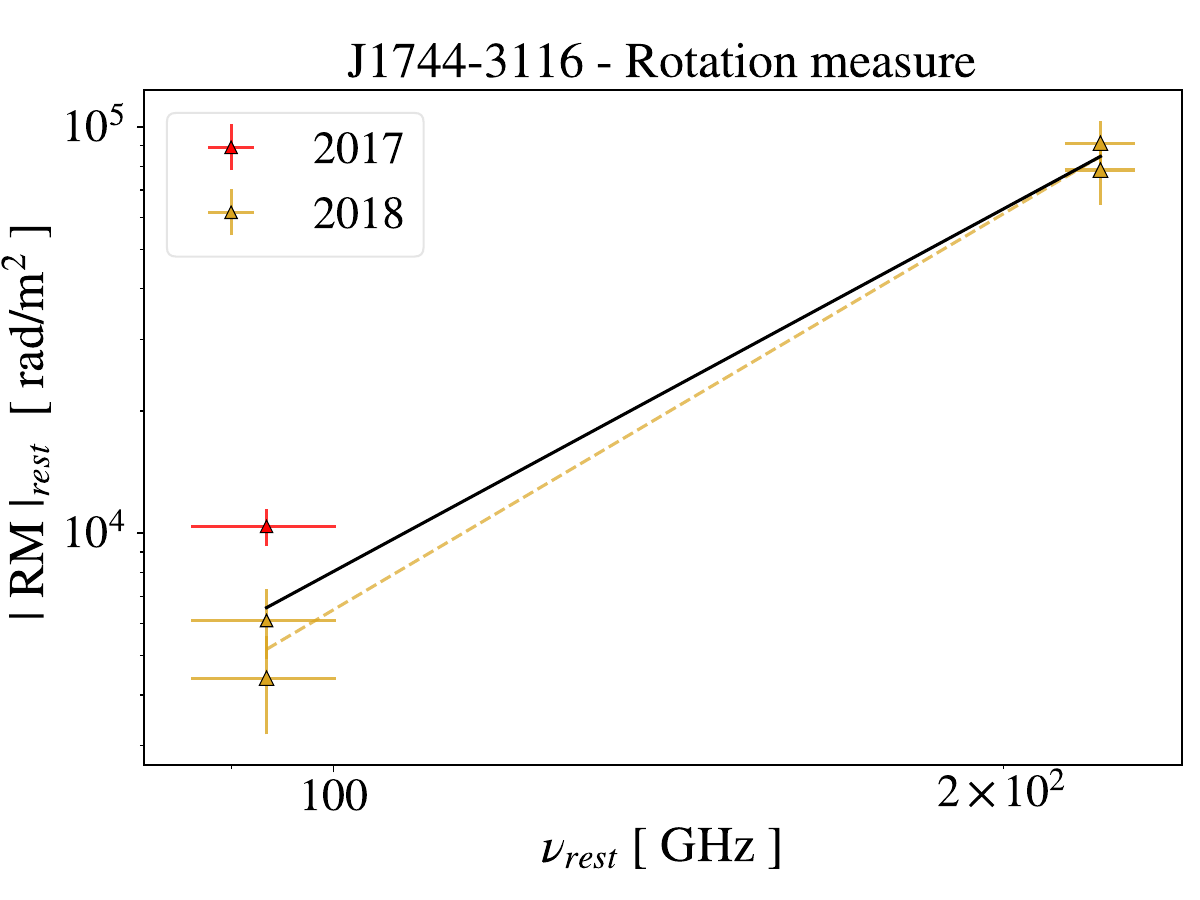}
     \caption{Plots of $|$RM$|$ as a function of frequency for the best-covered sources. Top: M87, Sgr A*, and 3C273. Center: 3C279, QSOB1921-293, and NRAO530. Bottom: 4C01.28, OJ287, and J1744-3116. The solid black line represents the fit using all data, while the dashed colored lines represent fitting using data from specific years. \label{fig:RM_freq}}
\end{figure*}

The power-law indices obtained from the millimeter/submillimeter ALMA observations lie in the range of $a\sim [0.7,\ 4.4]$, which is consistent with values from the literature at lower frequencies but covers a slightly wider range.
The values for M87 are consistent throughout the years $a_\textnormal{all}\simeq2.37\pm0.34$.
Sgr A* had an apparently lower index in 2023 ($0.66$) than in 2017 ($1.06$), but both are roughly consistent with the value obtained from fitting all years $a_\textnormal{all}=1.23\pm0.22$.
The index of 3C273 seems to have been much higher in 2017 ($\sim4$) than in 2021 or 2023. In 2021, 3C273 was observed in the three bands over multiple days, and we obtained a lower index \citep[$a_\textnormal{2021}=2.44\pm0.25$, closer to][]{Hovatta_2012} than when using all data ($a_\textnormal{all}=3.04\pm0.34$); the same was true for 2023, when it was observed only at 93 and 343 GHz $a_\textnormal{2023}=2.13\pm0.40$.
We obtained $a_\textnormal{all}=1.83\pm0.35$ for 3C279, but we note that the index seems to have been progressively decreasing over the years, from $2.56$ in 2017 to $1.36$ in 2023, when the source was also observed in the three bands.
\\

If the index $a$ is indeed linked to jet geometry, these results suggest that M87, 3C273, and 3C279 host outflows close to the canonical conical jet configuration ($a\sim2$).
Whether the decreasing index observed in 3C279 reflects genuine structural changes remains unclear, particularly in light of its major flare in 2021 and the transition to a more quiescent state in 2023.
Within this framework, the higher indices found for 4C01.28, OJ287, and J1744--3116 may indicate more rapidly expanding flows, while the lower values measured for NRAO530 and Sgr~A* would be consistent with more collimated outflows.
We find no clear association between $a$ and source class, as both FSRQs and BL~Lacs span the full range of measured indices.

The case of Sgr~A* is particularly intriguing. Although a jet has not yet been conclusively confirmed in this source, multiple lines of indirect evidence point to its existence, including large-scale, nonthermal radio structures aligned with the Galactic rotation axis \citep{Yusef_1986, Sofue_1989}, spectral-energy-distribution modeling \citep{Falcke_2000, Yuan_2002, Markoff_2007}, X-ray filaments aligned with these features \citep{Muno_2008, Li_2013}, and the discovery of the Fermi $\gamma$-ray bubbles \citep{Su_2010, Dobler_2010, Su_2012}. Our RM--frequency analysis could therefore provide an additional indirect indication of a highly collimated outflow in Sgr~A*, assuming the Faraday rotation originates in such a structure.

Alternatively, the accretion flow itself could act as the Faraday screen. This is argued in the works of \cite{Marrone_2006, Li_2015, Bower_2018} and is also supported by GRMHD simulations \citep{Ricarte_2020, Ressler_2023}.
If we revisit assumption (2) about the magnetic-field configuration and follow \cite{Marrone_2006} in adopting a radial magnetic-field profile tied to the electron density in the form of $B_\parallel\propto r^{-\frac{1}{2}(a+1)}$, then the expected behavior of the RM will also follow a power law of $|$RM$|\propto\nu^{a'}$, with $a'={\frac{1}{2}(\frac{3}{2}a+1)}$.
This yields a more restricted range of $a'\in[0.875,\,1.625]$, corresponding to convection-dominated and spherical or advection-dominated flows, respectively \citep{Blandford_Begelman_1999, Narayan_1994, Quataert_2000, Marrone_2006, Park_2019}.
This illustrates that a power-law RM dependence on frequency may not be a unique feature of jet-dominated Faraday rotation.
\\

Another important caveat is that the framework of \citet{Jorstad_2007} assumes that the polarized emission observed from the VLBI core at a given frequency, $\nu,$ originates predominantly from a single $\tau_\nu\sim1$ region in the jet. In practice, however, the core itself may contain unresolved substructure, and, particularly for sources viewed close to the line of sight, the observed polarization represents the integration of emission from multiple regions along the jet.
{Moreover, spatial variations in the magnetic field and electron density of the Faraday screen, as well as spectral evolution of the synchrotron emission, may modify the relative contributions of these regions to the observed polarization.}

The combined effect of these components on the observed $|{\rm RM}|\propto\nu^a$ relation is not yet well understood, as multiple Faraday screens can produce nonlinear EVPA--$\lambda^2$ behavior \citep{Sokoloff_1998,Sullivan2012}.
This issue is particularly relevant for ALMA observations, whose arcsecond-scale resolution blends emission that can be spatially resolved by VLBI experiments.
Nevertheless, we find no clear dependence of the $a$ index on source class. Quantifying the impact of unresolved Faraday components {and frequency-dependent polarized emission on the observed RM--frequency relation is beyond the scope of this work and will be investigated using the multiband polarization data presented here in a forthcoming publication.}

\section{Conclusions}\label{sec:conc}
We have presented a spectropolarimetric study of a sample of AGNs and Sgr~A* observed with ALMA during GMVA and EHT VLBI campaigns between 2017 and 2023.
The observations span several days over one to two weeks per year and cover three ALMA bands, enabling a characterization of the spectral and polarization properties of compact cores over the $86$--$345$\,GHz ($0.87$--$3.4$\,mm) range.
The full sample consists of 39 targets, including 29 quasars (22 FSRQs, 7 BL~Lacs, 1 unknown), nine active galaxies, and Sgr~A*. We analyzed the subset of best-sampled sources in detail and investigated their temporal and spectral evolution over the full observing period.

Our main results can be summarized as follows:

\begin{itemize}    
    \item On weekly timescales, most sources show limited variability in their total intensity and spectral index ($V_I < 10\%$), while more significant changes are commonly observed from year to year;
    
    \item Polarization properties often vary on much shorter timescales, with abrupt changes in polarization fractions and EVPA occurring from one day to the next. When considering the full 2017--2023 dataset, typical variability indices are $V_I \sim 60\%$ and $V_\mathrm{LP} \sim 80\%$;
    
    \item Large EVPA rotations observed during flaring states may be linked to the evolution of unresolved jet components. This interpretation can be tested through direct comparison with contemporaneous and future VLBI imaging observations;

    \item BLLac objects are systematically more polarized than FSRQs at millimeter and submillimeter wavelengths.
    Active galaxies show significantly lower polarization fractions than quasars and Sgr\~A*, with the latter reaching LP values of up to $\sim 9\%$ at 343 GHz, comparable to those measured in BLLacs;

    \item For most sources, the polarization fraction increases with observing frequency. FSRQs account for nearly all counterexamples, displaying as many episodes of inverse depolarization as regular depolarization. In particular, 3C273 exhibits inverse depolarization even during quiescent states, suggesting the presence of persistent and complex Faraday structures that require detailed modeling over a broad wavelength range;

    \item The large RMs observed ($\sim10^4$--$10^5$ rad m$^{-2}$) point to ordered magnetic-field structures close to the central engine. These RMs are generally stable within a given observing week but can change sign from one year to the next;
    
    \item In a few cases, rapid RM sign reversals or strong decreases in magnitude are observed over timescales of days, coinciding with episodes of rapid polarization variability, indicating a dynamically evolving Faraday screen.

    \item Using the multi-band ALMA coverage, we find evidence of a power-law dependence of $|\mathrm{RM}|$ on frequency, with indices in the $a \sim 0.7$--$4.4$ range. These values are consistent with Faraday rotation originating in a magnetized sheath surrounding the relativistic jet, allowing constraints to be placed on jet geometry. An origin in the accretion flow remains a viable alternative, particularly for Sgr~A*, for which a collimated jet has not been firmly established.  
    
\end{itemize}

\begin{acknowledgements}
The full acknowledgments are available in Appendix~\ref{ap:acknow}.
\end{acknowledgements}

%\bibliographystyle{aa}
%\bibliography{bibliography}
% Reference list as typeset in the published A&A version (bibtex output from
% bibliography.bib, hand-adjusted to the journal's format):

\begin{appendix}

%\nolinenumbers

\section{Acknowledgments}\label{ap:acknow}
\begin{acknowledgements}
We thank Rick Perley for his careful review of the manuscript, constructive suggestions, and valuable scientific discussions that helped improve the presentation and interpretation of our results.
This work used data from the Atacama Large Millimeter/submillimeter Array (facility: ALMA) and made use of the \texttt{CASA} software and the \texttt{astropy} Python package.

The Event Horizon Telescope Collaboration thanks the following organizations and programs: the Academia Sinica; the Research Council of Finland (project 362572); the Agencia Nacional de Investigaci\'{o}n y Desarrollo (ANID), Chile via NCN$19\_058$ (TITANs), Fondecyt 1221421 and 11251078, and BASAL FB210003; the Alexander von Humboldt Stiftung (including the Feodor Lynen Fellowship); an Alfred P. Sloan Research Fellowship; the ALMA North America Development Fund; the Astrophysics and High Energy Physics programme by MCIN (with funding from European Union NextGenerationEU, PRTR-C17I1); the Black Hole Initiative, which is funded by grants from the John Templeton Foundation (60477, 61497, 62286) and the Gordon and Betty Moore Foundation (Grants GBMF-8273, GBMF12987) -- although the opinions expressed in this work are those of the authors and do not necessarily reflect the views of these Foundations; the Brinson Foundation; the Canada Research Chairs (CRC) program; the University of Toronto Eric and Wendy Schmidt AI in Science Postdoctoral Fellowship Program, a program of Schmidt Sciences; the Natural Sciences \& Engineering Research Council of Canada (NSERC); the Ontario Research Fund - Research Excellence (Project Number RE012-045); Chandra DD7-18089X and TM6-17006X; the China Scholarship Council; the China Postdoctoral Science Foundation fellowships (2020M671266, 2022M712084); ANID through Fondecyt Postdoctorado (project 3250762); Conicyt through Fondecyt Postdoctorado (project 3220195); Consejo Nacional de Humanidades, Ciencia y Tecnología (CONAHCYT, Mexico, projects U0004-246083, U0004-259839, F0003-272050, M0037-279006, F0003-281692, 104497, 275201, 263356, CBF2023-2024-1102, 257435, Ph.D. Scholarship 963437); the Colfuturo Scholarship; the Consejo Superior de Investigaciones Cient\'{i}ficas (grant 2019AEP112); the Delaney Family via the Delaney Family John A. Wheeler Chair at Perimeter Institute; Dirección General de Asuntos del Personal Académico-Universidad Nacional Autónoma de México (DGAPA-UNAM, projects IN112820 and IN108324); the Dutch Research Council (NWO) for the VICI award (grant 639.043.513), the grant OCENW.KLEIN.113, and the Dutch Black Hole Consortium (with project No. NWA 1292.19.202) of the research programme the National Science Agenda; the Dutch National Supercomputers, Cartesius and Snellius  (NWO grant 2021.013); the EACOA Fellowship awarded by the East Asia Core Observatories Association, which consists of the Academia Sinica Institute of Astronomy and Astrophysics, the National Astronomical Observatory of Japan, Center for Astronomical Mega-Science, Chinese Academy of Sciences, and the Korea Astronomy and Space Science Institute; the European Research Council (ERC) Synergy Grant ``BlackHoleCam: Imaging the Event Horizon of Black Holes'' (grant 610058), Synergy Grant ``BlackHolistic:  Colour Movies of Black Holes: Understanding Black Hole Astrophysics from the Event Horizon to Galactic Scales'' (grant 10107164), and Horizon ERC Grants 2021 program under grant agreement No. 101040021; the European Union Horizon 2020 research and innovation programme under grant agreements BlackHolistic (No. 101071643), RadioNet (No. 730562), M2FINDERS (No. 101018682); the European Research Council for advanced grant ``JETSET: Launching, propagation and emission of relativistic jets from binary mergers and across mass scales'' (grant No. 884631); the European Horizon Europe staff exchange (SE) programme HORIZON-MSCA-2021-SE-01 grant NewFunFiCO (No. 10108625); the Horizon ERC Grants 2021 programme under grant agreement No. 101040021; the FAPESP (Funda\c{c}\~ao de Amparo \'a Pesquisa do Estado de S\~ao Paulo) under grant 2021/01183-8; the Fondes de Recherche Nature et Technologies (FRQNT); the Fondo CAS-ANID folio CAS220010; the Generalitat Valenciana (grants APOSTD/2018/177 and  ASFAE/2022/018) and GenT Program (project CIDEGENT/2018/021); the Hellenic Foundation for Research and Innovation (ELIDEK) under Grant No. 23698; the Gordon and Betty Moore Foundation (GBMF-3561, GBMF-5278, GBMF-10423); the Institute for Advanced Study; the ICSC – Centro Nazionale di Ricerca in High Performance Computing, Big Data and Quantum Computing, funded by European Union – NextGenerationEU; the Istituto Nazionale di Fisica Nucleare (INFN) sezione di Napoli, iniziative specifiche TEONGRAV; the Research Foundation -- Flanders (FWO) Postdoctoral Fellowship 1255226N; the International Max Planck Research School for Astronomy and Astrophysics at the Universities of Bonn and Cologne; the Italian Ministry of University and Research (MUR)– Project CUP F53D23001260001, funded by the European Union – NextGenerationEU; Deutsche Forschungsgemeinschaft (DFG, German Research Foundation) as part of the DFG Research Unit FOR5195 -– project number 443220636; Joint Columbia/Flatiron Postdoctoral Fellowship (research at the Flatiron Institute is supported by the Simons Foundation); the Japan Ministry of Education, Culture, Sports, Science and Technology (MEXT; grant JPMXP1020200109); the Japan Society for the Promotion of Science (JSPS) Grant-in-Aid for JSPS Research Fellowship (JP17J08829); the Joint Institute for Computational Fundamental Science, Japan; the Key Research Program of Frontier Sciences, Chinese Academy of Sciences (CAS, grants QYZDJ-SSW-SLH057, QYZDJSSW-SYS008, ZDBS-LY-SLH011); the Leverhulme Trust Early Career Research Fellowship; the Max-Planck-Gesellschaft (MPG); the Max Planck Partner Group of the MPG and the CAS; the MEXT/JSPS KAKENHI (grants 18KK0090, JP21H01137, JP18H03721, JP18K13594, 18K03709, JP19K14761, 18H01245, 25120007, 19H01943, 21H01137, 21H04488, 22H00157, 23K03453, 24KJ0773); the MICINN Research Projects PID2019-108995GB-C22, PID2022-140888NB-C22; the MIT International Science and Technology Initiatives (MISTI) Funds; the Ministry of Science and Technology (MOST) of Taiwan (103-2119-M-001-010-MY2, 105-2112-M-001-025-MY3, 105-2119-M-001-042, 106-2112-M-001-011, 106-2119-M-001-013, 106-2119-M-001-027, 106-2923-M-001-005, 107-2119-M-001-017, 107-2119-M-001-020, 107-2119-M-001-041, 107-2119-M-110-005, 107-2923-M-001-009, 108-2112-M-001-048, 108-2112-M-001-051, 108-2923-M-001-002, 109-2112-M-001-025, 109-2124-M-001-005, 109-2923-M-001-001, 110-2112-M-001-033, 110-2124-M-001-007 and 110-2923-M-001-001); the National Science and Technology Council (NSTC) of Taiwan (111-2124-M-001-005, 112-2124-M-001-014,  112-2112-M-003-010-MY3, and 113-2124-M-001-008); the Ministry of Education (MoE) of Taiwan Yushan Young Scholar Program; the Physics Division, National Center for Theoretical Sciences of Taiwan; the National Aeronautics and Space Administration (NASA, Fermi Guest Investigator grant 80NSSC23K1508, NASA Astrophysics Theory Program grant 80NSSC20K0527, NASA NuSTAR award 80NSSC20K0645); NASA Hubble Fellowship Program Einstein Fellowship; NASA Hubble Fellowship grants HST-HF2-51431.001-A, HST-HF2-51482.001-A, HST-HF2-51539.001-A, HST-HF2-51552.001A awarded by the Space Telescope Science Institute, which is operated by the Association of Universities for Research in Astronomy, Inc., for NASA, under contract NAS5-26555; the National Institute of Natural Sciences (NINS) of Japan; the National Key Research and Development Program of China (grant 2016YFA0400704, 2017YFA0402703, 2016YFA0400702); the National Science and Technology Council (NSTC, grants NSTC 111-2112-M-001 -041, NSTC 111-2124-M-001-005, NSTC 112-2124-M-001-014); the US National Science Foundation (NSF, grants AST-0096454, AST-0352953, AST-0521233, AST-0705062, AST-0905844, AST-0922984, AST-1126433, OIA-1126433, AST-1140030, DGE-1144085, AST-1207704, AST-1207730, AST-1207752, MRI-1228509, OPP-1248097, AST-1310896, AST-1440254, AST-1555365, AST-1614868, AST-1615796, AST-1715061, AST-1716327,  AST-1726637, OISE-1743747, AST-1743747, AST-1816420, AST-1935980, AST-1952099, AST-2034306,  AST-2205908, AST-2307887, AST-2407810, AST-2535855); NSF Astronomy and Astrophysics Postdoctoral Fellowship (AST-1903847); the NSF Graduate Research Fellowship (DGE 2140743); the Natural Science Foundation of China (grants 11650110427, 10625314, 11721303, 11725312, 11873028, 11933007, 11991052, 11991053, 12192220, 12192223, 12273022, 12325302, 12303021); the AWS Impact Computing Project at the Harvard Data Science Initiative (award no. A61166); the Natural Sciences and Engineering Research Council of Canada (NSERC); the Korea Aerospace Administration (KASA) (grant no. RS-2026-25587698); the National Research Foundation of Korea (the Global PhD Fellowship Grant: grants NRF-2015H1A2A1033752; the Korea Research Fellowship Program: NRF-2015H1D3A1066561; Brain Pool Program: RS-2024-00407499;  Basic Research Support Grant 2019R1F1A1059721, 2021R1A6A3A01086420, 2022R1C1C1005255, RS-2022-NR071771, RS-2025-16067786, RS-2025-02214038); the POSCO Science Fellowship of the POSCO TJ Park Foundation; the Global University 30 Project Fund of Kyungpook National University in 2026; the Global-Learning \& Academic research institution for Master's \textperiodcentered~PhD students, and Postdocs(G-LAMP) Program of the National Research Foundation of Korea (NRF) grant funded by the Ministry of Education (grants RS-2023-00301914, RS-2025-25442355); NOIRLab, which is managed by the Association of Universities for Research in Astronomy (AURA) under a cooperative agreement with the National Science Foundation; the A.G. Leventis Foundation; Onsala Space Observatory (OSO) national infrastructure, for the provisioning of its facilities/observational support (OSO receives funding through the Swedish Research Council under grant 2017-00648);  the Perimeter Institute for Theoretical Physics (research at Perimeter Institute is supported by the Government of Canada through the Department of Innovation, Science and Economic Development and by the Province of Ontario through the Ministry of Research, Innovation and Science); the Portuguese Foundation for Science and Technology (FCT, https://ror.org/00snfqn58) grants (Individual CEEC program – 5th edition,  FCT Multi-Annual Financing Program for R\&D Units Grants UID/04106/2025 (https://doi.org/10.54499/UID/04106/2025) and UID/PRR/04106/2025 (https://doi.org/10.54499/UID/PRR/04106/2025), PTDC/FISAST/3041/2020, CERN/FIS-PAR/0024/2021, 2022.04560.PTDC); the Princeton Gravity Initiative; the Spanish Ministerio de Ciencia, Innovaci\'{o}n  y Universidades (grants PID2022-140888NB-C21, PID2022-140888NB-C22, PID2023-147883NB-C21, RYC2023-042988-I); the Severo Ochoa grant CEX2021-001131-S funded by MICIU/AEI/10.13039/501100011033; The European Union’s Horizon Europe research and innovation program under grant agreement No. 101093934 (RADIOBLOCKS); The European Union “NextGenerationEU”, the Recovery, Transformation and Resilience Plan, the CUII of the Andalusian Regional Government and the Spanish CSIC through grant AST22\_00001\_Subproject\_10; ``la Caixa'' Foundation (ID 100010434) through fellowship codes LCF/BQ/DI22/11940027 and LCF/BQ/DI22/11940030 and ; the University of Pretoria for financial aid in the provision of the new Cluster Server nodes and SuperMicro (USA) for a SEEDING GRANT approved toward these nodes in 2020; the Shanghai Municipality orientation program of basic research for international scientists (grant no. 22JC1410600); the Shanghai Pilot Program for Basic Research, Chinese Academy of Science, Shanghai Branch (JCYJ-SHFY-2021-013); the Simons Foundation (grant 00001470); the Spanish Ministry for Science and Innovation grant CEX2021-001131-S funded by MCIN/AEI/10.13039/501100011033; the Spinoza Prize SPI 78-409; the South African Research Chairs Initiative, through the South African Radio Astronomy Observatory (SARAO, grant ID 77948),  which is a facility of the National Research Foundation (NRF), an agency of the Department of Science and Innovation (DSI) of South Africa; the Swedish Research Council (VR); the Taplin Fellowship; the Toray Science Foundation; the UK Science and Technology Facilities Council (grant no. ST/X508329/1); the US Department of Energy (USDOE) through the Los Alamos National Laboratory (operated by Triad National Security, LLC, for the National Nuclear Security Administration of the USDOE, contract 89233218CNA000001); and the YCAA Prize Postdoctoral Fellowship. This work was also supported by the National Research Foundation of Korea (NRF) grant funded by the Korea government(MSIT) (RS-2024-00449206). We acknowledge support from the Coordenação de Aperfeiçoamento de Pessoal de Nível Superior (CAPES) of Brazil through PROEX grant number 88887.845378/2023-00. We acknowledge financial support from Millenium Nucleus NCN23\_002 (TITANs) and Comité Mixto ESO-Chile.

We thank the staff at the participating observatories, correlation centers, and institutions for their enthusiastic support.
This paper makes use of the following ALMA data: \begingroup \scriptsize \almaproj{2016.1.00413.V}, \almaproj{2016.1.01114.V}, \almaproj{2016.1.01116.V}, \almaproj{2016.1.01154.V}, \almaproj{2016.1.01176.V}, \almaproj{2016.1.01198.V}, \almaproj{2016.1.01216.V}, \almaproj{2016.1.01290.V}, \almaproj{2016.1.01404.V}, \almaproj{2017.1.00795.V}, \almaproj{2017.1.00797.V}, \almaproj{2017.1.00841.V}, \almaproj{2017.1.00842.V}, \almaproj{2017.1.00985.V}, \almaproj{2017.1.00991.V}, \almaproj{2017.1.01339.V}, \almaproj{2017.1.01514.V}, \almaproj{2019.1.00183.V}, \almaproj{2019.1.00636.V}, \almaproj{2019.1.01218.V}, \almaproj{2019.1.01797.V}, \almaproj{2019.1.01812.V}, \almaproj{2021.1.00767.V}, \almaproj{2021.1.00889.V}, \almaproj{2021.1.00906.V}, \almaproj{2021.1.00910.V}, \almaproj{2021.1.01156.V}, \almaproj{2021.1.01324.V}, \almaproj{2021.1.01431.V}, \almaproj{2021.1.01458.V}, \almaproj{2021.1.01515.V}, and \almaproj{2022.1.01268.V}.
\endgroup ALMA is a partnership of the European Southern Observatory (ESO; Europe, representing its member states), NSF, and National Institutes of Natural Sciences of Japan, together with National Research Council (Canada), Ministry of Science and Technology (MOST; Taiwan), Academia Sinica Institute of Astronomy and Astrophysics (ASIAA; Taiwan), and Korea Astronomy and Space Science Institute (KASI; Republic of Korea), in cooperation with the Republic of Chile. The Joint ALMA Observatory is operated by ESO, Associated Universities, Inc. (AUI)/NRAO, and the National Astronomical Observatory of Japan (NAOJ).
The National Radio Astronomy Observatory (NRAO) and Green Bank Observatory are facilities of the U.S. National Science Foundation (NSF) operated under cooperative agreement by AUI.
This research used resources of the Oak Ridge Leadership Computing Facility at the Oak Ridge National Laboratory, which is supported by the Office of Science of the U.S. Department of Energy under contract No. DE-AC05-00OR22725; the ASTROVIVES FEDER infrastructure, with project code IDIFEDER-2021-086; the computing cluster of Shanghai VLBI correlator supported by the Special Fund for Astronomy from the Ministry of Finance in China; We also thank the Center for Computational Astrophysics, National Astronomical Observatory of Japan.

APEX is a collaboration between the Max-Planck-Institut f{\"u}r Radioastronomie (Germany), ESO, and the Onsala Space Observatory (Sweden). The SMA is a joint project between the SAO and ASIAA and is funded by the Smithsonian Institution and the Academia Sinica. The JCMT is operated by the East Asian Observatory on behalf of the NAOJ, ASIAA, and KASI, as well as the Ministry of Finance of China, Chinese Academy of Sciences, and the National Key Research and Development Program (No. 2017YFA0402700) of China and Natural Science Foundation of China grant 11873028.
Additional funding support for the JCMT is provided by the Science and Technologies Facility Council (UK) and participating universities in the UK and Canada.
The LMT is a project operated by the Instituto Nacional de Astr\'{o}fisica, \'{O}ptica, y Electr\'{o}nica (Mexico) and the University of Massachusetts at Amherst (USA).
The IRAM 30 m telescope on Pico Veleta, Spain and the NOEMA interferometer on Plateau de Bure, France are operated by IRAM and supported by CNRS (Centre National de la Recherche Scientifique, France), MPG (Max-Planck-Gesellschaft, Germany), and IGN (Instituto Geográfico Nacional, Spain).
The SMT is operated by the Arizona Radio Observatory, a part of the Steward Observatory of the University of Arizona, with financial support of operations from the State of Arizona and financial support for instrumentation development from the NSF.
Support for SPT participation in the EHT is provided by the National Science Foundation through award OPP-1852617 to the University of Chicago. Partial support is also provided by the Kavli Institute of Cosmological Physics at the University of Chicago. The SPT hydrogen maser was provided on loan from the GLT, courtesy of ASIAA.

This work used the Extreme Science and Engineering Discovery Environment (XSEDE), supported by NSF grant ACI-1548562, and CyVerse, supported by NSF grants DBI-0735191, DBI-1265383, and DBI-1743442. XSEDE Stampede2 resource at TACC was allocated through TG-AST170024 and TG-AST080026N. XSEDE JetStream resource at PTI and TACC was allocated through AST170028.
This research is part of the Frontera computing project at the Texas Advanced Computing Center through the Frontera Large-Scale Community Partnerships allocation AST20023. Frontera is made possible by National Science Foundation award OAC-1818253.
This research was done using services provided by the OSG Consortium~\citep{osg07,osg09}, which is supported by the National Science Foundation award Nos. 2030508 and 1836650.
Additional work used ABACUS2.0, which is part of the eScience center at Southern Denmark University, and the Kultrun Astronomy Hybrid Cluster (projects Conicyt Programa de Astronomia Fondo Quimal QUIMAL170001, Conicyt PIA ACT172033, Fondecyt Iniciacion 11170268, Quimal 220002).
Simulations were also performed on the SuperMUC cluster at the LRZ in Garching, on the LOEWE cluster in CSC in Frankfurt, on the HazelHen cluster at the HLRS in Stuttgart, and on the Pi2.0 and Siyuan Mark-I at Shanghai Jiao Tong University.
The computer resources of the Finnish IT Center for Science (CSC) and the Finnish Computing Competence Infrastructure (FCCI) project are acknowledged. This research was enabled in part by support provided by Compute Ontario (http://computeontario.ca), Calcul Quebec (http://www.calculquebec.ca), and the Digital Research Alliance of Canada (https://alliancecan.ca/en).

The EHTC has received generous donations of FPGA chips from Xilinx Inc., under the Xilinx University Program. The EHTC has benefited from technology shared under open-source license by the Collaboration for Astronomy Signal Processing and Electronics Research (CASPER). The EHT project is grateful to T4Science and Microsemi for their assistance with hydrogen masers. This research has made use of NASA's Astrophysics Data System. We gratefully acknowledge the support provided by the extended staff of the ALMA, from the inception of the ALMA Phasing Project through the observational campaigns of 2017 and 2018. We would like to thank A. Deller and W. Brisken for EHT-specific support with the use of DiFX. We thank Martin Shepherd for the addition of extra features in the Difmap software that were used for the CLEAN imaging results presented in this paper.
We acknowledge the significance that Maunakea, where the SMA and JCMT EHT stations are located, has for the indigenous Hawaiian people.
\end{acknowledgements}

\section{Tables and figures}
Here, we have some tables and figures cited in the main text. Table \ref{tab:srcs} lists all of the sources in the survey. Table \ref{tab:var_analysis} reports the results from the variability described in Sect. \ref{sec:variab}. Continuing Fig. \ref{fig:specs}, the spectral flux and polarization behavior of the sources is shown in Figures \ref{fig:specs2} and \ref{fig:specs3}.

\begin{table*}[h]
\centering
\caption{Sources observed during the GMVA and EHT campaigns from 2017 to 2023, including both VLBI-simultaneous and non-simultaneous observations (see Sect. \ref{sec:Data}).}\label{tab:srcs}
\begin{tabular}{ccccccccc}
\hline\hline
Source & J2000 name & RA & DEC & z$^{a}$ & $D_L$ & Class$^{b}$ & $\#$ Epochs$^c$ & Bands$^d$\\
 &  & (h:m:s) & (deg:m:s) &   & (Mpc) &  &  &  \\
\hline
3C279 & J1256-0547 & 12:56:11.167 & $-$5:47:21.525 & $0.536$ & $2957$ & B & 46 & B3,B6,B7\\
QSOB1921-293 & J1924-2914 & 19:24:51.056 & $-$29:14:30.121 & $0.353$ & $1798$ & F & 27 & B3,B6,B7\\
SgrA* & J1745.6-2900 & 17:45:40.036 & $-$29:0:28.170 & $-$ & $0.008$ & SgrA* & 26 & B3,B6,B7\\
M87 & J1230+1223 & 12:30:49.411 & 12:23:28.283 & $0.004$ & $17.66$ & G & 22 & B3,B6,B7\\
4C01.28 & J1058+0133 & 10:58:29.605 & 1:33:58.824 & $0.894$ & $5518$ & B & 17 & B3,B6,B7\\
NRAO530 & J1733-1304 & 17:33:2.706 & $-$13:4:49.548 & $0.899$ & $5559$ & F & 14 & B3,B6\\
3C273 & J1229+0203 & 12:29:6.700 & 2:3:8.600 & $0.158$ & $726.1$ & F & 17 & B3,B6,B7\\
OJ287 & J0854+2006 & 8:54:48.875 & 20:6:30.641 & $0.306$ & $1522$ & B & 9 & B3,B6\\
PKS1335-127 & J1337-1257 & 13:37:39.783 & $-$12:57:24.693 & $0.539$ & $2975$ & F & 8 & B6,B7\\
J1744-3116 & J1744-3116 & 17:44:23.578 & $-$31:16:36.292 & $-$ & $-$ & U & 7 & B3,B6\\
J1957-3845 & J1957-3845 & 19:57:59.819 & $-$38:45:6.356 & $0.626$ & $3567$ & F & 6 & B6\\
PKS1741-03 & J1743-0350 & 17:43:58.856 & $-$3:50:4.616 & $1.054$ & $6767$ & F & 6 & B6\\
J0510+1800 & J0510+1800 & 5:10:2.369 & 18:0:41.582 & $0.416$ & $2183$ & B & 4 & B3,B6\\
Cen A & J1325-4301 & 13:25:27.615 & $-$43:1:8.805 & $0.002$ & $7.511$ & G & 4 & B3,B6\\
PKS1243-072 & J1246-0730 & 12:46:4.232 & $-$7:30:46.575 & $1.286$ & $8655$ & F & 3 & B6,B7\\
PKS1510-089 & J1512-0905 & 15:12:50.533 & $-$9:5:59.830 & $0.360$ & $1842$ & F & 3 & B6,B7\\
3C84 & J0319+4130 & 3:19:48.160 & 41:30:42.103 & $0.018$ & $73.6$ & G & 3 & B6\\
M84 & J1225+1253 & 12:25:3.743 & 12:53:13.138 & $0.003$ & $13.98$ & G & 3 & B3,B6\\
J1224+0330 & J1224+0330 & 12:24:52.422 & 3:30:50.292 & $0.956$ & $6000$ & F & 2 & B3,B6\\
4C +04.42 & J1222+0413 & 12:22:22.550 & 4:13:15.776 & $0.966$ & $6078$ & F & 2 & B3,B6\\
3C345 & J1642+3948 & 16:42:58.810 & 39:48:36.993 & $0.593$ & $3342$ & F & 2 & B6\\
NRAO005 & J0006-0623 & 0:6:13.893 & $-$6:23:35.335 & $0.347$ & $1762$ & B & 2 & B6\\
OC $-150$ & J0132-1654 & 1:32:43.489 & $-$16:54:48.567 & $1.020$ & $6498$ & F & 2 & B6\\
4C09.57 & J1751+0939 & 17:51:32.819 & 9:39:0.729 & $0.322$ & $1617$ & B & 2 & B3\\
4C +29.45 & J1159+2914 & 11:59:31.834 & 29:14:43.826 & $0.725$ & $4265$ & F & 2 & B3,B6\\
OI280 & J0750+1231 & 7:50:52.046 & 12:31:4.828 & $0.889$ & $5484$ & F & 2 & B3,B6\\
J0837+2454 & J0837+2454 & 8:37:40.246 & 24:54:23.122 & $1.125$ & $7337$ & F & 2 & B6\\
J1215+1654 & J1215+1654 & 12:15:3.979 & 16:54:37.957 & $1.131$ & $7384$ & F & 2 & B6\\
NGC1052 & J0241-0815 & 2:41:4.799 & $-$8:15:20.752 & $0.005$ & $21.47$ & G & 2 & B6\\
NGC4261 & J1219+0549 & 12:19:23.216 & 5:49:29.701 & $0.007$ & $30.01$ & G & 2 & B3,B6\\
NGC4278 & J1220+2916 & 12:20:6.825 & 29:16:50.713 & $0.002$ & $8.912$ & G & 2 & B3,B6\\
3C275.1 & J1243+1622 & 12:43:57.649 & 16:22:53.393 & $0.555$ & $3084$ & F & 1 & B6\\
AP Librae & J1517-2422 & 15:17:41.813 & $-$24:22:19.476 & $0.049$ & $208.9$ & B & 1 & B6\\
3C454.3 & J2253+1608 & 22:53:57.748 & 16:8:53.561 & $0.859$ & $5256$ & F & 1 & B3\\
Mrk501 & J1653+3945 & 16:53:52.218 & 39:45:36.615 & $0.033$ & $139$ & F & 1 & B6\\
NGC4594 & J1239-1137 & 12:39:59.432 & $-$11:37:22.995 & $0.004$ & $14.97$ & G & 1 & B6\\
NGC5232 & J1336-0829 & 13:36:8.260 & $-$8:29:51.798 & $0.023$ & $94.21$ & G & 1 & B6\\
S4 1144+40 & J1146+3958 & 11:46:58.298 & 39:58:34.282 & $1.088$ & $7037$ & F & 1 & B7\\
PKS1124-186 & J1127-1857 & 11:27:4.392 & $-$18:57:17.442 & $1.052$ & $6751$ & F & 1 & B6\\
PKS0420-01 & J0423-0120 & 4:23:15.801 & $-$1:20:33.066 & $0.916$ & $5690$ & F & 1 & B6\\
\hline\hline
\end{tabular}
\tablefoot{\tablefootmark{a} Redshifts obtained from the Nasa Extragalactic Database (NED). \\
    \tablefootmark{b} Source classification based on \cite{Veron-Cetty_2006}, where we denote the letters F, B, G and U, to stand for FSRQs, BL Lacs, active galaxies and unknowns sources respectively. \\
    \tablefootmark{c} Number of epochs source was observed, including both VLBI and non-VLBI modes. \\
    \tablefootmark{d} Observations in ALMA bands B3 (93 GHz), B6 (221 GHz) and B7 (343 GHz).}
\end{table*}

\begin{table*}[h]
\centering
\caption{Results from variability analysis of observations at band 6 (221 GHz) from Sect. \ref{sec:variab}}\label{tab:var_analysis}
\begin{tabular}{cccccccc}
\hline\hline
Source & Data & $\bar{I}\pm\sigma_I$ $^a$ & $V_I$ & $\bar{\alpha}\pm\sigma_\alpha$ $^a$ & $\bar{LP}\pm\sigma_{\textnormal{LP}}$ $^a$ & $V_{\textnormal{LP}}$ & $\bar{\chi}\pm\sigma_\chi$ $^a$\\
 & & (Jy) & (\%) &   & (\%) & (\%) & (deg) \\
\hline
M87 & VAPOLA & $1.3\pm0.1$ & 17.3 & $-0.9\pm0.2$ & $1.9\pm0.5$ & 51.0 & $-0.4\pm4.4^b$\\
SgrA* & VAPOLA & $3.1\pm0.3$ & 16.9 & $0.07\pm0.15$ & $4.7\pm1.7$ & 62.9 & $-70.7\pm17.3^b$\\
\hline
3C273 & VAPOLA & $5.7\pm1.3$ & 38.5 & $-0.85\pm0.07$ & $2.3\pm1.0$ & 47.8 & $-50.3\pm17.7^b$\\
 & AMAPOLA & $4.7 \pm 1.7$ & 62.3 & $-$ & $3.8 \pm 1.4$ & 73.4 & $-37.9 \pm 24.8$\\
3C279 & VAPOLA & $9.8\pm1.6$ & 39.0 & $-0.62\pm0.10$ & $10.1\pm2.7$ & 41.4 & $16.5\pm36.9$\\
 & AMAPOLA & $6.3 \pm 2.7$ & 61.5 & $-$ & $7.3 \pm 3.4$ & 87.8 & $20.5 \pm 32.2$\\
QSOB1921-293 & VAPOLA & $3.6\pm1.3$ & 44.1 & $-0.73\pm0.19$ & $4.3\pm1.9$ & 88.6 & $-43.7\pm19.4^b$\\
 & AMAPOLA & $3.5 \pm 1.0$ & 60.7 & $-$ & $3.5 \pm 1.7$ & 90.5 & $-28.2 \pm 38.3$\\
NRAO530 & VAPOLA & $1.6\pm0.3$ & 35.2 & $-0.90\pm0.08$ & $1.4\pm2.4$ & 95.2 & $58.5\pm22.0$\\
 & AMAPOLA & $1.1 \pm 0.4$ & 58.1 & $-$ & $2.5 \pm 1.4$ & 96.0 & $6.0 \pm 41.0$\\
4C01.28 & VAPOLA & $2.8\pm0.5$ & 22.6 & $-0.8\pm0.1$ & $3.6\pm2.1$ & 74.6 & $-23.0\pm8.8^b$\\
 & AMAPOLA & $2.5 \pm 0.6$ & 49.9 & $-$ & $5.1 \pm 2.5$ & 96.7 & $-35.6 \pm 34.4$\\
OJ287 & VAPOLA & $3.5\pm1.0$ & 37.2 & $-0.94\pm0.21$ & $8.1\pm1.6$ & 29.0 &$-60.2\pm11.0$\\
 & AMAPOLA & $3.2 \pm 0.8$ & 43.7 & $-$ & $9.0 \pm 2.8$ & 76.9 & $-35.8 \pm 17.8$\\
PKS1335-127 & VAPOLA & $2.1\pm0.6$ & 47.0 & $-0.72\pm0.18$ & $4.8\pm4.5$ & 90.76 & $-6.0\pm19.7$\\
 & AMAPOLA & $2.3 \pm 0.8$ & 64.3 & $-$ & $5.3 \pm 3.0$ & 84.5 & $-5.2 \pm 40.4$\\
\hline
3C345 & AMAPOLA & $2.3 \pm 0.5$ & 52.2 & $-$ & $5.2 \pm 2.1$ & 81.9 & $-47.7 \pm 51.9$ \\
3C454.3 & AMAPOLA & $5.2 \pm 2.4$ & 70.7 & $-$ & $2.6 \pm 2.0$ & 93.8 & $-56.4 \pm 57.2$ \\
3C84 & AMAPOLA & $10.0 \pm 2.7$ & 50.8 & $-$ & $1.3 \pm 0.7$ & 90.7 & $-13.1 \pm 28.4$ \\
4C+29.45 & AMAPOLA & $1.6 \pm 1.1$ & 93.1 & $-$ & $3.1 \pm 1.4$ & 87.3 & $-27.6 \pm 47.5$ \\
4C09.57 & AMAPOLA & $1.7 \pm 0.5$ & 53.5 & $-$ & $3.6 \pm 1.7$ & 93.3 & $-35.7 \pm 48.9$ \\
APLibrae & AMAPOLA & $1.6 \pm 0.4$ & 62.2 & $-$ & $2.6 \pm 0.8$ & 82.4 & $-12.4 \pm 20.1$ \\
J0510+1800 & AMAPOLA & $1.1 \pm 0.7$ & 71.0 & $-$ & $7.6 \pm 2.9$ & 84.1 & $-67.8 \pm 60.3$\\
J1957-3845 & AMAPOLA & $2.3 \pm 1.5$ & 82.5 & $-$ & $2.9 \pm 2.0$ & 87.7 & $3.52 \pm 44.4$ \\
NRAO005 & AMAPOLA & $1.0 \pm 0.6$ & 70.2 & $-$ & $6.8 \pm 3.5$ & 90.9 & $24.8 \pm 18.2$ \\
OI280 & AMAPOLA & $0.89 \pm 0.27$ & 57.0 & $-$ & $3.5 \pm 1.5$ & 93.5 & $15.8 \pm 25.4$ \\
PKS1124-186 & AMAPOLA & $0.49 \pm 0.09$ & 46.5 & $-$ & $3.1 \pm 1.6$ & 91.8 & $36.4 \pm 53.8$ \\
PKS1510-089 & AMAPOLA & $1.8 \pm 0.5$ & 51.4 & $-$ & $3.3 \pm 1.3$ & 63.7 & $23.5 \pm 44.6$ \\
S4 1144+40 & AMAPOLA & $0.67 \pm 0.24$ & 60.4 & $-$ & $3.2 \pm 1.9$ & 91.9 & $-45.6 \pm 54.7$ \\
\hline\hline
\end{tabular}
\tablefoot{\tablefootmark{a} Median and standard deviation values across all observation epochs. The VAPOLA data spans at most a few dozen observations over five years, while the AMAPOLA data covers the same period with almost dayly cadence. \\
    \tablefootmark{b} Values excluding EVPA outliers.}
\end{table*}

\begin{figure*}[h]
    \centering
    \includegraphics[width=0.95\linewidth]{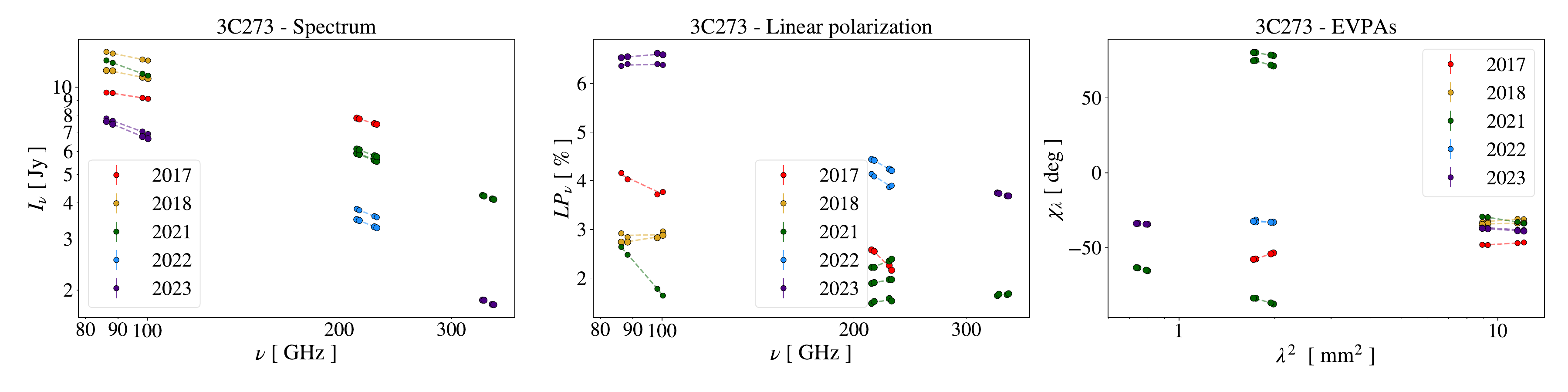}
    
    %\vspace{-6.5mm}
    \includegraphics[width=0.95\linewidth]{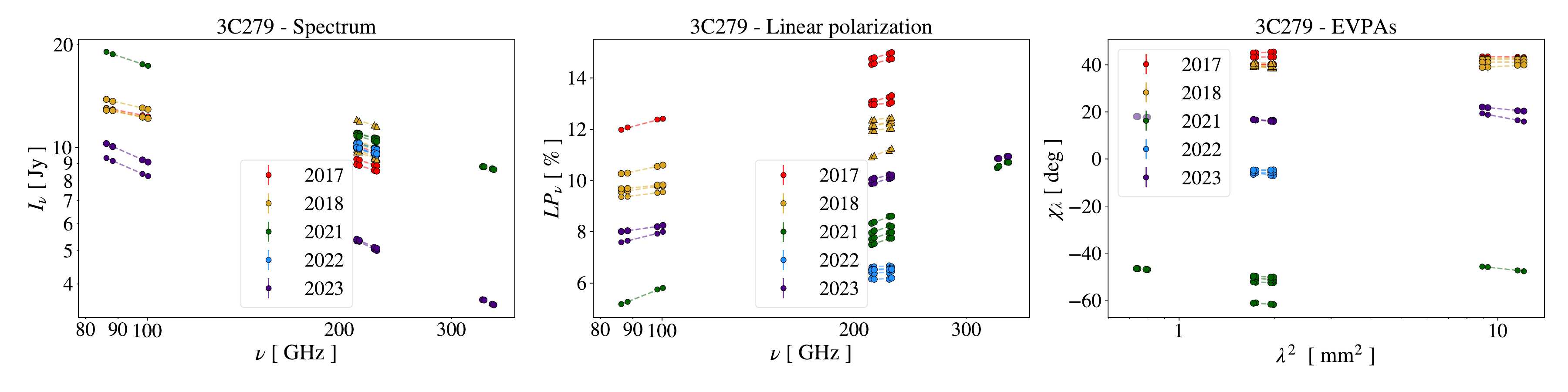}
    %\vspace{-6.5mm}
    \includegraphics[width=0.95\linewidth]{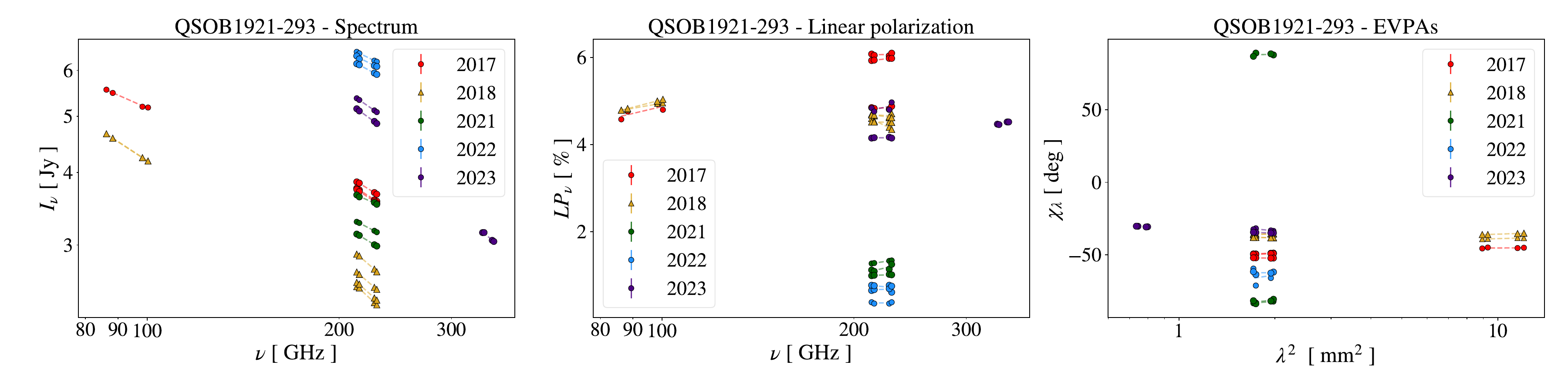}
    
    %\vspace{-6.5mm}
    \includegraphics[width=0.95\linewidth]{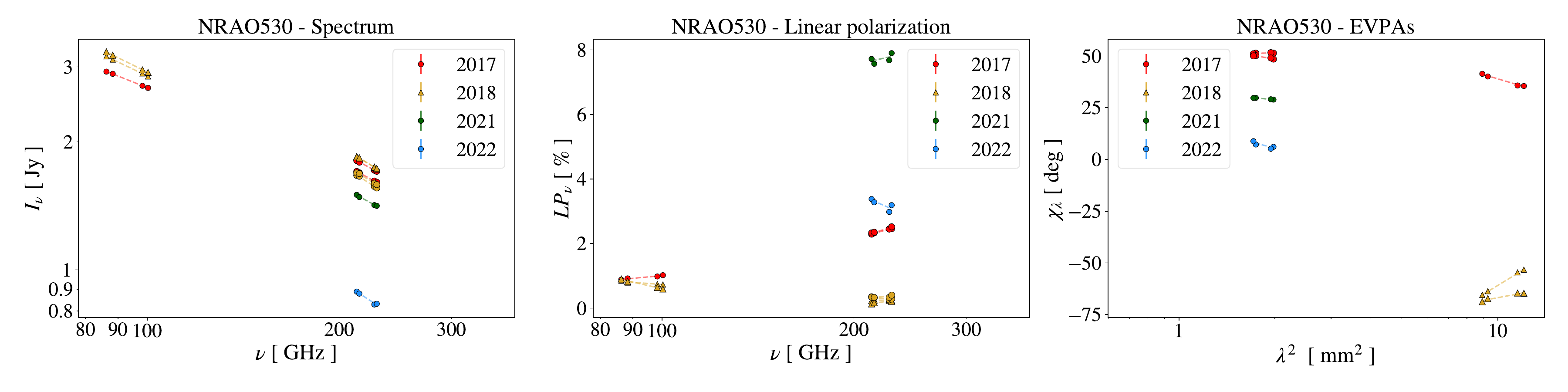}
    
    %\vspace{-6.5mm}
    \includegraphics[width=0.95\linewidth]{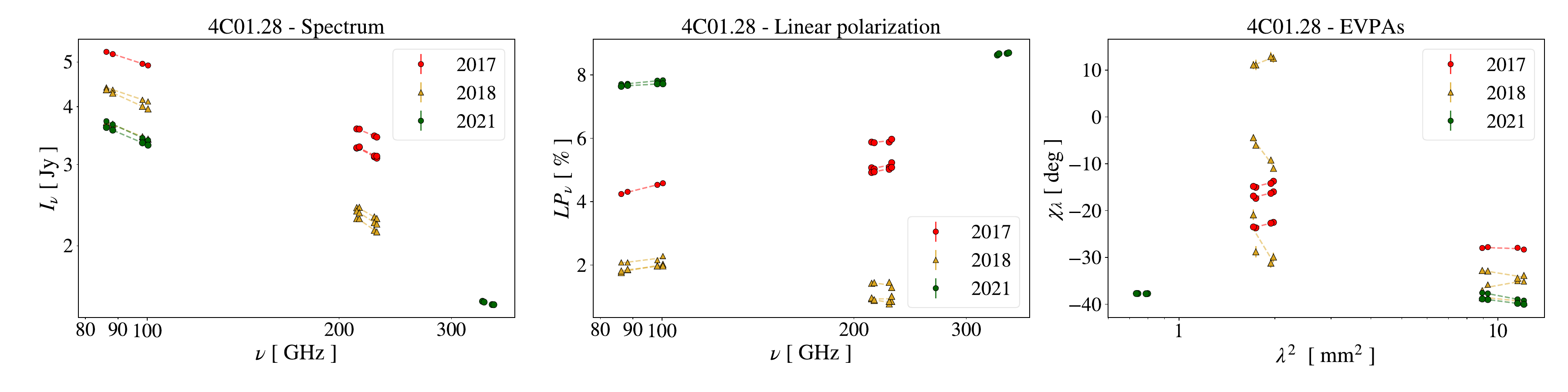}
    
    %\vspace{-6.5mm}
    \includegraphics[width=0.95\linewidth]{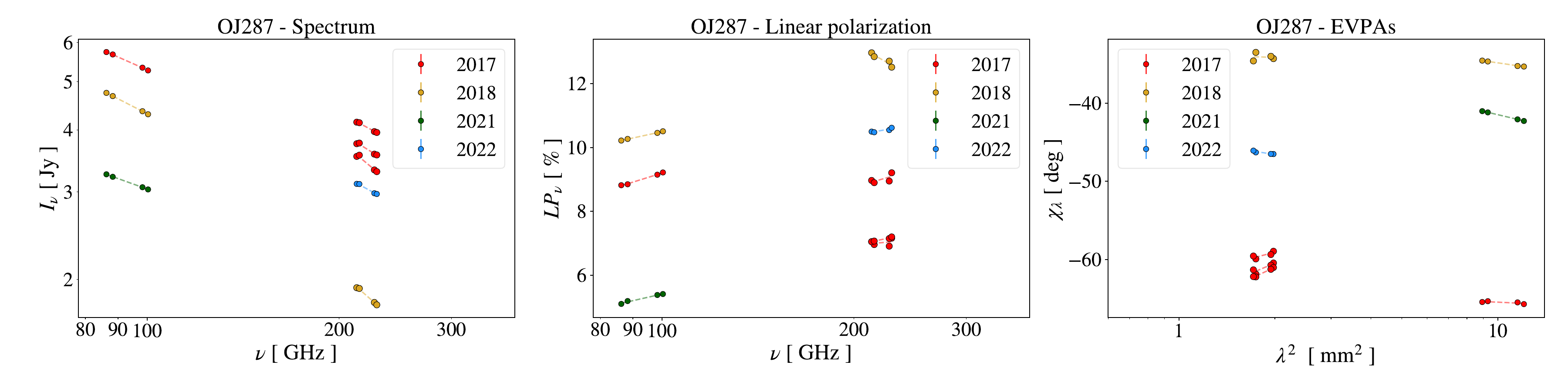}
    %\vspace{-2mm}
    \caption{Continuation of Fig. \ref{fig:specs} for sources: 3C273, 3C279, QSOB1921-293, NRAO530, 4C01.28 and OJ287.}
    \label{fig:specs2}
\end{figure*}
\begin{figure*}[h]
    \centering 
    \includegraphics[width=0.95\linewidth]{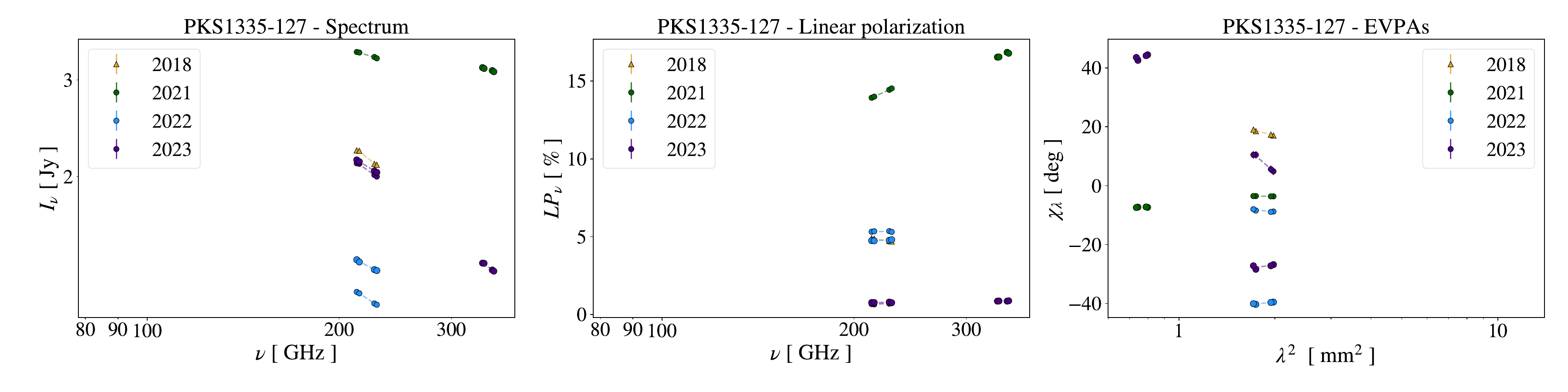}
    
    %\vspace{-6.5mm}
    \includegraphics[width=0.95\linewidth]{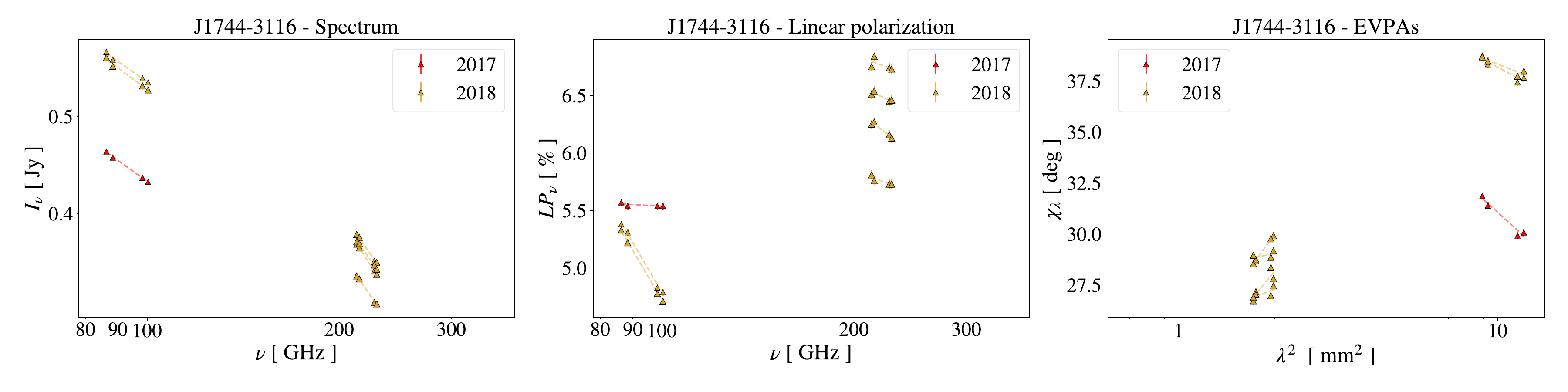}
    
    %\vspace{-2mm}
    \caption{Continuation of Fig. \ref{fig:specs} for sources: PKS1335-127 and J1744-3116.}\label{fig:specs3}
\end{figure*}

\section{Uncertainty budget}\label{ap:errors}
The uncertainties in the Stokes parameters include statistical errors from the visibility-domain fitting with UVMULTIFIT (Sect.~\ref{sec:data_polpar}) and systematic calibration contributions. For Stokes $Q$ and $U$; we adopt the systematic leakage of $0.03\%$ of Stokes $I$, while for Stokes $V$ we adopt $0.6\%$ of Stokes $I$. These values correspond to one third of the ALMA minimum detectable linear ($0.1\%$) and circular ($1.8\%$) polarization levels quoted at the $3\sigma$ level for compact sources (see ALMA Technical Handbook\footnote{\url{https://almascience.eso.org/documents-and-tools/cycle5/alma-technical-handbook}}).
The statistical and systematic contributions are added in quadrature.

The uncertainty in ALMA's absolute flux calibration, estimated at $5\%$ in Band 3 and $10\%$ in Bands 6 and 7, affects all Stokes parameters as a common multiplicative factor. We therefore do not include this contribution in the uncertainties of individual SPW measurements, since it is correlated across the simultaneously observed SPWs and Stokes parameters. Treating it as an independent error would artificially inflate the uncertainties in the derived spectral and polarization quantities, particularly the spectral index $\alpha$, LP, and EVPA. We include the absolute flux calibration uncertainty in the band-averaged Stokes $I$ measurements and image-based flux densities, for which the absolute flux scale is relevant.

The uncertainties in LP and EVPA are obtained by propagating those of the Stokes parameters, including the adopted systematic polarization calibration terms.
For the band-averaged LP and EVPA, we additionally include the standard deviation across the four SPWs to account for their observed frequency-dependent scatter. The uncertainties in fitted quantities such as $\alpha$, $D$, and RM are derived from the corresponding fits, using the individual SPW uncertainties. The statistical analysis and quality-control procedures are described in Appendix~\ref{ap:like}.

\section{Statistical analysis and quality control}
\label{ap:like}

We use likelihood analysis to evaluate the significance of relationships between measured quantities and to identify potentially unreliable determinations of $D$ and RM.

For a model $m^\theta$ with parameters $\theta$, we assume that the measurements $(x_i,y_i)$ satisfy
\begin{equation}
    y_i = m^\theta(x_i) + r_i,
\end{equation}
where the residuals $r_i$ are assumed to be independent Gaussian random variables with zero mean and standard deviations $\sigma_i$. The probability density for each measurement is then
\begin{equation}\label{eq:err_sample}
    f_Y(y_i|m^\theta) =
    \frac{1}{\sqrt{2\pi\sigma_i^2}}
    \exp\left[
    -\frac{(y_i-m^\theta(x_i))^2}{2\sigma_i^2}
    \right].
\end{equation}

The variance may include the measurement uncertainties in both variables, $e_{xi}$ and $e_{yi}$, as well as an intrinsic scatter $\sigma$:
\begin{equation}
    \sigma_i^2 =
    \sigma^2 + e_{yi}^2 +
    \left(m'_i e_{xi}\right)^2,
\end{equation}
where $m'_i$ is the derivative of the model evaluated at $x_i$.

Assuming independent measurements, the likelihood is the product of the individual probability densities:
\begin{equation}\label{eq:like}
    L(m^\theta|X,Y) =
    \prod_{i=1}^{N} f_Y(y_i|m^\theta).
\end{equation}

The corresponding log-likelihood is
\begin{equation}
    \mathcal{L} =
    \ln L =
    \sum_{i=1}^{N} \ln f_Y(y_i|m^\theta).
\end{equation}

To compare two models, $m_1^{\theta_1}$ and $m_2^{\theta_2}$, we use the likelihood-ratio test statistic
\begin{equation}
    {\rm TS} = 2(\mathcal{L}_1-\mathcal{L}_2).
\end{equation}
For nested models satisfying the regularity conditions of Wilks' theorem, TS approximately follows a $\chi^2$ distribution with $k=k_1-k_2$ degrees of freedom, where $k_1$ and $k_2$ are the numbers of free parameters in the two models. This allows us to estimate the statistical significance of the improvement in the fit.

In this work, we consider linear relations, $m(x)=ax+b$, including power-law relations fitted in logarithmic space. Their significance is evaluated relative to a constant null model, $m_{\rm null}(x)=c$, which represents the absence of a dependence between the two quantities.

\subsection{Outlier rejection}

For quantities derived from fits across the four SPWs, such as $D$ and RM, we apply an additional quality-control procedure to identify measurements potentially affected by anomalous SPWs.

For fits with a significance below $2\sigma$, we perform a leave-one-out test. Each SPW is excluded in turn, and a linear relation is fitted to the remaining three measurements.
We then compare the excluded measurement with the value predicted by this relation.

We reject the corresponding $D$ or RM determination if at least one SPW deviates from the prediction by more than four times its quoted measurement uncertainty. This procedure resulted in the rejection of 30 measurements among $D$ and RM, corresponding to approximately $7\%$ of the total.

\subsection{Reliability of in-band RM measurements}

The uncertainties used in the RM fits include both thermal noise and the systematic polarization calibration terms described in Appendix~\ref{ap:errors}, propagated to the EVPA measurements of each SPW.

After applying the quality-control procedure, most retained RM measurements are consistent with a linear EVPA--$\lambda^2$ relation within their quoted uncertainties, with a median reduced $\chi^2$ of 0.3.
The mean EVPA uncertainty is $\sigma_\chi\simeq0.25^\circ$ per SPW across all bands, comparable to values reported in previous ALMA polarization studies \citep[e.g.,][]{Nagai2016,Hull2020,Goddi_2021}.

The corresponding RM uncertainties are consistent with those expected from the wavelength-squared coverage of each ALMA band and the typical EVPA uncertainty, namely $\sim1.4\times10^3$, $1.6\times10^4$, and $7\times10^4$ rad\,m$^{-2}$ for B3, B6, and B7, respectively. The relatively low reduced $\chi^2$ values suggest that the adopted uncertainties may be conservative.
Nevertheless, the limited $\lambda^2$ coverage within each band makes the derived RM sensitive to small frequency-dependent calibration errors that may not be fully captured by the quoted uncertainties.

\section{Polarimetric tables}\label{ap:poltable}

The polarization properties of the compact cores are provided in two sets of tables. The first contains the full-Stokes parameters ($I$, $Q$, $U$, and $V$) and their derived polarization quantities for each individual SPW. The second reports band-averaged measurements of Stokes $I$, LP, and EVPA, together with derived spectral quantities such as $\alpha$,  $D$, and RM.

The quoted uncertainties follow the prescriptions described in Appendix~\ref{ap:errors}. Measurements of $D$ and RM are subject to the quality-control procedure described in Appendix~\ref{ap:like}.

As an example, we present the per-SPW and band-averaged measurements for M87 in Tables~\ref{poltab:pspw} and \ref{tab:M87resumed_avg}, respectively. Band-averaged measurements for the other main sources are provided in Tables~\ref{tab:SgrAresumed_avg}, \ref{tab:3C273resumed}, \ref{tab:3C279resumed}, \ref{tab:NRAO530resumed}, \ref{tab:QSOB1921resumed}, \ref{tab:J1744-3116resumed}, \ref{tab:4C01.28resumed}, \ref{tab:OJ287resumed} and \ref{tab:PKS1335-127resumed}.

The complete per-SPW and band-averaged tables for all sources are available through the downloads page of the VAPOLA archive\footnote{\url{https://vapola.ia2.inaf.it/download.html}} under \texttt{Tables/polarization\_tables/}.

\begin{table*}
\centering
\caption{Full-Stokes parameters and derived polarization quantities for the compact core of M87 in each SPW.}\label{poltab:pspw}
\begin{tabular}{cccccccc}
\hline\hline
Date & Frequency & $I$ & $Q$ & $U$ & $V$ & LP & EVPA\\
 & (GHz) & (Jy) & (mJy) & (mJy) & (mJy) & ($\%$) & (deg)\\
\hline
\multicolumn{8}{c}{...}\\
23/04/2021 & 86.3 & 1.8197 $\pm$ 0.0003 & 16.3 $\pm$ 0.6 & 5.9 $\pm$ 0.6 & 0 $\pm$ 11 & 0.95 $\pm$ 0.03 & 9.9 $\pm$ 0.9 \\
23/04/2021 & 88.3 & 1.8103 $\pm$ 0.0002 & 17.8 $\pm$ 0.5 & 5.3 $\pm$ 0.6 & 0 $\pm$ 11 & 1.03 $\pm$ 0.03 & 8.2 $\pm$ 0.8 \\
23/04/2021 & 98.3 & 1.7744 $\pm$ 0.0002 & 21.9 $\pm$ 0.5 & 11.6 $\pm$ 0.5 & -1 $\pm$ 11 & 1.40 $\pm$ 0.03 & 13.9 $\pm$ 0.6 \\
23/04/2021 & 100.3 & 1.7661 $\pm$ 0.0002 & 20.2 $\pm$ 0.5 & 12.8 $\pm$ 0.5 & -1 $\pm$ 11 & 1.35 $\pm$ 0.03 & 16.2 $\pm$ 0.6 \\
05/05/2023 & 86.3 & 2.138 $\pm$ 0.002 & 38.6 $\pm$ 0.7 & 14.6 $\pm$ 0.7 & -3 $\pm$ 13 & 1.93 $\pm$ 0.03 & 10.4 $\pm$ 0.5 \\
05/05/2023 & 88.3 & 2.128 $\pm$ 0.002 & 38.4 $\pm$ 0.7 & 11.1 $\pm$ 0.7 & -4 $\pm$ 13 & 1.88 $\pm$ 0.03 & 8.0 $\pm$ 0.5 \\
05/05/2023 & 98.3 & 2.089 $\pm$ 0.001 & 45.9 $\pm$ 0.7 & 4.1 $\pm$ 0.7 & -5 $\pm$ 12 & 2.20 $\pm$ 0.03 & 2.5 $\pm$ 0.4 \\
05/05/2023 & 100.3 & 2.080 $\pm$ 0.001 & 47.2 $\pm$ 0.7 & 4.3 $\pm$ 0.7 & -6 $\pm$ 12 & 2.28 $\pm$ 0.03 & 2.6 $\pm$ 0.4 \\
\multicolumn{8}{c}{...}\\
\hline
\multicolumn{8}{c}{...}\\
25/04/2018 & 213.1 & 1.1159 $\pm$ 0.0001 & 7.5 $\pm$ 0.3 & 20.3 $\pm$ 0.3 & -2 $\pm$ 7 & 1.94 $\pm$ 0.03 & 34.8 $\pm$ 0.5 \\
25/04/2018 & 215.1 & 1.1157 $\pm$ 0.0001 & 7.1 $\pm$ 0.3 & 21.1 $\pm$ 0.3 & -2 $\pm$ 7 & 2.00 $\pm$ 0.03 & 35.7 $\pm$ 0.4 \\
25/04/2018 & 227.1 & 1.0485 $\pm$ 0.0001 & 4.9 $\pm$ 0.3 & 24.4 $\pm$ 0.3 & -2 $\pm$ 6 & 2.37 $\pm$ 0.03 & 39.3 $\pm$ 0.4 \\
25/04/2018 & 229.1 & 1.0402 $\pm$ 0.0001 & 4.6 $\pm$ 0.3 & 24.9 $\pm$ 0.3 & -2 $\pm$ 6 & 2.43 $\pm$ 0.03 & 39.8 $\pm$ 0.3 \\
18/04/2021 & 213.1 & 1.4748 $\pm$ 0.0001 & 26.0 $\pm$ 0.4 & -0.1 $\pm$ 0.4 & -7 $\pm$ 9 & 1.76 $\pm$ 0.03 & -0.1 $\pm$ 0.5 \\
18/04/2021 & 215.1 & 1.4666 $\pm$ 0.0001 & 26.0 $\pm$ 0.4 & 0.1 $\pm$ 0.4 & -7 $\pm$ 9 & 1.77 $\pm$ 0.03 & 0.1 $\pm$ 0.5 \\
18/04/2021 & 227.1 & 1.4167 $\pm$ 0.0001 & 28.2 $\pm$ 0.4 & 1.9 $\pm$ 0.4 & -7 $\pm$ 8 & 2.00 $\pm$ 0.03 & 2.0 $\pm$ 0.4 \\
18/04/2021 & 229.1 & 1.4084 $\pm$ 0.0001 & 29.2 $\pm$ 0.4 & 2.4 $\pm$ 0.4 & -8 $\pm$ 8 & 2.08 $\pm$ 0.03 & 2.4 $\pm$ 0.4 \\
22/03/2022 & 213.1 & 1.2137 $\pm$ 0.0003 & 15.9 $\pm$ 0.4 & -0.7 $\pm$ 0.4 & -2 $\pm$ 7 & 1.31 $\pm$ 0.03 & -1.3 $\pm$ 0.7 \\
22/03/2022 & 215.1 & 1.2021 $\pm$ 0.0003 & 15.4 $\pm$ 0.4 & -0.9 $\pm$ 0.4 & -2 $\pm$ 7 & 1.29 $\pm$ 0.03 & -1.7 $\pm$ 0.7 \\
22/03/2022 & 227.1 & 1.1435 $\pm$ 0.0002 & 14.8 $\pm$ 0.3 & -1.8 $\pm$ 0.4 & -2 $\pm$ 7 & 1.30 $\pm$ 0.03 & -3.5 $\pm$ 0.7 \\
22/03/2022 & 229.1 & 1.1359 $\pm$ 0.0002 & 15.9 $\pm$ 0.3 & -2.5 $\pm$ 0.3 & -1 $\pm$ 7 & 1.42 $\pm$ 0.03 & -4.4 $\pm$ 0.6 \\
\hline\hline
\end{tabular}
\tablefoot{The quoted uncertainties include statistical errors for Stokes $I$, and both statistical and systematic polarization calibration errors for Stokes $Q$, $U$, and $V$. The absolute flux calibration uncertainty is not included for any Stokes parameter (see Appendix~\ref{ap:errors}). The full tables, including all sources and epochs, are available for download at \url{https://vapola.ia2.inaf.it/download.html} under \texttt{Tables/polarization\_tables/.}}
\end{table*}

\begin{table*}
\begin{center}
\caption{Frequency-averaged polarization properties of M87 in each observed ALMA band, all years. See Sect. \ref{sec:Res_M87}.}\label{tab:M87resumed_avg}
\begin{tabular}{cccccccc}
\hline\hline
Date & $I$ & $\alpha$ & LP & EVPA & $\chi_0$ & RM & Depol. \\
 & (Jy) &  & ($\%$) & (deg) & (deg) & ($10^5$ rad m$^{-2}$) & ($10^{-4}$ GHz$^{-1}$) \\
\hline
\multicolumn{8}{c}{B3: 93.3 GHz}\\
15/04/2018 & $1.400 \pm 0.070 $ & $-0.11 \pm 0.01$ & $1.0 \pm 0.2$ & $8 \pm 5$ & $-28.5 \pm 3.4 $ & $0.603 \pm 0.059 $ & $3.77 \pm 0.25 $ \\
23/04/2021 & $1.793 \pm 0.090 $ & $-0.19 \pm 0.01$ & $1.2 \pm 0.2$ & $12 \pm 3$ & $35.2 \pm 2.8 $ & $-0.386 \pm 0.049 $ & $3.12 \pm 0.25 $ \\
05/05/2023 & $2.11 \pm 0.10 $ & $-0.18 \pm 0.01$ & $2.1 \pm 0.2$ & $6 \pm 3$ & $-20.0 \pm 1.7 $ & $0.431 \pm 0.029 $ & $2.75 \pm 0.26 $ \\
09/05/2023 & $2.08 \pm 0.10 $ & $-0.28 \pm 0.01$ & $2.0 \pm 0.2$ & $3 \pm 1$ & $-5.9 \pm 1.6 $ & $0.142 \pm 0.028 $ & $3.87 \pm 0.25 $ \\
\hline
\multicolumn{8}{c}{B6: 221.1 GHz}\\
05/04/2017 & $1.32 \pm 0.13 $ & $-1.17 \pm 0.03$ & $2.44 \pm 0.03$ & $-7.8 \pm 0.9$ & $-20.7 \pm 2.8 $ & $1.22 \pm 0.26 $ & $0.19 \pm 0.21 $ \\
06/04/2017 & $1.28 \pm 0.13 $ & $-1.17 \pm 0.02$ & $2.15 \pm 0.06$ & $-8 \pm 1$ & $-24.6 \pm 3.1 $ & $1.61 \pm 0.30 $ & $0.67 \pm 0.21 $ \\
10/04/2017 & $1.35 \pm 0.14 $ & $-1.18 \pm 0.03$ & $2.74 \pm 0.04$ & $-0.1 \pm 0.5$ & $4.8 \pm 2.5 $ & $-0.46 \pm 0.23 $ & $-0.39 \pm 0.21 $ \\
11/04/2017 & $1.34 \pm 0.13 $ & $-1.14 \pm 0.01$ & $2.71 \pm 0.08$ & $-1.2 \pm 0.5$ & $3.9 \pm 2.5 $ & $-0.48 \pm 0.24 $ & $1.05 \pm 0.21 $ \\
21/04/2018 & $1.04 \pm 0.10 $ & $-1.07 \pm 0.07$ & $2.24 \pm 0.08$ & $28 \pm 3$ & $73.6 \pm 3.0 $ & $-4.33 \pm 0.29 $ & $0.98 \pm 0.21 $ \\
22/04/2018 & $1.16 \pm 0.12 $ & $-0.97 \pm 0.01$ & $2.1 \pm 0.2$ & $25 \pm 2$ & $60.4 \pm 3.2 $ & $-3.39 \pm 0.31 $ & $3.36 \pm 0.21 $ \\
25/04/2018 & $1.08 \pm 0.11 $ & $-1.03 \pm 0.07$ & $2.2 \pm 0.2$ & $37 \pm 2$ & $71.3 \pm 3.1 $ & $-3.21 \pm 0.30 $ & $3.11 \pm 0.21 $ \\
13/04/2021 & $1.34 \pm 0.13 $ & $-0.67 \pm 0.03$ & $1.73 \pm 0.10$ & $2 \pm 3$ & $40.6 \pm 3.9 $ & $-3.66 \pm 0.37 $ & $1.33 \pm 0.21 $ \\
17/04/2021 & $1.47 \pm 0.15 $ & $-0.61 \pm 0.01$ & $2.0 \pm 0.1$ & $3 \pm 2$ & $31.9 \pm 3.4 $ & $-2.72 \pm 0.32 $ & $1.81 \pm 0.21 $ \\
18/04/2021 & $1.44 \pm 0.14 $ & $-0.64 \pm 0.01$ & $1.9 \pm 0.1$ & $1 \pm 1$ & $18.1 \pm 3.5 $ & $-1.61 \pm 0.34 $ & $1.93 \pm 0.21 $ \\
22/03/2022 & $1.17 \pm 0.12 $ & $-0.92 \pm 0.01$ & $1.33 \pm 0.06$ & $-3 \pm 1$ & $-23.0 \pm 5.2 $ & $1.91 \pm 0.49 $ & $0.50 \pm 0.22 $ \\
27/03/2022 & $1.13 \pm 0.11 $ & $-0.86 \pm 0.02$ & $0.89 \pm 0.04$ & $4 \pm 4$ & $-56.3 \pm 7.6 $ & $5.67 \pm 0.71 $ & $-0.14 \pm 0.21 $ \\
\hline
\multicolumn{8}{c}{B7: 342.6 GHz}\\
19/04/2021 & $0.970 \pm 0.097 $ & $-1.24 \pm 0.02$ & $2.80 \pm 0.05$ & $12.6 \pm 0.4$ & $8.5 \pm 5.0 $ & $0.9 \pm 1.1 $ & $-0.45 \pm 0.28 $ \\
\hline\hline\\
\end{tabular}
\tablefoot{The quoted Stokes $I$ uncertainties include both statistical and absolute flux calibration errors. Uncertainties in the remaining quantities follow the prescriptions described in Appendix~\ref{ap:errors}.
    The full tables, including all sources and epochs, are available for download at \url{https://vapola.ia2.inaf.it/download.html} under \texttt{Tables/polarization\_tables/}.}
\end{center}
\end{table*}

\begin{table*}
\begin{center}
\caption{Frequency-averaged polarization properties of Sgr A* in each observed ALMA band, all years. See Sect. \ref{sec:Res_SgrA}}\label{tab:SgrAresumed_avg}
%\vspace{-0.4cm}
\begin{tabular}{cccccccc}
\hline\hline
Date & $I$ & $\alpha$ & LP & EVPA & $\chi_0$ & RM & Depol. \\
 & (Jy) &  & ($\%$) & (deg) & (deg) & ($10^5$ rad m$^{-2}$) & ($10^{-4}$ GHz$^{-1}$) \\
\hline
\multicolumn{8}{c}{B3: 93.3 GHz}\\
03/04/2017 & $2.43 \pm 0.12 $ & $-0.14 \pm 0.05$ & $0.8 \pm 0.4$ & $10 \pm 15$ & $127.3 \pm 4.8 $ & $-1.959 \pm 0.089 $ & $5.78 \pm 0.25 $ \\
14/04/2018 & $2.41 \pm 0.12 $ & $0.33 \pm 0.01$ & $0.17 \pm 0.08$ & $76 \pm 17$ & $-$ & $-$ & $-$ \\
17/04/2018 & $2.48 \pm 0.12 $ & $0.38 \pm 0.02$ & $0.05 \pm 0.04$ & $45 \pm 28$ & $-$ & $-$ & $-$ \\
\hline
\multicolumn{8}{c}{B6: 221.1 GHz}\\
06/04/2017 & $2.84 \pm 0.28 $ & $0.07 \pm 0.02$ & $6.8 \pm 0.3$ & $-66 \pm 3$ & $-14.09 \pm 0.99 $ & $-4.889 \pm 0.094 $ & $3.74 \pm 0.21 $ \\
07/04/2017 & $2.63 \pm 0.26 $ & $-0.03 \pm 0.01$ & $7.1 \pm 0.3$ & $-66 \pm 3$ & $-16.26 \pm 0.94 $ & $-4.662 \pm 0.090 $ & $3.67 \pm 0.21 $ \\
11/04/2017 & $2.63 \pm 0.26 $ & $-0.17 \pm 0.02$ & $7.0 \pm 0.2$ & $-50 \pm 3$ & $-9.10 \pm 0.96 $ & $-3.869 \pm 0.091 $ & $2.13 \pm 0.21 $ \\
21/04/2018 & $3.11 \pm 0.31 $ & $-0.00 \pm 0.06$ & $3.0 \pm 0.3$ & $-75 \pm 4$ & $-9.1 \pm 2.3 $ & $-6.24 \pm 0.22 $ & $3.94 \pm 0.21 $ \\
22/04/2018 & $2.87 \pm 0.29 $ & $-0.11 \pm 0.04$ & $3.8 \pm 0.5$ & $-71 \pm 1$ & $-55.0 \pm 1.8 $ & $-1.56 \pm 0.17 $ & $7.24 \pm 0.21 $ \\
24/04/2018 & $3.20 \pm 0.32 $ & $0.01 \pm 0.03$ & $4.4 \pm 0.1$ & $74 \pm 3$ & $121.0 \pm 1.5 $ & $-4.48 \pm 0.15 $ & $1.52 \pm 0.21 $ \\
25/04/2018 & $3.22 \pm 0.32 $ & $0.03 \pm 0.07$ & $3.4 \pm 0.2$ & $67 \pm 3$ & $118.5 \pm 2.0 $ & $-4.88 \pm 0.19 $ & $3.50 \pm 0.21 $ \\
14/04/2021 & $3.63 \pm 0.36 $ & $0.25 \pm 0.01$ & $4.7 \pm 0.2$ & $79 \pm 4$ & $140.0 \pm 1.4 $ & $-5.73 \pm 0.14 $ & $2.43 \pm 0.21 $ \\
15/04/2021 & $3.69 \pm 0.37 $ & $0.25 \pm 0.09$ & $5.7 \pm 0.6$ & $-31.7 \pm 0.3$ & $-28.7 \pm 1.2 $ & $-0.28 \pm 0.11 $ & $8.62 \pm 0.21 $ \\
17/04/2021 & $3.46 \pm 0.35 $ & $0.16 \pm 0.02$ & $5.6 \pm 0.$ & $-89 \pm 4$ & $147.6 \pm 1.2 $ & $-5.33 \pm 0.11 $ & $0.92 \pm 0.21 $ \\
20/03/2022 & $2.71 \pm 0.27 $ & $0.11 \pm 0.03$ & $2.8 \pm 0.3$ & $-68 \pm 6$ & $17.8 \pm 2.4 $ & $-8.13 \pm 0.23 $ & $4.17 \pm 0.21 $ \\
22/03/2022 & $3.07 \pm 0.31 $ & $0.0 \pm 0.1$ & $5.7 \pm 0.2$ & $-73 \pm 5$ & $0.2 \pm 1.2 $ & $-6.92 \pm 0.11 $ & $3.41 \pm 0.21 $ \\
23/03/2022& $2.84 \pm 0.28 $ & $-0.09 \pm 0.01$ & $5.9 \pm 0.2$ & $-59 \pm 3$ & $-12.8 \pm 1.1 $ & $-4.37 \pm 0.11 $ & $2.18 \pm 0.21 $ \\
16/04/2023 & $3.16 \pm 0.32 $ & $0.20 \pm 0.02$ & $1.7 \pm 0.2$ & $-75 \pm 5$ & $8.0 \pm 3.9 $ & $-7.84 \pm 0.38 $ & $2.87 \pm 0.21 $ \\
18/04/2023 & $3.36 \pm 0.34 $ & $0.27 \pm 0.01$ & $2.9 \pm 0.1$ & $-83 \pm 4$ & $-19.4 \pm 2.3 $ & $-6.00 \pm 0.22 $ & $1.50 \pm 0.21 $ \\
\hline
\multicolumn{8}{c}{B7: 342.6 GHz}\\
15/04/2023 & $3.35 \pm 0.34 $ & $-0.04 \pm 0.04$ & $8.78 \pm 0.08$ & $-43 \pm 1$ & $-2.7 \pm 1.4 $ & $-9.10 \pm 0.31 $ & $1.20 \pm 0.25 $ \\
\hline\hline\\
\end{tabular}
\end{center}
\end{table*}

\begin{table*}
\centering
\caption{Frequency-averaged polarization properties of 3C273 in each observed ALMA band, all years. See Sect. \ref{sec:Res_3C273}.}\label{tab:3C273resumed}
\begin{tabular}{cccccccc}
\hline\hline
Date & $I$ & $\alpha$ & LP & EVPA & $\chi_0$ & RM & Depol. \\
 & (Jy) &  & ($\%$) & (deg) & (deg) & ($10^5$ rad m$^{-2}$) & ($10^{-4}$ GHz$^{-1}$) \\
\hline
\multicolumn{8}{c}{B3: 93.3 GHz}\\
04/04/2017 & $9.36 \pm 0.47 $ & $-0.34 \pm 0.02$ & $3.9 \pm 0.2$ & $-47.3 \pm 0.8$ & $-52.79 \pm 0.86 $ & $0.091 \pm 0.014 $ & $-2.85 \pm 0.25 $ \\
14/04/2018 & $12.78 \pm 0.64 $ & $-0.45 \pm 0.01$ & $2.89 \pm 0.07$ & $-31.6 \pm 0.9$ & $-37.8 \pm 1.1 $ & $0.103 \pm 0.019 $ & $0.12 \pm 0.25 $ \\
15/04/2018 & $11.08 \pm 0.55 $ & $-0.44 \pm 0.03$ & $2.80 \pm 0.07$ & $-33.8 \pm 0.5$ & $-36.3 \pm 1.2 $ & $0.042 \pm 0.020 $ & $0.99 \pm 0.25 $ \\
23/04/2021 & $11.64 \pm 0.58 $ & $-0.80 \pm 0.01$ & $2.1 \pm 0.4$ & $-31 \pm 2$ & $-16.3 \pm 1.7 $ & $-0.250 \pm 0.027 $ & $-7.05 \pm 0.25 $ \\
05/05/2023 & $7.35 \pm 0.37 $ & $-0.81 \pm 0.01$ & $6.38 \pm 0.03$ & $-37.3 \pm 0.8$ & $-31.49 \pm 0.52 $ & $-0.097 \pm 0.009 $ & $0.07 \pm 0.25 $ \\
09/05/2023 & $7.12 \pm 0.36 $ & $-0.91 \pm 0.01$ & $6.57 \pm 0.04$ & $-38.0 \pm 0.8$ & $-31.93 \pm 0.51 $ & $-0.101 \pm 0.008 $ & $0.49 \pm 0.25 $ \\
\hline
\multicolumn{8}{c}{B6: 221.1 GHz}\\
06/04/2017 & $7.63 \pm 0.76 $ & $-0.68 \pm 0.01$ & $2.4 \pm 0.2$ & $-56 \pm 2$ & $-85.9 \pm 2.9 $ & $2.87 \pm 0.27 $ & $-2.54 \pm 0.21 $ \\
13/04/2021 & $5.76 \pm 0.58 $ & $-0.85 \pm 0.02$ & $2.30 \pm 0.08$ & $73 \pm 2$ & $99.3 \pm 2.9 $ & $-2.48 \pm 0.28 $ & $1.07 \pm 0.21 $ \\
15/04/2021 & $5.73 \pm 0.57 $ & $-0.85 \pm 0.01$ & $1.93 \pm 0.04$ & $79 \pm 1$ & $94.1 \pm 3.5 $ & $-1.42 \pm 0.33 $ & $0.46 \pm 0.21 $ \\
18/04/2021 & $5.95 \pm 0.59 $ & $-0.85 \pm 0.01$ & $1.53 \pm 0.05$ & $-85 \pm 2$ & $-57.9 \pm 4.4 $ & $-2.59 \pm 0.42 $ & $0.37 \pm 0.21 $ \\
22/03/2022 & $3.68 \pm 0.37 $ & $-0.91 \pm 0.02$ & $4.0 \pm 0.1$ & $-32.4 \pm 0.8$ & $-21.9 \pm 1.7 $ & $-0.99 \pm 0.16 $ & $-1.60 \pm 0.21 $ \\
27/03/2022 & $3.39 \pm 0.34 $ & $-0.91 \pm 0.01$ & $4.3 \pm 0.1$ & $-32.7 \pm 0.3$ & $-29.6 \pm 1.6 $ & $-0.29 \pm 0.15 $ & $-1.47 \pm 0.21 $ \\
\hline
\multicolumn{8}{c}{B7: 342.6 GHz}\\
19/04/2021 & $4.17 \pm 0.42 $ & $-0.78 \pm 0.01$ & $1.66 \pm 0.03$ & $-64 \pm 1$ & $-38.6 \pm 7.3 $ & $-5.8 \pm 1.7 $ & $0.13 \pm 0.25 $ \\
15/04/2023 & $1.81 \pm 0.18 $ & $-0.99 \pm 0.09$ & $3.72 \pm 0.04$ & $-34.0 \pm 0.3$ & $-27.1 \pm 3.4 $ & $-1.58 \pm 0.77 $ & $-0.42 \pm 0.26 $ \\
\hline\hline\\
\end{tabular}
\end{table*}

\begin{table*}
\begin{center}
\caption{Frequency-averaged polarization properties of 3C279 in each observed ALMA band, all years. See Sect. \ref{sec:Res_3c279}}\label{tab:3C279resumed}
\begin{tabular}{cccccccc}
\hline\hline
Date & I & $\alpha$ & LP & EVPA & $\chi_0$ & RM & Depol. \\
 & (Jy) &  & ($\%$) & (deg) & (deg) & ($10^5$ rad m$^{-2}$) & ($10^{-4}$ GHz$^{-1}$) \\
\hline
\multicolumn{8}{c}{B3: 93.3 GHz}\\
04/04/2017 & $12.71 \pm 0.64 $ & $-0.38 \pm 0.01$ & $12.2 \pm 0.2$ & $43.5 \pm 0.1$ & $44.50 \pm 0.27 $ & $-0.017 \pm 0.005 $ & $3.04 \pm 0.25 $ \\
13/04/2018 & $12.57 \pm 0.63 $ & $-0.41 \pm 0.01$ & $9.5 \pm 0.1$ & $42.9 \pm 0.1$ & $42.64 \pm 0.35 $ & $0.004 \pm 0.006 $ & $1.41 \pm 0.25 $ \\
14/04/2018 & $12.58 \pm 0.63 $ & $-0.41 \pm 0.01$ & $9.7 \pm 0.1$ & $42.3 \pm 0.1$ & $42.20 \pm 0.34 $ & $0.002 \pm 0.006 $ & $1.70 \pm 0.25 $ \\
15/04/2018 & $12.55 \pm 0.63 $ & $-0.38 \pm 0.03$ & $9.8 \pm 0.1$ & $41.2 \pm 0.1$ & $40.91 \pm 0.34 $ & $0.005 \pm 0.006$ & $1.15 \pm 0.25 $ \\
17/04/2018 & $13.39 \pm 0.67 $ & $-0.43 \pm 0.01$ & $10.4 \pm 0.1$ & $39.5 \pm 0.5$ & $35.97 \pm 0.32 $ & $0.058 \pm 0.005$ & $2.38 \pm 0.25 $ \\
23/04/2021 & $18.18 \pm 0.91 $ & $-0.63 \pm 0.01$ & $5.5 \pm 0.3$ & $-46.6 \pm 0.9$ & $-39.96 \pm 0.60 $ & $-0.11\pm 0.01 $ & $4.62 \pm 0.25 $ \\
05/05/2023 & $8.79 \pm 0.44 $ & $-0.81 \pm 0.01$ & $7.8 \pm 0.2$ & $18 \pm 1$ & $29.00 \pm 0.44 $ & $-0.189 \pm 0.007 $ & $2.87 \pm 0.25 $ \\
09/05/2023 & $9.67 \pm 0.48 $ & $-0.83 \pm 0.01$ & $8.1 \pm 0.1$ & $21.2 \pm 0.8$ & $27.13 \pm 0.41 $ & $-0.098 \pm 0.007 $ & $1.70 \pm 0.25 $ \\
\hline
\multicolumn{8}{c}{B6: 221.1 GHz}\\
05/04/2017 & $9.91 \pm 0.99 $ & $-0.61 \pm 0.01$ & $13.2 \pm 0.1$ & $45.3 \pm 0.2$ & $42.35 \pm 0.51 $ & $0.277 \pm 0.048 $ & $1.41 \pm 0.21 $ \\
06/04/2017 & $9.78 \pm 0.98 $ & $-0.58 \pm 0.03$ & $13.0 \pm 0.1$ & $43.4 \pm 0.1$ & $42.66 \pm 0.52 $ & $0.066 \pm 0.049 $ & $0.53 \pm 0.21 $ \\
10/04/2017 & $9.06 \pm 0.91 $ & $-0.61 \pm 0.01$ & $14.6 \pm 0.1$ & $40.2 \pm 0.1$ & $39.31 \pm 0.46 $ & $0.088 \pm 0.043 $ & $1.37 \pm 0.21 $ \\
11/04/2017 & $8.74 \pm 0.87 $ & $-0.61 \pm 0.01$ & $14.9 \pm 0.1$ & $40.1 \pm 0.1$ & $40.60 \pm 0.45 $ & $-0.048 \pm 0.043 $ & $1.46 \pm 0.21 $ \\
21/04/2018 & $9.62 \pm 0.96 $ & $-0.81 \pm 0.01$ & $12.0 \pm 0.1$ & $39.2 \pm 0.2$ & $42.78 \pm 0.56 $ & $-0.342 \pm 0.053 $ & $0.59 \pm 0.21 $ \\
22/04/2018 & $10.2 \pm 1.0 $ & $-0.71 \pm 0.01$ & $12.2 \pm 0.1$ & $39.1 \pm 0.3$ & $44.10 \pm 0.55 $ & $-0.478 \pm 0.052 $ & $1.00 \pm 0.21 $ \\
24/04/2018 & $10.1 \pm 1.0 $ & $-0.71 \pm 0.01$ & $12.2 \pm 0.1$ & $39.2 \pm 0.4$ & $44.91 \pm 0.55 $ & $-0.541 \pm 0.052 $ & $0.92 \pm 0.21 $ \\
25/04/2018 & $9.50 \pm 0.95 $ & $-0.73 \pm 0.06$ & $12.4 \pm 0.1$ & $39.5 \pm 0.2$ & $42.21 \pm 0.54 $ & $-0.252 \pm 0.051 $ & $0.56 \pm 0.21 $ \\
27/04/2018 & $11.8 \pm 1.2 $ & $-0.60 \pm 0.06$ & $11.1 \pm 0.1$ & $40.8 \pm 0.1$ & $41.48 \pm 0.61 $ & $-0.065 \pm 0.058 $ & $1.96 \pm 0.21 $ \\
13/04/2021 & $10.8 \pm 1.1 $ & $-0.51 \pm 0.01$ & $7.6 \pm 0.1$ & $-61.4 \pm 0.3$ & $-56.86 \pm 0.88 $ & $-0.432 \pm 0.084 $ & $1.65 \pm 0.21 $ \\
15/04/2021 & $10.1 \pm 1.0 $ & $-0.54 \pm 0.01$ & $7.9 \pm 0.1$ & $-52.4 \pm 0.2$ & $-50.20 \pm 0.86 $ & $-0.205 \pm 0.081 $ & $1.68 \pm 0.21 $ \\
17/04/2021 & $10.8 \pm 1.1 $ & $-0.51 \pm 0.01$ & $8.1 \pm 0.1$ & $-50.9 \pm 0.5$ & $-43.91 \pm 0.83 $ & $-0.660 \pm 0.079 $ & $1.64 \pm 0.21 $ \\
18/04/2021 & $10.6 \pm 1.1 $ & $-0.54 \pm 0.01$ & $8.5 \pm 0.1$ & $-49.9 \pm 0.2$ & $-46.85 \pm 0.79 $ & $-0.293 \pm 0.075 $ & $1.80 \pm 0.21 $ \\
19/03/2022 & $9.77 \pm 0.98 $ & $-0.65 \pm 0.01$ & $6.66 \pm 0.04$ & $-6.1 \pm 0.4$ & $-1.4 \pm 1.0 $ & $-0.440 \pm 0.096 $ & $0.17 \pm 0.21 $ \\
22/03/2022 & $9.77 \pm 0.98 $ & $-0.65 \pm 0.01$ & $6.41 \pm 0.04$ & $-6.3 \pm 0.9$ & $-$ & $-$ & $0.29 \pm 0.21 $ \\
23/03/2022 & $10.1 \pm 1.0 $ & $-0.62 \pm 0.06$ & $6.17 \pm 0.04$ & $-5.6 \pm 0.2$ & $-6.3 \pm 1.1 $ & $0.07 \pm 0.10 $ & $0.16 \pm 0.21 $ \\
26/03/2022 & $9.74 \pm 0.97 $ & $-0.63 \pm 0.01$ & $6.55 \pm 0.04$ & $-4.7 \pm 0.1$ & $-5.0 \pm 1.0 $ & $0.027 \pm 0.097 $ & $0.26 \pm 0.21 $ \\
27/03/2022 & $9.82 \pm 0.98 $ & $-0.59 \pm 0.01$ & $6.53 \pm 0.03$ & $-4.7 \pm 0.1$ & $-5.1 \pm 1.0 $ & $0.045 \pm 0.097 $ & $0.13 \pm 0.21 $ \\
16/04/2023 & $5.18 \pm 0.52 $ & $-0.92 \pm 0.05$ & $10.0 \pm 0.1$ & $16.5 \pm 0.2$ & $19.06 \pm 0.68 $ & $-0.238 \pm 0.064 $ & $1.45 \pm 0.21 $ \\
18/04/2023& $5.24 \pm 0.52 $ & $-0.86 \pm 0.01$ & $10.2 \pm 0.1$ & $16.4 \pm 0.3$ & $21.52 \pm 0.66 $ & $-0.488 \pm 0.063 $ & $1.16 \pm 0.21 $ \\
\hline
\multicolumn{8}{c}{B7: 342.6 GHz}\\
19/04/2021 & $8.73 \pm 0.87 $ & $-0.48 \pm 0.04$ & $10.6 \pm 0.1$ & $-46.7 \pm 0.2$ & $-41.5 \pm 1.1 $ & $-1.20 \pm 0.26 $ & $1.55 \pm 0.25 $ \\
15/04/2023 & $3.54 \pm 0.35 $ & $-0.88 \pm 0.05$ & $10.9 \pm 0.1$ & $17.9 \pm 0.2$ & $22.8 \pm 1.1 $ & $-1.12 \pm 0.25 $ & $0.61 \pm 0.25 $ \\
\hline\hline\\
\end{tabular}
\end{center}
\end{table*}

\begin{table*}
\begin{center}
\caption{Frequency-averaged polarization properties of NRAO530 in each observed ALMA band, all years. See Sect. \ref{sec:Res_NRAO530}}\label{tab:NRAO530resumed}
%\vspace{-0.4cm}
\begin{tabular}{cccccccc}
\hline\hline
Date & $I$ & $\alpha$ & LP & EVPA & $\chi_0$ & RM & Depol. \\
 & (Jy) &  & ($\%$) & (deg) & (deg) & ($10^5$ rad m$^{-2}$) & ($10^{-4}$ GHz$^{-1}$) \\
\hline
\multicolumn{8}{c}{B3: 93.3 GHz}\\
03/04/2017 & $2.80 \pm 0.14 $ & $-0.59 \pm 0.01$ & $0.95 \pm 0.06$ & $38 \pm 3$ & $58.1 \pm 3.5 $ & $-0.332 \pm 0.059 $ & $0.89 \pm 0.25 $ \\
14/04/2018 & $3.01 \pm 0.15 $ & $-0.71 \pm 0.01$ & $0.78 \pm 0.06$ & $-59 \pm 5$ & $-100.3 \pm 4.4 $ & $0.685 \pm 0.072 $ & $-0.75 \pm 0.25 $ \\
17/04/2018 & $3.08 \pm 0.15 $ & $-0.74 \pm 0.01$ & $0.7 \pm 0.1$ & $-66 \pm 2$ & $-79.0 \pm 4.9 $ & $0.211 \pm 0.077 $ & $-2.08 \pm 0.25 $ \\
\hline
\multicolumn{8}{c}{B6: 221.1 GHz}\\
06/04/2017 & $1.75 \pm 0.17 $ & $-0.80 \pm 0.01$ & $2.38 \pm 0.08$ & $51.4 \pm 0.4$ & $49.2 \pm 2.9 $ & $0.21 \pm 0.27 $ & $1.07 \pm 0.21 $ \\
07/04/2017 & $1.65 \pm 0.17 $ & $-0.82 \pm 0.01$ & $2.41 \pm 0.08$ & $49.4 \pm 0.8$ & $60.0 \pm 2.8 $ & $-1.00 \pm 0.27 $ & $1.08 \pm 0.21 $ \\
21/04/2018 & $1.79 \pm 0.18 $ & $-0.93 \pm 0.02$ & $0.25 \pm 0.03$ & $71 \pm 5$ & $-$ & $-$ & $-$ \\
22/04/2018 & $1.79 \pm 0.18 $ & $-0.90 \pm 0.03$ & $0.19 \pm 0.05$ & $73 \pm 6$ & $-$ & $-$ & $-$ \\
24/04/2018 & $1.61 \pm 0.16 $ & $-0.97 \pm 0.03$ & $0.30 \pm 0.05$ & $70 \pm 5$ & $-$ & $-$ & $-$ \\
25/04/2018 & $1.64 \pm 0.16 $ & $-0.91 \pm 0.04$ & $0.35 \pm 0.04$ & $65 \pm 3$ & $-$ & $-$ & $-$ \\
17/04/2021 & $1.45 \pm 0.14 $ & $-0.82 \pm 0.04$ & $7.7 \pm 0.1$ & $29.4 \pm 0.4$ & $35.1 \pm 1.0 $ & $-0.541 \pm 0.096 $ & $1.03 \pm 0.25 $ \\
20/03/2022 & $0.857 \pm 0.086 $ & $-0.98 \pm 0.09$ & $3.2 \pm 0.2$ & $7 \pm 1$ & $25.3 \pm 4.4 $ & $-1.74 \pm 0.42 $ & $-1.64 \pm 0.45 $ \\
\hline\hline\\
\end{tabular}
\end{center}
\end{table*}

\begin{table*}
\begin{center}
\caption{Frequency-averaged polarization properties of QSOB1921-293 in each observed ALMA band, all years. See Sect. \ref{sec:Res_QSOB1921-293}}\label{tab:QSOB1921resumed}
%\vspace{-0.4cm}
\begin{tabular}{cccccccc}
\hline\hline
Date & $I$ & $\alpha$ & LP & EVPA & $\chi_0$ & RM & Depol. \\
 & (Jy) &  & ($\%$) & (deg) & (deg) & ($10^5$ rad m$^{-2}$) & ($10^{-4}$ GHz$^{-1}$) \\
\hline
\multicolumn{8}{c}{B3: 93.3 GHz}\\
03/04/2017 & $5.35 \pm 0.27 $ & $-0.48 \pm 0.02$ & $4.8 \pm 0.1$ & $-45.2 \pm 0.3$ & $-45.60 \pm 0.71 $ & $0.006 \pm 0.012 $ & $1.58 \pm 0.25 $ \\
14/04/2018 & $4.41 \pm 0.22 $ & $-0.71 \pm 0.01$ & $4.87 \pm 0.08$ & $-38.7 \pm 0.4$ & $-41.10 \pm 0.68 $ & $0.040 \pm 0.011 $ & $1.20 \pm 0.25 $ \\
17/04/2018 & $4.42 \pm 0.22 $ & $-0.72 \pm 0.01$ & $4.9 \pm 0.1$ & $-35.6 \pm 0.4$ & $-38.69 \pm 0.67 $ & $0.051 \pm 0.011 $ & $1.74 \pm 0.25 $ \\
\hline
\multicolumn{8}{c}{B6: 221.1 GHz}\\
06/04/2017 & $3.66 \pm 0.37 $ & $-0.72 \pm 0.01$ & $6.07 \pm 0.04$ & $-49.1 \pm 0.3$ & $-53.3 \pm 1.1 $ & $0.39 \pm 0.10 $ & $0.03 \pm 0.21 $ \\
07/04/2017 & $3.64 \pm 0.36 $ & $-0.74 \pm 0.02$ & $5.96 \pm 0.04$ & $-49.4 \pm 0.3$ & $-52.7 \pm 1.1 $ & $0.32 \pm 0.11 $ & $0.32 \pm 0.21 $ \\
11/04/2017 & $3.76 \pm 0.38 $ & $-0.68 \pm 0.01$ & $4.84 \pm 0.04$ & $-52.2 \pm 0.2$ & $-50.6 \pm 1.4 $ & $-0.15 \pm 0.13 $ & $0.08 \pm 0.21 $ \\
21/04/2018 & $2.80 \pm 0.28 $ & $-0.96 \pm 0.02$ & $4.67 \pm 0.05$ & $-35.6 \pm 0.2$ & $-34.2 \pm 1.4 $ & $-0.13 \pm 0.14 $ & $-0.11 \pm 0.21 $ \\
22/04/2018 & $2.46 \pm 0.25 $ & $-1.02 \pm 0.04$ & $4.5 \pm 0.1$ & $-35.8 \pm 0.5$ & $-43.5 \pm 1.5 $ & $0.73 \pm 0.14 $ & $-1.43 \pm 0.21 $ \\
24/04/2018 & $2.50 \pm 0.25 $ & $-0.98 \pm 0.02$ & $4.52 \pm 0.03$ & $-38.0 \pm 0.3$ & $-33.9 \pm 1.5 $ & $-0.39 \pm 0.14 $ & $-0.04 \pm 0.21 $ \\
25/04/2018 & $2.60 \pm 0.26 $ & $-0.95 \pm 0.01$ & $4.65 \pm 0.04$ & $-38.1 \pm 0.2$ & $-37.0 \pm 1.5 $ & $-0.11 \pm 0.14 $ & $-0.38 \pm 0.21 $ \\
14/04/2021 & $3.22 \pm 0.32 $ & $-0.57 \pm 0.01$ & $1.30 \pm 0.04$ & $-82.7 \pm 0.8$ & $-90.2 \pm 5.2 $ & $0.71 \pm 0.49 $ & $0.46 \pm 0.21 $ \\
15/04/2021 & $3.59 \pm 0.36 $ & $-0.48 \pm 0.04$ & $1.15 \pm 0.07$ & $-82 \pm 1$ & $-92.2 \pm 5.8 $ & $0.97 \pm 0.55 $ & $0.69 \pm 0.21 $ \\
17/04/2021 & $3.06 \pm 0.31 $ & $-0.67 \pm 0.01$ & $1.00 \pm 0.03$ & $88 \pm 1$ & $85.8 \pm 6.7 $ & $0.22 \pm 0.64 $ & $0.12 \pm 0.21 $ \\
20/03/2022 & $6.33 \pm 0.63 $ & $-0.54 \pm 0.01$ & $0.37 \pm 0.03$ & $-71 \pm 3$ & $-$ & $-$ & $-$ \\
22/03/2022 & $6.21 \pm 0.62 $ & $-0.55 \pm 0.06$ & $0.68 \pm 0.07$ & $-65 \pm 5$ & $-$ & $-$ & $-$ \\
23/03/2022 & $6.03 \pm 0.60 $ & $-0.59 \pm 0.01$ & $0.74 \pm 0.04$ & $-62 \pm 1$ & $-64.2 \pm 9.1 $ & $0.18 \pm 0.85 $ & $-0.29 \pm 0.21 $ \\
16/04/2023 & $5.23 \pm 0.52 $ & $-0.76 \pm 0.01$ & $4.84 \pm 0.08$ & $-32.7 \pm 0.7$ & $-22.9 \pm 1.4 $ & $-0.92 \pm 0.13 $ & $0.62 \pm 0.21 $ \\
18/04/2023 & $5.00 \pm 0.50 $ & $-0.79 \pm 0.02$ & $4.16 \pm 0.03$ & $-34.7 \pm 0.3$ & $-31.4 \pm 1.6 $ & $-0.32 \pm 0.15 $ & $0.06 \pm 0.21 $ \\
\hline
\multicolumn{8}{c}{B7: 342.6 GHz}\\
15/04/2023 & $3.10 \pm 0.31 $ & $-0.93 \pm 0.09$ & $4.49 \pm 0.04$ & $-30.5 \pm 0.3$ & $-24.6 \pm 2.7 $ & $-1.35 \pm 0.61 $ & $0.45 \pm 0.25 $ \\
\hline\hline\\
\end{tabular}
\end{center}
\end{table*}

\begin{table*}
\begin{center}
\caption{Frequency-averaged polarization properties of J1744-3116 in each observed ALMA band, all years. See Sect. \ref{sec:res_J1744}}\label{tab:J1744-3116resumed}
%\vspace{-0.4cm}
\begin{tabular}{cccccccc}
\hline\hline
Date & $I$ & $\alpha$ & LP & EVPA & $\chi_0$ & RM & Depol. \\
 & (Jy) &  & ($\%$) & (deg) & (deg) & ($10^5$ rad m$^{-2}$) & ($10^{-4}$ GHz$^{-1}$) \\
\hline
\multicolumn{8}{c}{B3: 93.3 GHz}\\
03/04/2017 & $0.445 \pm 0.022 $ & $-0.45 \pm 0.01$ & $5.55 \pm 0.04$ & $30.8 \pm 0.9$ & $37.05 \pm 0.66 $ & $-0.104 \pm 0.011 $ & $-0.13 \pm 0.27 $ \\
14/04/2018 & $0.559 \pm 0.028 $ & $-0.44 \pm 0.03$ & $5.1 \pm 0.3$ & $38.0 \pm 0.5$ & $41.71 \pm 0.72 $ & $-0.061 \pm 0.012 $ & $-4.44 \pm 0.27 $ \\
17/04/2018 & $0.550 \pm 0.028 $ & $-0.47 \pm 0.04$ & $5.0 \pm 0.3$ & $38.2 \pm 0.4$ & $40.86 \pm 0.73 $ & $-0.044 \pm 0.012 $ & $-4.44 \pm 0.27 $ \\
\hline
\multicolumn{8}{c}{B6: 221.1 GHz}\\
21/04/2018 & $0.369 \pm 0.037 $ & $-0.93 \pm 0.05$ & $6.77 \pm 0.06$ & $29.2 \pm 0.6$ & $19.6 \pm 1.3 $ & $0.91 \pm 0.12 $ & $-0.36 \pm 0.28 $ \\
22/04/2018 & $0.360 \pm 0.036 $ & $-0.96 \pm 0.01$ & $6.49 \pm 0.05$ & $28.9 \pm 0.2$ & $27.5 \pm 1.3 $ & $0.13 \pm 0.13 $ & $-0.46 \pm 0.28 $ \\
24/04/2018 & $0.336 \pm 0.034 $ & $-0.92 \pm 0.04$ & $6.20 \pm 0.07$ & $27.1 \pm 0.3$ & $24.4 \pm 1.4 $ & $0.26 \pm 0.13 $ & $-0.79 \pm 0.28 $ \\
\hline\hline\\
\end{tabular}
\end{center}
\end{table*}

\begin{table*}
\begin{center}
\caption{Frequency-averaged polarization properties of 4C01.28 in each observed ALMA band, all years. See Sect. \ref{sec:Res_4C}}\label{tab:4C01.28resumed}
%\vspace{-0.4cm}
\begin{tabular}{cccccccc}
\hline\hline
Date & $I$ & $\alpha$ & LP & EVPA & $\chi_0$ & RM & Depol. \\
 & (Jy) &  & ($\%$) & (deg) & (deg) & ($10^5$ rad m$^{-2}$) & ($10^{-4}$ GHz$^{-1}$) \\
\hline
\multicolumn{8}{c}{B3: 93.3 GHz}\\
03/04/2017 & $5.08 \pm 0.25 $ & $-0.45 \pm 0.01$ & $4.4 \pm 0.1$ & $-28.0 \pm 0.3$ & $-27.03 \pm 0.75 $ & $-0.016 \pm 0.013 $ & $2.31 \pm 0.25 $ \\
14/04/2018 & $4.25 \pm 0.21 $ & $-0.47 \pm 0.01$ & $2.15 \pm 0.09$ & $-35.8 \pm 0.9$ & $-41.1 \pm 1.5 $ & $0.089 \pm 0.026 $ & $1.19 \pm 0.25 $ \\
15/04/2018 & $3.55 \pm 0.18 $ & $-0.56 \pm 0.02$ & $1.9 \pm 0.1$ & $-33.5 \pm 0.8$ & $-28.7 \pm 1.8 $ & $-0.080 \pm 0.029 $ & $1.64 \pm 0.25 $ \\
16/04/2018 & $4.15 \pm 0.21 $ & $-0.63 \pm 0.01$ & $1.90 \pm 0.07$ & $-39.0 \pm 0.6$ & $-35.9 \pm 1.7 $ & $-0.052 \pm 0.029 $ & $1.09 \pm 0.25 $ \\
23/04/2021 & $3.55 \pm 0.18 $ & $-0.63 \pm 0.01$ & $7.77 \pm 0.06$ & $-38.4 \pm 0.7$ & $-32.74 \pm 0.43 $ & $-0.094 \pm 0.007 $ & $0.93 \pm 0.25 $ \\
26/04/2021 & $3.45 \pm 0.17 $ & $-0.59 \pm 0.01$ & $7.69 \pm 0.04$ & $-39.4 \pm 0.5$ & $-35.61 \pm 0.43 $ & $-0.064 \pm 0.007 $ & $0.54 \pm 0.25 $ \\
\hline
\multicolumn{8}{c}{B6: 221.1 GHz}\\
05/04/2017 & $3.18 \pm 0.32 $ & $-0.75 \pm 0.09$ & $5.90 \pm 0.05$ & $-23.1 \pm 0.5$ & $-30.7 \pm 1.1 $ & $0.72 \pm 0.11 $ & $0.44 \pm 0.21 $ \\
10/04/2017 & $3.20 \pm 0.32 $ & $-0.67 \pm 0.09$ & $5.10 \pm 0.08$ & $-16.6 \pm 0.6$ & $-24.1 \pm 1.3 $ & $0.71 \pm 0.12 $ & $-$ \\
11/04/2017 & $3.51 \pm 0.35 $ & $-0.58 \pm 0.04$ & $4.99 \pm 0.07$ & $-14.4 \pm 0.5$ & $-22.1 \pm 1.4 $ & $0.72 \pm 0.13 $ & $0.91 \pm 0.21 $ \\
22/04/2018 & $2.22 \pm 0.22 $ & $-0.97 \pm 0.07$ & $0.86 \pm 0.07$ & $-28 \pm 4$ & $-$ & $-$ & $-0.66 \pm 0.22 $ \\
25/04/2018 & $2.36 \pm 0.24 $ & $-0.79 \pm 0.07$ & $0.93 \pm 0.07$ & $12 \pm 1$ & $-0.0 \pm 7.5 $ & $1.13 \pm 0.70 $ & $-$ \\
27/04/2018 & $2.30 \pm 0.23 $ & $-0.90 \pm 0.01$ & $1.40 \pm 0.07$ & $-8 \pm 3$ & $31.4 \pm 4.9 $ & $-3.70 \pm 0.46 $ & $-$ \\
\hline
\multicolumn{8}{c}{B7: 342.6 GHz}\\
19/04/2021 & $1.50 \pm 0.15 $ & $-0.42 \pm 0.03$ & $8.67 \pm 0.04$ & $-37.7 \pm 0.1$ & $-37.4 \pm 1.5 $ & $-0.08 \pm 0.35 $ & $0.36 \pm 0.27 $ \\
\hline\hline\\
\end{tabular}
\end{center}
\end{table*}

\begin{table*}
\begin{center}
\caption{Frequency-averaged polarization properties of OJ287 in each observed ALMA band, all years. See Sect. \ref{sec:res_OJ287}}\label{tab:OJ287resumed}
%\vspace{-0.4cm}
\begin{tabular}{cccccccc}
\hline\hline
Date & $I$ & $\alpha$ & LP & EVPA & $\chi_0$ & RM & Depol. \\
 & (Jy) &  & ($\%$) & (deg) & (deg) & ($10^5$ rad m$^{-2}$) & ($10^{-4}$ GHz$^{-1}$) \\
\hline
\multicolumn{8}{c}{B3: 93.3 GHz}\\
03/04/2017 & $5.51 \pm 0.28 $ & $-0.57 \pm 0.01$ & $9.0 \pm 0.2$ & $-65.5 \pm 0.2$ & $-64.64 \pm 0.37 $ & $-0.014 \pm 0.006 $ & $2.86 \pm 0.25 $ \\
16/04/2018 & $4.52 \pm 0.23 $ & $-0.66 \pm 0.01$ & $10.4 \pm 0.1$ & $-35.0 \pm 0.3$ & $-32.37 \pm 0.32 $ & $-0.043 \pm 0.005 $ & $1.99 \pm 0.25 $ \\
26/04/2021 & $3.14 \pm 0.16 $ & $-0.46 \pm 0.01$ & $5.3 \pm 0.1$ & $-41.6 \pm 0.6$ & $-37.45 \pm 0.63 $ & $-0.070 \pm 0.010 $ & $2.15 \pm 0.25 $ \\
\hline
\multicolumn{8}{c}{B6: 221.1 GHz}\\
05/04/2017 & $3.43 \pm 0.34 $ & $-1.1 \pm 0.1$ & $9.0 \pm 0.1$ & $-61.1 \pm 0.6$ & $-68.67 \pm 0.74 $ & $0.718 \pm 0.071 $ & $-$ \\
10/04/2017 & $3.66 \pm 0.37 $ & $-0.80 \pm 0.09$ & $7.0 \pm 0.1$ & $-61.7 \pm 0.6$ & $-69.98 \pm 0.96 $ & $0.787 \pm 0.090 $ & $-$ \\
11/04/2017 & $4.05 \pm 0.41 $ & $-0.69 \pm 0.04$ & $7.1 \pm 0.1$ & $-59.4 \pm 0.4$ & $-64.00 \pm 0.94 $ & $0.431 \pm 0.089 $ & $0.82 \pm 0.21 $ \\
27/04/2018 & $1.85 \pm 0.18 $ & $-1.13 \pm 0.06$ & $12.8 \pm 0.2$ & $-34.1 \pm 0.4$ & $-$ & $-$ & $-2.21 \pm 0.22 $ \\
26/03/2022 & $3.04 \pm 0.30 $ & $-0.70 \pm 0.05$ & $10.5 \pm 0.1$ & $-46.3 \pm 0.2$ & $-43.75 \pm 0.64 $ & $-0.246 \pm 0.060 $ & $0.72 \pm 0.21 $ \\
\hline\hline\\
\end{tabular}
\end{center}
\end{table*}

\begin{table*}
\begin{center}
\caption{Frequency-averaged polarization properties of PKS1335-127 in each observed ALMA band, all years. See Sect. \ref{sec:res_PKS1335}}\label{tab:PKS1335-127resumed}
%\vspace{-0.4cm}
\begin{tabular}{cccccccc}
\hline\hline
Date & $I$ & $\alpha$ & LP & EVPA & $\chi_0$ & RM & Depol. \\
 & (Jy) &  & ($\%$) & (deg) & (deg) & ($10^5$ rad m$^{-2}$) & ($10^{-4}$ GHz$^{-1}$) \\
\hline
\multicolumn{8}{c}{B6: 221.1 GHz}\\
22/04/2018 & $2.17 \pm 0.22 $ & $-0.90 \pm 0.06$ & $4.80 \pm 0.10$ & $18.0 \pm 0.8$ & $30.7 \pm 1.4 $ & $-1.20 \pm 0.14 $ & $-1.27 \pm 0.21 $ \\
15/04/2021 & $3.33 \pm 0.33 $ & $-0.36 \pm 0.02$ & $14.2 \pm 0.3$ & $-3.56 \pm 0.08$ & $-2.80 \pm 0.47 $ & $-0.072 \pm 0.045 $ & $3.69 \pm 0.21 $ \\
19/03/2022 & $1.20 \pm 0.12 $ & $-0.76 \pm 0.01$ & $5.33 \pm 0.04$ & $-8.5 \pm 0.4$ & $-3.5 \pm 1.3 $ & $-0.48 \pm 0.12 $ & $0.01 \pm 0.21 $ \\
26/03/2022 & $1.38 \pm 0.14 $ & $-0.62 \pm 0.03$ & $4.77 \pm 0.04$ & $-39.9 \pm 0.4$ & $-44.5 \pm 1.4 $ & $0.43 \pm 0.13 $ & $0.32 \pm 0.21 $ \\
16/04/2023 & $2.06 \pm 0.21 $ & $-0.80 \pm 0.04$ & $0.69 \pm 0.04$ & $8 \pm 3$ & $48.5 \pm 9.9 $ & $-3.85 \pm 0.94 $ & $0.40 \pm 0.21 $ \\
18/04/2023 & $2.09 \pm 0.21 $ & $-0.72 \pm 0.01$ & $0.77 \pm 0.03$ & $-27 \pm 1$ & $-33.6 \pm 8.8 $ & $0.59 \pm 0.83 $ & $0.07 \pm 0.21 $ \\
\hline
\multicolumn{8}{c}{B7: 342.6 GHz}\\
19/04/2021 & $3.13 \pm 0.31 $ & $-0.38 \pm 0.05$ & $16.7 \pm 0.1$ & $-7.31 \pm 0.09$ & $-8.15 \pm 0.77 $ & $0.19 \pm 0.17 $ & $2.21 \pm 0.26 $ \\
15/04/2023 & $1.37 \pm 0.14 $ & $-0.88 \pm 0.06$ & $0.86 \pm 0.03$ & $44 \pm 1$ & $27 \pm 15 $ & $3.9 \pm 3.3 $ & $0.06 \pm 0.26 $ \\
\hline\hline\\
\end{tabular}
\end{center}
\end{table*}

\section{History plots}\label{ap:histplots}
\begin{figure*}[h]
    \centering
    \includegraphics[width=0.97\linewidth]{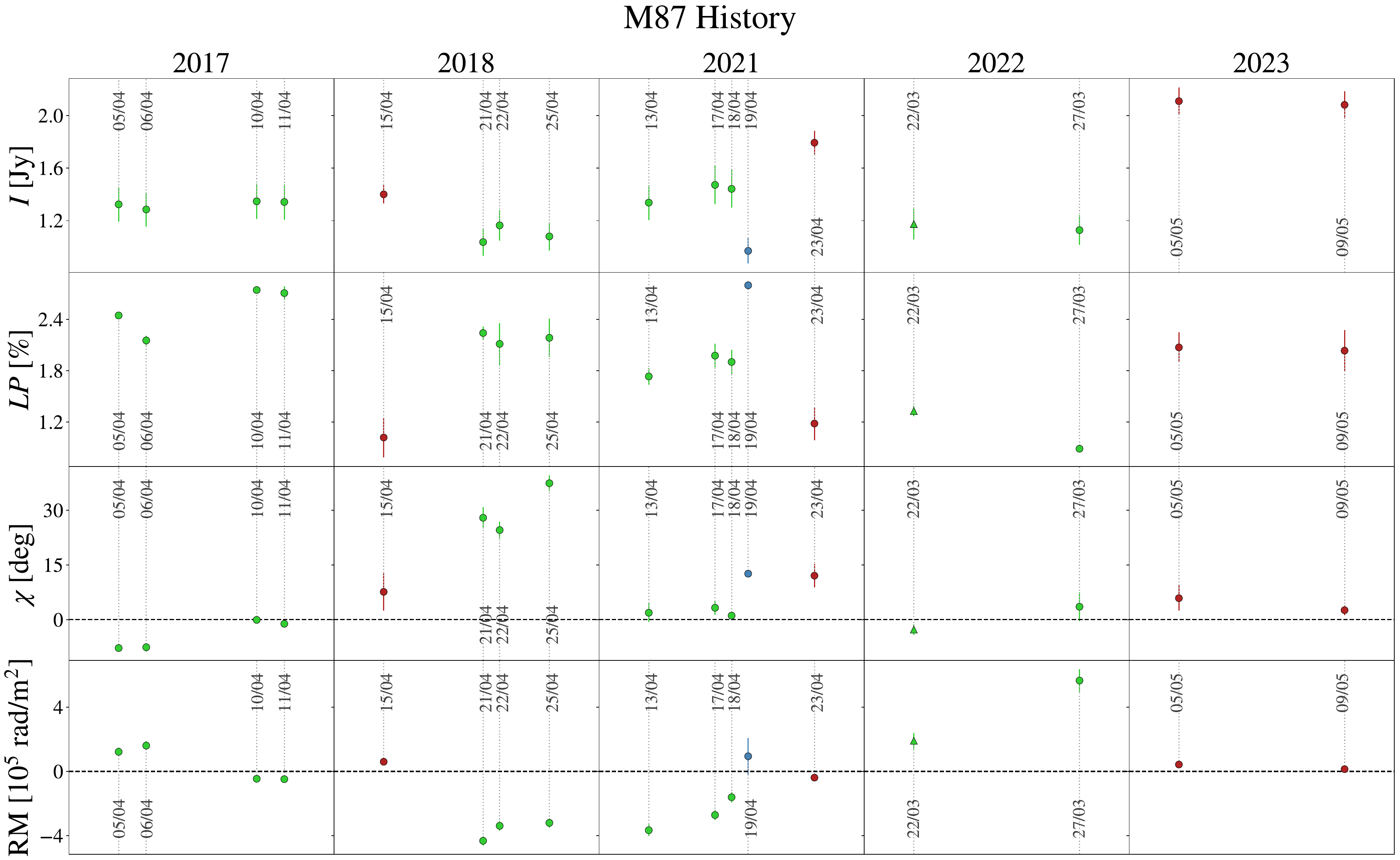}
    \caption{History of ALMA measurements of M87 during the VLBI campaigns (Sect. \ref{sec:Res_M87}).
    Each row displays the time evolution of some parameter, from top to bottom: Stokes $I$, linear polarization fraction (LP), EVPAs ($\chi$) and Rotation Measure (RM).
    Each column refers to the observing campaign of each year.
    The data points are labeled with the day/month of the observation, and their color and shapes represent the band (red: B3, green: B6, blue: B7) and observation mode (circle: VLBI scans, triangle: non-VLBI scans). Plots are available at \protect\url{https://vapola.ia2.inaf.it/download.html} under \texttt{Tables/history/}.}
    \label{fig:hist_M87}
    %\vspace{0.2cm}
    \includegraphics[width=0.97\linewidth]{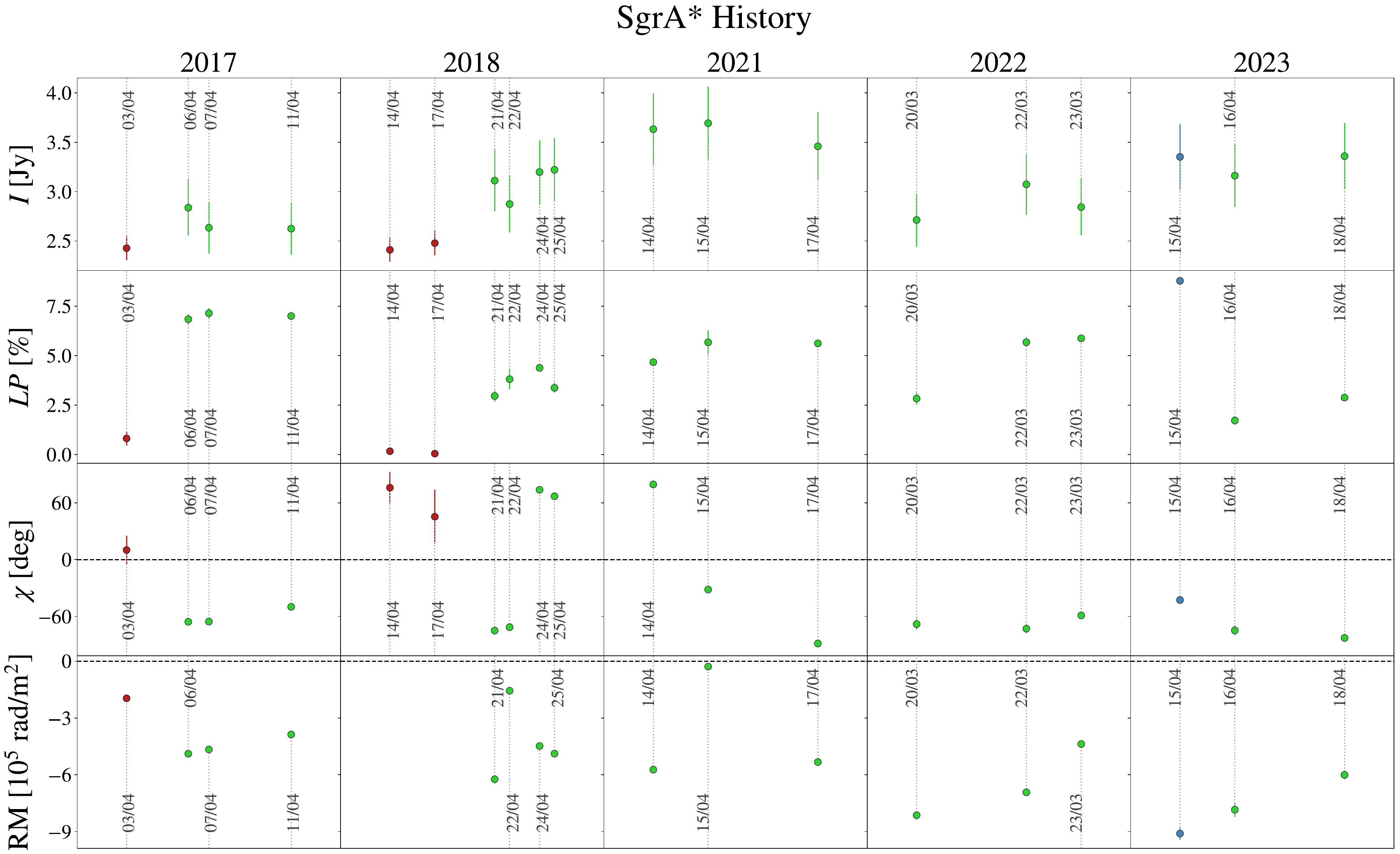}
    \caption{History of ALMA measurements of Sgr A* during the VLBI campaigns (Sect. \ref{sec:Res_SgrA}). Color scheme is the same as in Fig. \ref{fig:hist_M87}.}
    \label{fig:hist_Sgra}
\end{figure*}

\begin{figure*}[h]
    \centering
    \includegraphics[width=\linewidth]{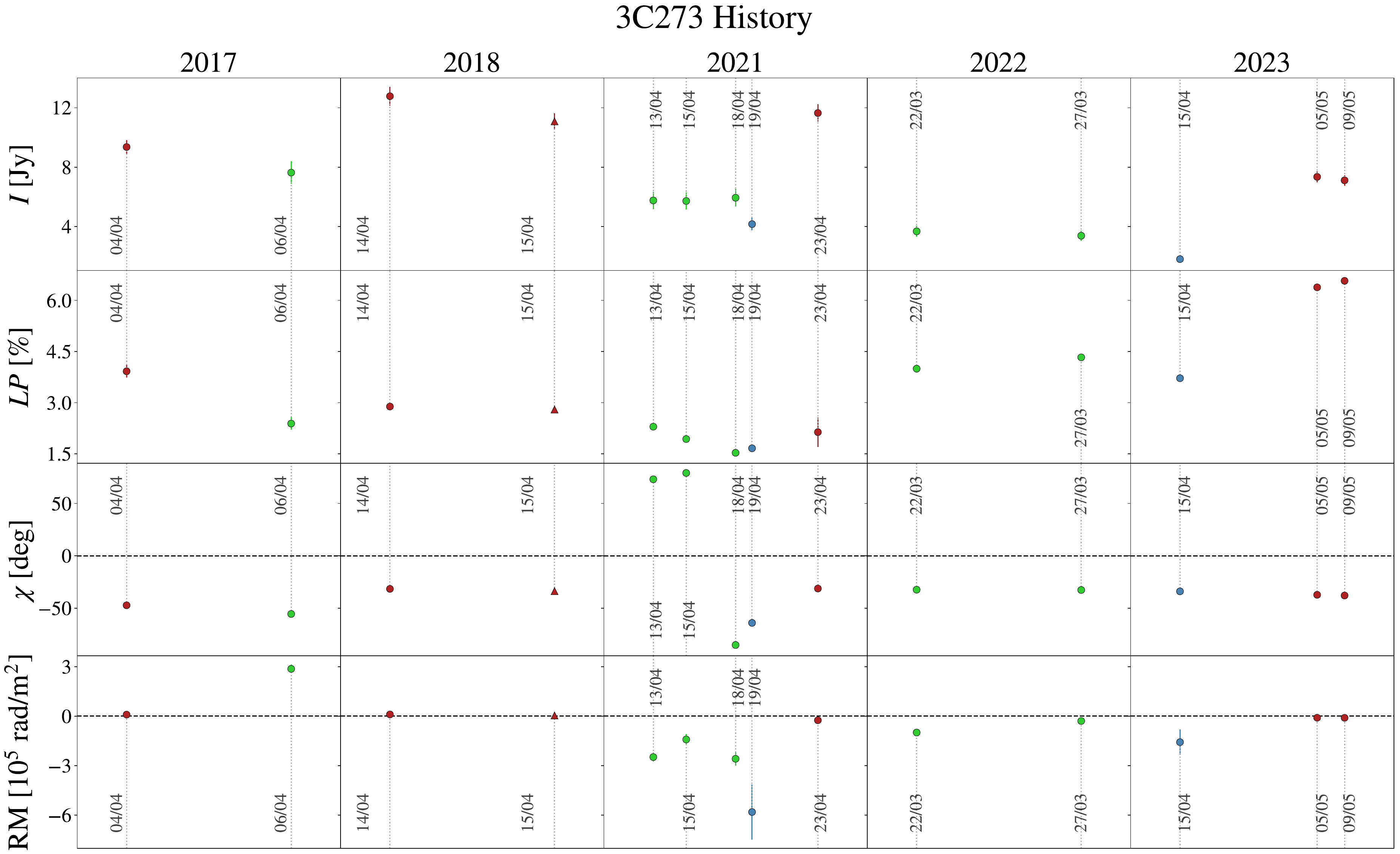}
    \caption{History of ALMA measurements of 3C273 during the VLBI campaigns (Sect. \ref{sec:Res_3C273}). Formatting is the same as in Fig. \ref{fig:hist_M87}.}
    \label{fig:hist_3C273}
    %\vspace{0.3cm}
    \includegraphics[width=\linewidth]{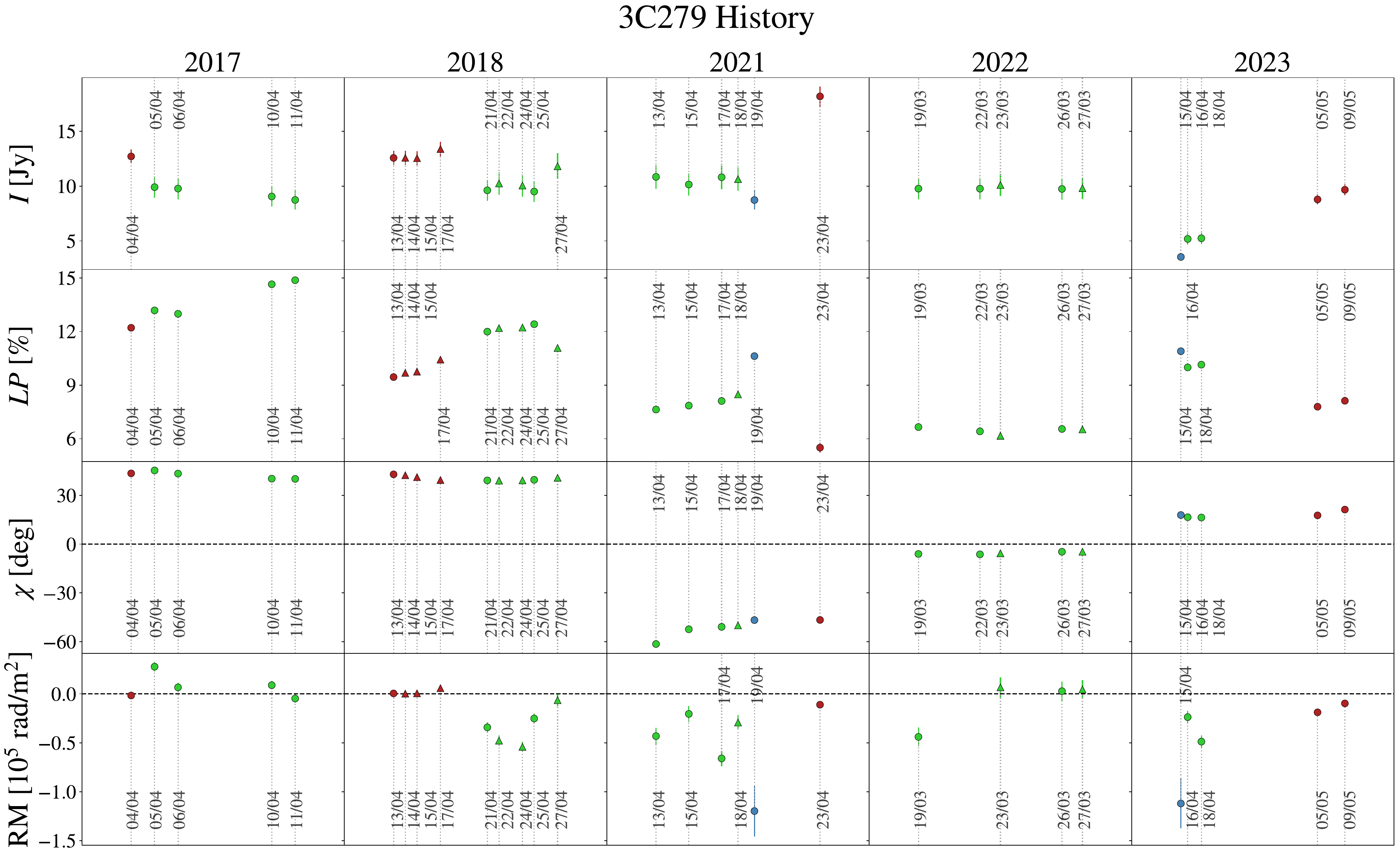}
    \caption{History of ALMA measurements of 3C279 during the VLBI campaigns (Sect. \ref{sec:Res_3c279}). Formatting is the same as in Fig. \ref{fig:hist_M87}.}
    \label{fig:hist_3C279}
\end{figure*}

Here we show the historical evolution of a few spectral and polarimetric parameters, namely Stokes $I$, LP, EVPA $\chi$ and RM. This is done for M87 (Fig. \ref{fig:hist_M87}), SgrA* (Fig. \ref{fig:hist_Sgra}), 3C273 (Fig. \ref{fig:hist_3C273}) and 3C279 (Fig. \ref{fig:hist_3C279}).
Plots for these and other sources from Sect. \ref{sec:individual} can be viewed and downloaded from the VAPOLA archive in \texttt{Tables/history/}.

\section{AMAPOLA plots}\label{ap:amaplots}
Here we show the trends in Stokes $I$, LP and EVPA for the sources covered by the AMAPOLA GS which are listed in Table \ref{tab:var_analysis}.

\begin{figure*}[h!]
    \centering
    %\vspace*{-0.2cm}
    \includegraphics[width=0.32\linewidth]{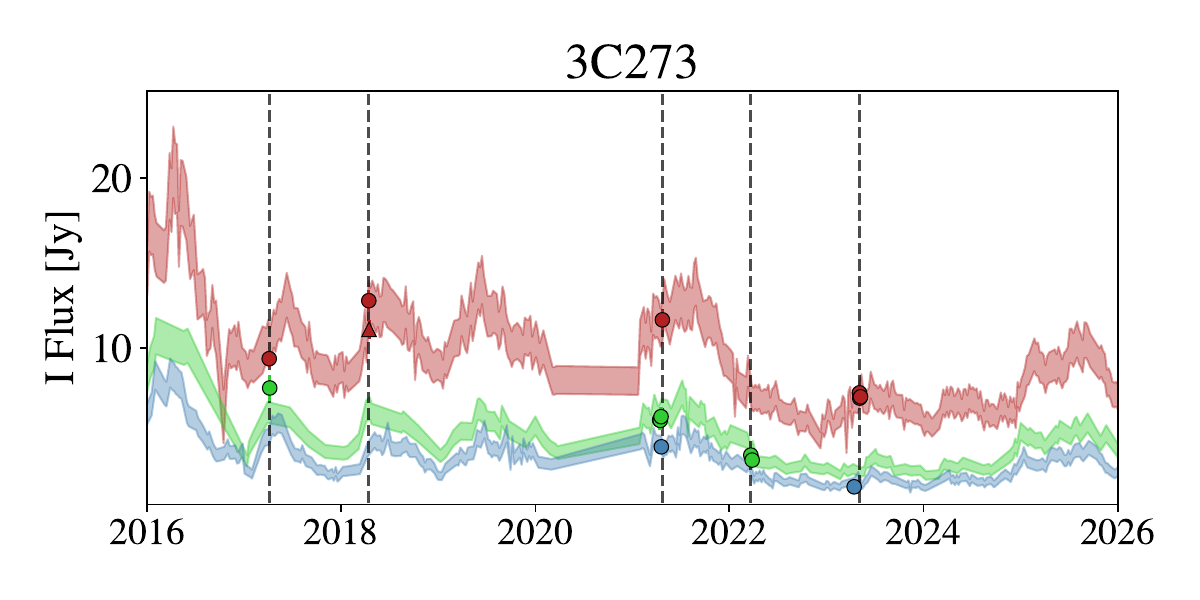}
    \includegraphics[width=0.32\linewidth]{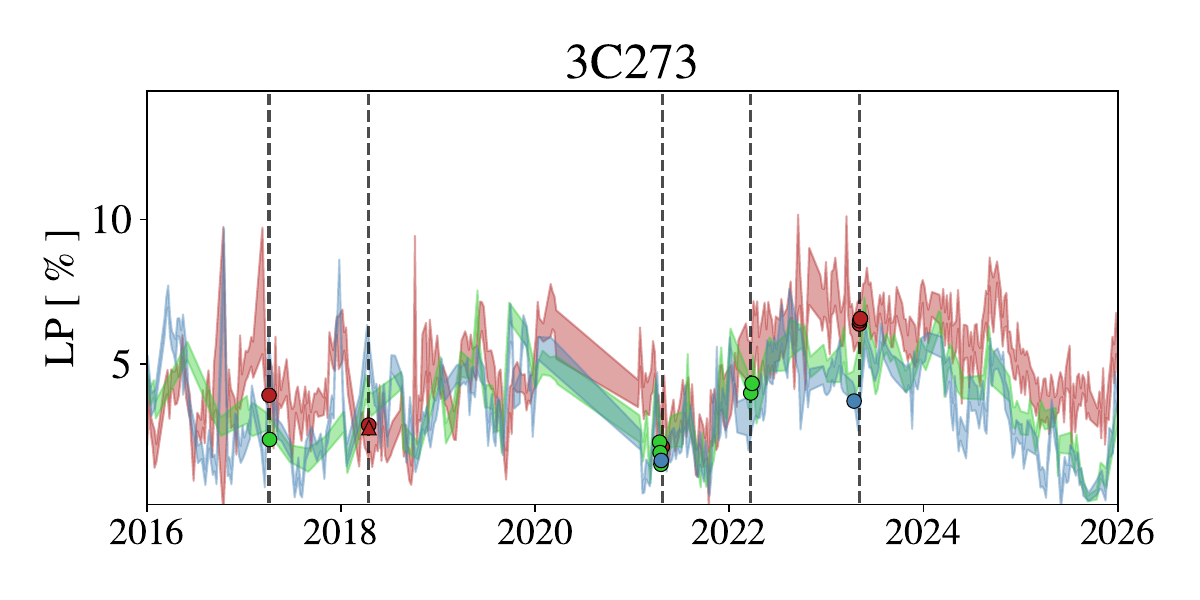}
    \includegraphics[width=0.32\linewidth]{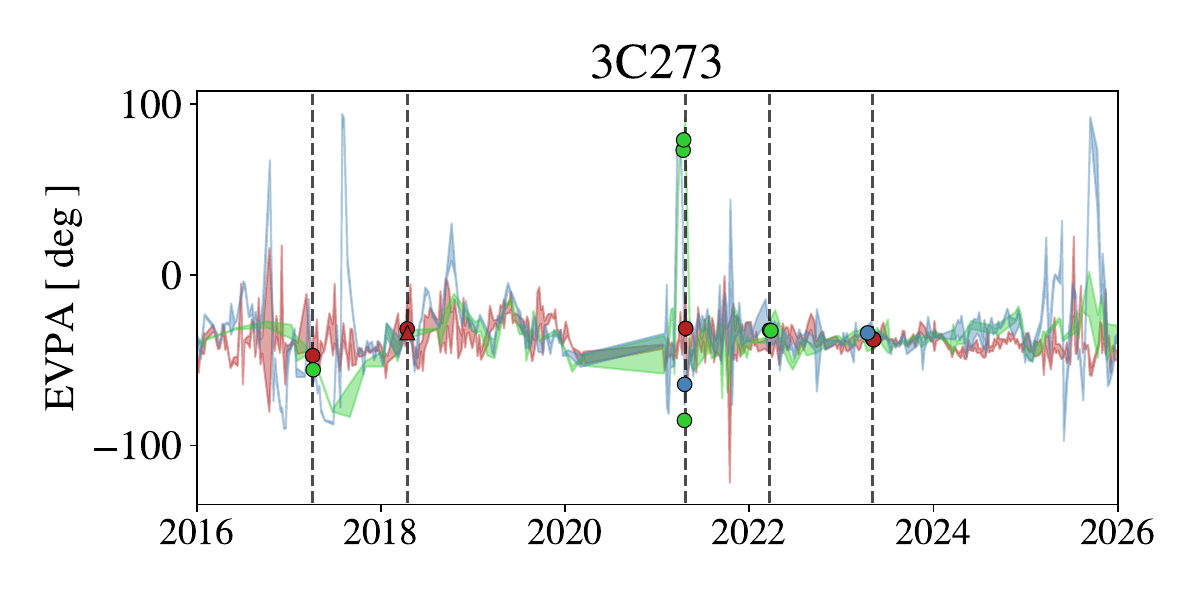}
    
    %\vspace*{-0.2cm}
    \includegraphics[width=0.32\linewidth]{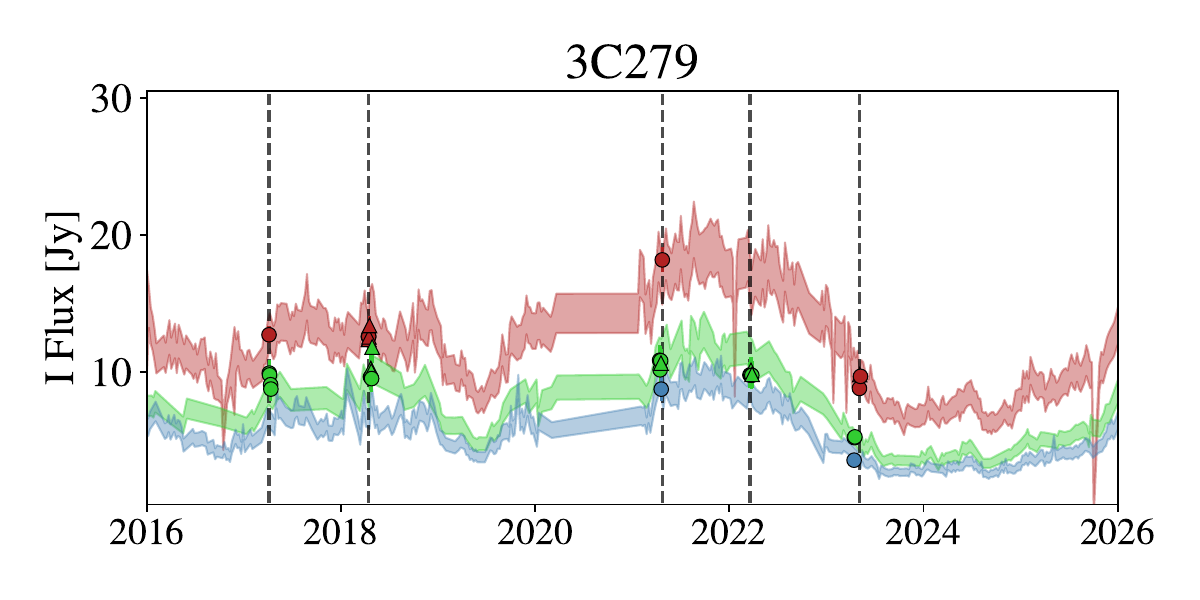}
    \includegraphics[width=0.32\linewidth]{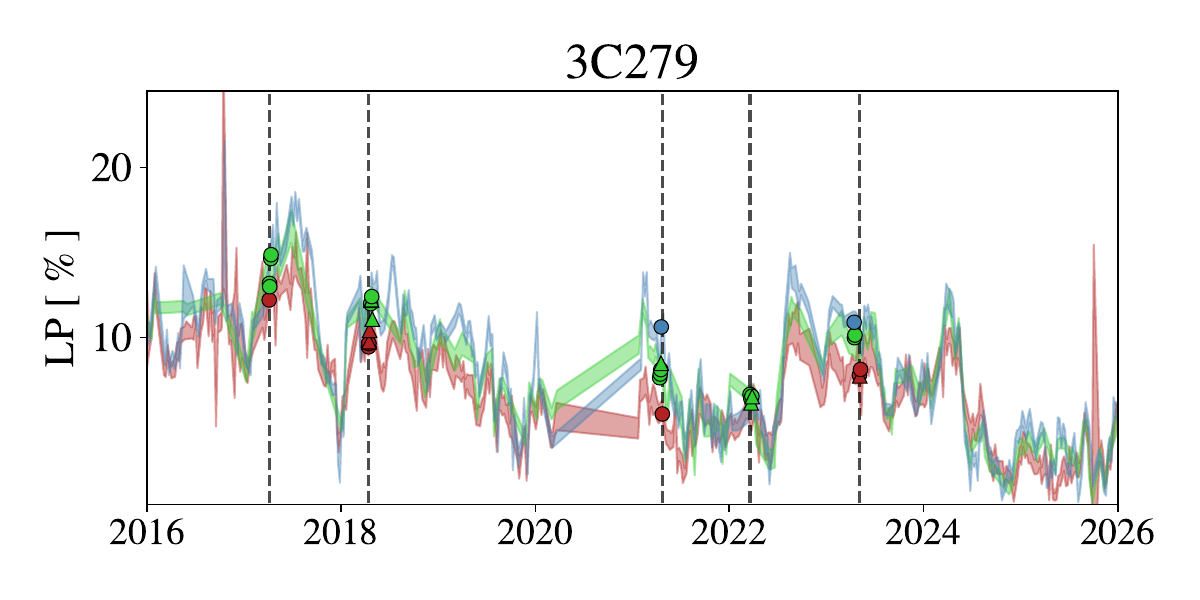}
    \includegraphics[width=0.32\linewidth]{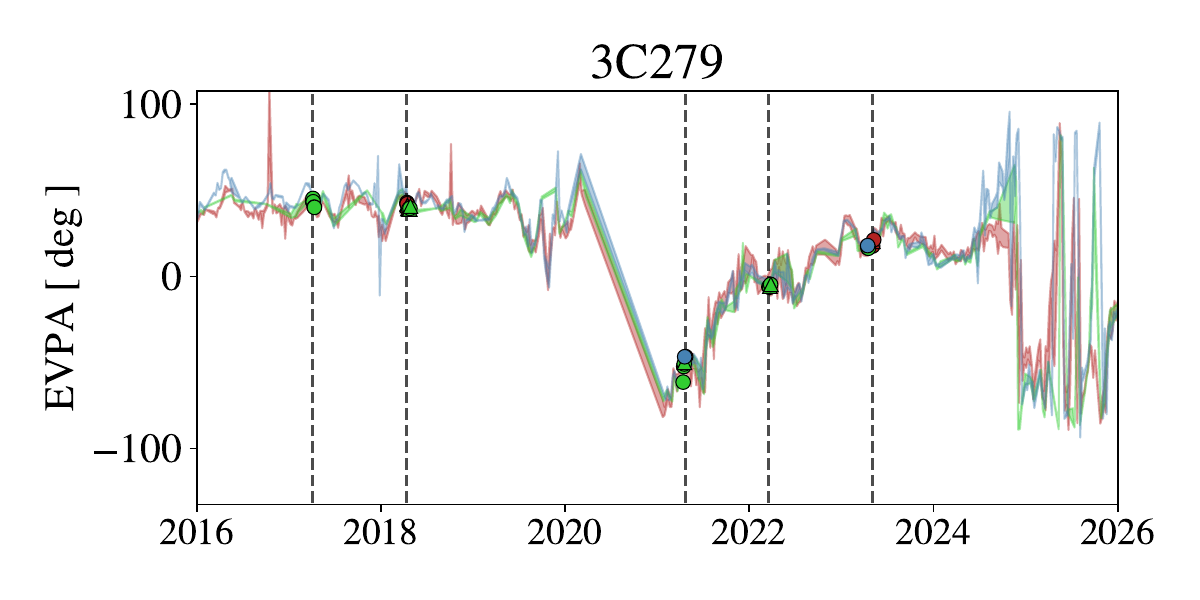}
    
    %\vspace*{-0.2cm}
    \includegraphics[width=0.325\linewidth]{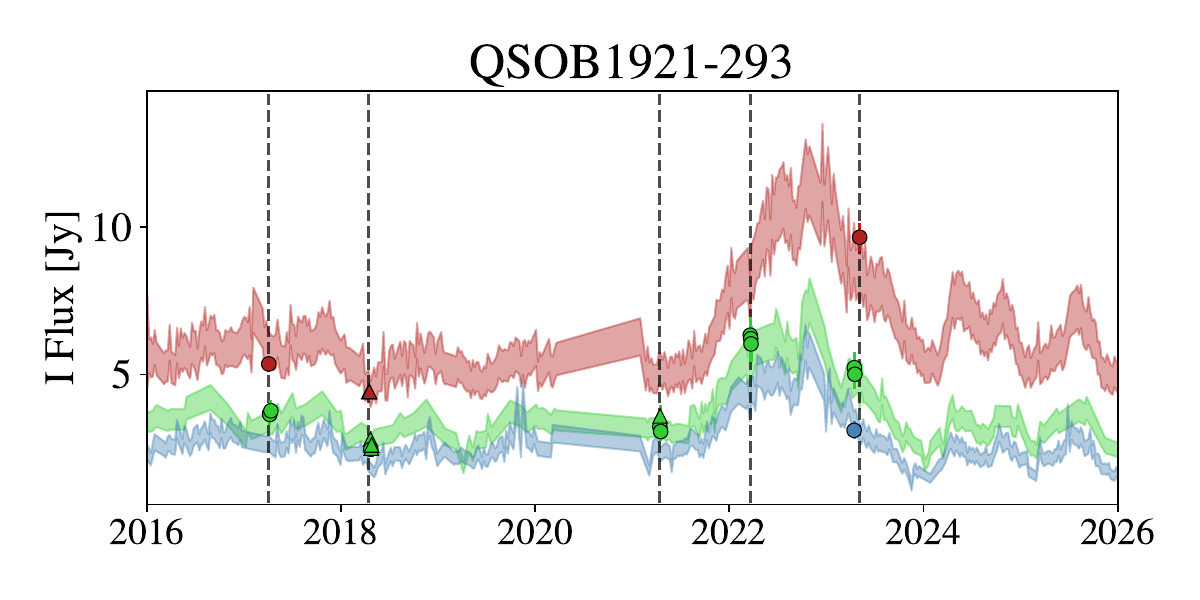}
    \includegraphics[width=0.325\linewidth]{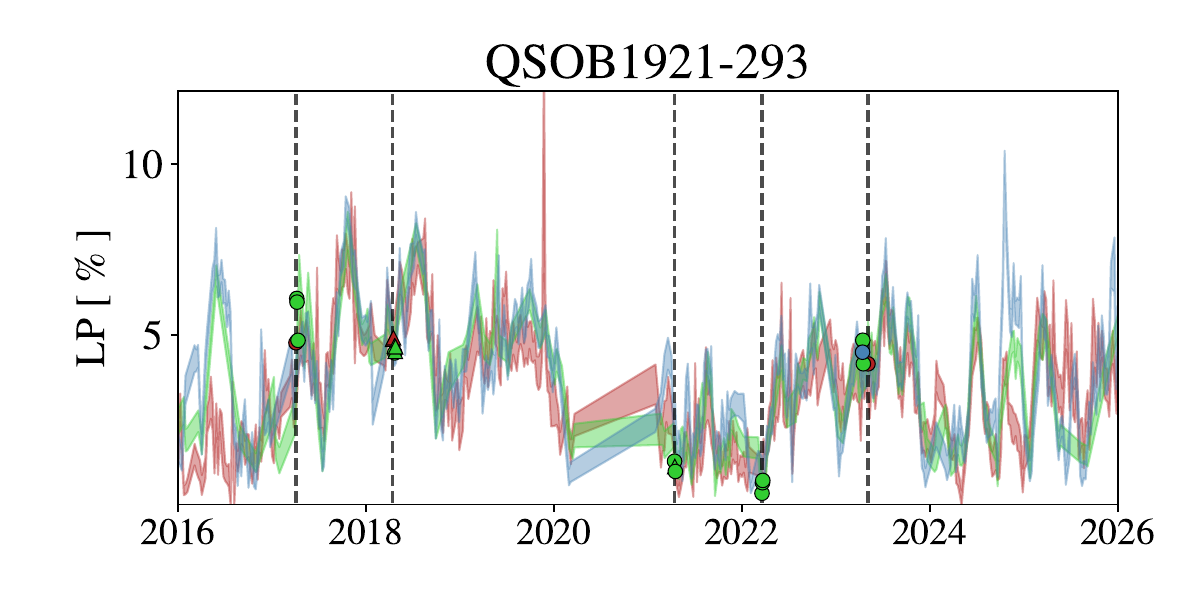}
    \includegraphics[width=0.325\linewidth]{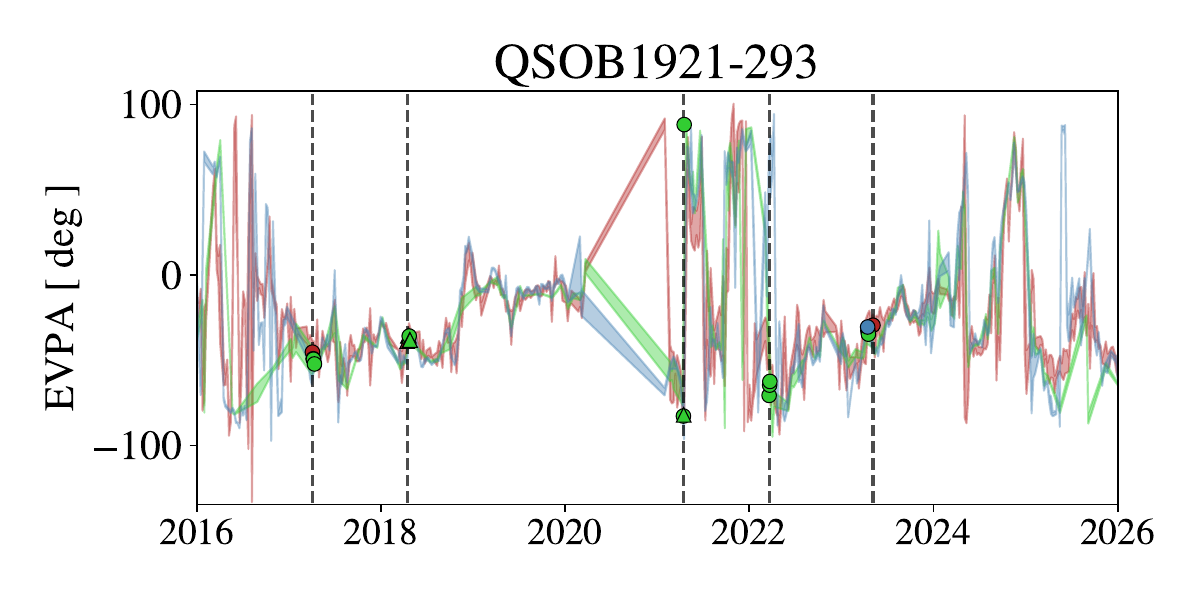}
    
    %\vspace*{-0.2cm}
    \includegraphics[width=0.32\linewidth]{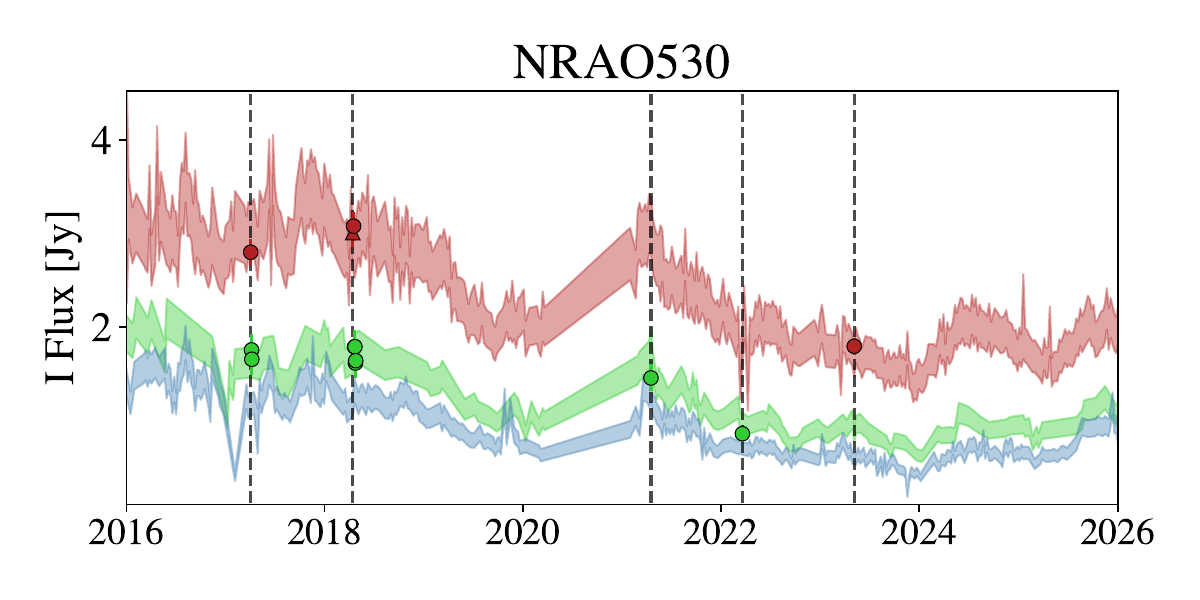}
    \includegraphics[width=0.32\linewidth]{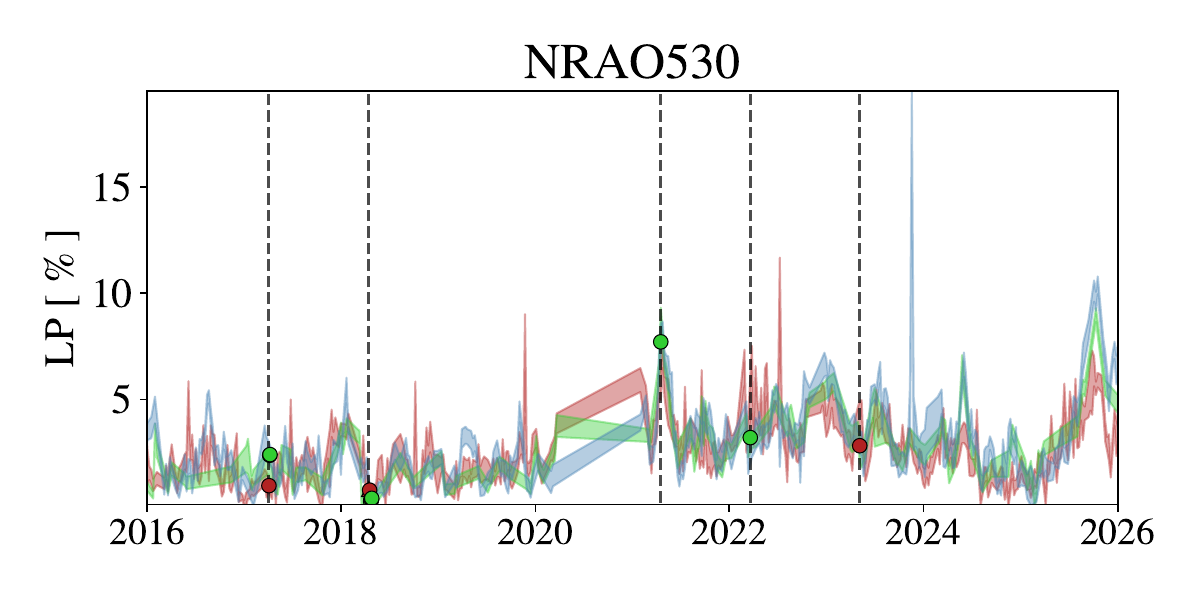}
    \includegraphics[width=0.32\linewidth]{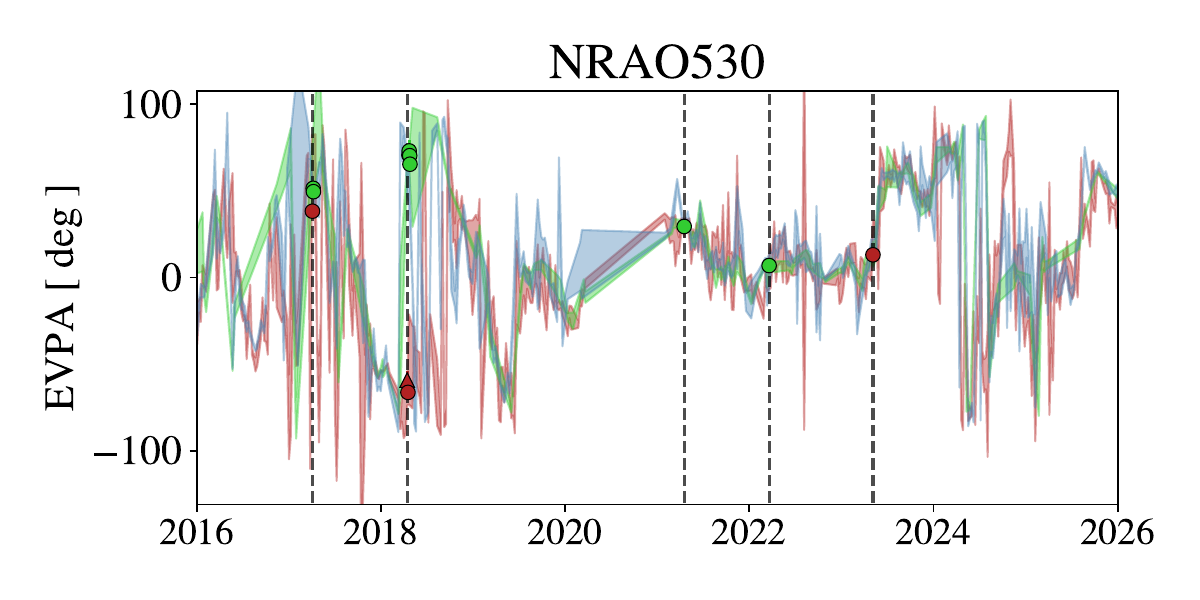}
    
    %\vspace*{-0.2cm}
    \includegraphics[width=0.325\linewidth]{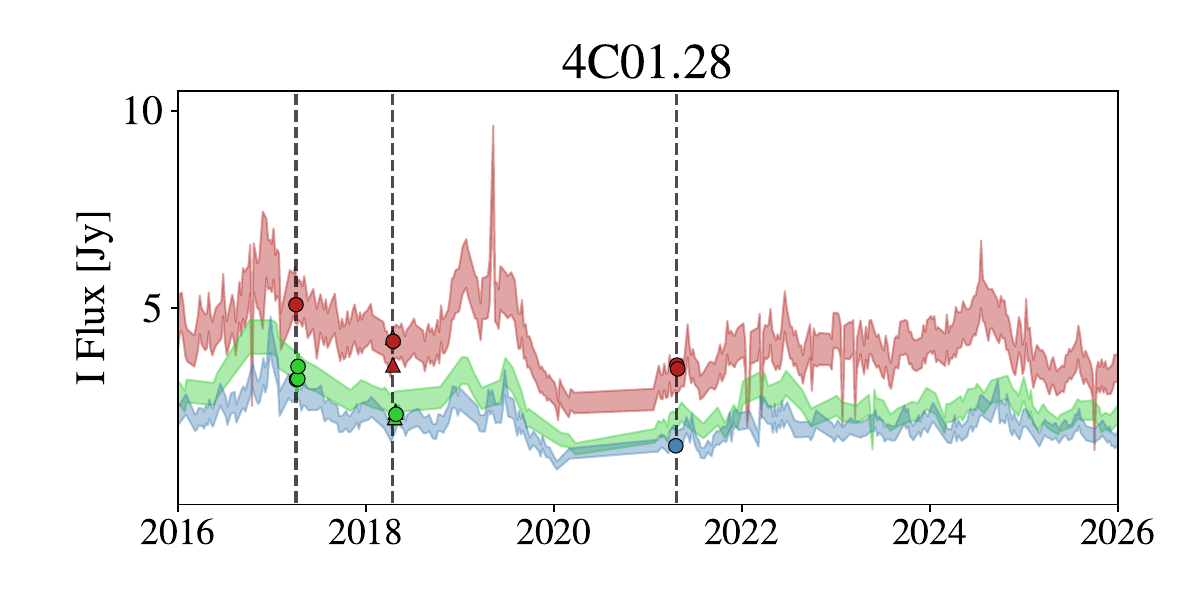}
    \includegraphics[width=0.325\linewidth]{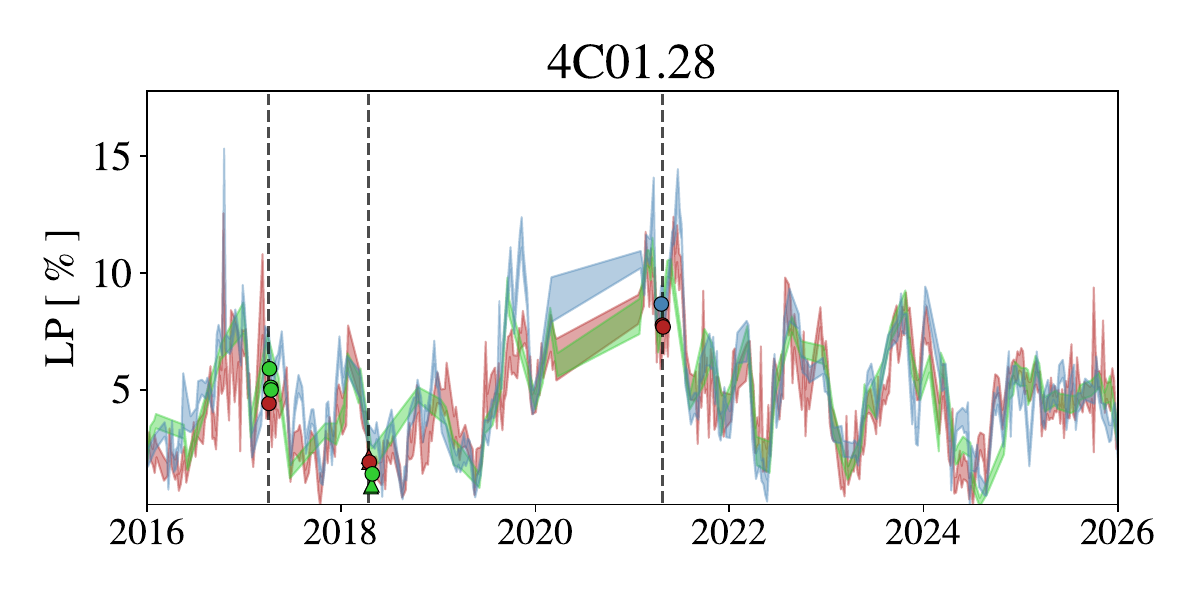}
    \includegraphics[width=0.325\linewidth]{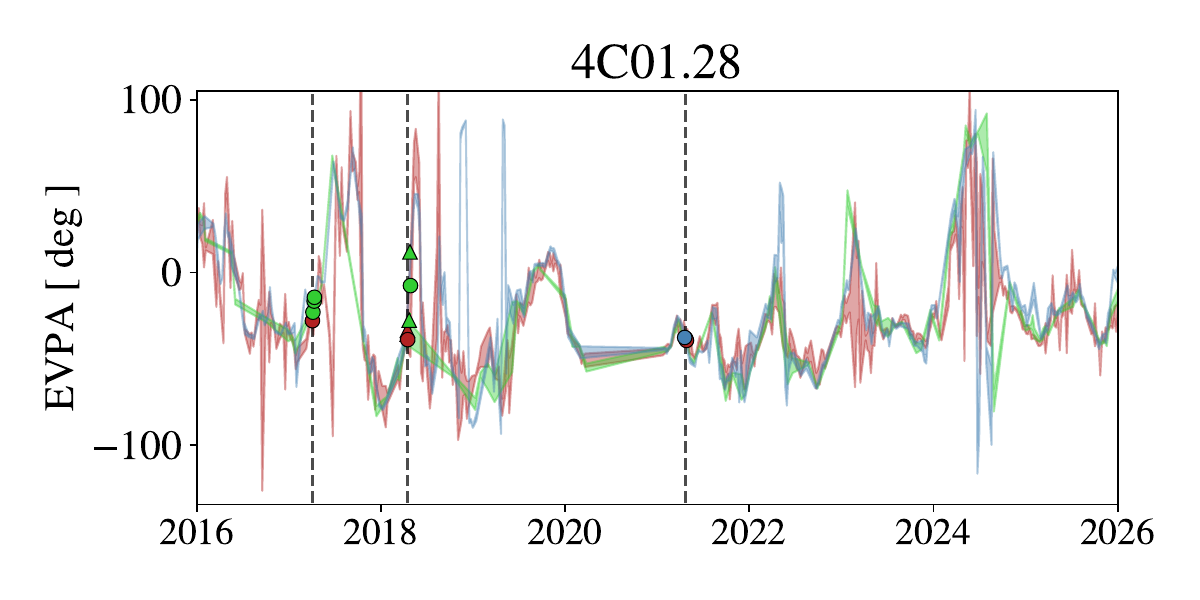}
    
    %\vspace*{-0.2cm}
    \includegraphics[width=0.32\linewidth]{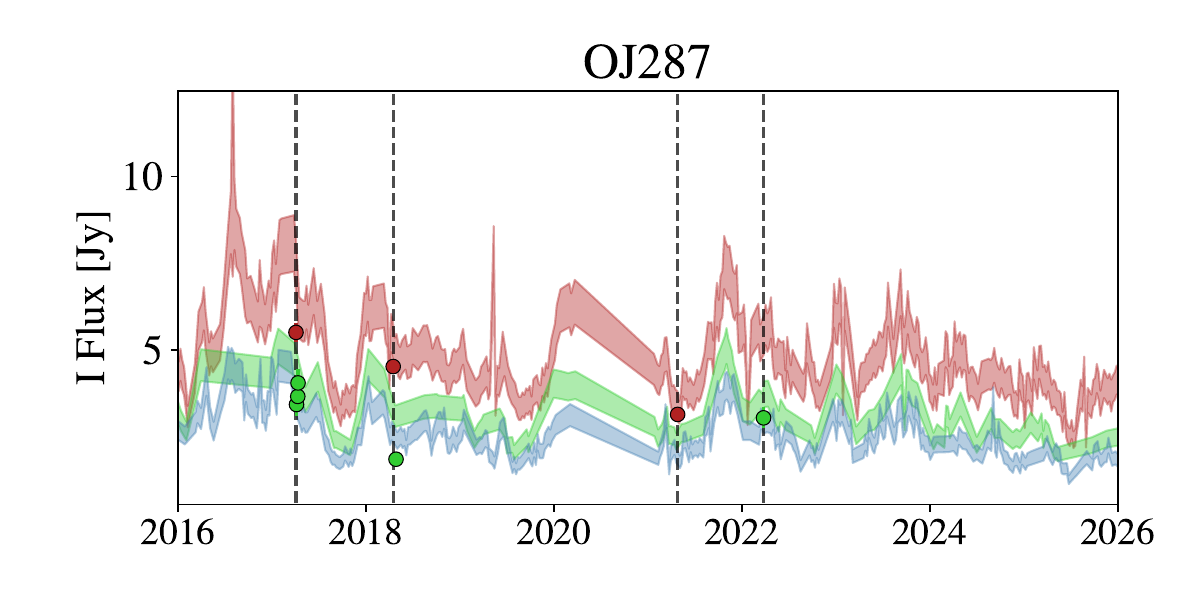}
    \includegraphics[width=0.32\linewidth]{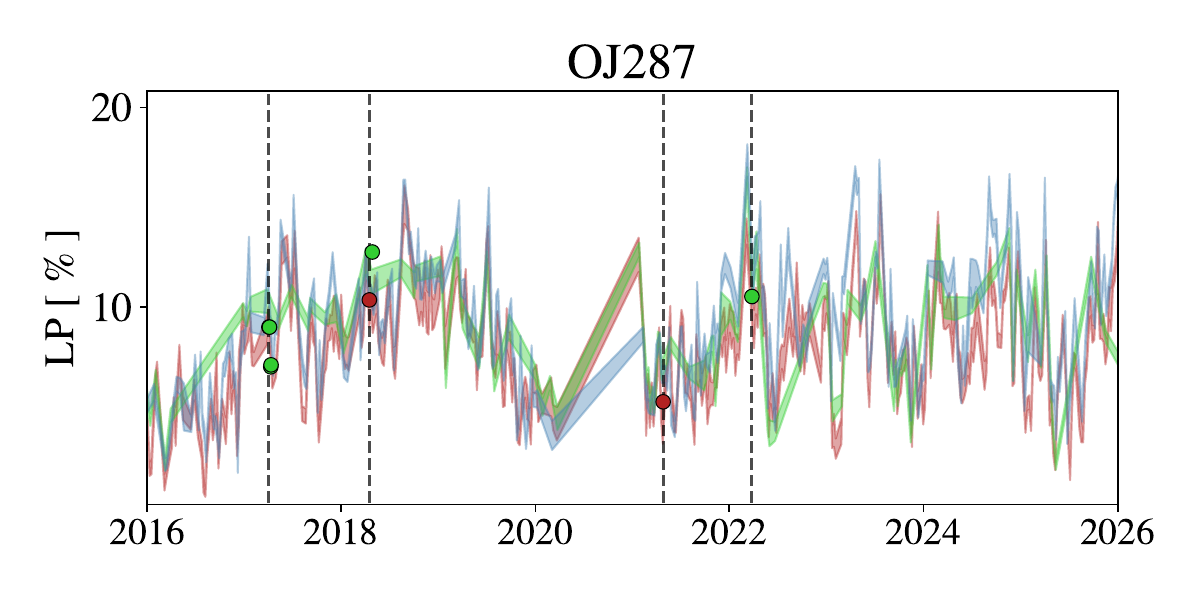}
    \includegraphics[width=0.32\linewidth]{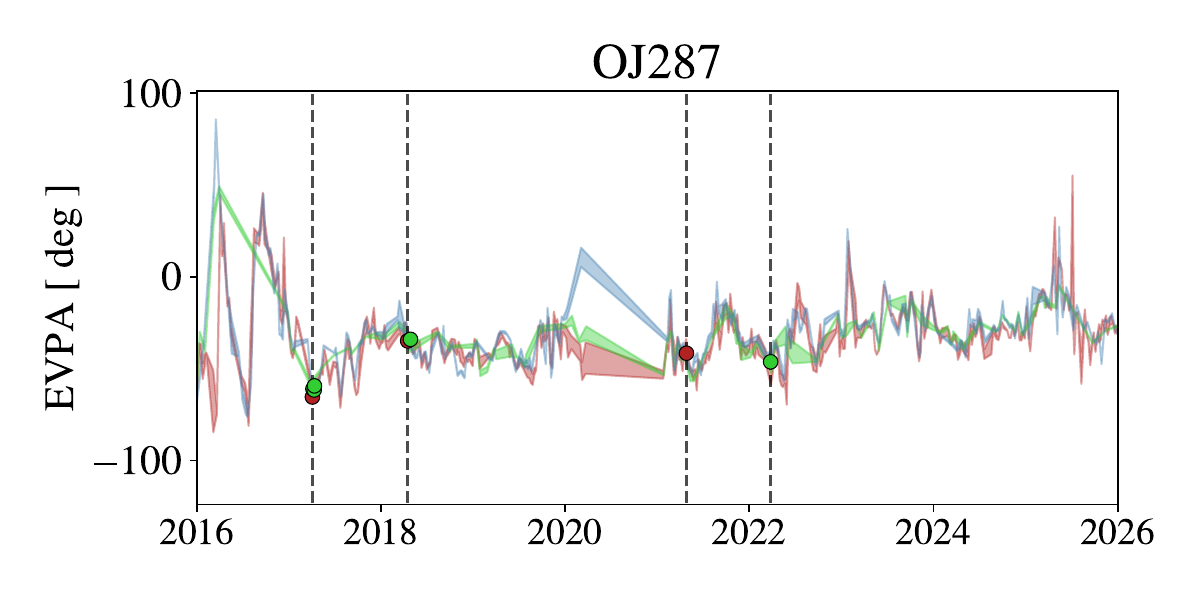}
    
    %\vspace*{-0.2cm}
    \includegraphics[width=0.32\linewidth]{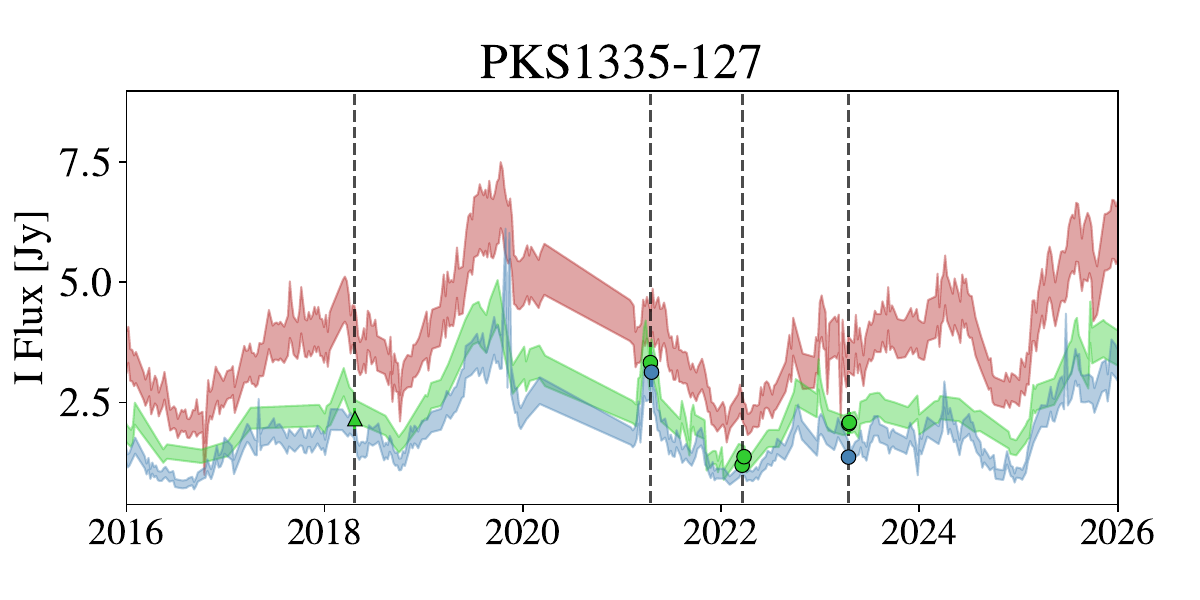}
    \includegraphics[width=0.32\linewidth]{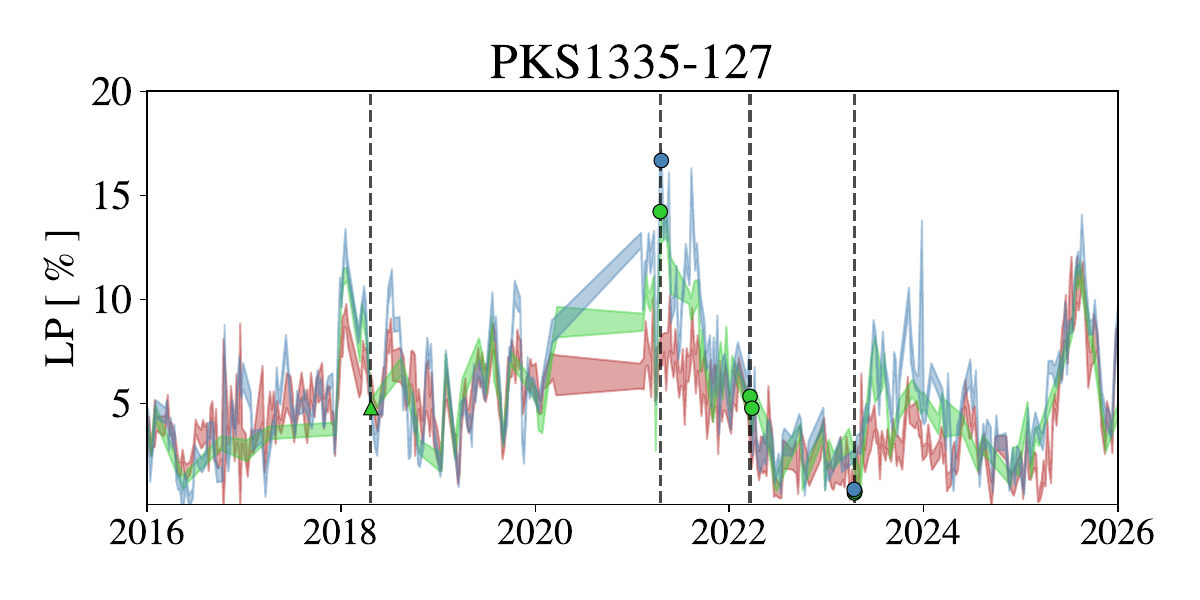}
    \includegraphics[width=0.32\linewidth]{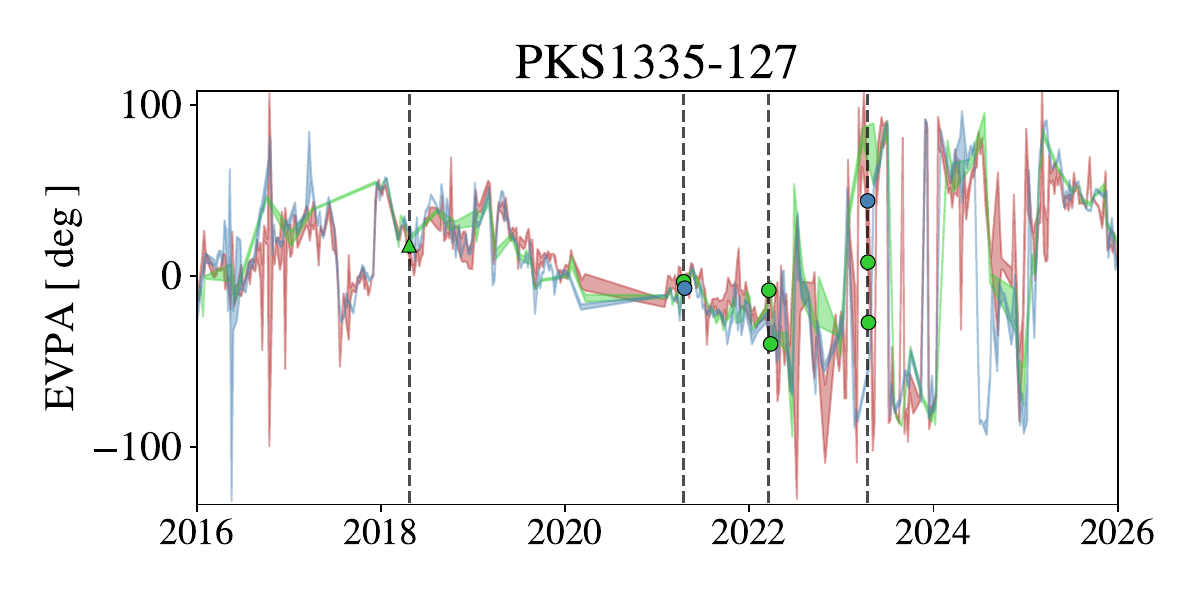}
    %\vspace*{-0.2cm}

\caption{AMAPOLA measurements for the most frequently observed sources over the 2016--2026 period.
Each row corresponds to a different source, and each column to a different parameter: left, Stokes $I$; center, LP; right, EVPA.
The colored bands indicate the $1\sigma$ interval of the AMAPOLA measurements, while the colored circles represent the corresponding VAPOLA measurements.
Colors indicate observing frequency: red, 98 GHz; green, 230 GHz; and cyan, 343 GHz.
The vertical black dashed lines indicate the dates of the VLBI campaigns.}
    \label{fig:amaplots}
\end{figure*}

\begin{figure*}[h!]
    \centering
    %\vspace*{-0.2cm}
    \includegraphics[width=0.32\linewidth]{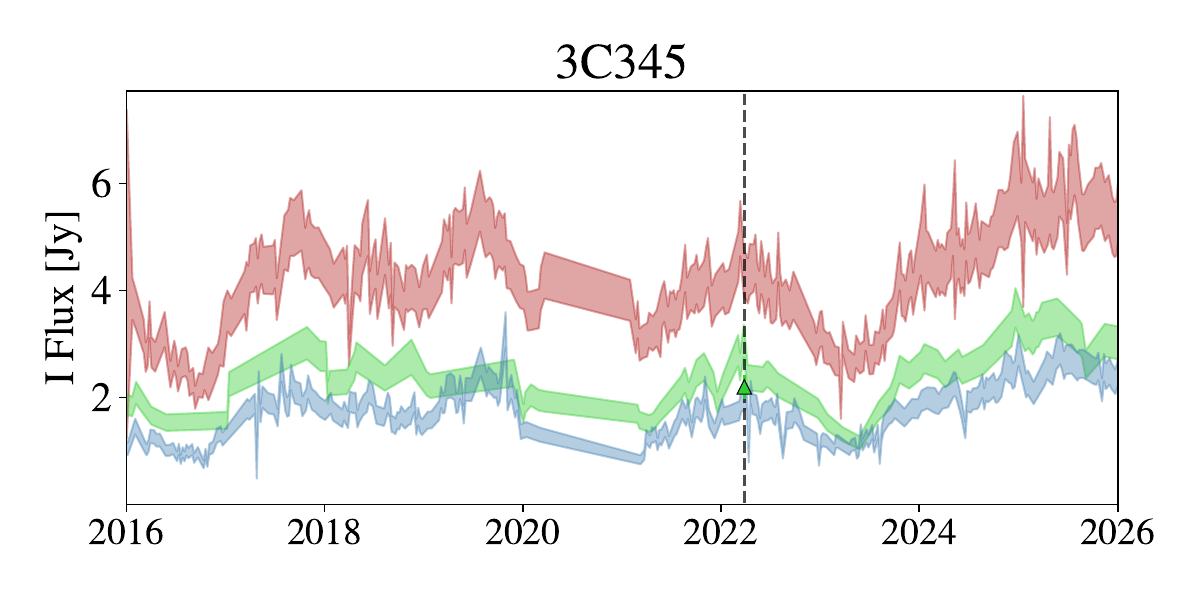}
    \includegraphics[width=0.32\linewidth]{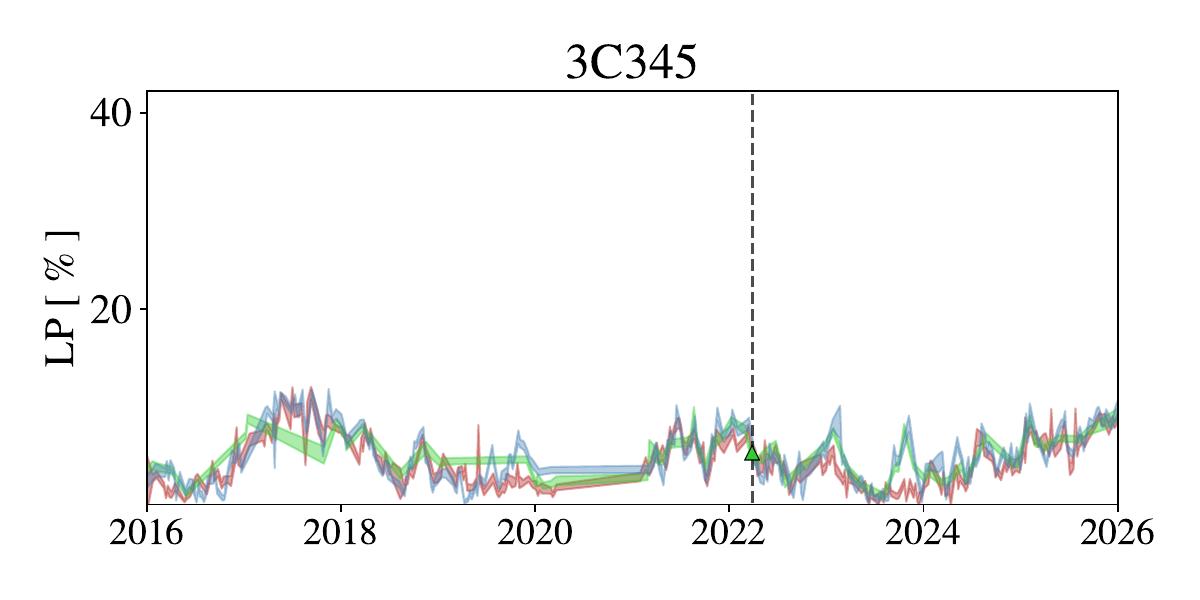}
    \includegraphics[width=0.32\linewidth]{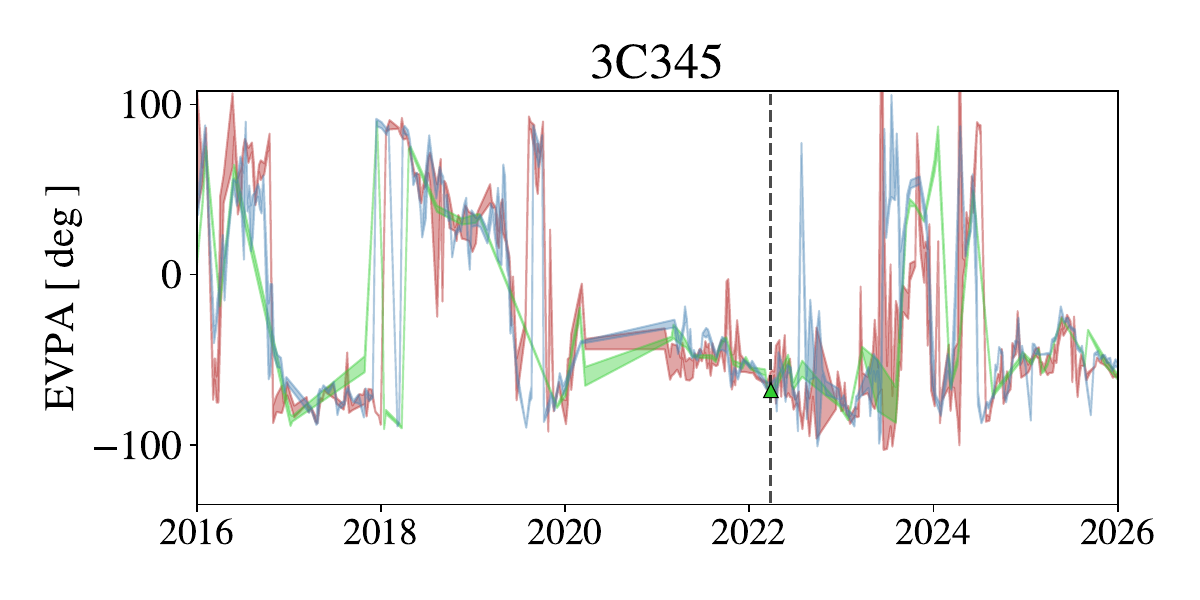}
    
    %\vspace*{-0.2cm}
    \includegraphics[width=0.32\linewidth]{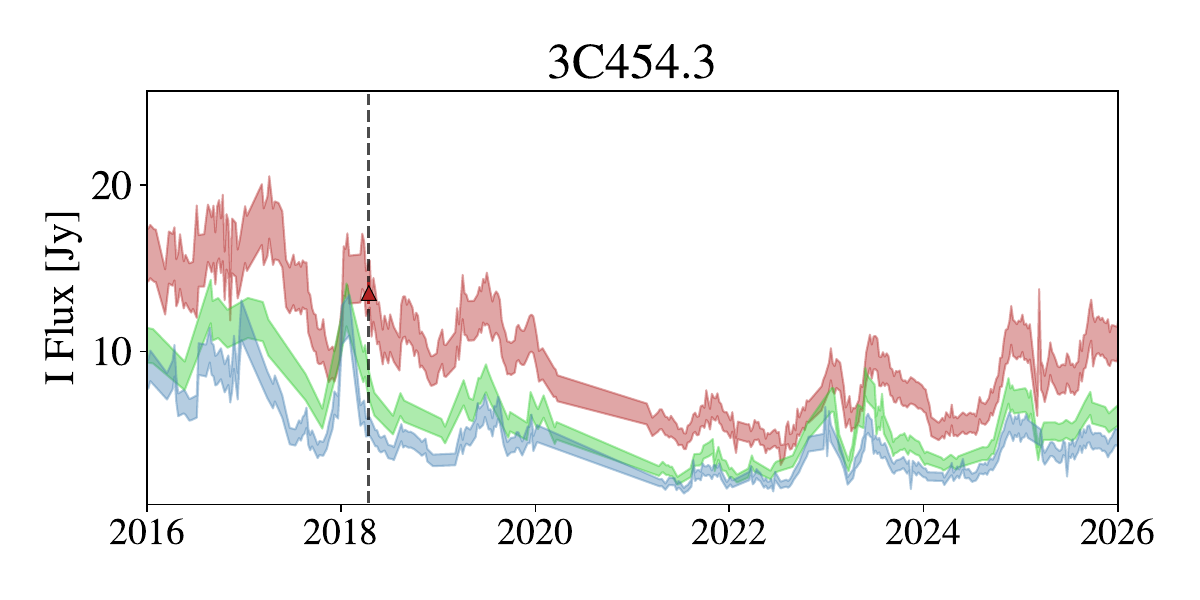}
    \includegraphics[width=0.32\linewidth]{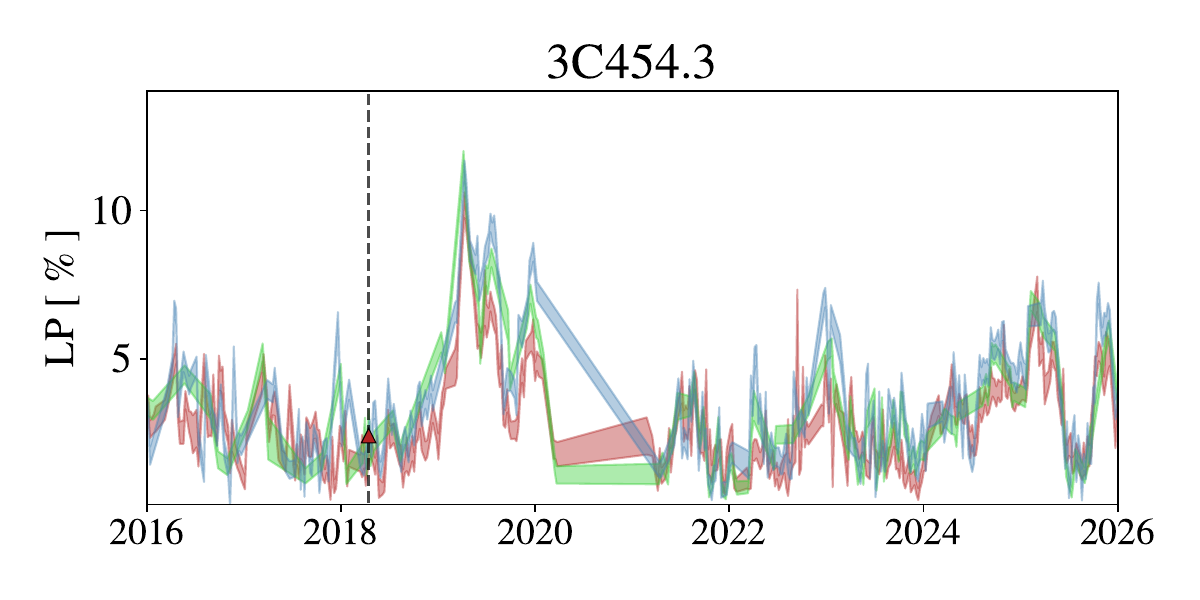}
    \includegraphics[width=0.32\linewidth]{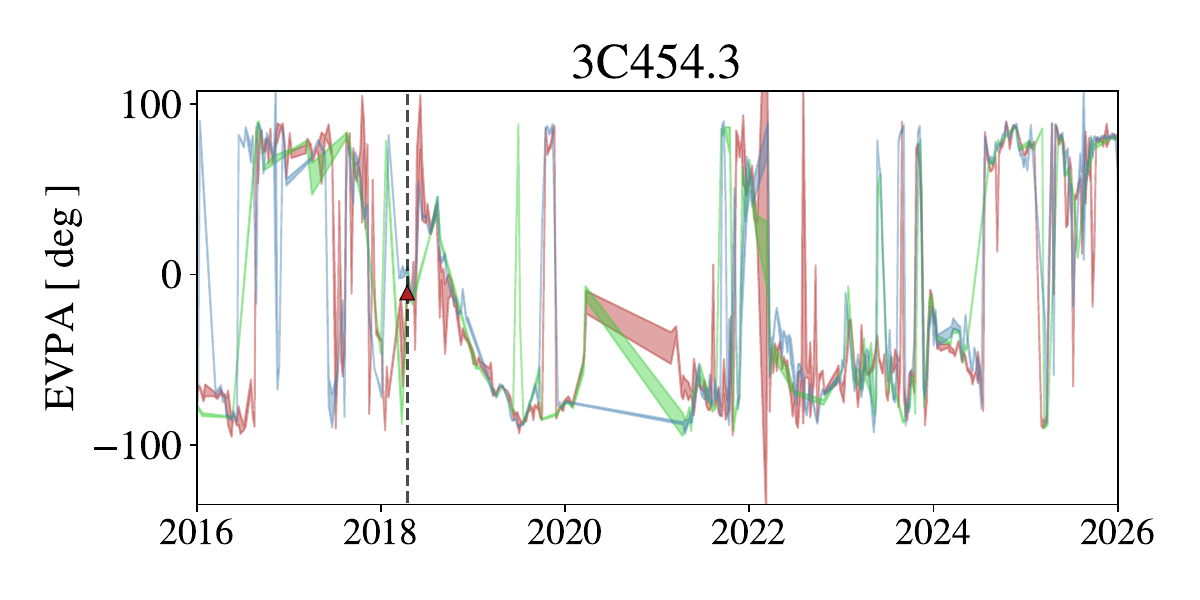}
    
    %\vspace*{-0.2cm}
    \includegraphics[width=0.32\linewidth]{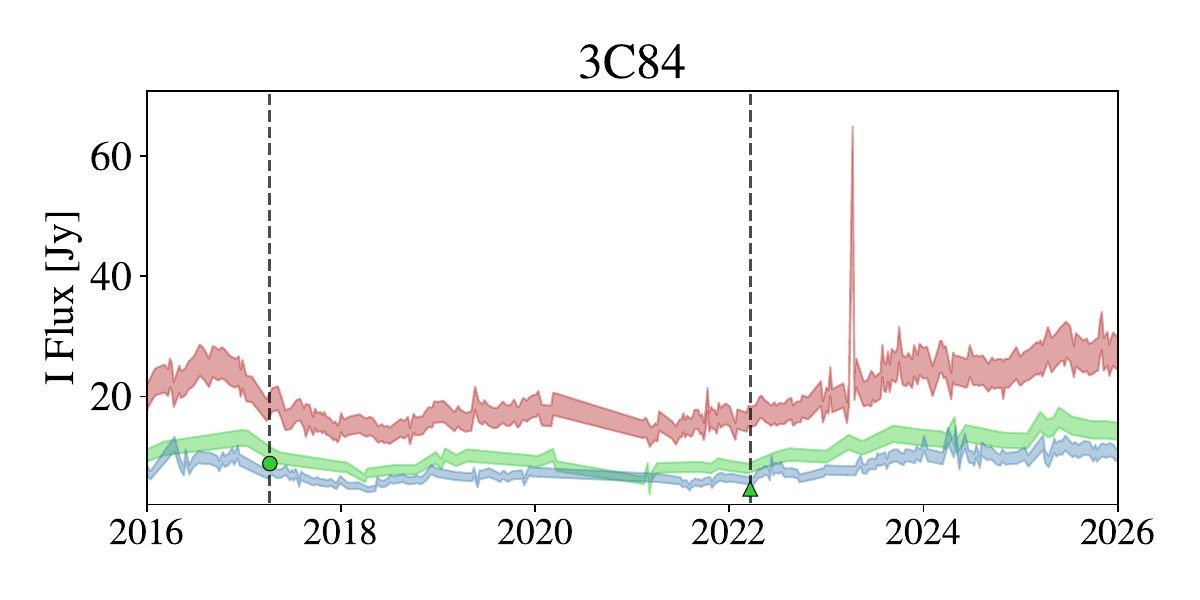}
    \includegraphics[width=0.32\linewidth]{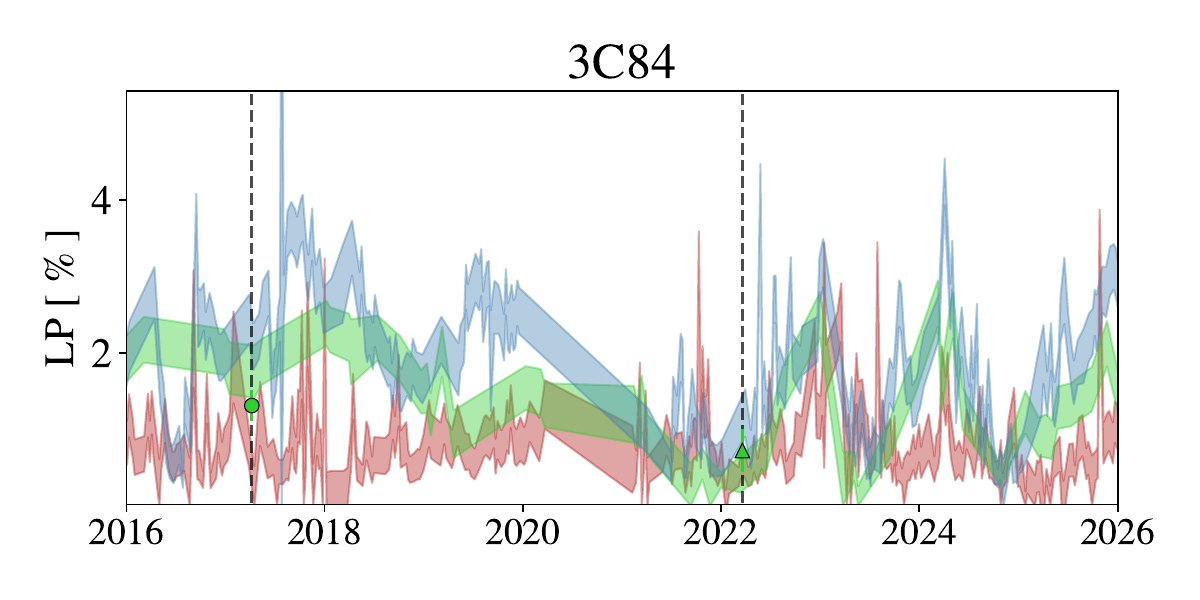}
    \includegraphics[width=0.32\linewidth]{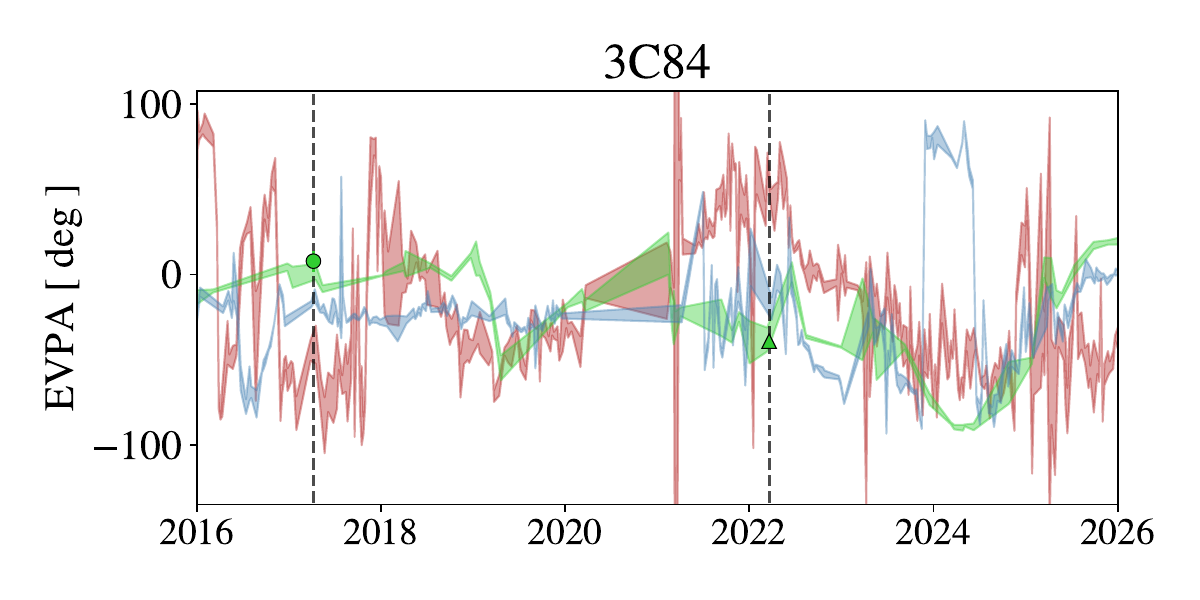}
    
    %\vspace*{-0.2cm}
    \includegraphics[width=0.32\linewidth]{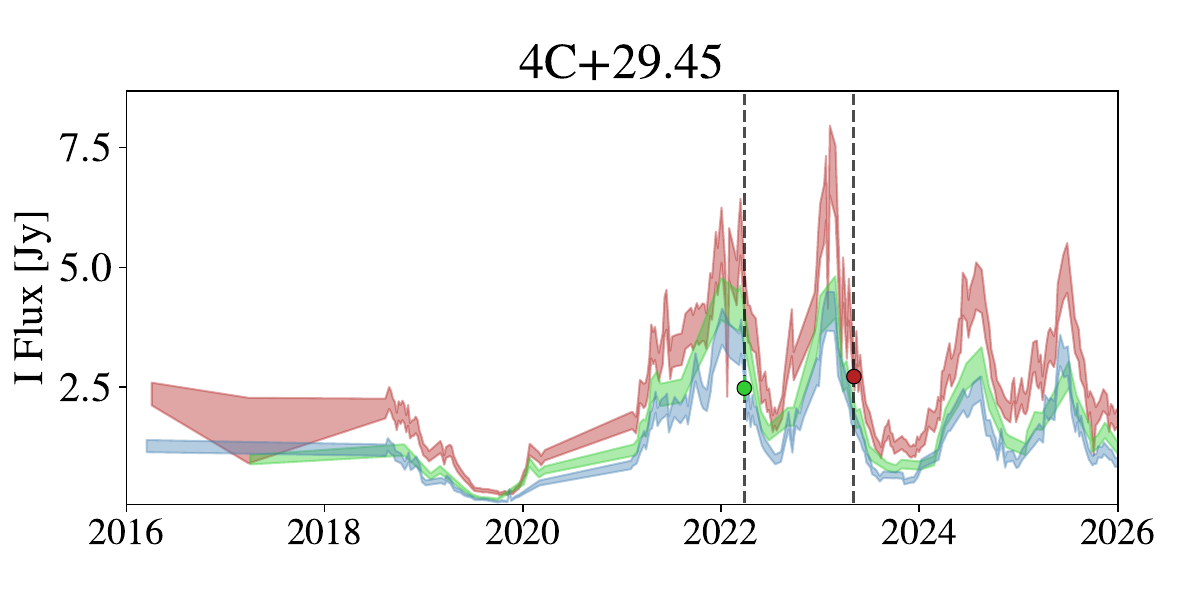}
    \includegraphics[width=0.32\linewidth]{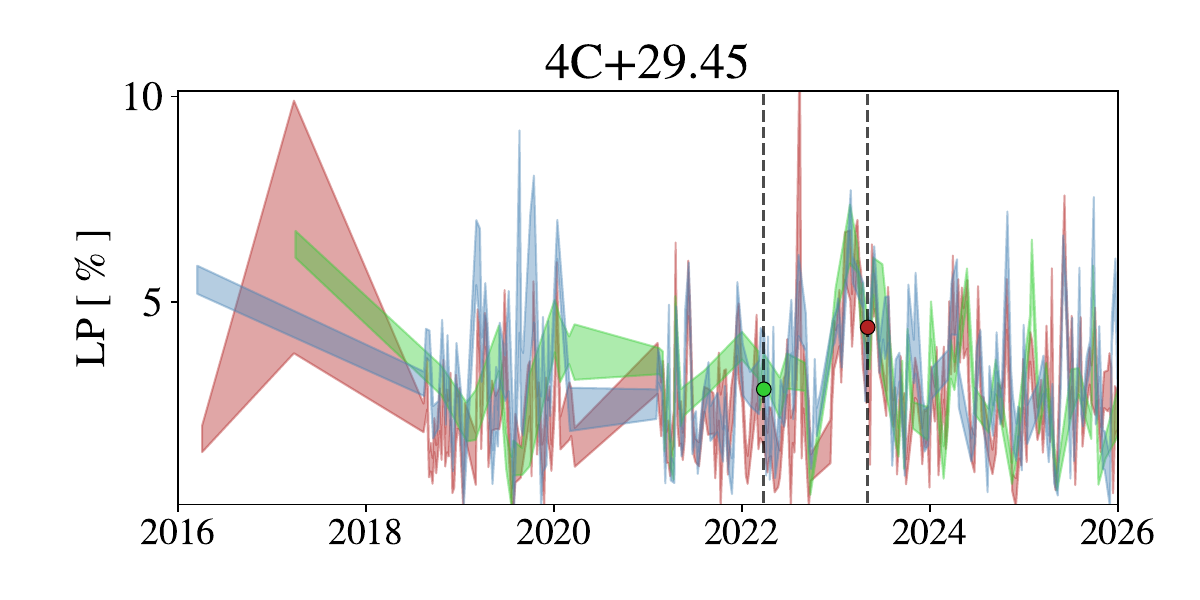}
    \includegraphics[width=0.32\linewidth]{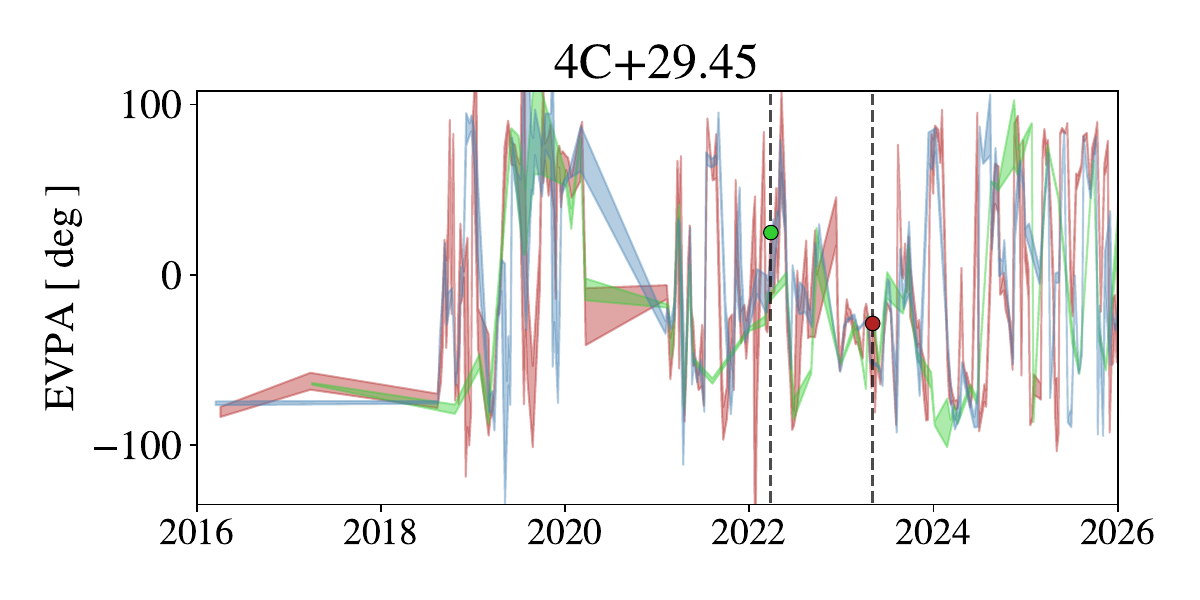}
    
    %\vspace*{-0.2cm}
    \includegraphics[width=0.32\linewidth]{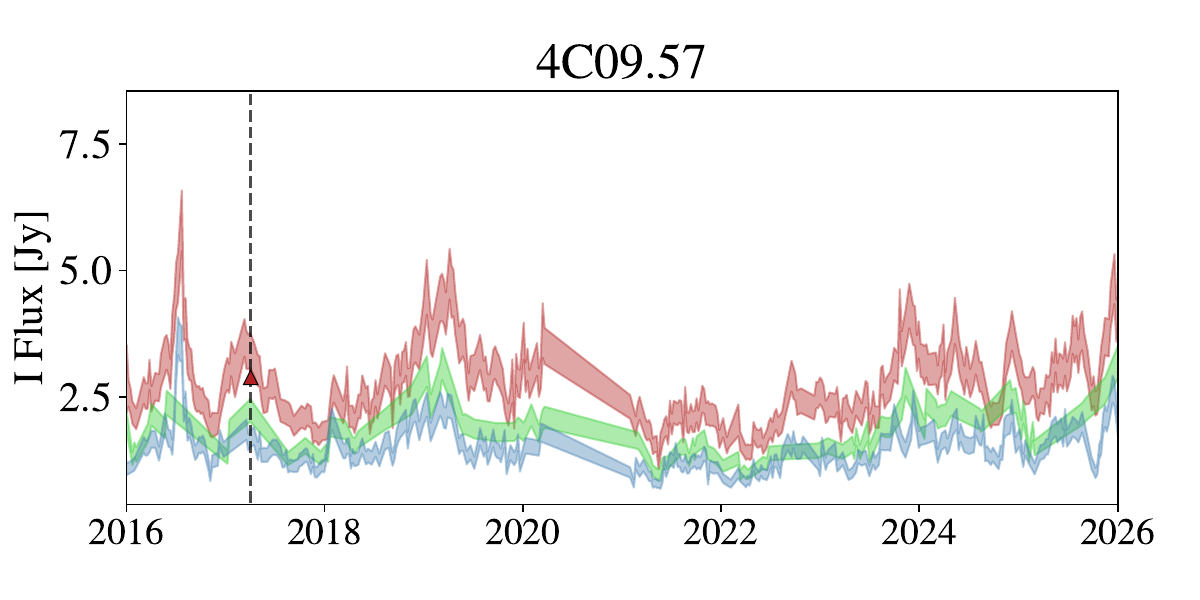}
    \includegraphics[width=0.32\linewidth]{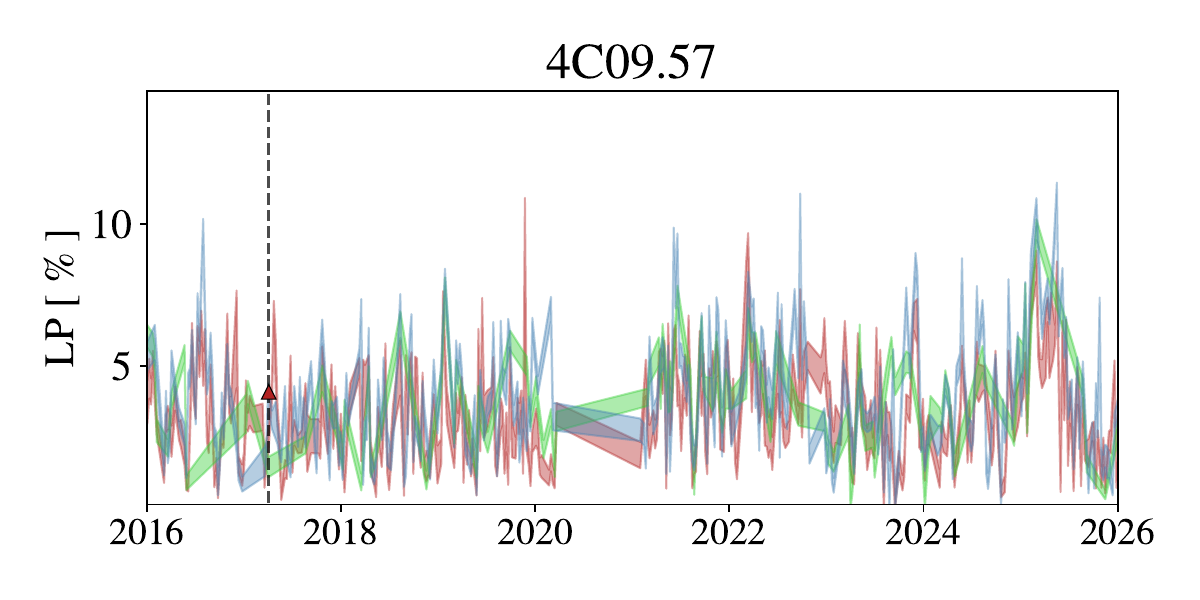}
    \includegraphics[width=0.32\linewidth]{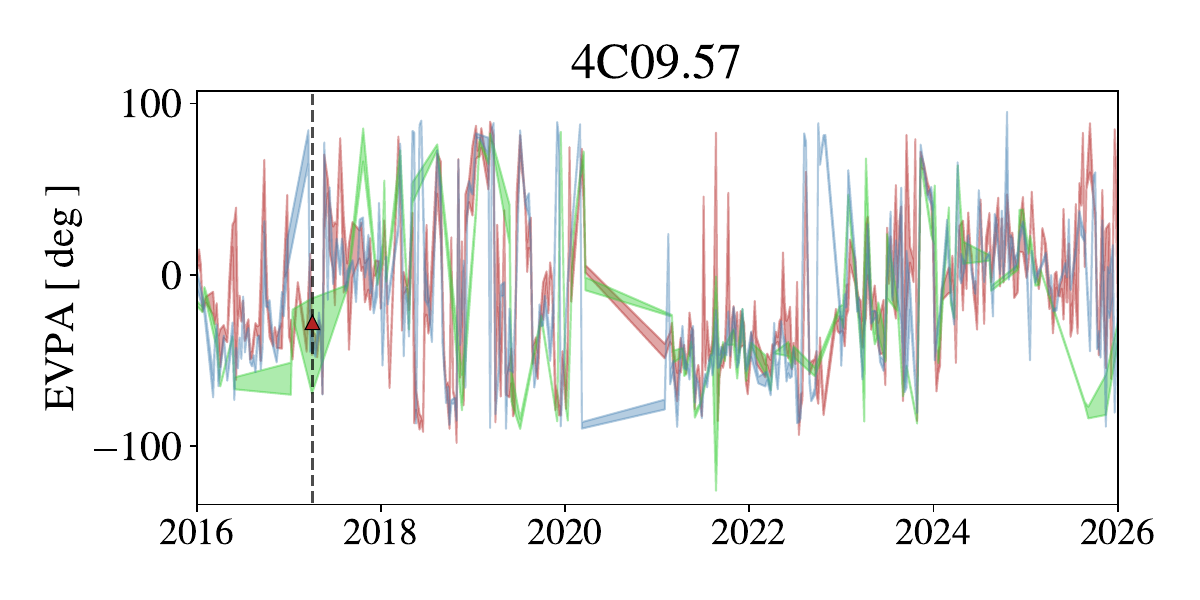}
    
    %\vspace*{-0.2cm}
    \includegraphics[width=0.32\linewidth]{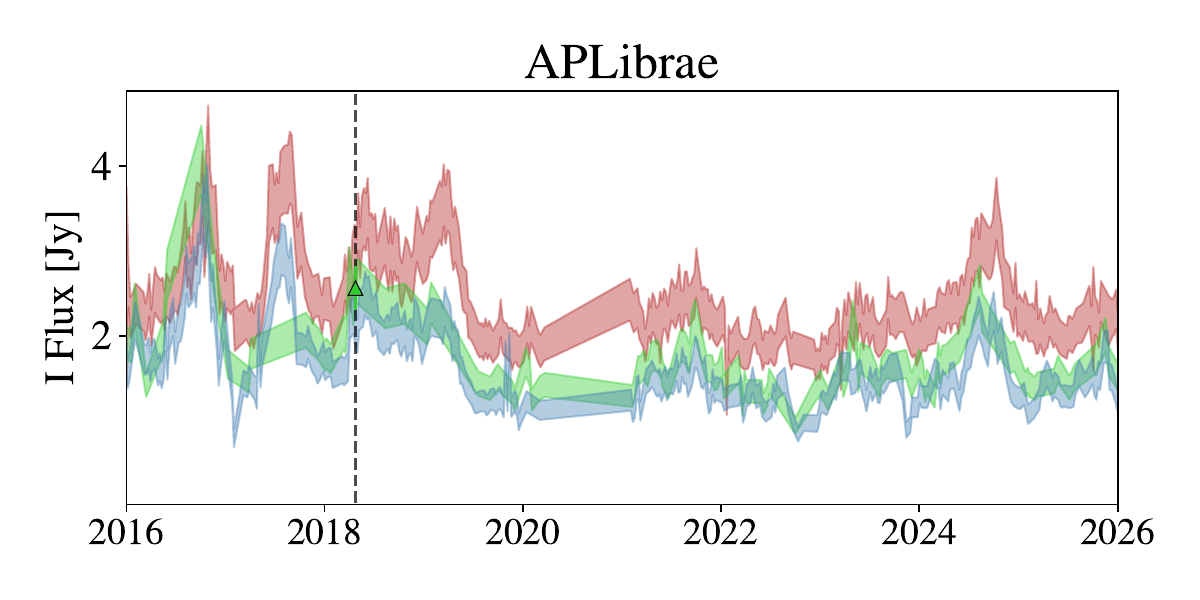}
    \includegraphics[width=0.32\linewidth]{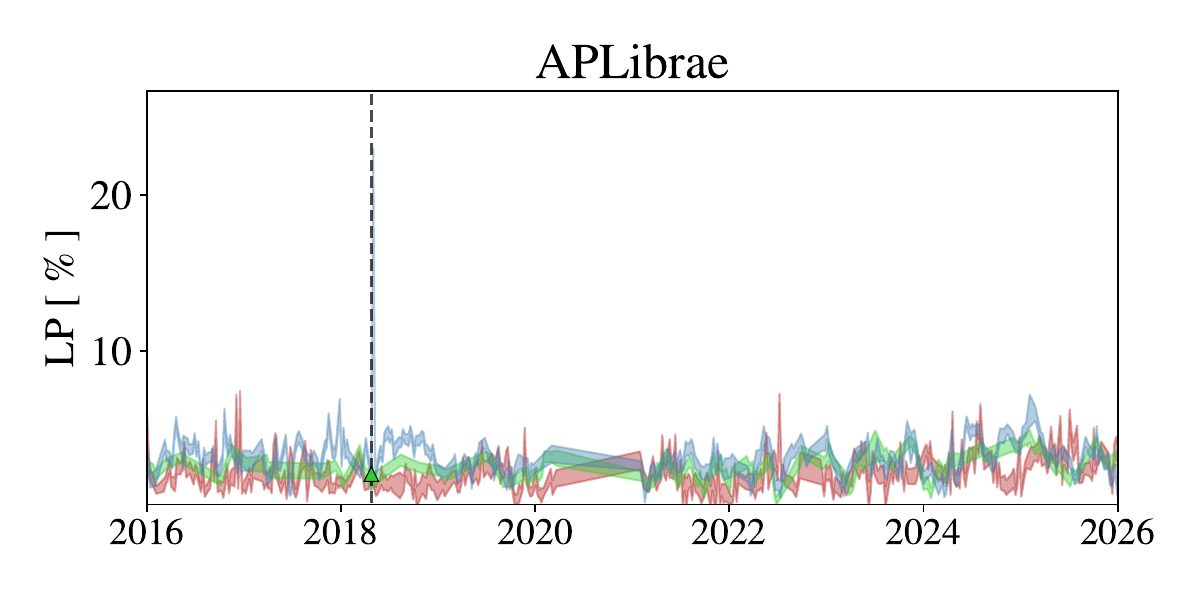}
    \includegraphics[width=0.32\linewidth]{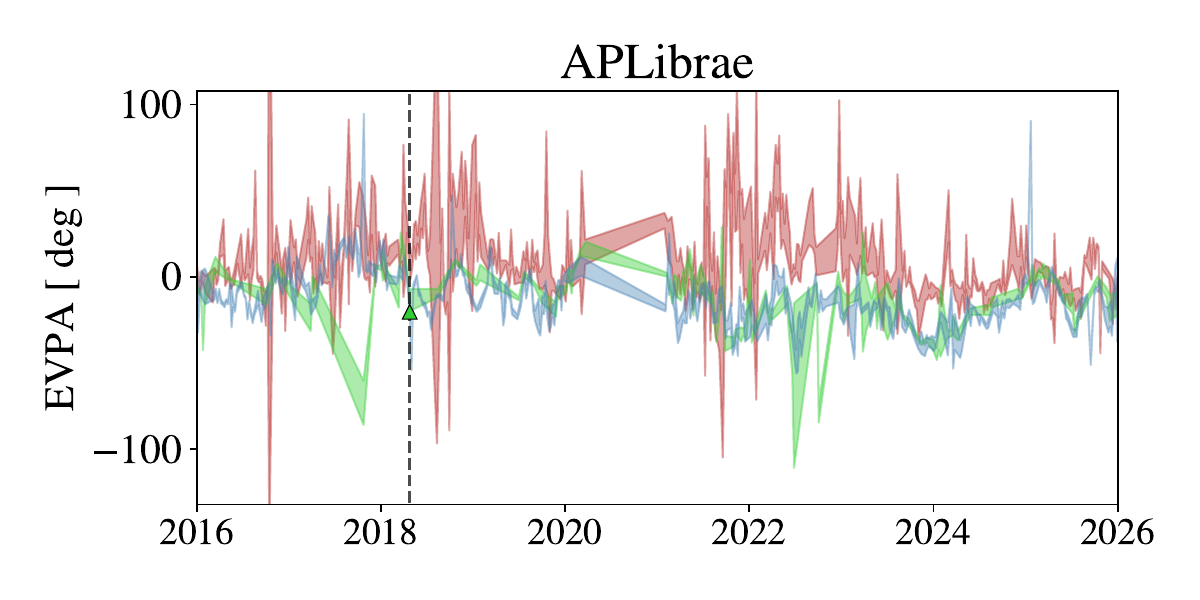}
    
    %\vspace*{-0.2cm}
    \includegraphics[width=0.32\linewidth]{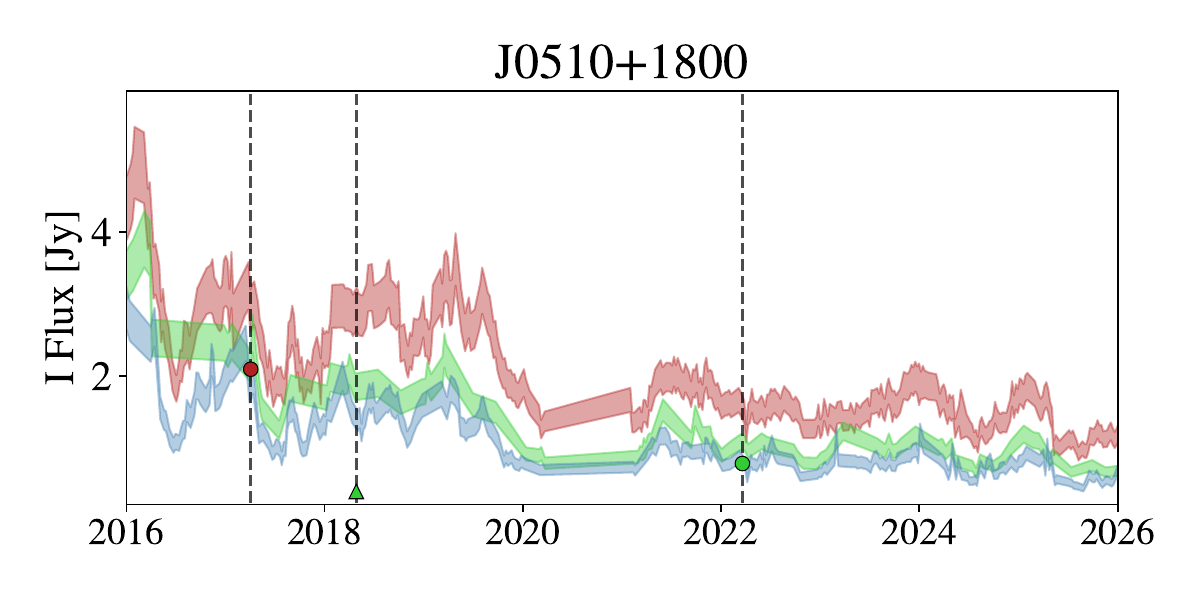}
    \includegraphics[width=0.32\linewidth]{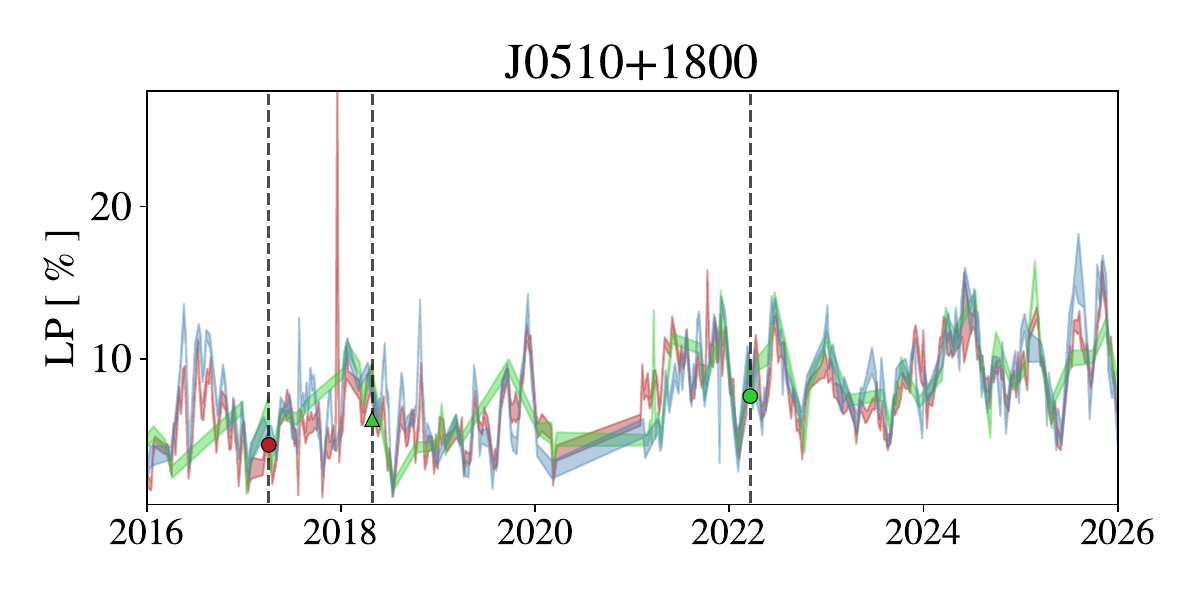}
    \includegraphics[width=0.32\linewidth]{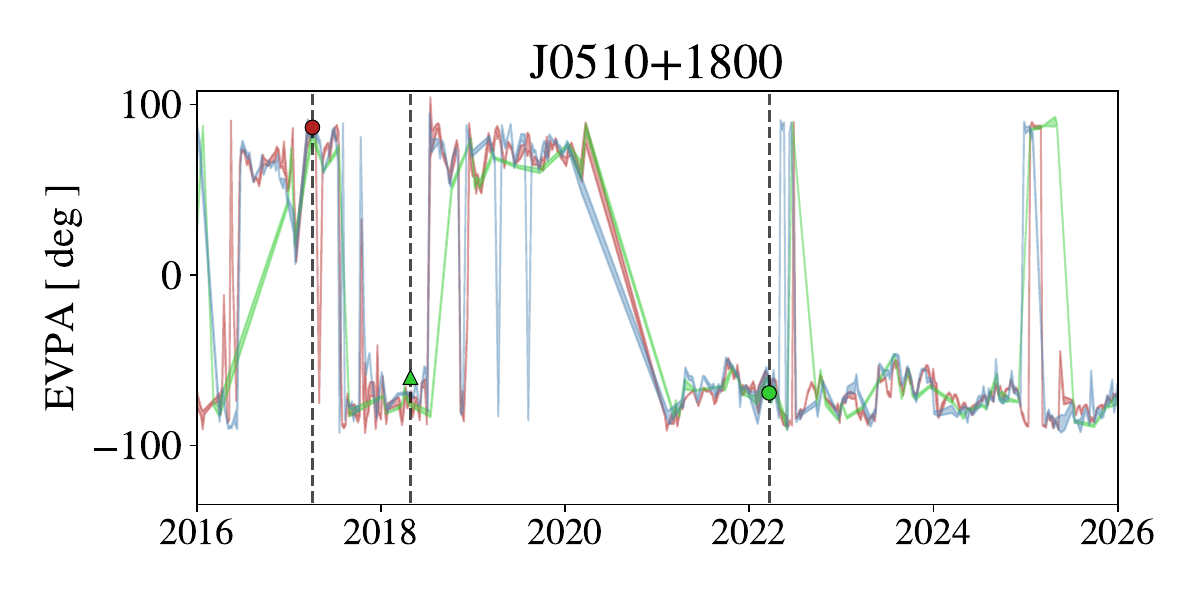}
    
    %\vspace*{-0.2cm}
    \includegraphics[width=0.32\linewidth]{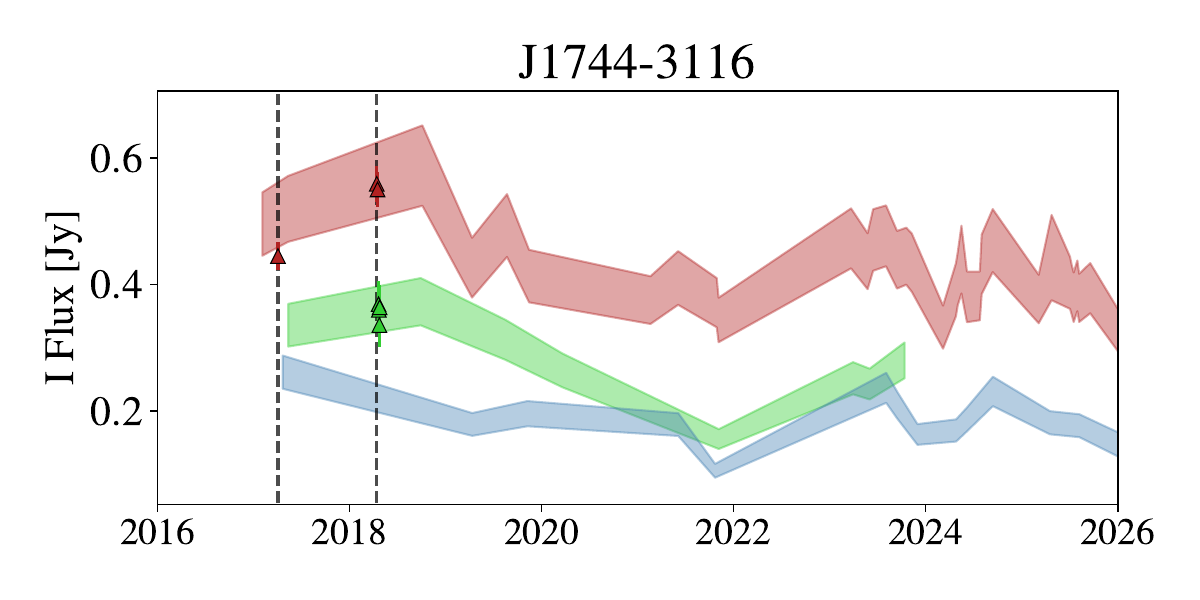}
    \includegraphics[width=0.32\linewidth]{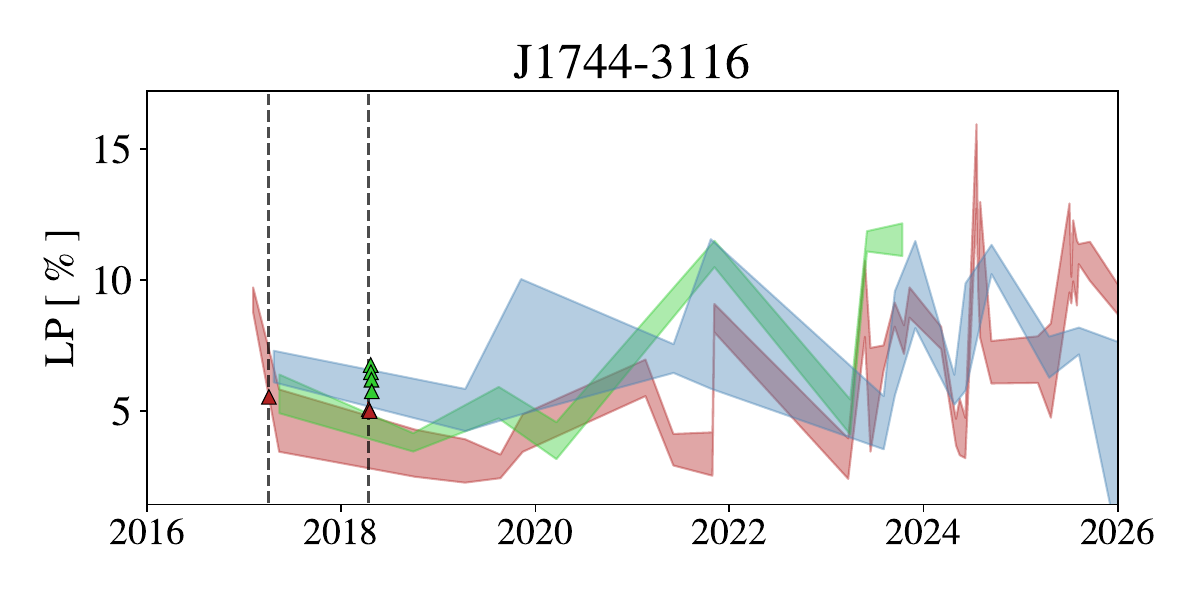}
    \includegraphics[width=0.32\linewidth]{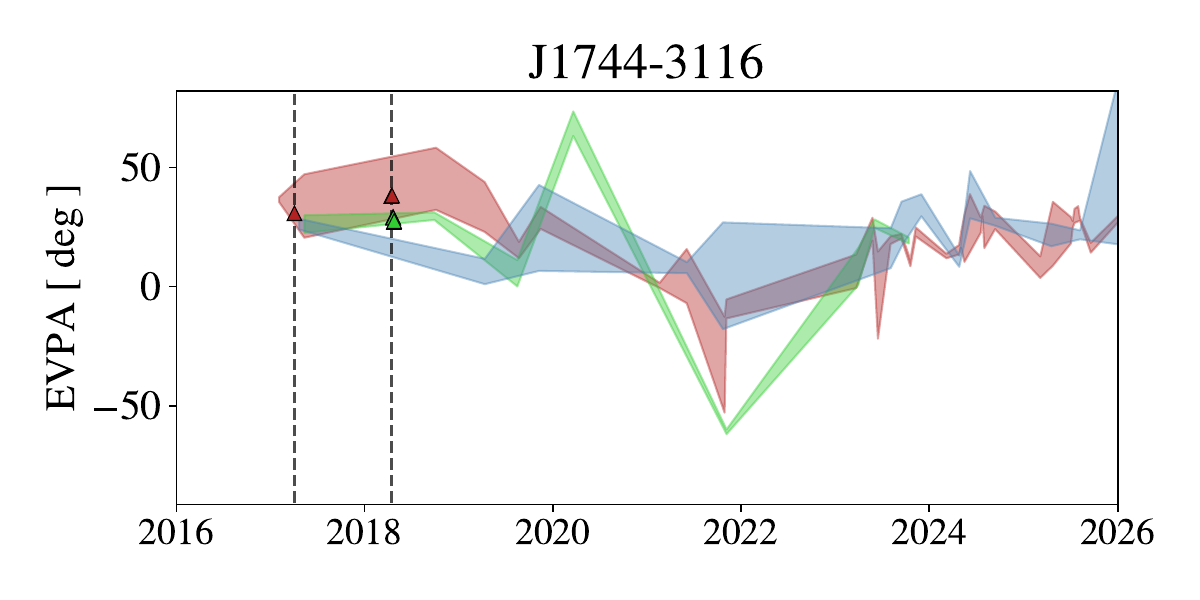}
    \caption{Same as Fig. \ref{fig:amaplots} for other sources.}\label{fig:amaplots2}
\end{figure*}

\begin{figure*}[h!]
    \centering
    \includegraphics[width=0.32\linewidth]{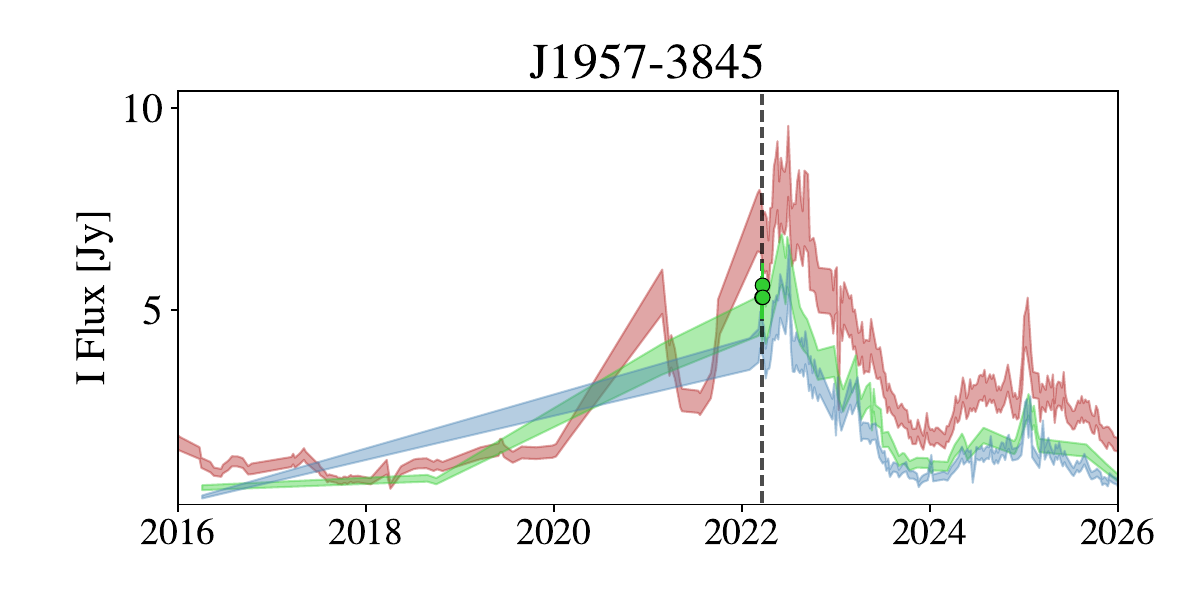}
    \includegraphics[width=0.32\linewidth]{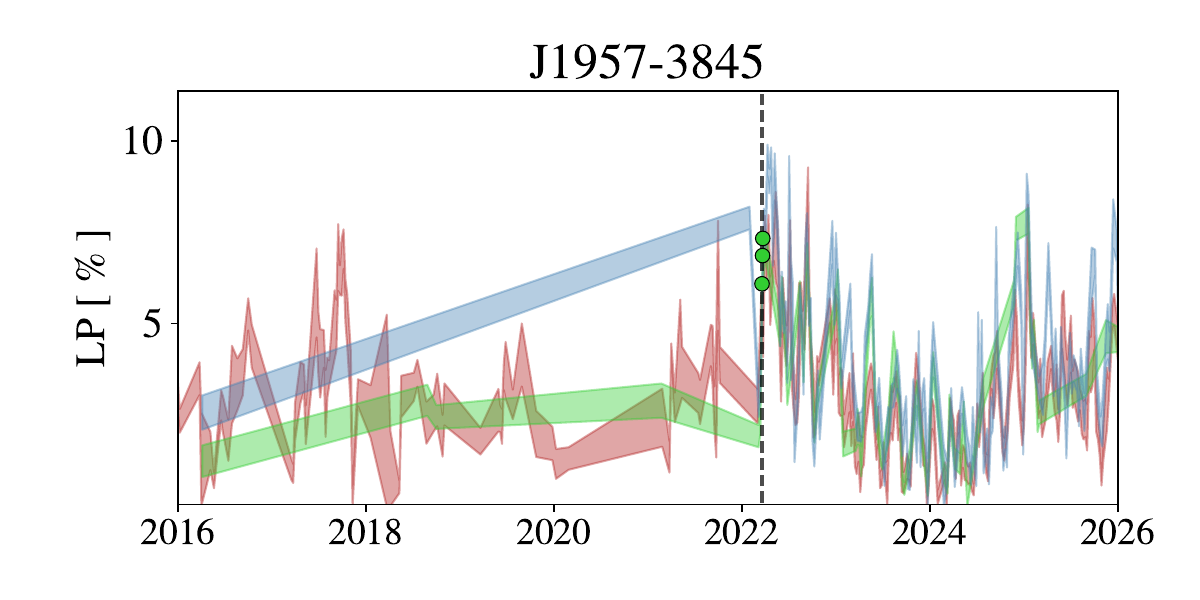}
    \includegraphics[width=0.32\linewidth]{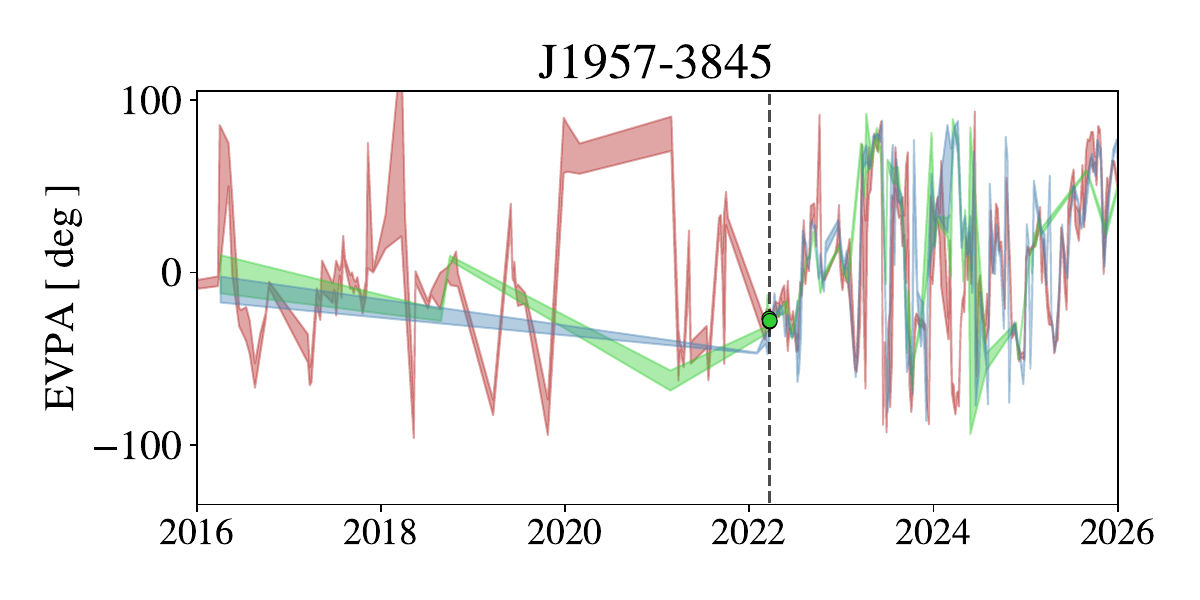}
    
    %\vspace*{-0.2cm}
    \includegraphics[width=0.32\linewidth]{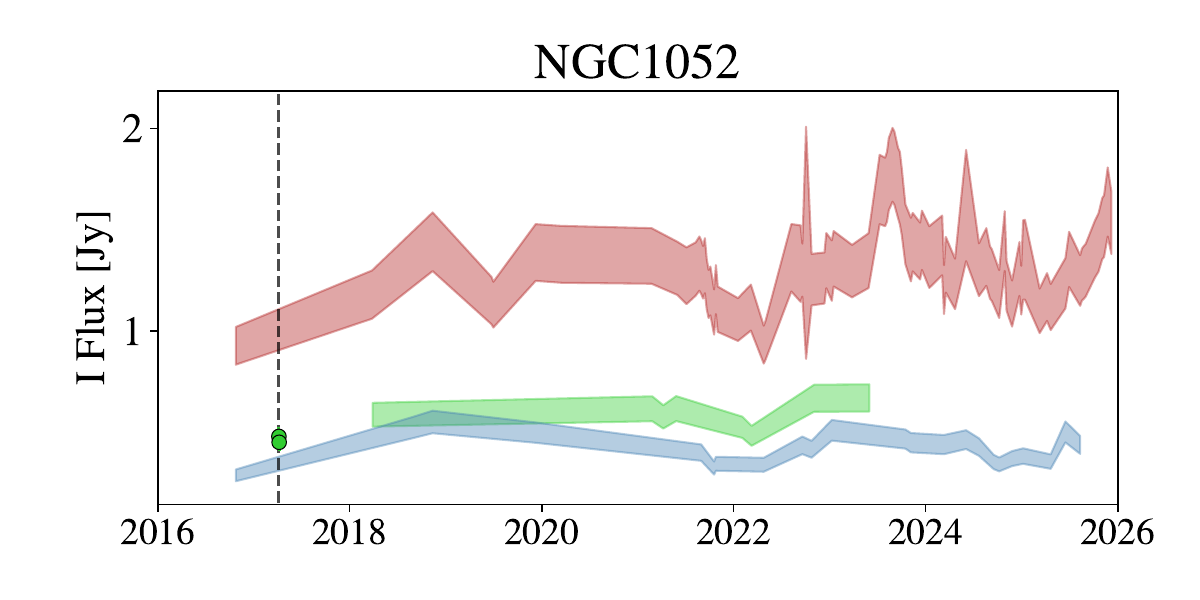}
    \includegraphics[width=0.32\linewidth]{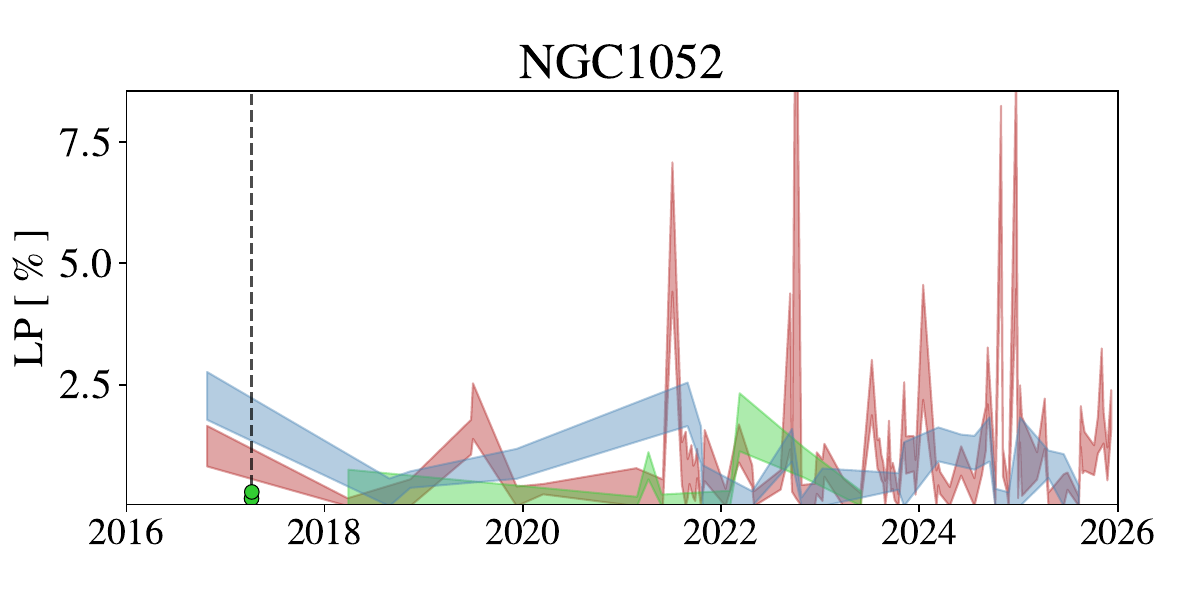}
    \includegraphics[width=0.32\linewidth]{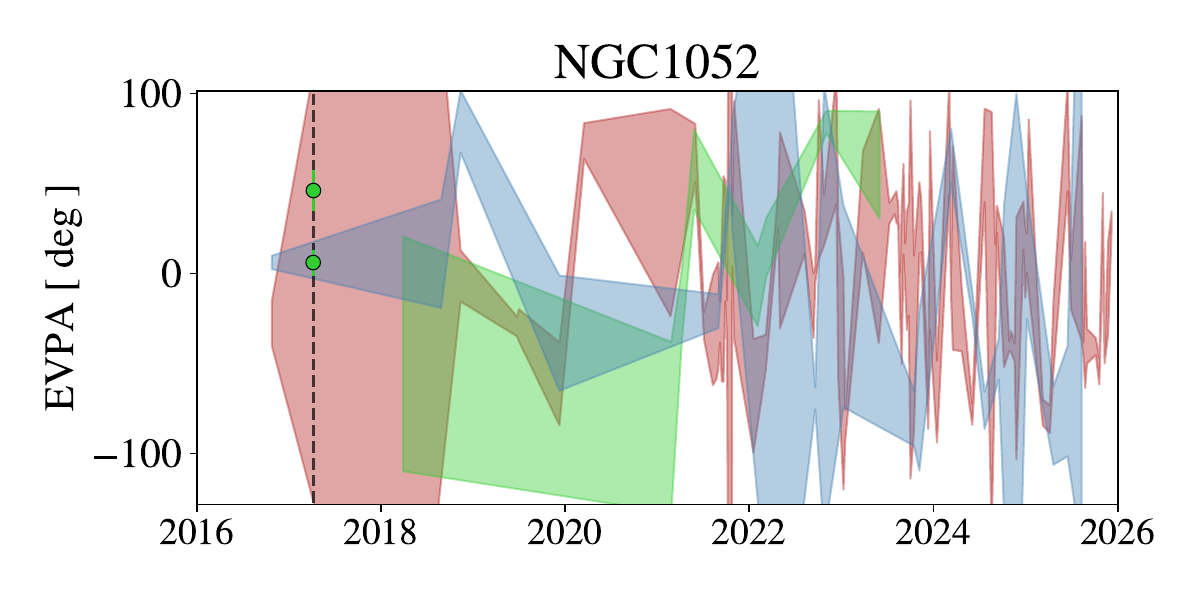}
    
    %\vspace*{-0.2cm}
    \includegraphics[width=0.32\linewidth]{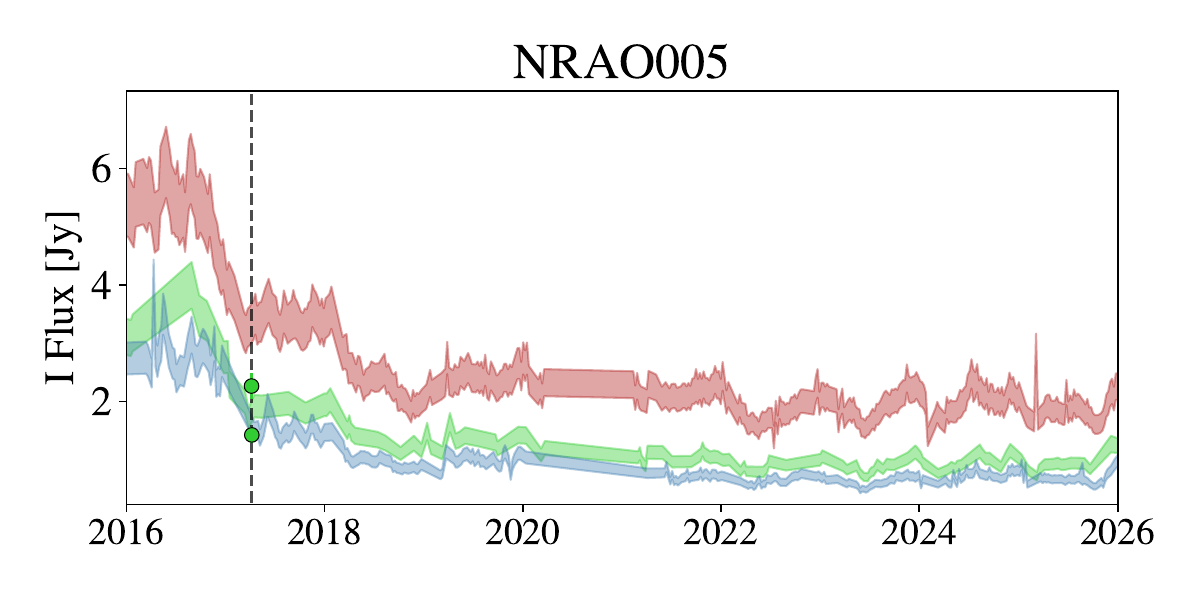}
    \includegraphics[width=0.32\linewidth]{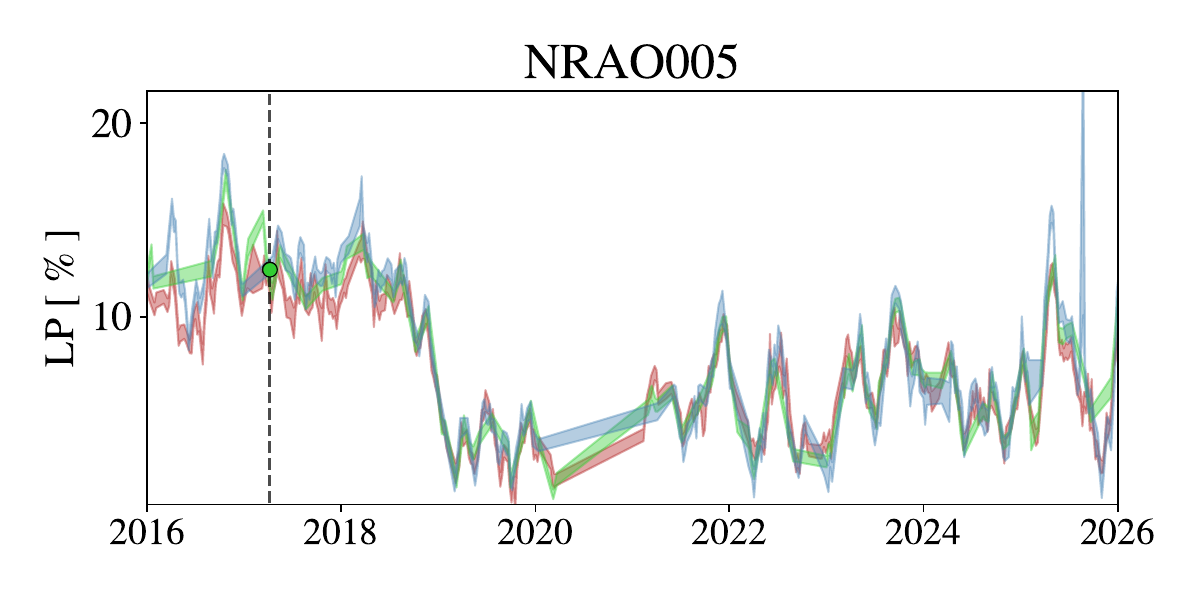}
    \includegraphics[width=0.32\linewidth]{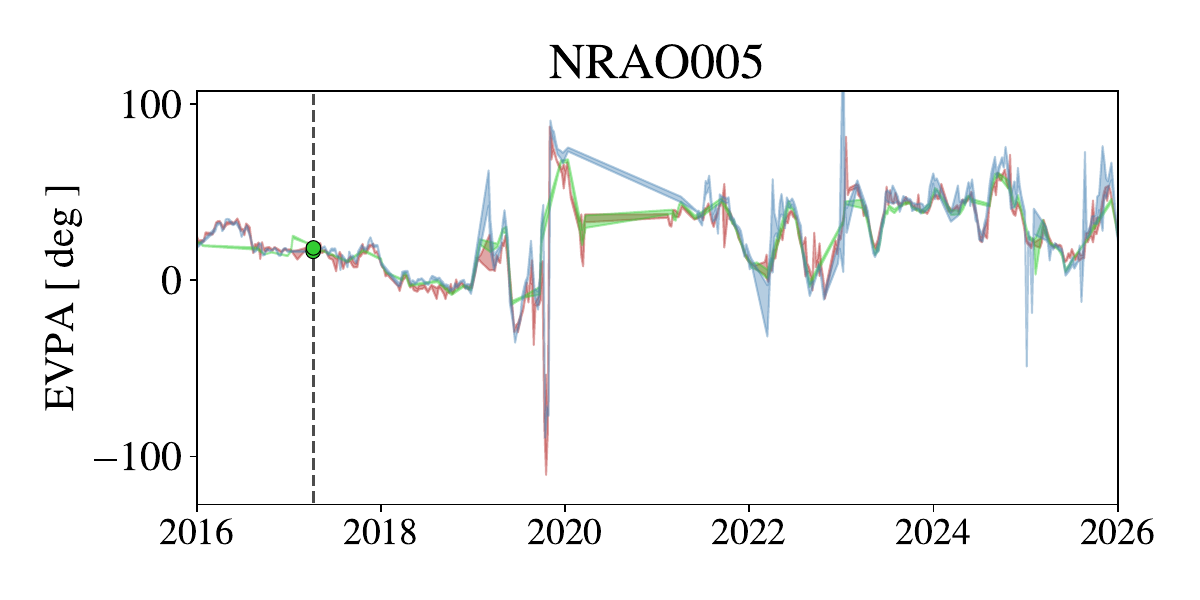}
    
    %\vspace*{-0.2cm}
    \includegraphics[width=0.32\linewidth]{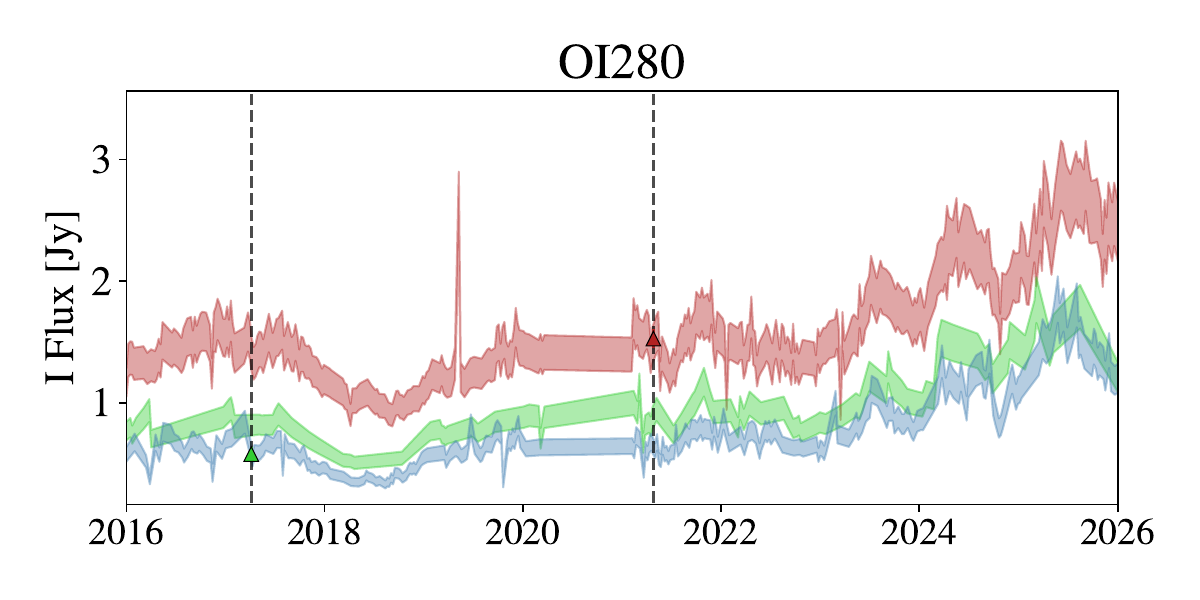}
    \includegraphics[width=0.32\linewidth]{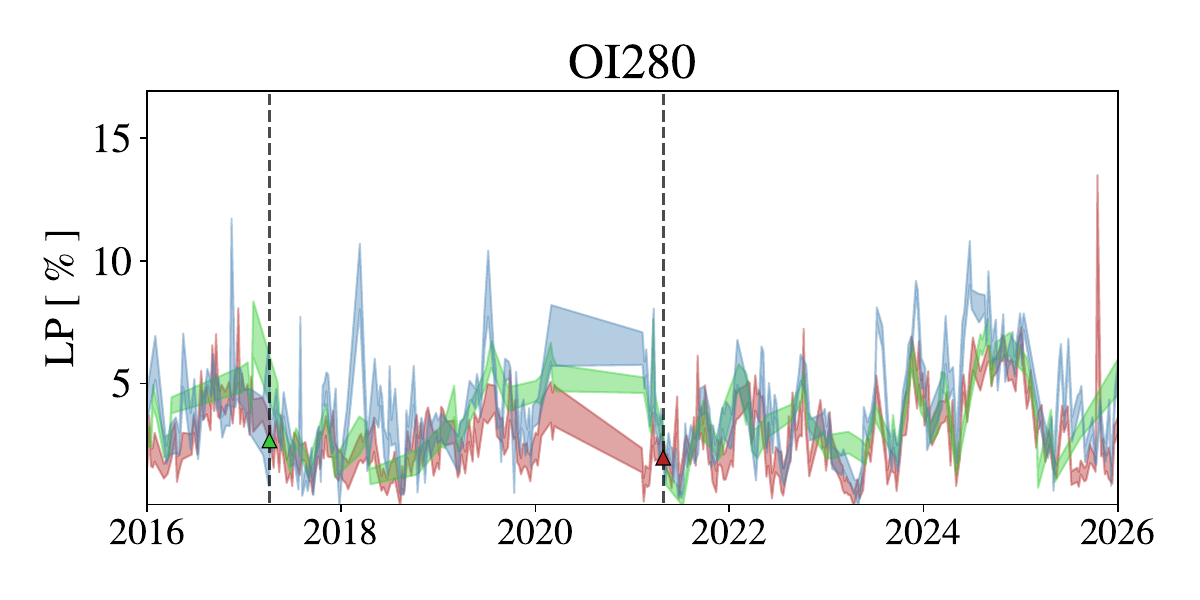}
    \includegraphics[width=0.32\linewidth]{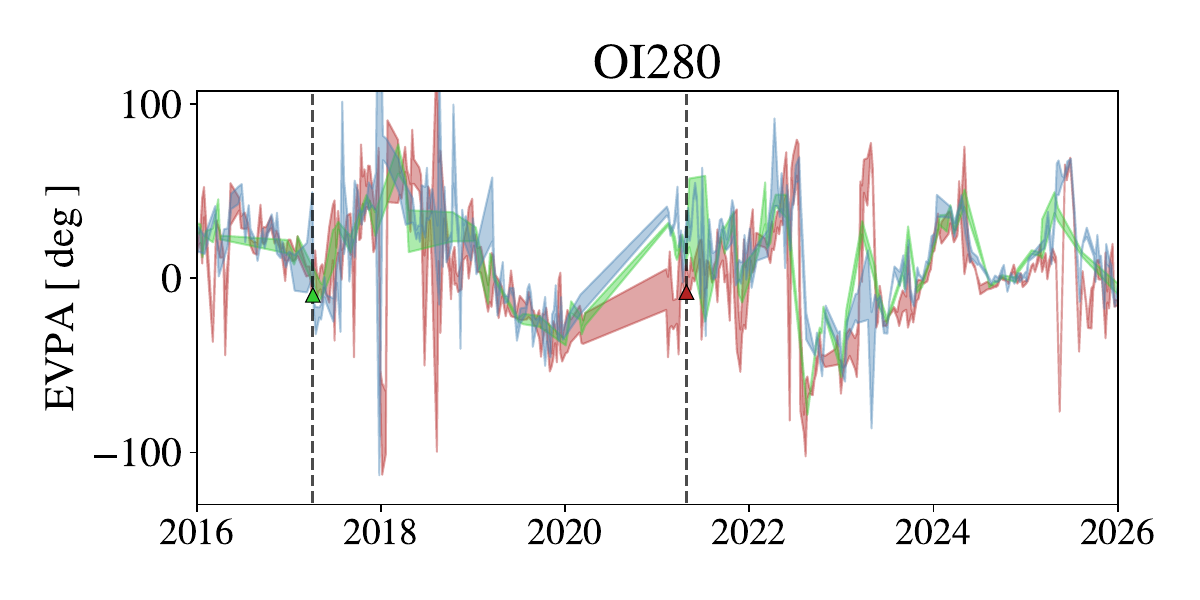}
    
    %\vspace*{-0.2cm}
    \includegraphics[width=0.32\linewidth]{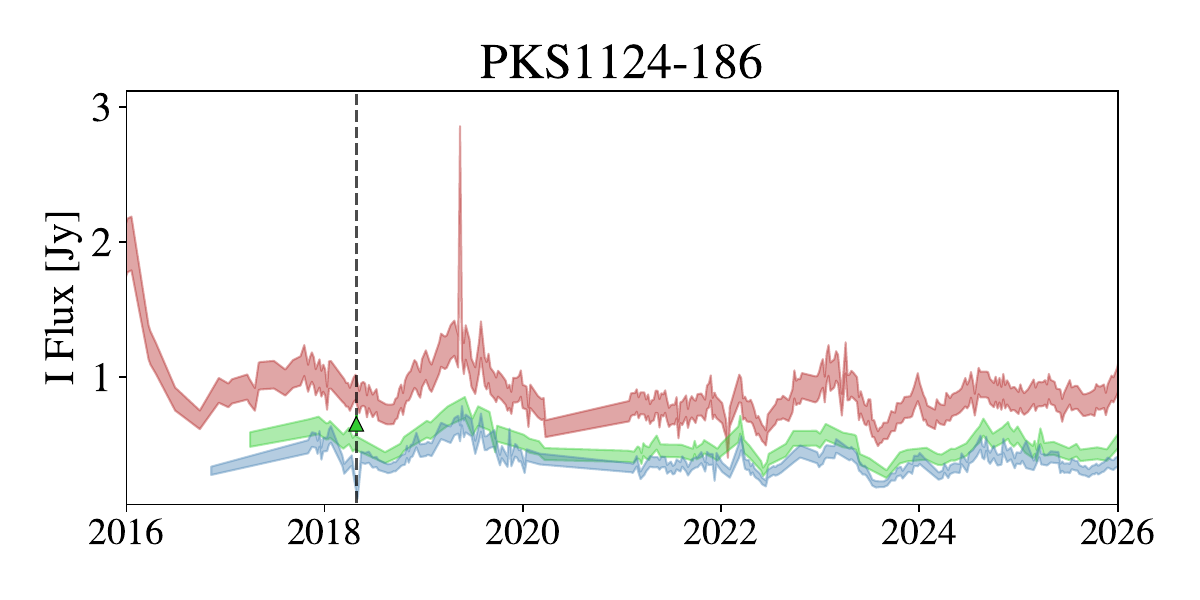}
    \includegraphics[width=0.32\linewidth]{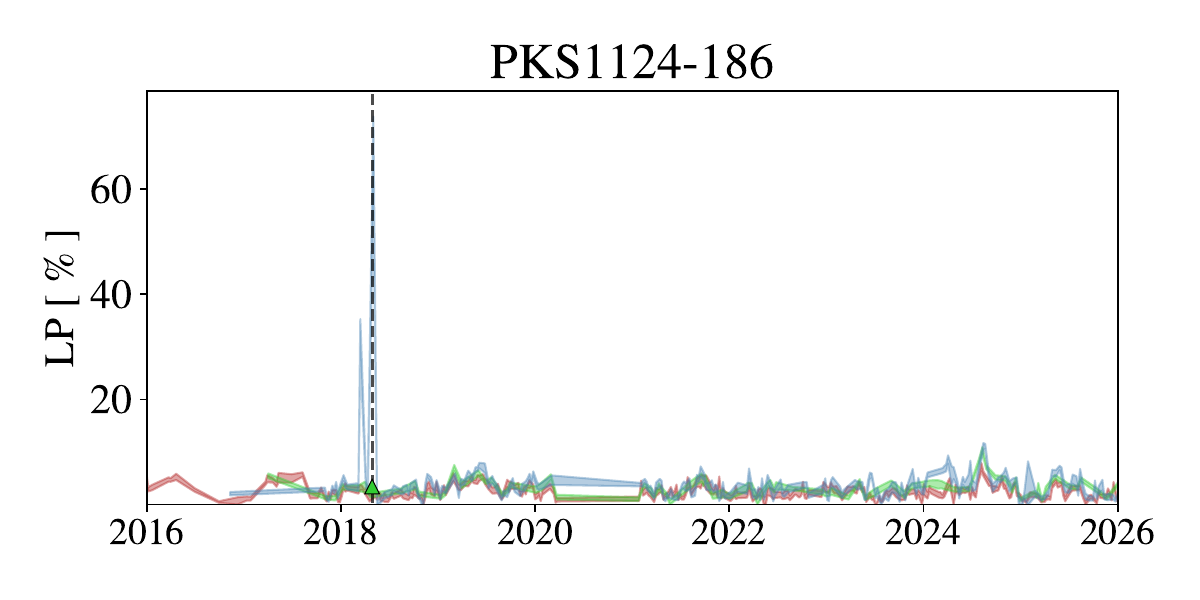}
    \includegraphics[width=0.32\linewidth]{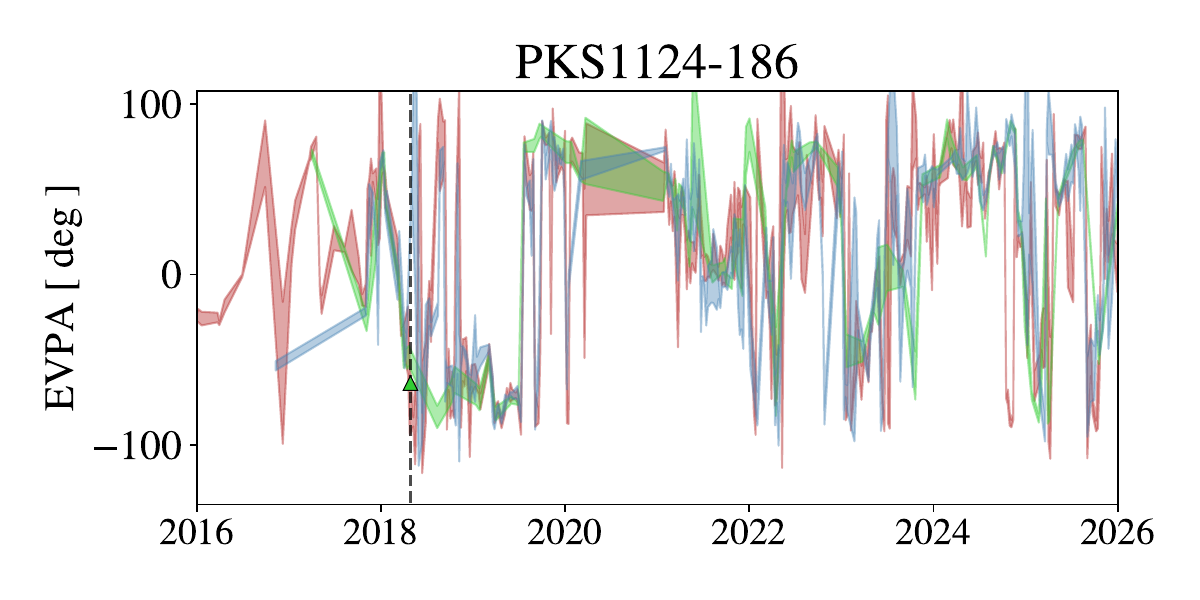}
    
    %\vspace*{-0.2cm}
    \includegraphics[width=0.32\linewidth]{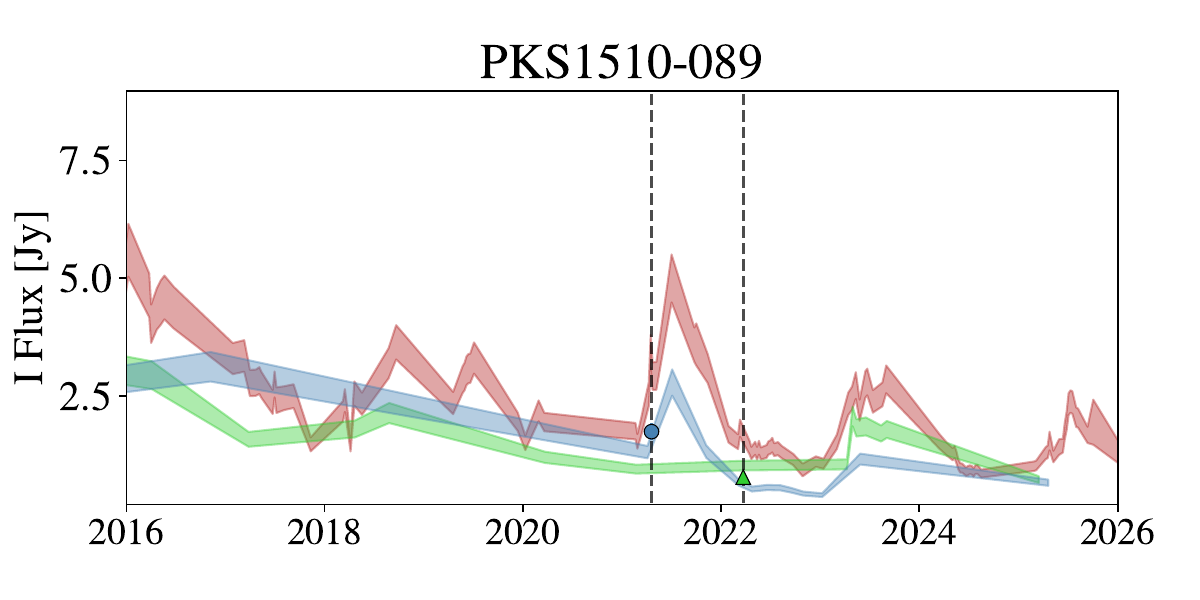}
    \includegraphics[width=0.32\linewidth]{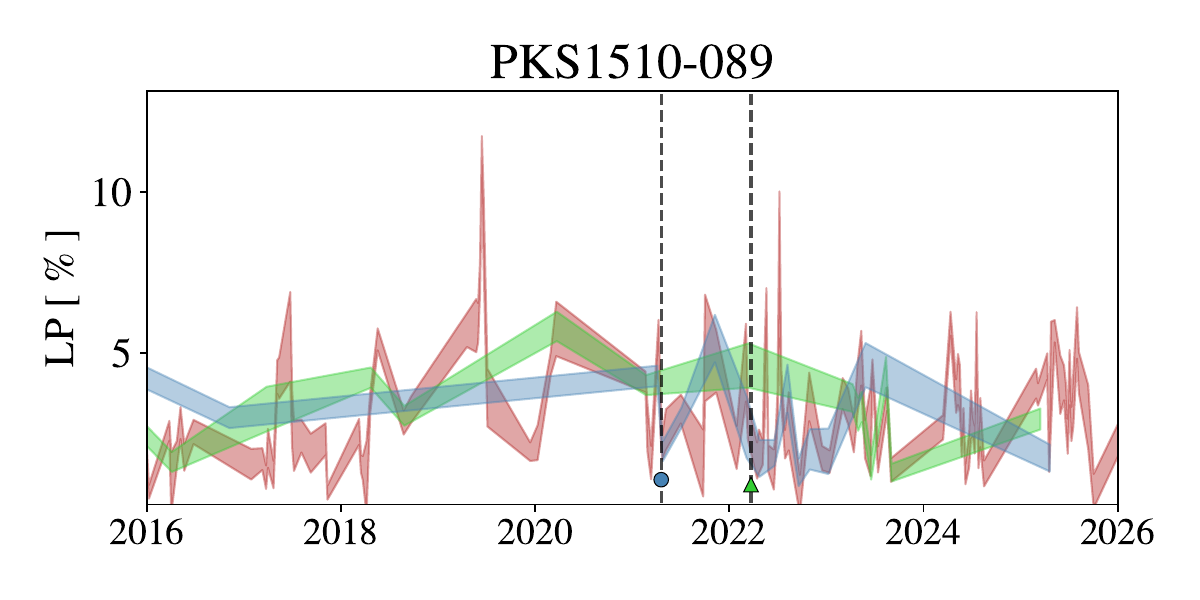}
    \includegraphics[width=0.32\linewidth]{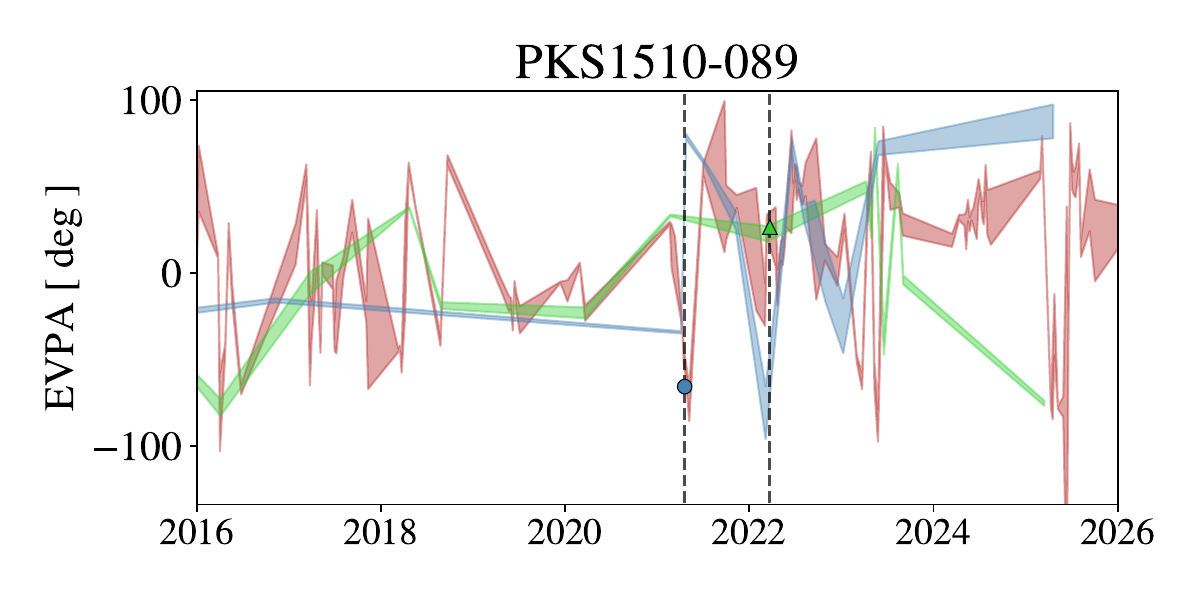}
    
    %\vspace*{-0.2cm}
    \includegraphics[width=0.32\linewidth]{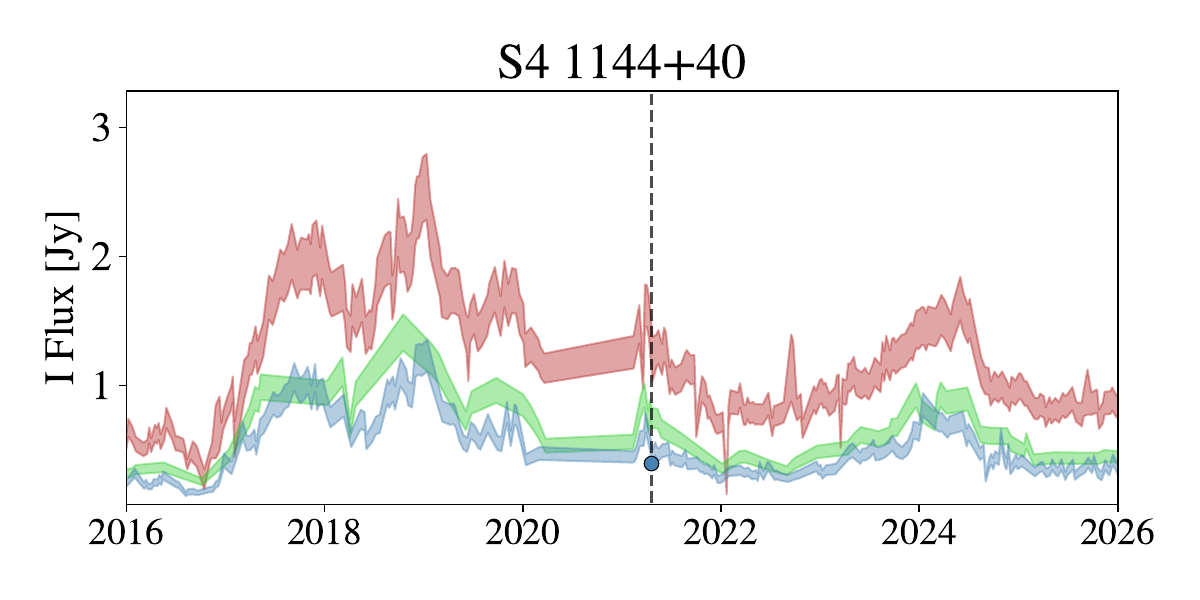}
    \includegraphics[width=0.32\linewidth]{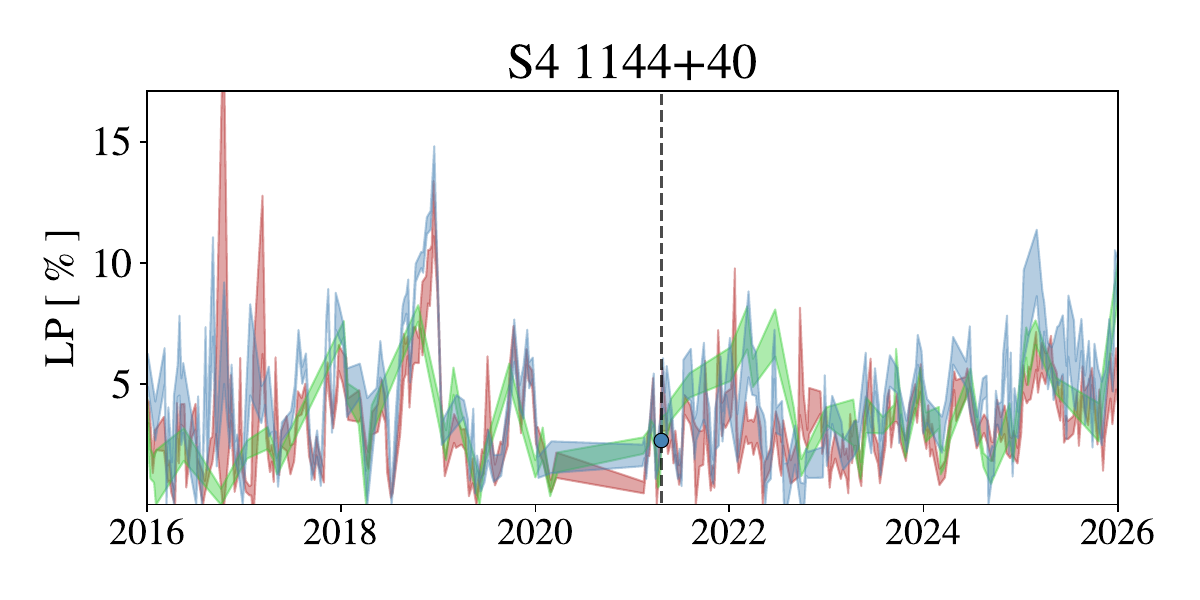}
    \includegraphics[width=0.32\linewidth]{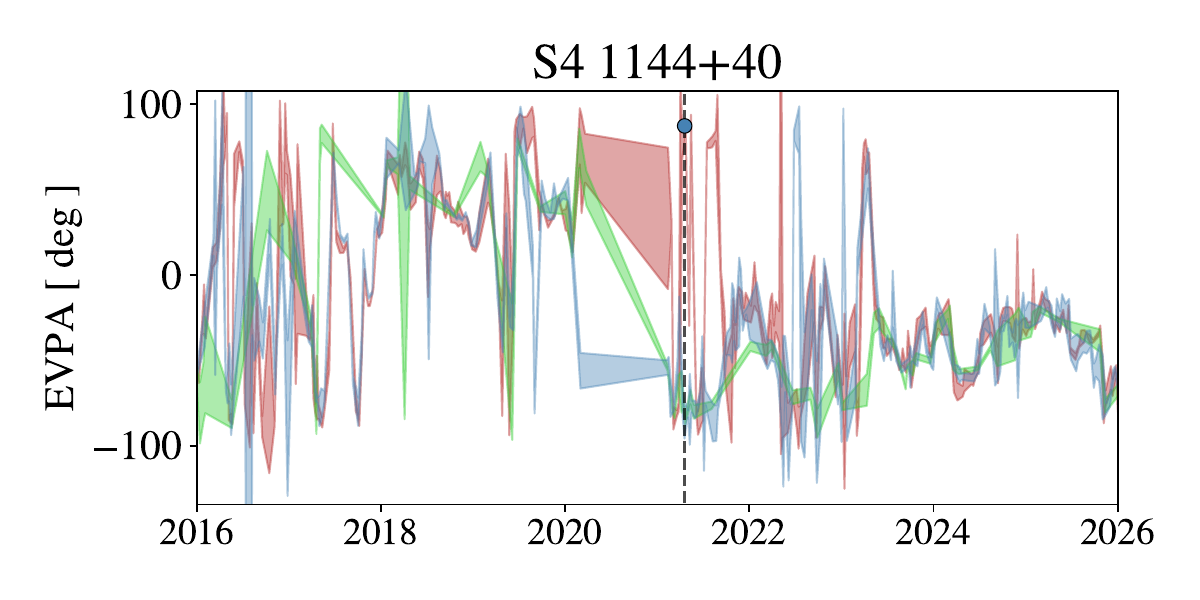}
    \caption{Same as Fig. \ref{fig:amaplots} for other sources.}
    \label{fig:amaplots3}
\end{figure*}

\end{appendix}

\end{document}